\documentclass[12pt,letterpaper]{article}

\usepackage{epsfig}
\usepackage{amsmath}
\usepackage{amssymb}
\usepackage{amsthm}
\usepackage{mathrsfs}
\usepackage{bbm}
\usepackage{bm}
\usepackage[T1]{fontenc}

\usepackage[utf8]{inputenc}

\usepackage{graphicx}
\usepackage[percent]{overpic}
\usepackage{indentfirst}
\usepackage{xspace}
\usepackage{multirow}
\usepackage{hyperref}
\usepackage{xcolor}
\definecolor{darkred}{rgb}{0.5,0.15,0.15}
\hypersetup{colorlinks=true,urlcolor=darkred,linkcolor=darkred,citecolor=darkred}
\usepackage{verbatim}
\usepackage[letterpaper,margin=1in,headheight=15pt]{geometry}
\usepackage{mathpazo}
\usepackage{authblk}
\usepackage{tikz-cd}
\usepackage{capt-of}
\usepackage{silence}
\usepackage{enumerate}
\usepackage{tikz}
\usetikzlibrary{calc}   
\usetikzlibrary{topaths}
\usetikzlibrary{decorations}
\usetikzlibrary{decorations.pathmorphing}

\usepackage{csquotes}
\PassOptionsToPackage{%
backend=bibtex8,bibencoding=ascii,%
language=auto,%
style=alphabetic, 
sorting=nyt, 
maxbibnames=10, 
natbib=true 
}{biblatex}
\usepackage{biblatex}
\numberwithin{equation}{section}

\theoremstyle{definition}

\newtheorem{theorem*}{Theorem}

\newtheorem{example}{Example}

\newcommand{\tops}{\texorpdfstring}

\newcommand{\fg}{{\mathfrak g}}

\newcommand{\fX}{{\mathfrak X}}

\newcommand{\cA}{\ensuremath{\mathcal A}}
\newcommand{\cB}{\ensuremath{\mathcal B}}

\newcommand{\cD}{\ensuremath{\mathcal D}}
\newcommand{\cL}{\ensuremath{\mathcal L}}
\newcommand{\cS}{\ensuremath{\mathcal S}}
\newcommand{\cF}{\ensuremath{\mathcal F}}

\newcommand{\cO}{\ensuremath{\mathcal O}}

\newcommand{\cX}{\ensuremath{\mathcal X}}

\newcommand{\cW}{\ensuremath{\mathcal W}}

\newcommand{\scrW}{\ensuremath{\mathscr W}}
\newcommand{\scrT}{\ensuremath{\mathscr T}}
\newcommand{\scrM}{\ensuremath{\mathscr M}}
\newcommand{\scrB}{\ensuremath{\mathscr B}}
\newcommand{\scrI}{\ensuremath{\mathscr I}}

\newcommand{\scrL}{\ensuremath{\mathscr L}}

\newcommand{\R}{\ensuremath{\mathbb R}}
\newcommand{\C}{\ensuremath{\mathbb C}}
\newcommand{\PP}{\ensuremath{\mathbb P}}
\newcommand{\Z}{\ensuremath{\mathbb Z}}

\newcommand{\N}{{\mathcal N}}

\newcommand{\RR}{{\mathcal R}}

\newcommand{\kahler}{k\"ahler\xspace}
\newcommand{\hk}{hyperk\"ahler\xspace}

\newcommand{\I}{{\mathrm i}}
\newcommand{\E}{{\mathrm e}}
\newcommand{\de}{\mathrm{d}}

\newcommand{\FN}{\textrm{\tiny FN}}
\newcommand{\lang}{\textrm{\tiny L}}

\DeclareMathOperator{\Tr}{Tr}

\newcommand{\nablaab}{\nabla^{\text{ab}}}

\newcommand{\Weff}{\widetilde{\mathcal{W}}^{\text{eff}}}
\newcommand{\Mdol}{\mathscr{M}_{\text{Dol}}}

\newcommand{\Setminus}{\!\setminus\!}

\definecolor{Blue}{RGB}{50,50,255}
\definecolor{Orange}{RGB}{253,127,0}
\definecolor{Red}{RGB}{255,0,0}
\definecolor{Green}{RGB}{0,175,0}
\definecolor{ToDo}{RGB}{150,0,0}

\begin{document}

\title{A geometric recipe for twisted superpotentials\footnote{Based
    on lectures given by the first author at the {\sl Pre-StringMath
      Summer School}, 17--21 July 2017, and at \emph{StringMath 2017}, 24--28 July 2017, Hamburg University.}}
\date{{{\tiny \color{gray} \tt}} 
{{\tiny \color{gray} \tt}}} 
\author[1]{Lotte Hollands}
\author[2]{Philipp R\"uter}
\author[3]{Richard J. Szabo}
\affil[1,2,3]{\small Department of Mathematics, Heriot-Watt
  University}
\affil[1,2,3] {\small Maxwell Institute for Mathematical Sciences, Edinburgh}
\affil[1,3]{\small Higgs Centre for Theoretical Physics, Edinburgh}

\maketitle

\begin{abstract}
\noindent
We give a pedagogical introduction to spectral networks and abelianization, as well as their relevance to $\N=2$ supersymmetric field theories in four dimensions. Motivated by a conjecture of Nekrasov-Rosly-Shatashvili, we detail a geometric recipe for computing the effective twisted superpotential for $\N=2$ field theories of class $\cS$ as a generating function of the brane of opers, with respect to the spectral coordinates found from abelianization. We present two new examples, the simplest Argyres-Douglas theory and the pure $SU(2)$ gauge theory, while we conjecture the $\epsilon$-expansion of the effective twisted superpotential for the $E_6$ Minahan-Nemeschansky theory.
\end{abstract}

\vspace{2cm}

\begin{flushright}
{\sf\small EMPG--20--02}
\end{flushright}

\newpage
{\baselineskip=12pt
\tableofcontents
}
\newpage

\section{Introduction}
\noindent
In these notes we give a pedagogical introduction to
four-dimensional $\N=2$ supersymmetric field theories of class $\cS$ and certain novel geometric
methods that can be used to study them.

\subsection{Background}
\noindent
The origin of class $\cS$ theories lies in six dimensions in the $\N=(2,0)$ superconformal field theory with gauge algebra $\fg$. This theory is rather mysterious and elusive: it does not admit a Lagrangian description and it might not even exist as a classical theory.\footnote{It is possible to find the field content and a Lagrangian description in the specific case of the abelian gauge algebra $\fg=\mathfrak{u}(1)$ \cite{Claus:1997cq}.} However, it can be studied indirectly after compactification on lower-dimensional submanifolds where it gives rise to a host of new structures in both quantum field theory and geometry. In fact, it has broadened our fundamental understanding of what a quantum field theory really is.

The first evidence for the six-dimensional $(2,0)$-theory traces back to the seminal work by Nahm in his classification of the simple Lie superalgebras which can occur as the symmetry algebra of a superconformal quantum field theory. The highest dimensions in which a superconformal field theory can exist was found to be six with the superconformal algebra $\mathfrak{osp}(2,6|2k)$; the theory with the highest amount of supercharges in six dimensions is the $\N=(2,0)$ theory \cite{Nahm:1977tg}. At the time there was no known way to actually construct this theory, but later on constructions of this six-dimensional superconformal field theory in terms of string theory were realized independently by Witten~\cite{Witten:1995zh} and Strominger~\cite{Strominger:1995ac}.

One way to construct this theory is in type IIB string theory on an
ADE singularity. This is defined in the ten-dimensional spacetime
which is the direct product of $\mathbb{R}^{5,1}$ with the resolution
of an {orbifold singularity}. An orbifold singularity is a
quotient $\mathbb{C}^2/\Gamma$ by the natural action of a finite subgroup
$\Gamma\subset SU(2)$ (with $\C^2$ viewed as the fundamental
representation of $SU(2)$). By the McKay correspondence, $\Gamma$ can be
identified with the Dynkin diagram of type ADE of a simply-laced Lie algebra $\fg=\fg(\Gamma)$. By wrapping a D3-brane on a vanishing two-cycle of the resolution of $\C^2/\Gamma$, the localized dynamics at the tip $0\in\C^2/\Gamma$ describes Nahm's six-dimensional superconformal field theory. The resulting theory on $\mathbb{R}^{5,1}$ is labeled by the corresponding Lie algebra as $\fX=\fX[\fg]$.

The theory $\fX$ also has a realization in M-theory as the worldvolume
theory of M5-branes. Consider a collection of M5-branes, whose
interactions are mediated by open M2-branes ending on them. The
$(2,0)$-theory is then realized in the limit where the M5-branes
coincide; for instance, $K$ flat coincident M5-branes correspond to
the $(2,0)$-theory $\fX[K]$ at an $A_{K-1}$-singularity (with $\fg=\mathfrak{su}(K)$). 
For further reading about the six-dimensional $(2,0)$-theory we suggest \cite{moore2012lecture}.

The $(2,0)$-theory can be compactified on a (possibly punctured)
Riemann surface $C$. In these notes we will study the theory
$\fX[\fg]$ on the six-dimensional backgrounds
\begin{align}
C\times\R^{3,1} \ . 
\end{align}
After
a partial topological twist of the theory, in the limit where $C$ collapses to a
point this leads to a four-dimensional $\N=2$ supersymmetric field theory on
$\R^{3,1}$ which depends only on the complex structure of the
punctured Riemann surface $C$, together with certain
specified singularity data $\cD$ at the punctures of $C$. We denote this
theory by ${\sf T}_{\fg}[C,\cD]$, or simply by $\sf T$. This is called
a {theory of class $\cS$}~\cite{Gaiotto:2009we,gaiotto2013wall}.\footnote{The $\cS$ stands for `six',
  or alternatively for `S-duality' which acts in a particularly nice
  way on the Riemann surface $C$~\cite{tachikawa2015review}.} 
  
Class $\cS$
theories can be studied from the perspective offered by the geometry
of the Riemann surface $C$, and this is the perspective we will take
in these notes. One advantage of such a geometric approach is that it
is not only useful for understanding the conventional four-dimensional
field theories, where a Lagrangian description allows the usual
methods of quantum field theory to be employed, but also for more
unconventional theories where a Lagrangian description is not
available and other methods fail. The theories $\sf T$ of class $\cS$
naturally encompass both types of field theories.

\subsection{Summary and Outline}
\noindent
The goal of these notes is to define and compute the effective twisted superpotential for any four-dimensional $\N=2$ theory $\sf T$ of class $\cS$ in the $\frac12\Omega$-background, following a conjecture by Nekrasov, Rosly and Shatashvili~\cite{nekrasov2011darboux}.\footnote{For readers more familiar with topological string theory, we mention that this superpotential equals the refined topological string free energy $\cF^{\textrm{\tiny NS}}$ in the two-dimensional Nekrasov-Shatashvili limit.} The two main ingredients are the ``brane of opers'', which is a holomorphic Lagrangian submanifold in the moduli space of complexified flat connections on $C$, and a suitable system $\{x_i,y^i\}$ of holomorphic Darboux coordinates on this moduli space. With these choices the effective twisted superpotential \smash{$\Weff$} may be extracted as the generating function
\begin{align}
y^i = \frac1\epsilon \, \frac{\partial\Weff(x;\epsilon)}{\partial x_i}
\end{align}
of the brane of opers with respect to the Darboux coordinates $\{x_i,y^i\}$, where $\epsilon$ is the deformation parameter of the $\frac12\Omega$-background.

These notes have grown out of a series of lectures by the first author at the {Pre-StringMath Summer School 2017}. Paralleling these lectures, we have tried to keep these notes accessible and self-contained by including extensive background material in the first few sections. In particular, we define spectral networks on the Riemann surface $C$ and point out how they encode the four-dimensional BPS particle spectrum of the theory $\sf T$. We introduce the central ideas of Seiberg-Witten theory and motivate the importance of $\N=2$ theories of class $\cS$. We also review the abelianization method for constructing Darboux coordinate systems on the moduli space of complexified flat connections on $C$, which we relate to the exact WKB method as well as TBA-like equations. Along the way we introduce the Hitchin system and flat oper connections. All of these elements play an important role in reaching our final goal: a geometric recipe for computing the effective twisted superpotential \smash{$\Weff$} for any theory $\sf T$ of class $\cS$.

These notes are naturally structured in four parts. Most of our new results are part of Section~\ref{sec:newNRS}, and the expert reader may wish to skip over the earlier sections. 

In Section~\ref{WKBSNs} we present some of the geometry, introducing (Fenchel-Nielsen type) spectral networks, abelianization and (Fenchel-Nielsen type) spectral coordinates, while emphasizing their generalization to Riemann surfaces $C$ which have irregular punctures. Throughout we illustrate all of the
formalism with many explicit examples (which will be useful in Section~\ref{sec:newNRS}).

In Section~\ref{sec:classS} we
move on to the physics side of the story. After a brief introduction
to Seiberg-Witten theory, we introduce four-dimensional $\N=2$ theories
of class $\cS$ and, in particular, our main examples. These are the simplest Argyres-Douglas theory, the pure $SU(2)$ gauge theory, the conformal $SU(2)$ gauge theory coupled to four fundamental hypermultiplets, and the $E_6$ Minahan-Nemeschansky theory. As one application of the theory of spectral networks, we review how to determine the BPS particle spectrum of any theory of class $\cS$.

In Section~\ref{sec:NRS}, we
combine the geometry and physics perspectives through the richness of the Hitchin
system. We introduce the $\frac12\Omega$-background, along with the effective twisted superpotential of an $\N=2$ theory of class $\cS$ in this background. We further motivate the brane of opers as a quantization of the Coulomb branch of the $\N=2$ theory, while we relate the abelianization method, applied to flat oper connections, to the exact WKB method. This shows that the spectral coordinates from Section~\ref{WKBSNs} have good WKB asymptotics when evaluated on flat oper connections.

In Section~\ref{sec:newNRS}, we explain the Nekrasov-Rosly-Shatashvili conjecture which relates the generating function of opers in Fenchel-Nielsen type coordinates to the effective twisted superpotential of an $\N=2$ theory of class $\cS$ in the $\frac12\Omega$-background.
We then formulate our geometric recipe, and test it by computing the effective twisted superpotential \smash{$\Weff(x;\epsilon)$} --- in a perturbative expansion in the coupling or ultraviolet scale, which is exact in $\epsilon$ --- in two new examples: the simplest Argyres-Douglas theory and the pure $SU(2)$ gauge theory; for the treatment of the conformal $SU(2)$ theory we refer to~\cite{hollands2018higher}. 

For the pure $SU(2)$ theory this requires computing the asymptotically small solutions of the Mathieu equation in terms of the Mathieu functions; we are able to do this in a perturbation expansion in the ultraviolet scale, which is exact in $\epsilon$,  following the approach of~\cite{hollands2018higher} for the Heun oper. For the $E_6$ Minahan-Nemeschansky theory, we content ourselves with conjecturing the $\epsilon$-expansion of the effective twisted superpotential. We conclude by describing the string theory magic behind the geometric recipe, and by giving an interpretation to the generating function of opers computed with respect to other types of spectral coordinates.

In the time elapsed between delivering and writing up these lecture notes, various other relevant works have appeared. We have tried to incorporate them into these notes, either in the main text or in Section~\ref{sec:Open} below. Our presentation in these notes is distinguished by deriving the familiar effective twisted superpotential from scratch, without using any ingredients that are not known for instrinsically strongly coupled theories (such as the Matone relation~\cite{Matone:1995rx}). In particular, for the weakly coupled pure $SU(2)$ gauge theory we simply obtain the superpotential from the expression
\begin{align}
\Weff = \frac12\,\int\,\de x \, \log \frac{\textrm{Wr}(\psi_1',\psi_1'')}{\textrm{Wr}(\psi_1,\psi_1'')} \, \frac{\textrm{Wr}(\psi_1,\psi_2'')}{\textrm{Wr}(\psi_1',\psi_2'')} \ ,
\end{align}
where ${\rm Wr}$ denotes the Wronskian and the $\psi$'s are certain solutions to the Mathieu equation \eqref{eq:recipeMathieu}; see Section~\ref{ex:recipepure} for details.

\subsection{Open Questions}\label{sec:Open}
\noindent
These notes do not by any means close the door on this topic. There are various open computational as well as conceptual questions to be addressed. Let us mention a few here:
\begin{itemize}
\item Can one extend the computations of the effective twisted superpotential for the pure $SU(2)$ theory to higher order, or perhaps even to all orders?
\item Is it possible to find the exact form of the effective twisted superpotential for the $E_6$ Minahan-Nemeshansky theory? Can one verify its relation to the corresponding topological string partition function?
\item There are many more interesting examples to apply the recipe to, for instance more general Argyres-Douglas theories. Is there some new physics encoded in the corresponding effective twisted superpotentials?
\item Is it possible to find a proof of the geometric recipe (in one of the examples or even in general), perhaps taking inspiration from~\cite{Jeong:2018qpc}?
\item Can one develop a better physical framework for studying the relation between $\N=2$ boundary conditions and spectral coordinates?
\item Any flat connection $\nabla$ can be rewritten as an oper connection, possibly with adjacent singularities. In these notes we only discuss the generating function for the brane of opers without adjacent singularities. Can one define more general generating functions? What is the physical relevance of these? (This may compute the $\N=2$ superpotential of an $\N=2$ theory with added surface defects.)
\item Can one lift the geometric recipe to three dimensions, to compute the effective twisted superpotentials of five-dimensional $\N=1$ theories in the background $\R_\epsilon^2\times\R^2\times S^1$?
\item Could we also extend the recipe to four dimensions, to compute the full Nekrasov instanton partition function of the $\N=2$ theory in the $\Omega$-background $\R_{\epsilon_1}^2\times\R_{\epsilon_2}^2$ (see~\cite{TeschnerOber} for some initial results)? And likewise to five dimensions?
\item In~\cite{Coman:2018uwk,Coman:2020qgf} the dual topological string (or free fermion) partition function in the self-dual $\Omega$-background $\epsilon_1=-\epsilon_2=\epsilon$ is characterized as an isomonodromic tau-function with respect to the same types of spectral coordinates, found using either abelianization or the exact WKB method. Can we understand this result from the perspective of the present paper? Aside from the difference between the self-dual and Nekrasov-Shatashvili limits, the isomonodromic tau-function is formulated with respect to opers with apparent singularities. A gauge theoretic understanding of this relation has been recently
found in~\cite{Nekrasov:2020qcq,Jeong:2020uxz}, see also~\cite{bonelli2016painlev} for earlier work in this direction.
\item Equally intriguing are the non-perturbative approaches to topological string theory formulated in~\cite{Grassi:2014zfa,Codesido:2015dia,Ito:2018eon}, which are centred around the spectral determinant of the five-dimensional mirror curve. This spectral determinant is compared to the exact WKB method for the four-dimensional pure $SU(2)$ theory in~\cite{Grassi:2019coc}, and for the $SU(2)$ theory with one flavour in~\cite{Grassi:2021wpw}, in the $\frac12\Omega$-background. What is the role of the spectral determinant in our set-up?
\item Finally, another closely related avenue should be the mathematical programme outlined in~\cite{bridgeland2019riemann,Bridgeland:2017vbr,Barbieri:2019yya,Bridgelandcubic}, which studies a class of Riemann-Hilbert problems arising in Donaldson-Thomas theory (inspired by~\cite{Gaiotto:2008cd,gaiotto2013wall}).\footnote{Very relevant here as well is the series of papers by Allegretti
(see for instance~\cite{Allegretti:2018kvc,Allegretti:2019dij}) providing a mathematically rigorous
framework for understanding the conformal limit of the spectral
coordinates introduced by Gaiotto, Moore and Neitzke.} This theory is worked out in the example of the hypergeometric differential equation on the three-punctured sphere in~\cite{Iwaki:2020efz,Iwaki:2021zif}, producing the corresponding four-dimensional Nekrasov partition function in the self-dual limit. On the physics side this programme has inspired a better understanding of the non-perturbative topological string partition function~\cite{Alim:2020tpw,Alim:2021ukq,Alim:2021mhp}.
\end{itemize}

\paragraph{Acknowledgments.}
LH wishes to thank Andrew Neitzke for years of discussions and collaboration on the topics of these notes, and the organisors of the Pre-StringMath Summer School and StringMath 2017 in Hamburg for the invitation to deliver these lectures, and for hospitality during the school and conference. We thank Andrew Neitzke for his Mathematica program {\sl swn-plotter}~\cite{swn-plotter}. The work of LH is supported by a Royal Society Dorothy Hodgkin Fellowship. The work of PR is supported by a James Watt Scholarship from Heriot-Watt University. The work of RJS was supported by
the Consolidated Grant ST/P000363/1 
from the Science and Technology Facilities Council.

\clearpage
\newpage
\section{Spectral Networks and Abelianization}\label{WKBSNs}
\noindent
In this section we introduce spectral networks and explain how each
spectral network gives a coordinate system on the moduli space of flat
connections. Spectral networks were first defined in
\cite{gaiotto2012spectral}, where it was shown that they are a
fundamental ingredient in the story of BPS states and wall-crossing
for $\N=2$ gauge theories of class
$\cS$~\cite{Gaiotto:2009we,gaiotto2013wall}. In
\cite{gaiotto2012spectral} it was also observed that spectral networks
can be used to construct coordinate systems for moduli spaces of flat
connections. This method, called abelianization, was further developed in
\cite{hollands2016spectral}. In this section we follow the exposition
of~\cite{hollands2016spectral}, while extending the abelianization
method, as well as the notion of a Fenchel-Nielsen network, to Riemann
surfaces with irregular punctures.\footnote{We have used the Mathematica program {\sl swn-plotter} from Andrew Neitzke~\cite{swn-plotter} for many of the spectral network plots in these notes; we encourage the reader to play around with this her/himself.}

\subsection{WKB Spectral Networks}
\label{sec:WKBnetworks}
\noindent
A spectral network is a certain kind of {network} that can be drawn
on a (possibly punctured) Riemann surface $C$. Suppose that $C$ has $n$
punctures at positions $z_i$ and write
$C=\overline{C}\,\Setminus\{z_1,\dots,z_n\}$, where $\overline{C}$ is compact. Our initial input is a fixed meromorphic quadratic differential $p_2$ on $C$, which can locally be written as
\begin{equation}\label{phi2loc}
p_2(z) = u(z)\ \de z\otimes \de z \ .
\end{equation}
The differential $p_2$ may be singular at the punctures. A singular point is {regular} if $u(z)$
  has a pole of order~$2$ there; otherwise it is {irregular}. In a
neighbourhood of a {regular puncture} $z_i$,\footnote{ We use the terminology {regular
    pole}, {regular singularity}, and {regular puncture}
  interchangeably to mean the same thing.} the
function $u(z)$ may be brought to the form
\begin{equation}
u(z)\simeq\frac{m_i^2}{(z-z_i)^2} \ ,
\end{equation}
where $m_i$ are called the {mass parameters}.\footnote{We will discuss {irregular punctures} later on and for the moment assume that $p_2$ has only regular singularities at all punctures.}

Consider the square root $\lambda=\sqrt{p_2}$. This is a
one-form on $C$ which is locally given by $\lambda(z) = \sqrt{u(z)}\,\de
z$. The one-form $\lambda$ is \emph{a priori} multi-valued on $C$, but
we will remedy this by choosing appropriate branch cuts on $C$. Fix a phase $\vartheta\in\mathbb{R}/2\pi\,\mathbb{Z}$. A $\vartheta${-trajectory} is then a real curve $\gamma$ on $C$ such that 
\begin{equation}\label{thetaTrajectory}
\E^{-\I\,\vartheta}\,\sqrt{p_2}(v) \ \in \ \mathbb{R}^\times=\R\setminus\{0\}
\end{equation}
for any tangent vector $v$ to $\gamma$. This means that $\E^{-\I\,\vartheta}\,\sqrt{p_2}$ restricts to a real and non-vanishing one-form on $\gamma$.
We can write this condition more concretely by choosing a
parametrization $\gamma(t)$ of the curve and using the local form of $p_2$ as in (\ref{phi2loc}). Then (\ref{thetaTrajectory}) says
\begin{equation}
\E^{-\I\,\vartheta}\,\sqrt{u(\gamma(t))}\, \frac{\de\gamma(t)}{\de t} \ \in \ \mathbb{R}^\times \ .
\end{equation}

The set of all $\vartheta$-trajectories form the leaves of a foliation
of $C'=C\Setminus\{z|p_2(z)=0\}$. To see this, consider local
coordinates $w(z)$ defined around any point $z_0\in C'$ by
\begin{equation}
w(z)=\int_{z_0}^z\, \sqrt{u(z')}\, \de z' \ .
\end{equation}
Then $\sqrt{p_2(z)}=\de w$ and the $\vartheta$-trajectories are
just straight line segments of inclination $\vartheta$. With respect
to the coordinates $w$ we thus get a foliation whose leaves are
straight lines of inclination $\vartheta$.\footnote{Note also that $\de w$ is non-vanishing on $C'$ and thus defines a nowhere vanishing vector field on $C'$.}

\begin{figure}[h!]
\small
\centering
\begin{overpic}
[width=0.55\textwidth]{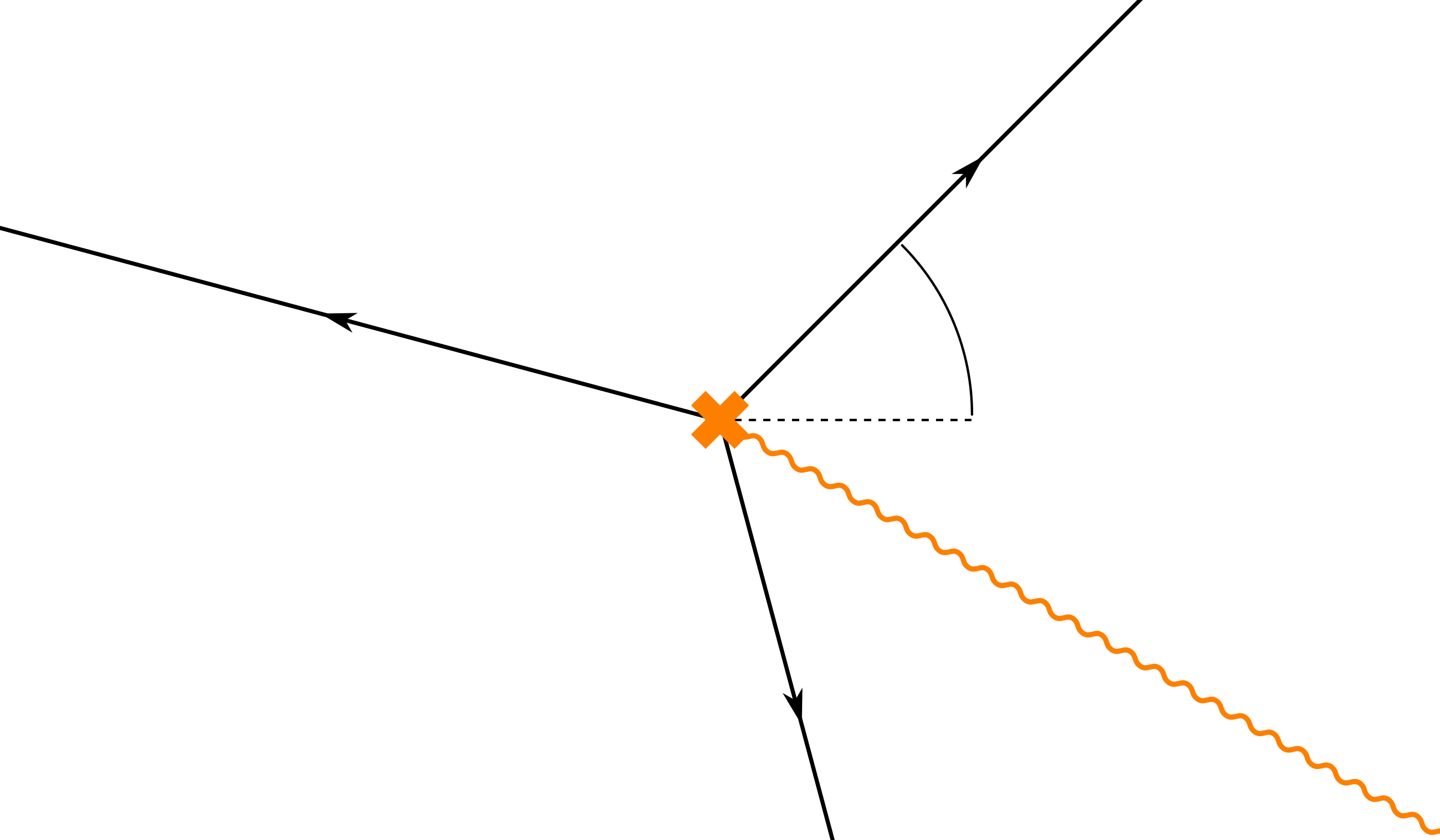}
\put(57,32){$\frac{2}{3}\vartheta$}
\end{overpic}
\caption{\small Illustration of the three $\vartheta$-trajectories
  emitted from a simple zero of $p_2$. The orange cross depicts
  the position of the zero, whereas the wavy line represents a choice
  of branch cut.}
\label{TPTrajectories}
\normalsize
\end{figure}

At a simple zero of $p_2$ the foliation by
$\vartheta$-trajectories becomes singular. In a neighbourhood of the
simple zero we can choose a local coordinate $z$ in which the zero is
located at $z=0$; in this coordinate, locally $p_2(z)\simeq z\,\de z\otimes \de z$. Consider
a trajectory starting at $z=0$ as $\gamma(t)=\E^{\,\I\,\alpha}\,t$ for $t>0$. Then
\begin{equation}
\sqrt{p_2}(v)=\sqrt{\gamma(t)}\,\frac{\de\gamma}{\de t}=\E^{\frac32\, \I\,\alpha}\, t^{1/2} \ .
\end{equation}
The condition for $\gamma$ to be a $\vartheta$-trajectory is then
\begin{equation}
\E^{-\I\,\vartheta}\,\E^{\frac32\,\I\,\alpha}\, t^{1/2} \ \in \ \mathbb{R}^\times
\end{equation}
which yields
\begin{equation}
\alpha=\frac{2\,\vartheta}{3}+k\,\frac{2\pi}{3} \qquad \mbox{for} \quad k=0,1,2 \ .
\end{equation}
Thus there are three $\vartheta$-trajectories emitted from every simple
zero of $p_2$ as shown in Figure~\ref{TPTrajectories}.

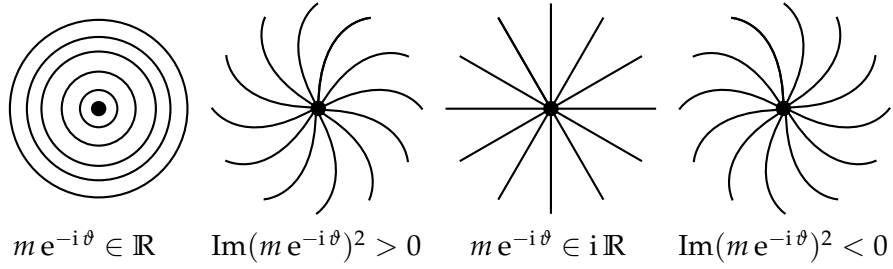
\begin{figure}[h!]
\small
\centering
\begin{overpic}
[width=0.80\textwidth]{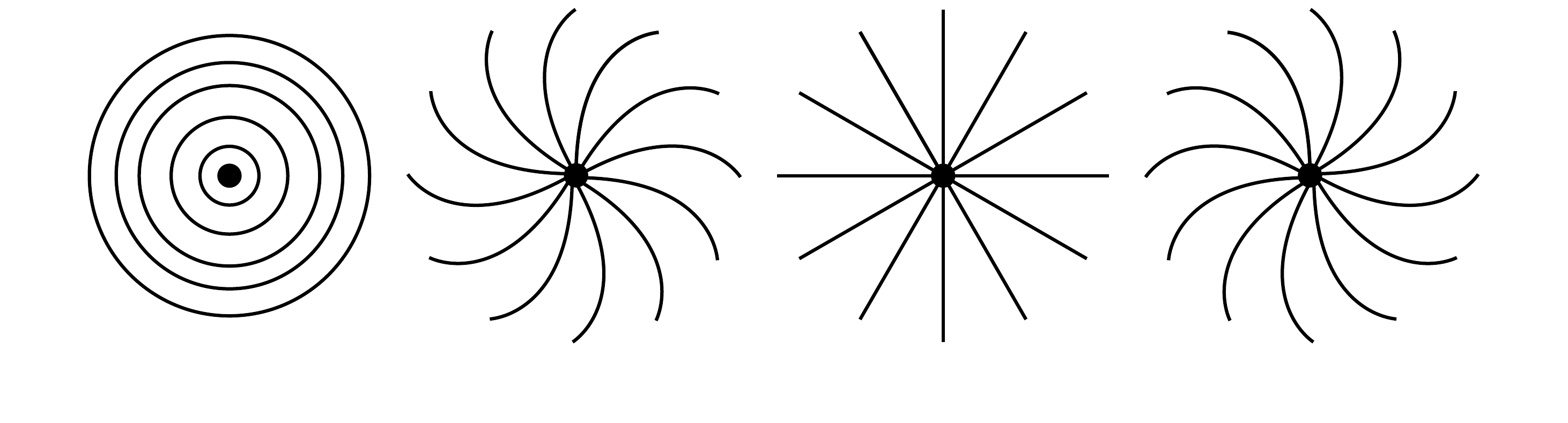}
\put(6,1){$m\,\E^{-\I\,\vartheta}\in\mathbb{R}$}
\put(26,1){$\text{Im}(m\,\E^{-\I\,\vartheta})^2>0$}
\put(52,1){$m\,\E^{-\I\,\vartheta}\in \I\,\mathbb{R}$}
\put(73,1){$\text{Im}(m\,\E^{-\I\,\vartheta})^2<0$}
\end{overpic}
\caption{\small Behaviour of $\vartheta$-trajectories near a puncture
  (depicted as a solid circle) depending on the phase of $m\,\E^{-\I\,\vartheta}$.}
\label{TrajectoriesPuncture}
\normalsize
\end{figure}

On the other hand, around regular poles of $p_2$ the behaviour of $\vartheta$-trajectories depends on the value of the corresponding mass parameter $m$, as depicted in Figure~\ref{TrajectoriesPuncture}.

A generic $\vartheta$-trajectory has both endpoints at punctures of
$C$. We call a trajectory {critical} if one or both of its
endpoints are zeroes of $p_2$. We then define the
{critical graph} to be the union of all critical trajectories
together with the zeroes of $p_2$. The trajectories may be 
oriented as outgoing from the zeroes, and, together with a choice of
branch cuts of $\sqrt{p_2}$, labeled by either $12$ or by $21$
if the sign of $\sqrt{p_2}(v)$ is positive or negative,
respectively. We call the critical graph together with orientations a
{WKB spectral network} on $C$ and denote it
by~$\scrW(p_2,\vartheta)$.  

\begin{example}\label{ex:CAD2}
Consider $C=\PP^1\Setminus\{\infty\}=\C$ with the quadratic differential
\begin{equation}
p_2(z)=(z^2+m)\,\de z\otimes\de z \ .
\end{equation}
This differential has an irregular singularity at $z=\infty$ of type
$L=4$ (i.e. a pole of order six), since $p_2$ behaves as $z^2\,\de z\otimes\de z$ near there.
For $m>0$ and $\vartheta=(\pi/2)^+$, we get the spectral network $\scrW$
shown in Figure \ref{AD2jump}~A.\footnote{In Figure~\ref{AD2jump} we have
  left out the branch cuts at infinity, which we rectify in
  Example~\ref{ex:AD2abelianisation} below.}
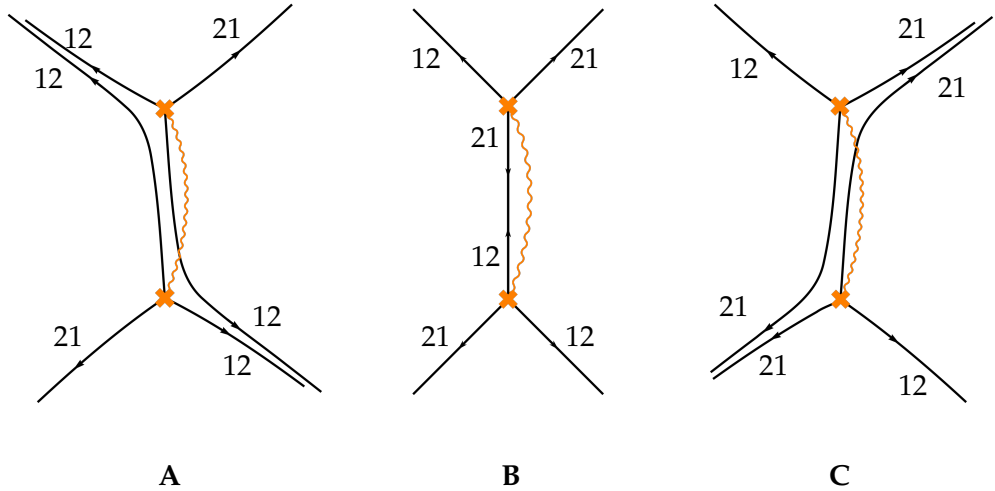
\begin{figure}[h!]
\small
\centering
\begin{overpic}
[width=0.80\textwidth]{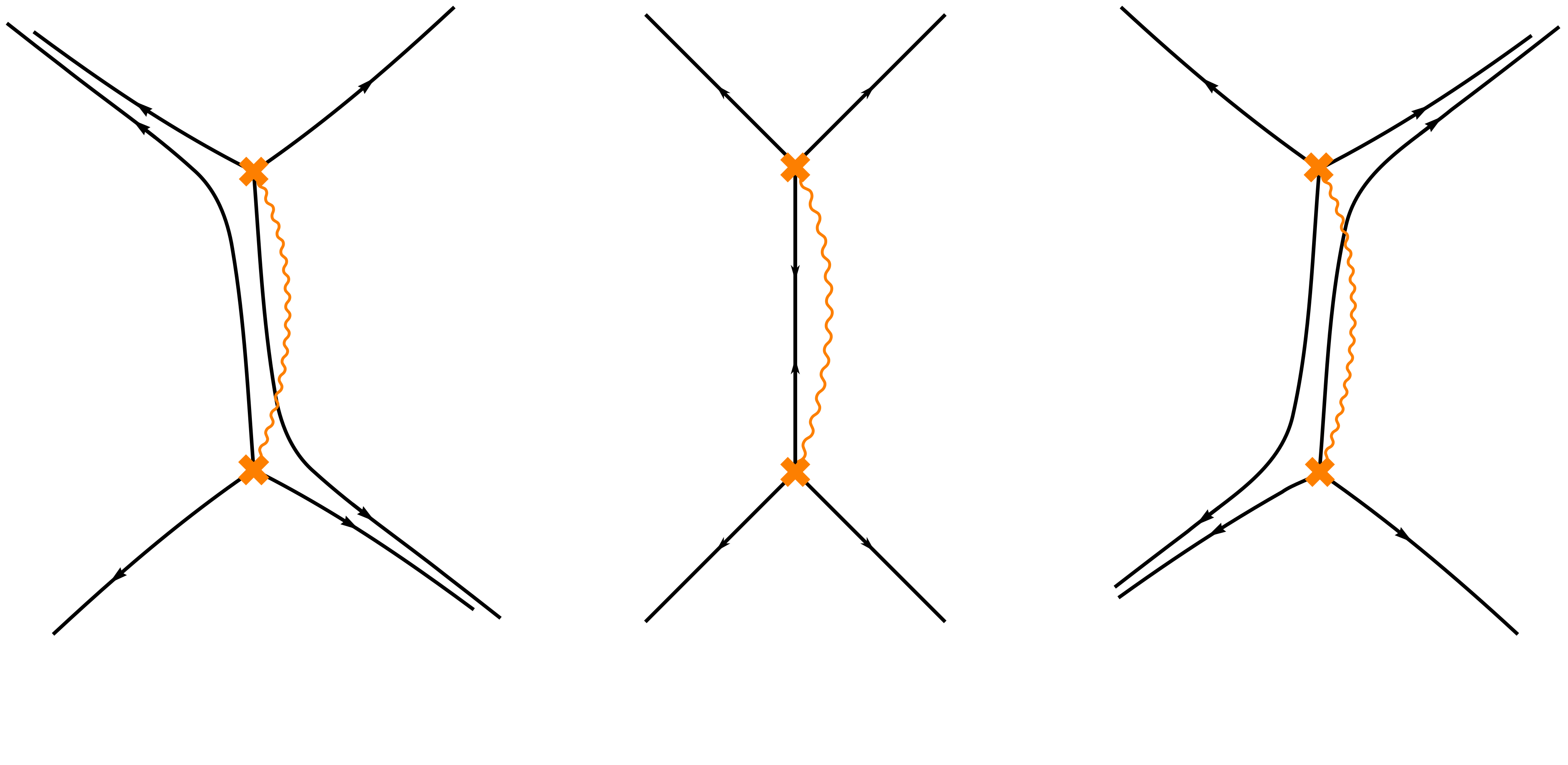}
\put(5,14){$21$}
\put(21,45){$21$}
\put(3,40){$12$}
\put(6,44){$12$}
\put(22,11){$12$}
\put(25,16){$12$}
\put(15.5,0){$\mathbf{A}$}
\put(42,14){$21$}
\put(56.5,14){$12$}
\put(47,22){$12$}
\put(47,34){$21$}
\put(41,42){$12$}
\put(57,42){$21$}
\put(50,0){$\mathbf{B}$}
\put(72,17){$21$}
\put(76,11){$21$}
\put(90,45){$21$}
\put(94,39){$21$}
\put(90,9){$12$}
\put(73,41){$12$}
\put(83,0){$\mathbf{C}$}
\end{overpic}
\caption{\small Spectral networks $\scrW(p_2,\vartheta)$ for $m>0$ as $\vartheta$ is varied.}
\label{AD2jump}
\normalsize
\end{figure}
Increasing the phase $\vartheta$ 
does not change the isotopy class of the network $\scrW$, but decreasing $\vartheta$ to
$\vartheta_{\rm c}=\pi/2$ does. At $\vartheta=\vartheta_{\rm c}$ the
two trajectories in between the branch points overlap and form a {saddle}, as shown in Figure~\ref{AD2jump}~B.

By decreasing $\vartheta$ further to $\vartheta=(\pi/2)^-$ we end up with the network in Figure~\ref{AD2jump}~C. That is, if we vary $\vartheta$ from $(\pi/2)^+$ to $(\pi/2)^-$ the
isotopy class of $\scrW$ changes, with
a saddle at the critical value $\vartheta_{\rm c}=\pi/2$. We call such a
topology change a {flip}. We will explain in
Section~\ref{4dBPS} why this physically describes the appearence of a
BPS hypermultiplet of mass~$m$.
\end{example}

\begin{example}
Let $C$ be any punctured Riemann surface, and take $p_2$ to be a
generic differential with regular singularities at the
punctures. For generic values of $\vartheta$ every trajectory 
ends on a puncture and the foliation defined by $p_2$ looks
locally as in Figure~\ref{FolCell}.
\begin{figure}[h!]
\centering
\small
\begin{overpic}
[width=0.60\textwidth]{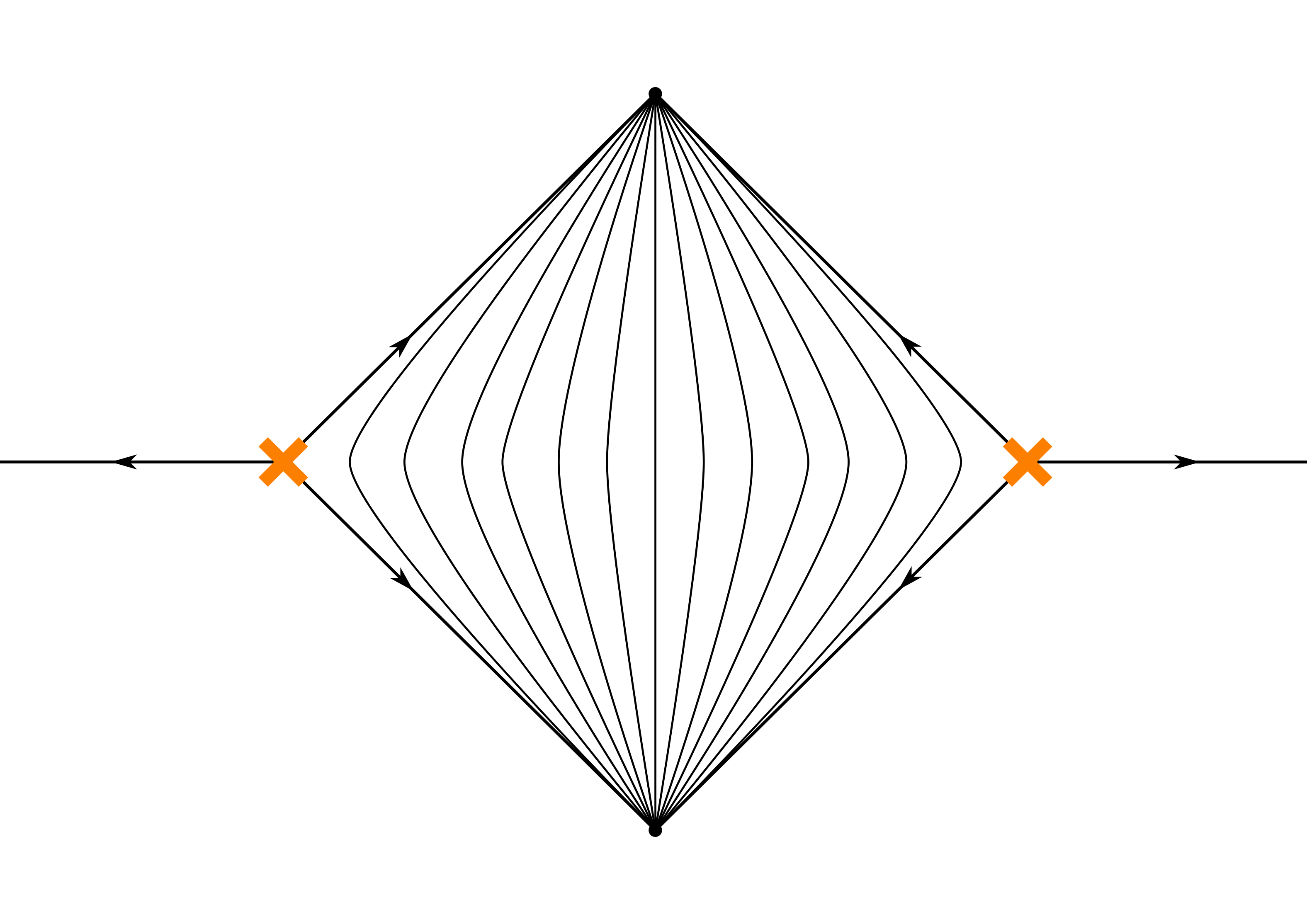}
\end{overpic}
\caption{\small Topology of a cell of the foliation. The cell is
  bounded by critical trajectories running from a branch point to a
  puncture and it includes an infinite family of generic trajectories that run between the punctures.}
\label{FolCell}
\normalsize
\end{figure}
We call the resulting spectral network a {Fock-Goncharov
  network} (in the terminology of~\cite{hollands2016spectral}).
For a Fock-Goncharov network we can use the cells of the foliation to define an {ideal triangulation} $\scrT=\scrT(p_2,\vartheta)$ of $C$. The vertices of $\scrT$ are the punctures of $C$ and the edges are obtained by choosing one generic trajectory in each cell~\cite{hollands2016spectral,gaiotto2013wall}. An example on the four-punctured sphere is shown in Figure~\ref{Nf4SN}.
\begin{figure}[h!]
\centering
\small
\begin{overpic}
[width=0.60\textwidth]{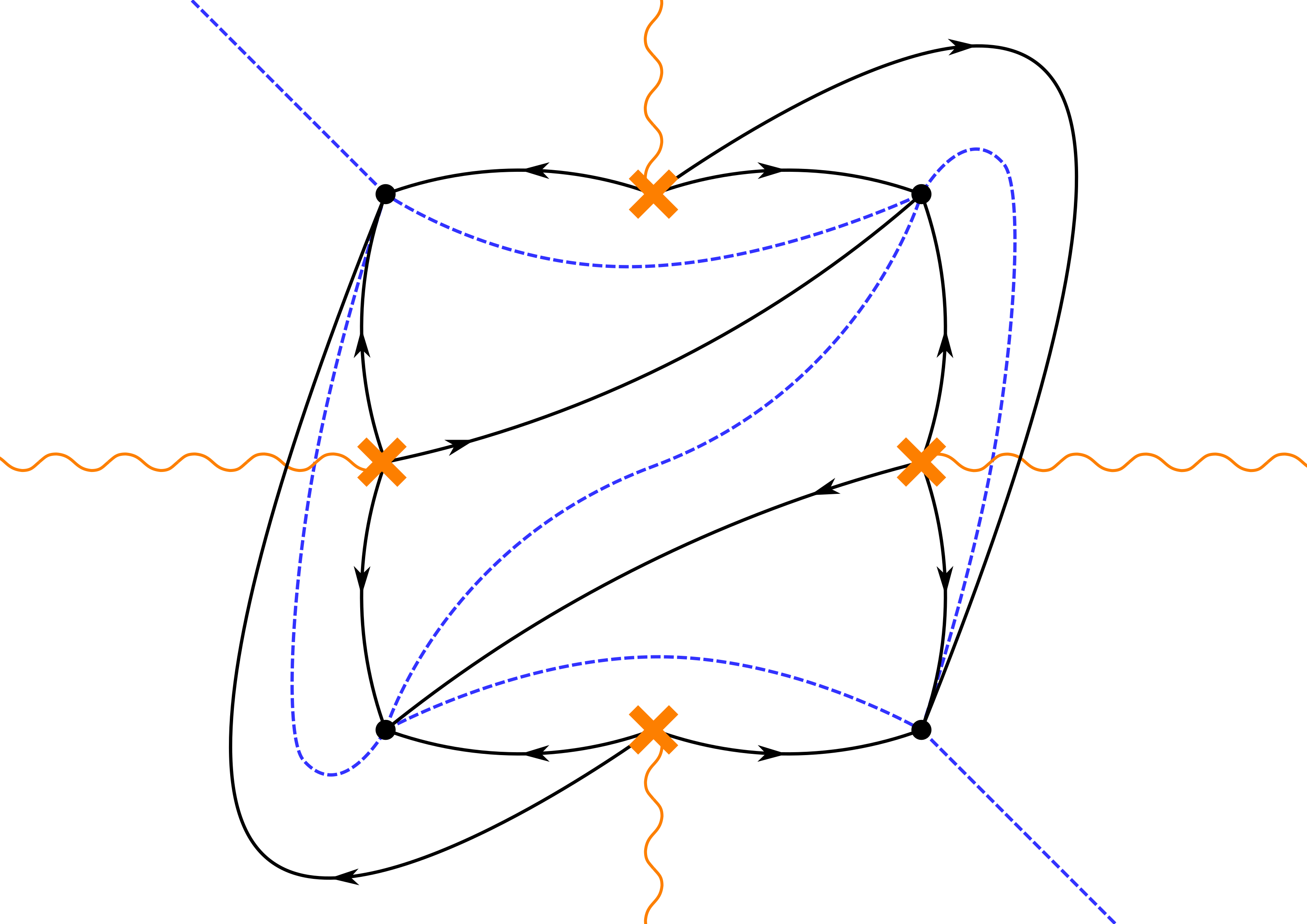}
\end{overpic}
\caption{\small The four-punctured sphere (viewed as the plane with
  the point at infinity omitted) with a Fock-Goncharov network $\scrW$. The edges of an ideal triangulation $\scrT$ are indicated in blue.}
\label{Nf4SN}
\normalsize
\end{figure}
\end{example}

\begin{example}\label{ex:purenetworks}
Consider $C=\C^\times=\C\setminus\{0\}$ with the quadratic differential
\begin{align}\label{eq:pureSU2p2}
p_2(z) = \Big(\frac1{z^3}+\frac{9}{4\,z^2}+\frac1z\Big)\, \de
  z\otimes\de z \ .
\end{align}
This differential has irregular singularities at $z=0$ and $z=\infty$,
since the behaviour of $p_2$ is given by $\frac1{\tilde z^3}\,\de\tilde
z\otimes\de\tilde z$ near there (in the local coordinate $\tilde z=z$ and
$\tilde z=\frac1z$, respectively). Near $\vartheta=\vartheta_{\rm
  c}=\pi/2$, something interesting happens: Close to this critical phase,
some trajectories start to wind around the cylinder $C=\C^\times$, see
Figure~\ref{PureSU2Wjuggle}. If one studies the winding for $\vartheta<\vartheta_{\rm c}$
(and the unwinding for $\vartheta>\vartheta_{\rm c}$), one finds that it
is really an infinite sequence of flips; this is explained in detail
in~\cite[Section~5.9]{gaiotto2013wall}. At $\vartheta=\vartheta_{\rm
  c}$ the network changes topology. At this phase a family of closed
trajectories emerges that surround the cylinder. This is illustrated
in Figure~\ref{PureSU2Wjuggle}, and is called a {juggle} (see
also~\cite[Section~6.6.3]{gaiotto2013wall}).
\begin{figure}[h!]
\centering
\small
\begin{overpic}
[width=0.80\textwidth]{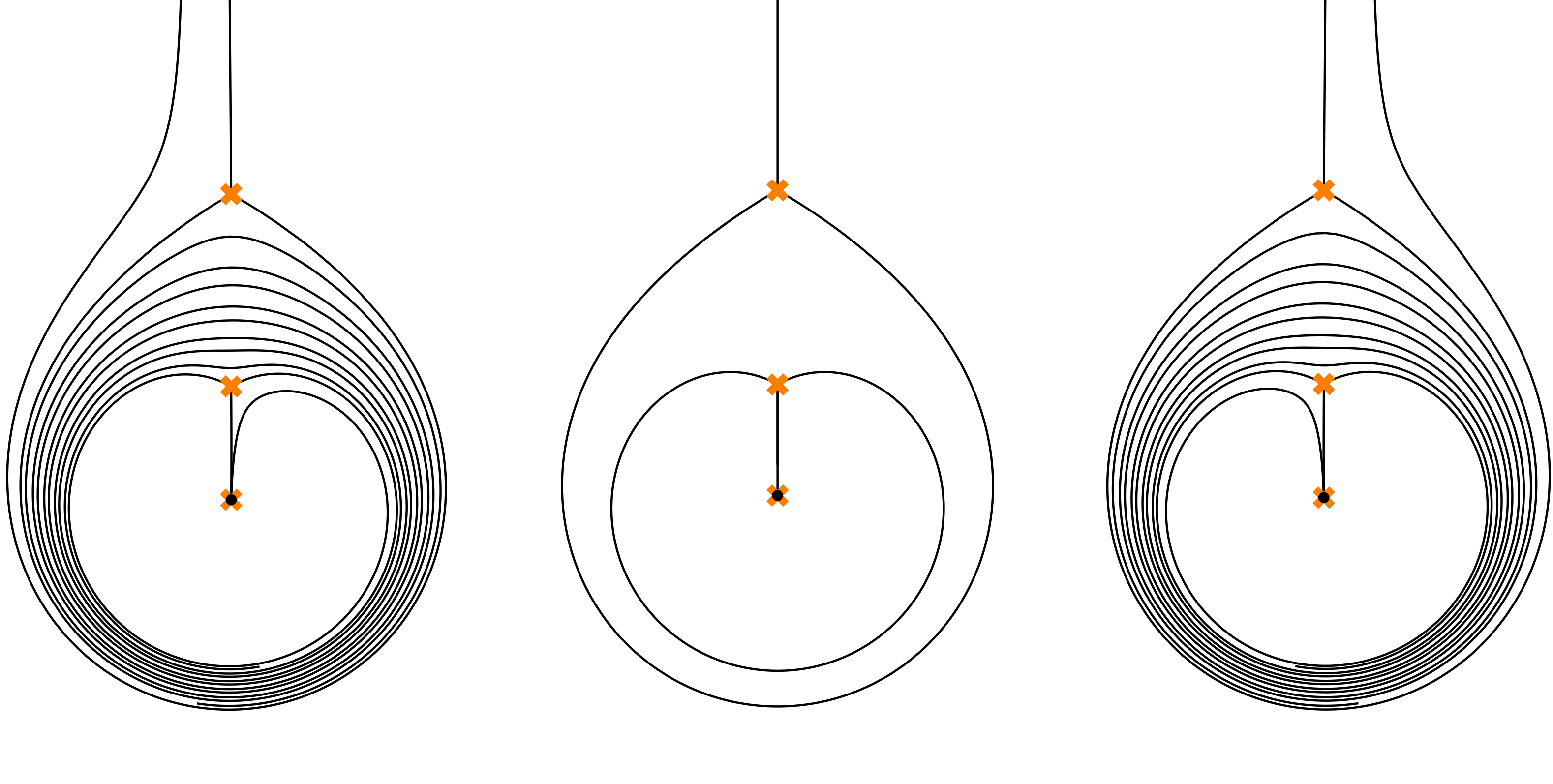}
\end{overpic}
\caption{\small Spectral networks $\scrW(p_2,\vartheta)$, rotated through $-90^\circ$, for the quadratic differential \eqref{eq:pureSU2p2} and phases $\vartheta_{\rm c}-\delta$, $\vartheta_{\rm c}$, $\vartheta_{\rm c}+\delta$, respectively, with $\vartheta_{\rm c}=\frac\pi2$ and $\delta$ small. The transition of the spectral network from $\vartheta_{\rm c}-\delta$ to $\vartheta_{\rm c}+\delta$ is called a juggle.}
\label{PureSU2Wjuggle}
\normalsize
\end{figure}
As we will see in Section~\ref{4dBPS}, a juggle corresponds to
a BPS vector multiplet in the corresponding supersymmetric field
theory. The two limits $\vartheta\to\vartheta_{\rm c}^\pm$ are called the {resolutions} of the spectral network at the critical phase; they are illustrated schematically in Figure~\ref{PureSU2resolutions}.
\begin{figure}[h!]
\centering
\small
\begin{overpic}
[width=0.80\textwidth]{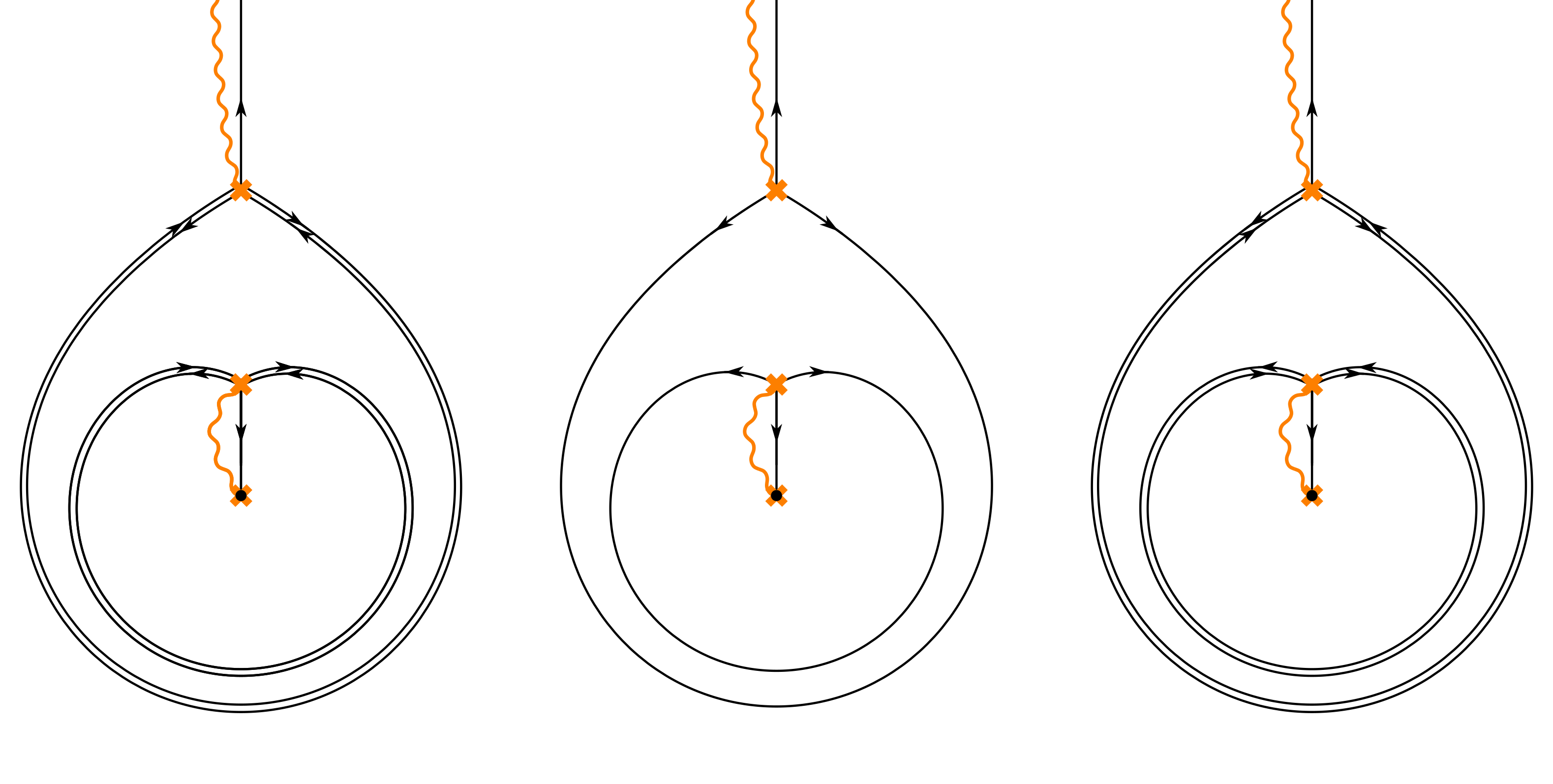}
\put(13,0){$\scrW^-$}
\put(48.5,0){$\scrW_{\rm c}$}
\put(83,0){$\scrW^+$}
\put(16,43){$21$}
\put(51,43){$21$}
\put(85.5,43){$21$}
\put(16,20){$21$}
\put(51,20){$21$}
\put(85.5,20){$21$}
\put(20,37){$12$}
\put(53,37){$12$}
\put(89,37){$21$}
\put(17,33){$12$}
\put(86.5,33){$21$}
\put(9,37){$12$}
\put(44,37){$21$}
\put(78,37){$21$}
\put(12,33){$21$}
\put(81,33){$21$}
\put(17,27.5){$12$}
\put(52,27.5){$12$}
\put(86.5,27.5){$21$}
\put(17,23){$21$}
\put(86.5,23){$12$}
\put(10,27.5){$21$}
\put(45,27.5){$21$}
\put(79,27.5){$12$}
\put(10,23){$21$}
\put(79,23){$12$}
\end{overpic}
\caption{\small In the middle: The spectral network $\scrW_{\rm c}=\scrW(p_2,\vartheta_{\rm c})$ for the quadratic differential \eqref{eq:pureSU2p2} and critical phase $\vartheta_{\rm c}=\frac\pi2$. On either side: Its resolutions $\scrW^{\pm}=\scrW(p_2,\vartheta_{\rm c}^\pm)$.}
\label{PureSU2resolutions}
\normalsize
\end{figure}
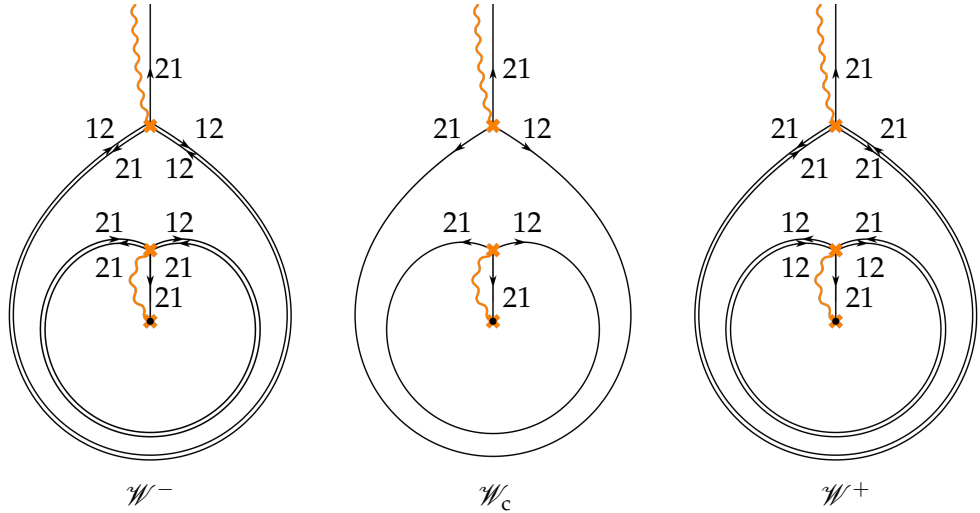
\end{example}

\begin{example}
If $C$ is any (possibly punctured) Riemann surface with regular
punctures, we may consider a special class of quadratic differentials
$\E^{-2\,\I\,\vartheta}\,p_2$ for which all trajectories are
compact and the critical graph is built out of saddles. These
differentials are known as {Strebel differentials}. In
particular, given any pants decomposition of $C$ along with a choice
of ``length'' parameters $l_k$, there is a unique Strebel differential $\E^{-2\,\I\,\vartheta}\,p_2$ respecting this pants decomposition and satisfying
\begin{align}
\E^{-\I\,\vartheta}\, \oint_{\alpha_k} \, \sqrt{p_2} = l_k
\end{align}
for all pants curves $\alpha_k$. The
corresponding spectral network is dual to the pants decomposition of
$C$. We call such a network a {Fenchel-Nielsen network} (for
more details see~\cite{hollands2016spectral}). See
Figures~\ref{MolOne} and~\ref{Mols} for examples of Fenchel-Nielson
networks on the three-punctured and four-punctured sphere,
respectively.
\begin{figure}[h!]
\centering
\small
\begin{overpic}
[width=0.40\textwidth]{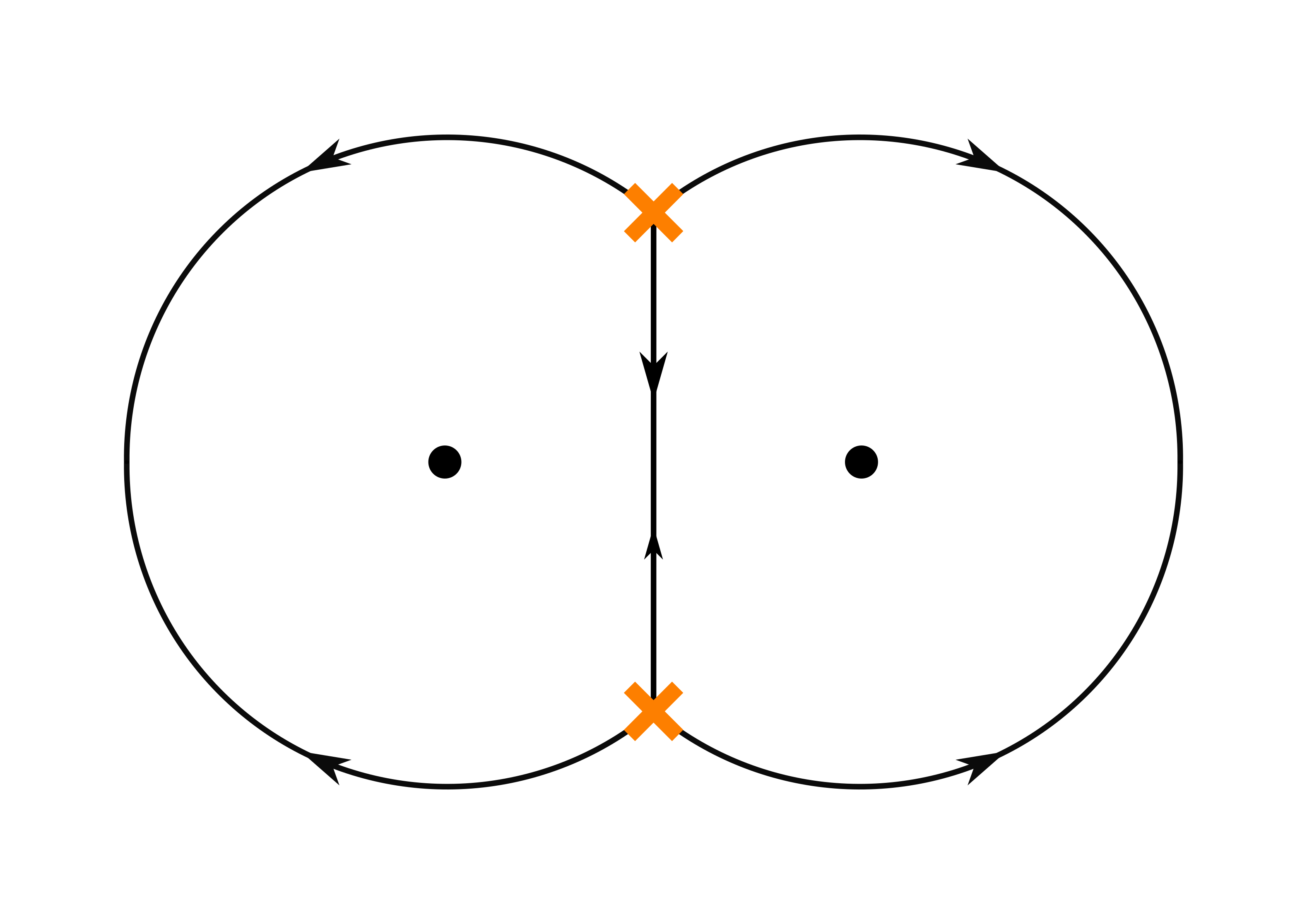}
\end{overpic}
\caption{\small One of two Fenchel-Nielsen network topologies on the three-punctured sphere with the third puncture being a regular puncture at infinity. This is equivalent to a pair of pants, if the punctures are viewed as boundaries instead.}
\label{MolOne}
\normalsize
\end{figure}
\begin{figure}[h!]
\centering
\small
\begin{overpic}
[width=0.60\textwidth]{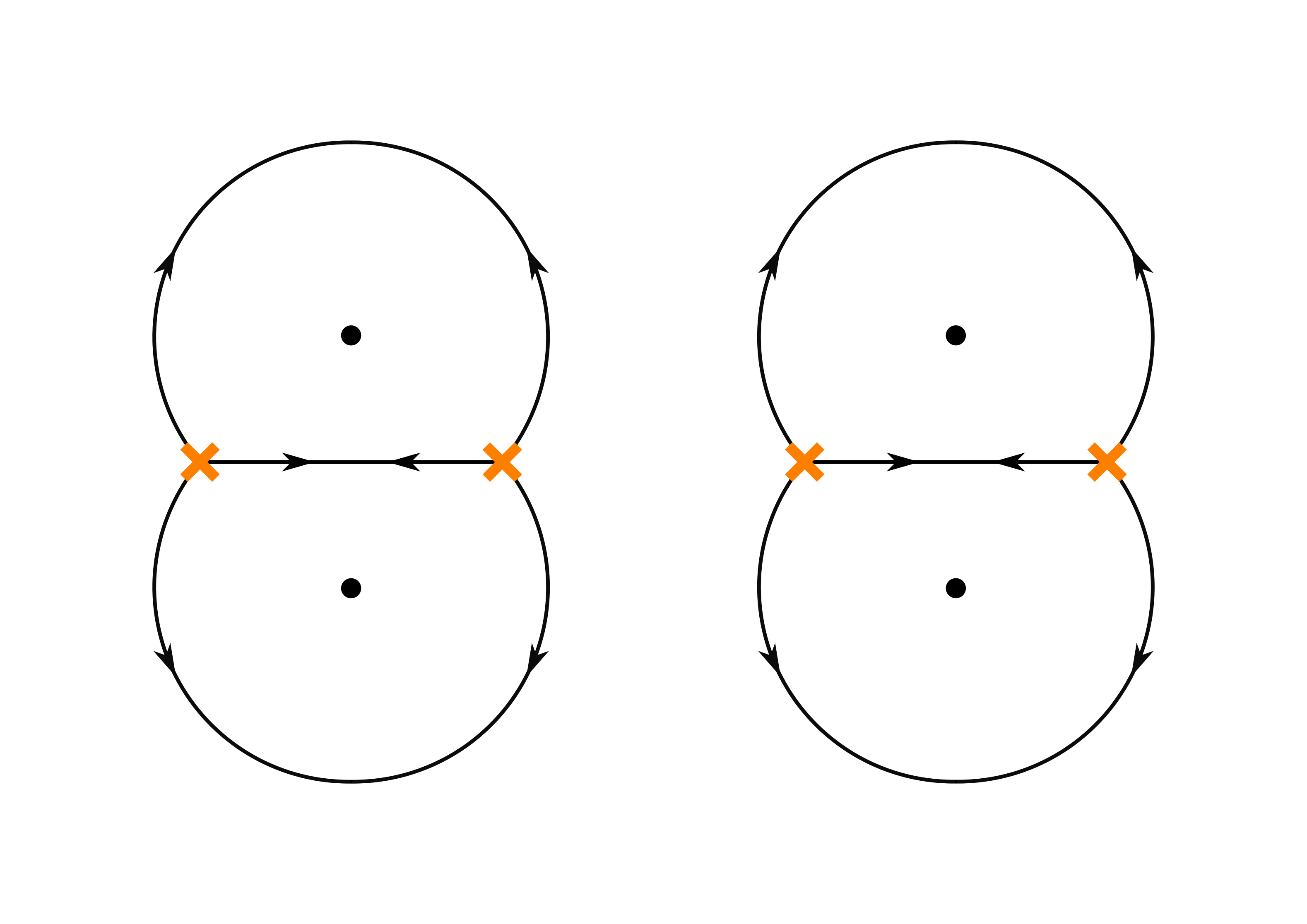}
\end{overpic}
\caption{\small A Fenchel-Nielsen network on the four-punctured sphere
  glued together from two pairs of pants.}
\label{Mols}
\normalsize
\end{figure}

If the Riemann surface $C$ has irregular singularities, the Strebel condition is too strong (there are always non-compact trajectories in this case). Yet we would like to call a spectral network $\scrW(p_2,\vartheta)$ of Fenchel-Nielsen type if there is a maximal number of non-degenerate ring domains (see also~\cite{Coman:2020qgf}). This is the case, for instance, at the critical phase $\vartheta_{\rm c}=\frac\pi2$ in Example~\ref{ex:purenetworks}. We can thus consider the middle network in Figure~\ref{PureSU2Wjuggle} as
another example of a Fenchel-Nielsen network.
\end{example}

\subsection{Higher Rank Generalization}\label{sec:SpectralHigherRank}
\noindent
So far we considered spectral networks defined by a single quadratic
differential $p_2$. This may be generalized to a higher rank
version of a WKB spectral network by considering a tuple of differentials
$p=(p_2,\dots,p_K)$ on $C$, where each $p_k$
is a meromorphic $k$-differential on $C$, possibly singular at the
punctures $z_i$. In this case we call a singularity {regular} if
the order of the pole of each $p_k$ is at most $k$, and
{irregular} otherwise. As we will see later on, regular punctures
may be classified by Young diagrams with $K$ boxes and at most $K-1$
rows.

The tuple of differentials $(p_2,\dots,p_K)$ defines a
(possibly branched) $K$-fold covering $\Sigma\subset T^*C$ of $C$, by
the equation\footnote{In \eqref{Sigma} we have changed the sign in front of the coefficients $p_k$ to match our conventions from Section~\ref{sec:WKBnetworks}. This agrees with the conventions of~\cite{gaiotto2013wall}, but unfortunately not with those of~\cite{gaiotto2012spectral}. In particular, to verify the networks below, one needs to introduce an additional minus sign in {\sl swn-plotter}~\cite{swn-plotter}.}
\begin{equation}\label{Sigma}
\Sigma\colon \quad \lambda^K-\sum_{k=2}^{K}\,p_k\, \lambda^{K-k}=0 \ ,
\end{equation}
where $\lambda = w\,\de z$ is the tautological one-form on the
cotangent bundle $T^*C$. The curve
$\Sigma$ is called the {spectral curve}, and we will encounter
it again in Section~\ref{sec:classS} when we explain how spectral
networks are relevant to $\N=2$ field theories. In the following we
assume that the covering $\Sigma\to C$ has only simple branch points,
where exactly two sheets come together; this is the generalization of
the requirement that $p_2$ has only simple zeroes for $K=2$. 

Let $\lambda_i$ be the restriction of the tautological one-form
$\lambda$ to the $i$-th sheet. 
We define an $ij${-trajectory} on $C$ for $i\neq j$ to be
a real curve $\gamma$ on $C$ such that
\begin{equation}
\E^{-\I\,\vartheta}\,(\lambda_i-\lambda_j)(v) \ \in \ \mathbb{R}^\times
\end{equation}
for any non-zero tangent vector $v$ to $\gamma$. We call such a
trajectory {critical} if at least one of its endpoints is a
branch point. The {critical graph} is again defined as the union of all
critical trajectories, and we add an orientation and a label $ij$ to
each $ij$-trajectory to get a spectral network
$\scrW(p,\vartheta)$ on $C$. Sometimes we will call
$\scrW(p,\vartheta)$ a spectral network {subordinate} to
the cover $\Sigma\to C$.

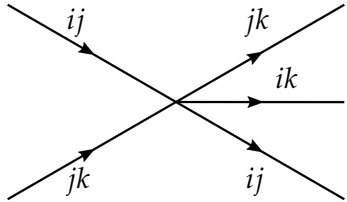
\begin{figure}[h!]
\centering
\small
\begin{overpic}
[width=0.80\textwidth]{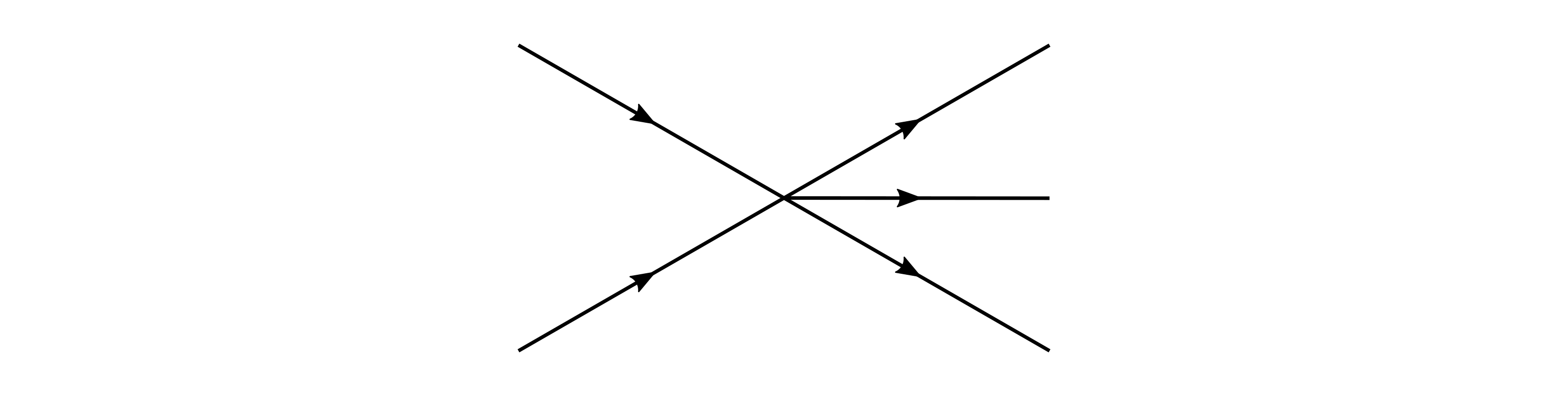}
\put(39,20){$ij$}
\put(39,4){$jk$}
\put(57,20){$jk$}
\put(57,4){$ij$}
\put(60,14){$ik$}
\end{overpic}
\caption{\small The crossing of walls with labels $ij$ and $jk$ results in a new wall with label $ik$ which is ``born'' at the intersection.}
\label{Intersectionijk}
\normalsize
\end{figure}

There is a new phenomenon for critical trajectories in higher rank
networks for which $K>2$: trajectories with different labels (say $ij$ and
$jk$) may cross paths and form a {junction}. For instance, the
crossing of a $12$ and a $23$ trajectory will lead to a new trajectory
of type $13$, see Figure~\ref{Intersectionijk}. In this way new trajectories can be
``born''. The possibility of such crossings vastly increases the
complexity of the higher rank networks. In fact, they are largely
unexplored apart from the cases of higher rank generalizations of
Fock-Goncharov and Fenchel-Nielsen networks~\cite{gaiotto2012spectral,
  Hollands:2016kgm, hollands2016spectral}.

\begin{example}\label{ex:higherFN}
Let $K=3$ and $C=\PP^1_{0,1,\infty}=\PP^1\setminus\{0,1,\infty\}$ be the three-punctured sphere with the differentials
\begin{align}\label{eq:E6differentials}
p_2 &= \frac{c_\infty \, z^2 -
            (c_0-c_1+c_\infty)\,z+c_0}{z^2\,(z-1)^2} \, (\de
            z)^{\otimes 2} \ , \nonumber \\[4pt]
p_3 &= \frac{d_\infty\,z^3 + u\,z^2 + (d_0+d_1-d_\infty-u)\,z -
            d_0}{z^3\,(z-1)^3} \, (\de z)^{\otimes 3} \ ,
\end{align}
where $u\in\C$ is a free parameter, while
\begin{align}
c_l=-m_{l,1}^2-m_{l,1}\,m_{l,2}-m_{l,2}^2 \qquad \mbox{and} \qquad d_l
  = m_{l,1}\,m_{l,2}\,(m_{l,1}+m_{l,2}) \ ,
\end{align}
and $m_{l,1}\neq m_{l,2}$. The residues of these differentials at the regular punctures
$z_l=l$ are given by $\{m_{l,1},m_{l,2},-m_{l,1}-m_{l,2}\}$, respectively, and
the spectral curve $\Sigma$ is a three-fold branched covering of $C$
with six simple branch points. This implies that $\Sigma$
is a punctured Riemann surface of genus one. In the limit $m_{l,j}\to0$, the branch points of the covering $\Sigma\to C$ move towards the punctures and the topology of the corresponding spectral networks $\scrW(u,\vartheta)$ depends only on the phase of the quantity $\E^{-3\,\I\,\vartheta} \, u$. For generic phase the network seems to be ``wild'', that is, it is dense in at least some parts of $C$. For non-generic phase the network is compact, with
\begin{align}
\E^{-\I\,\vartheta} \, \oint_\gamma \, \lambda \ \in \ \R \ ,
\end{align}
for some one-cycle $\gamma$ on $\Sigma$. 

These non-generic compact networks were studied in~\cite{Hollands:2016kgm}. They are labelled by two coprime integers $p$ and $q$. The three simplest topologies $\scrW_{[1,0]}$, $\scrW_{[1,2]}$ and $\scrW_{[1,3]}$ are illustrated in Figure~\ref{E6Networks}. 
\begin{figure}[h!]
\centering
\small
\begin{overpic}
[width=\textwidth]{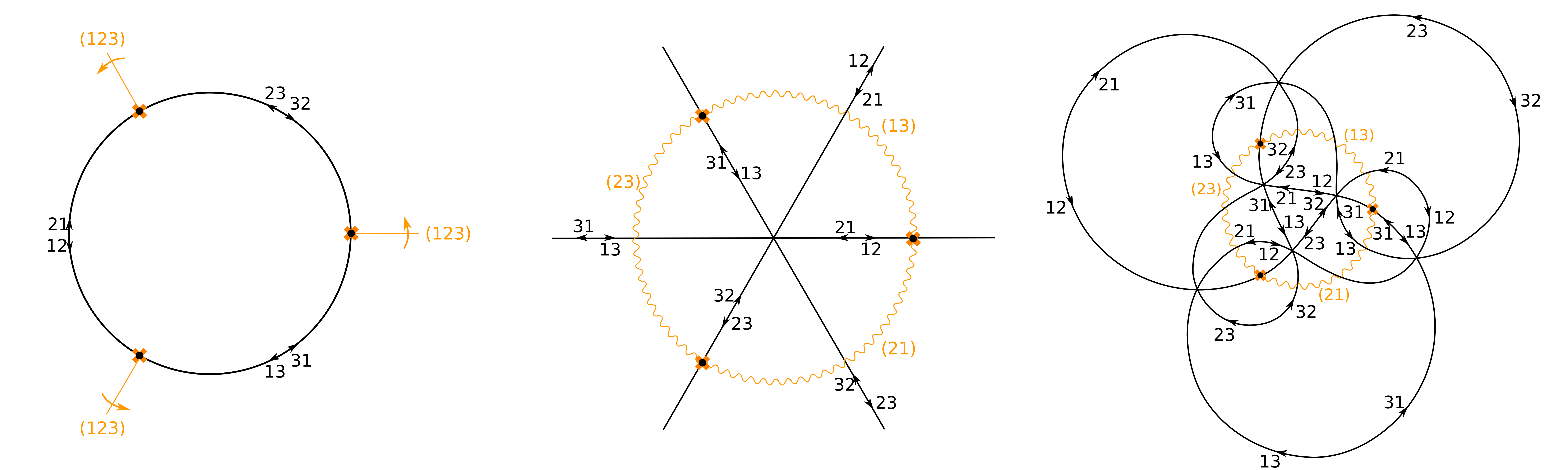}
\put(14,-4){$\mathbf{A}$}
\put(50,-4){$\mathbf{B}$}
\put(83,-4){$\mathbf{C}$}
\end{overpic}
\bigskip
\caption{\small Three examples of higher rank Fenchel-Nielsen networks on the three-punctured sphere (taken from~\cite{Hollands:2016kgm}): {\bf A} \ $\scrW_{[1,0]}$; {\bf B} \ $\scrW_{[1,2]}$; {\bf C} \ $\scrW_{[1,3]}$. Here we have moved the three punctures to $z=1,\omega,\omega^2$ with \smash{$\omega=\E^{\,2\pi\,\I/3}$}.}
\label{E6Networks}
\normalsize
\end{figure}
These networks are rather degenerate: If we were to perturb $\vartheta$ slightly away from the critical value, we would see that there is infinite winding around the saddles. For this reason we call these networks of higher rank Fenchel-Nielsen type.

As we will learn in Section~\ref{4dBPS}, each network $\scrW_{[q,p]}$ encodes all BPS states of electromagnetic charge $(q,p)$ in the $E_6$ Minahan-Nemeschansky theory.
\end{example}

\subsection{Abelianization}\label{sec:abelianization}
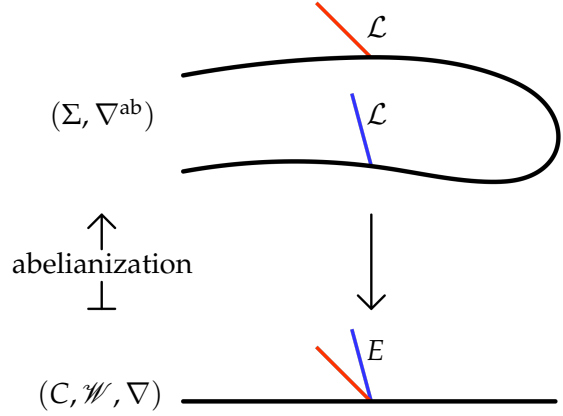
\begin{figure}[h!]
\centering
\small
\begin{overpic}
[width=0.50\textwidth]{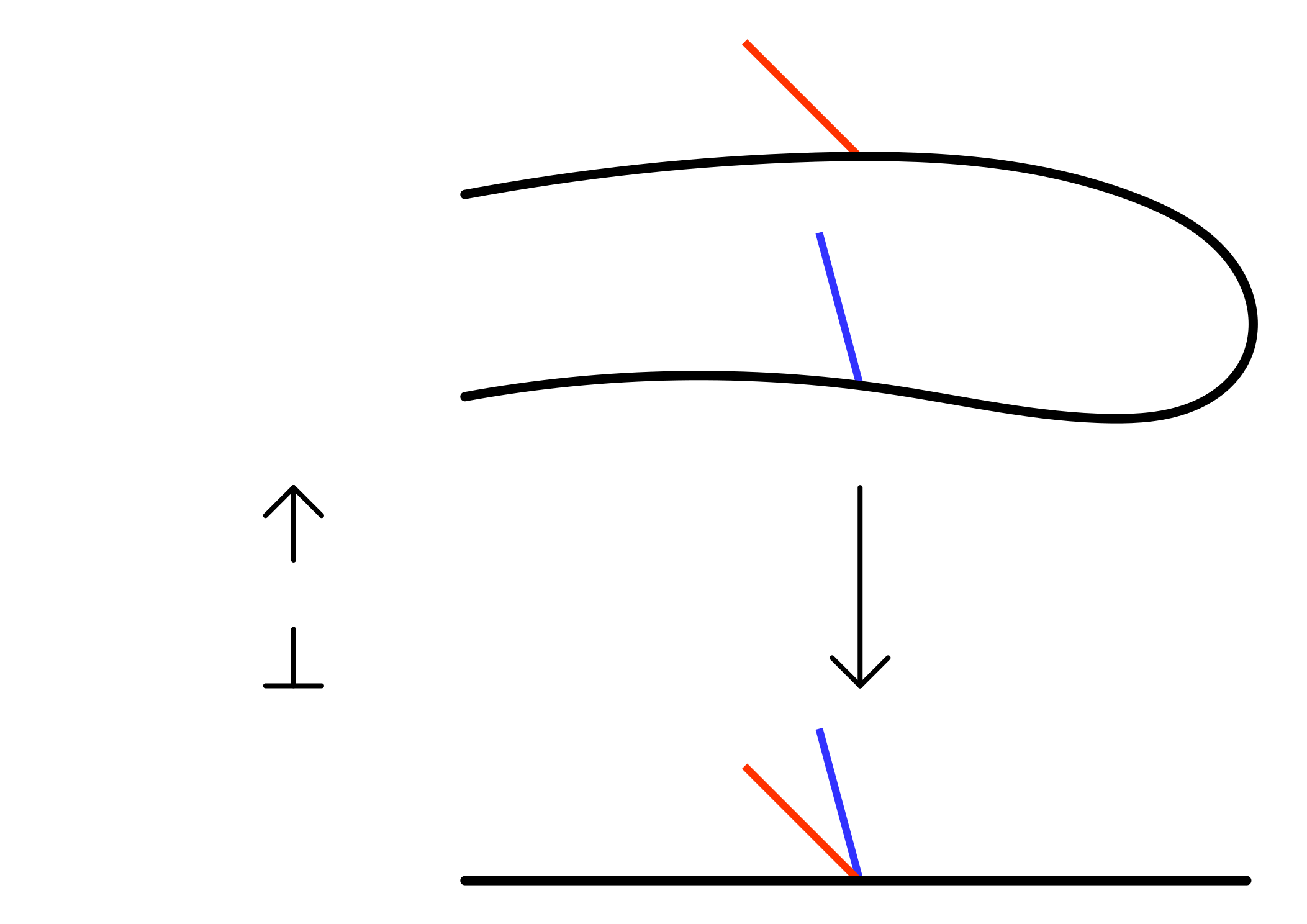}
\put(14,48){$(\Sigma,\nablaab)$}
\put(12,3){$(C,\scrW,\nabla)$}
\put(65,62){$\cL$}
\put(65,48){$\cL$}
\put(65,10){$E$}
\put(8,24){abelianization}
\end{overpic}
\caption{\small Given a spectral
  network $\scrW$ and a generic flat $SL(K,\C)$ connection $\nabla$ on the
  vector bundle $E$ over $C$, abelianization is a way of bringing
  $\nabla$ into an almost diagonal form, such that it may be lifted to
  a flat $GL(1,\C)$ connection $\nablaab$ on the line bundle $\cL$
  over the spectral cover $\Sigma'$.}
\label{fig:abelianization}
\normalsize
\end{figure}

\noindent
In the following we fix a branched $K$-fold cover
$\pi\colon\Sigma\rightarrow C$ as in (\ref{Sigma}), and a spectral
network $\scrW$ on $C$ subordinate to the covering. Let $\Sigma'$
denote $\Sigma$ with the branch points removed. The general
idea of {abelianization} is to lift non-abelian structures on a given rank~$K$ vector
bundle $E$ over $ C$ to construct corresponding abelian structures on a line bundle $\mathcal{L}$
over the spectral cover $\Sigma'$. In particular, starting from a flat
$SL(K,\mathbb{C})$ connection $\nabla$ on $C$, we construct a
flat $GL(1,\mathbb{C})$ connection $\nablaab$ on $\Sigma'$, using the
data of the spectral network $\scrW$. This is illustrated in
Figure~\ref{fig:abelianization}.
We first consider $\nabla$ on the connected components, or cells, of
$C\!\setminus\!\scrW$. The cells are either contractible or tubular,
and the construction works for either case~\cite{hollands2016spectral}. On these we can find a gauge which diagonalizes $\nabla$. We then glue this patchwise diagonalization of $\nabla$ together by assigning unipotent gauge transformations to the trajectories of~$\scrW$.

Concretely, in each cell we look for a basis of sections $(s_1,\dots,s_K)$ of $E$ with respect to which $\nabla$ is diagonal,
\begin{equation}
\nabla s_i=d_i\otimes s_i \ ,
\end{equation}
where $d_i$ are closed one-forms on $C$ for $i=1,\dots,K$. 
On crossing a trajectory with label $ij$ we require a unipotent transformation
\begin{align}\label{smaps}
s_i\longmapsto s_i+c_{ij}\, s_j=:s_i'  \qquad \mbox{and} \qquad
s_k\longmapsto s_k \quad \text{for} \quad k\neq i 
\ ,
\end{align}
for some function $c_{ij}$. 
On crossing a branch cut of type $(ij)$ we require that the sections
on either side are related by a permutation matrix as
\begin{align}\label{eq:sbranchmaps}
s_i\longmapsto s_j \ , \quad
s_j\longmapsto -s_i \qquad \mbox{and} \qquad
s_k\longmapsto s_k\quad \text{for} \quad k\neq i,j \ .
\end{align}

If we can find such a gauge, we may lift $\nabla$ on $C$ to  a
$GL(1,\C)$ connection $\nablaab$ on the spectral cover $\Sigma'$ as follows. On
$\Sigma\!\setminus\!\pi^{-1}(\scrW)$ we define $\nablaab$ on the $i$-th sheet by the diagonal entries $d_i$ of $\nabla$:
\begin{equation}
\nablaab(s_i)=d_i\otimes s_i \ .
\end{equation}
To show that the unipotent gauge transformations (\ref{smaps}) extend
$\nablaab$ across the trajectories to all of $\Sigma$, we need to show that
$\nabla s_i' = d_i\otimes s_i'$ for all $i=1,\dots,K$.\footnote{The
  connection $\nablaab$ can be
  extended to a connection on all of $\Sigma$ which is {almost flat}:
  its holonomy around a simple branch point is $-\mathbb{1}$.}  Indeed, since
$\nabla$ is an $SL(K,\C)$ connection, on
crossing a trajectory with label $ij$, $\nabla s_i=d_i\otimes s_i$ is sent to
\begin{align}
\nabla s_i'=d_i'\otimes s_i'=d_i'\otimes s_i+d_i'\otimes c_{ij}\,s_j 
\end{align}
with
\begin{align}
\nabla s_i'=\nabla(s_i+c_{ij}\,s_j) = \nabla s_i + \de c_{ij}\otimes s_j +
  c_{ij}\,\nabla s_j = d_i\otimes s_i+(\de c_{ij}+c_{ij}\,d_j)\otimes s_j \ ,
\end{align}
showing that $d_i'=d_i$ as desired (and $\de c_{ij} =
c_{ij}\,(d_i-d_j)$, so the functions $c_{ij}$ are bicovariantly
constant). Note that, since $\nabla$ is an $SL(K,\C)$ connection, the
connection $\nablaab$ carries some extra structure: parallel transport
of the sections $(s_1,\dots,s_K)$ along a path in $C\setminus\scrW$
(not crossing any branch cuts) is given by a diagonal matrix with
determinant equal to~$1$. We say that $\nablaab$ is
{equivariant}, see~\cite[Section~4.2]{hollands2016spectral}
and~\cite[Section~5.3]{hollands2018higher}. 

To help us find such a gauge, and to show it is unique, we may need
some additional discrete choices on $\nabla$; this is called a
{framing} of $\nabla$. 

\begin{example}
Fix a Fock-Goncharov network $\scrW$ 
and consider a flat $SL(2,\mathbb{C})$ connection
$\nabla$ on $C$. Locally $C\setminus \scrW$ contains
cells of the form shown in Figure~\ref{AbelCells}~A. 
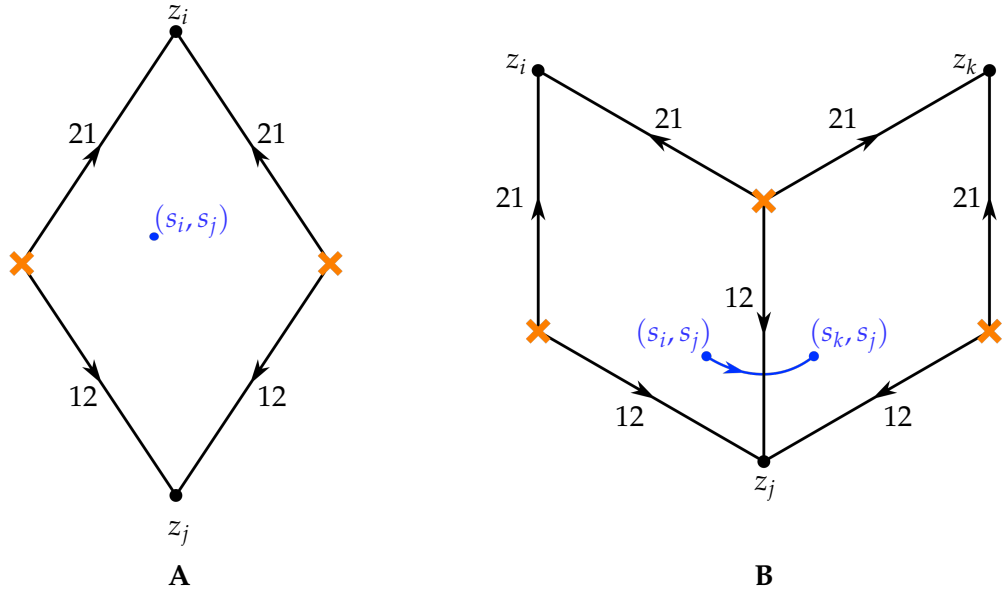
\begin{figure}[h!]
\centering
\small
\begin{overpic}
[width=0.80\textwidth]{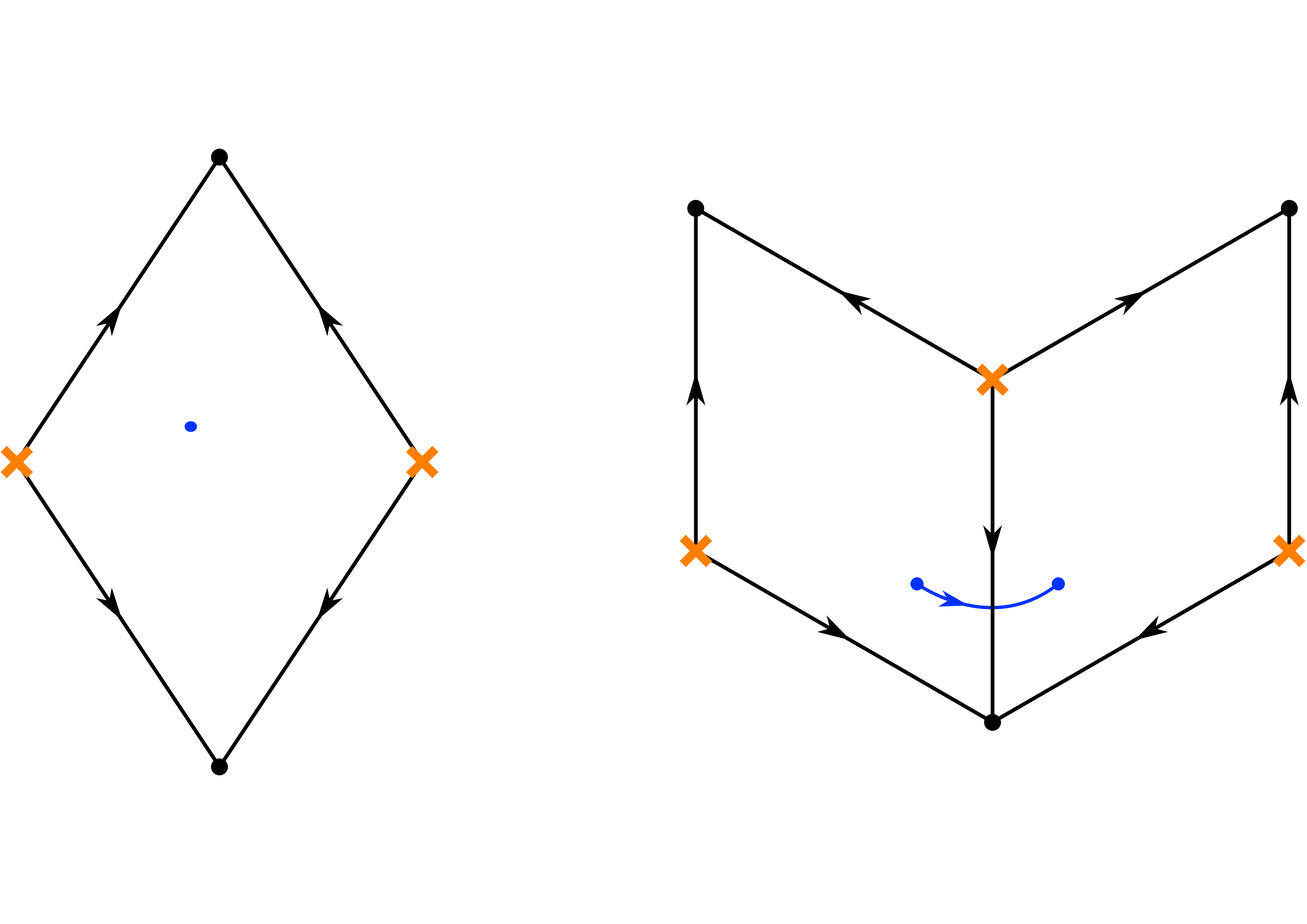}
\put(14.5,39){\textcolor{Blue}{$(s_i,s_j)$}}
\put(16,60){$z_i$}
\put(16,8){$z_j$}
\put(50,55){$z_i$}
\put(75,12.5){$z_j$}
\put(95,55){$z_k$}
\put(63,27.5){\textcolor{Blue}{$(s_i,s_j)$}}
\put(80.5,27.5){\textcolor{Blue}{$(s_k,s_j)$}}
\put(6,47.5){$21$}
\put(25,47.5){$21$}
\put(6,21){$12$}
\put(25,21){$12$}
\put(49,41){$21$}
\put(95,41){$21$}
\put(65,49){$21$}
\put(82.5,49){$21$}
\put(61,19){$12$}
\put(88,19){$12$}
\put(72,31){$12$}
\put(16,3){\textbf{A}}
\put(75,3){\textbf{B}}
\end{overpic}
\caption{\small \textbf{A:} A single cell of a Fock-Goncharov network, with basis of sections $(s_i,s_j)$. \textbf{B:} Adjacent cells with bases related by transport over a trajectory.}
\label{AbelCells}
\normalsize
\end{figure}
For any puncture $z_i$, we defined the framing of $\nabla$ at $z_i$ to
be a choice of a $\nabla$-invariant line sub-bundle $\ell_i$ of $E$
in a neighbourhood of $z_i$ with the following condition. If, for each
cell with two punctures $z_i$ and $z_j$, we parallel transport the
respective line bundles $\ell_i$ and $\ell_j$ to a common point $z$ in
the cell, we require that $\ell_i(z)\neq\ell_j(z)$. We call the 
connection $\nabla$ together with this framing data a
{$\scrW$-framed connection}.

Note that the condition above is automatically satisfied for a generic
connection $\nabla$. Moreover, for generic $\nabla$ there are exactly
$2^n$ possible $\scrW$-framings, where $n$ is the number of
punctures. Indeed the monodromy around each puncture has two distinct
eigenspaces, and a $\scrW$-framing corresponds to choosing just one of
them. 

How do we abelianize the $\scrW$-framed connection $\nabla$? Let us
locally trivialize the cover $\Sigma$ over a cell with two
punctures $z_i$ and $z_j$. Suppose that puncture $z_i$ has incoming
trajectories of type $21$, whereas puncture $z_j$ has incoming trajectories
of type $12$. The basis $(s_i,s_j)$ is then
obtained by choosing $(s_i\in \ell_i,s_j\in \ell_j)$ for each cell. 

The local picture for two neighbouring cells is shown in Figure
\ref{AbelCells} B. Each of the cells has basis $(s_i,s_j)$ defined as
above, and crossing the trajectory between them gives a unipotent transformation that
leaves one of the sections unchanged and modifies the other one. A
choice of local bases of sections like this thus defines a
$\scrW$-abelianization of $\nabla$. This can be shown to give a canonical one-to-one
correspondence between $\scrW$-framings of $\nabla$ and $\scrW$-abelianizations of
$\nabla$~\cite[Section~5.2]{hollands2016spectral}.
\end{example}

\begin{example}\label{ex:AD2abelianisation}
  Let us revisit Example~\ref{ex:CAD2} with $C=\C$ and the spectral
  network $\scrW=\scrW(p_2,\vartheta_{\rm c})$ with
  \begin{align}
p_2(z) = (z^2+m)\, \de z\otimes\de z \qquad \mbox{and} \qquad
    \vartheta_{\rm c}=\frac\pi2 \ .
  \end{align}
Since $p_2$ has an
irregular singularity at $z=\infty$ (a pole of order six), we consider
flat $SL(2,\C)$ connections $\nabla$ with a corresponding irregular
singularity at $z=\infty$. Any such connection $\nabla$ experiences
the {Stokes phenomenon} at $z=\infty$. That is, it is impossible
to find a single, well-defined section $s$ that is asymptotically
small as $z\to\infty$. Instead, for this particular singularity, we
require four sections $\tilde s_1$, $\tilde s_2$, $\tilde s_3$ and $\tilde s_4$ that each become
asymptotically small as $z\to\infty$ along what is called a {Stokes
  ray}.
(In
Section~\ref{sec:WKB} we will write down explicit expressions for
the sections $\tilde s_i$ when $\nabla$ is an {oper} connection.)

To keep track of the angular information, we consider the
{blow-up} of the singularity at infinity and replace the puncture at
$z=\infty$ with a small circle $S^1$ bounding an infinitesimal disc $D_\infty$. We then mark
four points $\tilde z_i$ on this circle corresponding to the Stokes rays, and
to these points we assign the sections $\tilde s_i$ that become
asymptotically small along the corresponding rays. Each section $\tilde s_i$
has good asymptotics in the angular regions adjacent to the
Stokes ray labeled by $\tilde s_i$. On the overlap of the regions where two
sections are well-defined, they are related by a {Stokes
  matrix}~\cite{boalch2014geometry}.

To frame the connection $\nabla$ at infinity, we similarly replace the
puncture at $z=\infty$ by a small circle $S^1$ with four marked points
$z_i$ corresponding to the four incoming trajectories. The framing of
$\nabla$ then corresponds to a choice of $\nabla$-invariant sub-bundle
$\ell_i$ for every marked point $z_i$, again with the constraint that
if two marked points $z_i$ and $z_j$ are connected by a path which
does not cross any trajectories, and we parallel transport $\ell_i$
and $\ell_j$ to a common point $z$ on this path, then
$\ell_i(z)\neq\ell_j(z)$. More precisely, consider the network $\scrW(p_2,\vartheta)$ for either
$\vartheta=(\pi/2)^+$ or $\vartheta=(\pi/2)^-$; these are sometimes
called the two {resolutions} of the (Fenchel-Nielsen type) network
$\scrW$, and we denote them by $\scrW^\pm$ respectively.\footnote{It
  is not possible to abelianize precisely at the critical phase
  $\vartheta_{\rm c}=\pi/2$, we have to make a choice of
  resolution.} Then the choice of framing $\ell_1,\ell_2,\ell_3,\ell_4$ determines a
unique $\scrW^\pm$-abelianization of $\nabla$, whose basis of sections in each
cell is shown in Figure~\ref{AD2sections}, with $s_i\in\ell_i$. Note that the two bases on either side of a
trajectory are indeed related by a unipotent transformation of the
type \eqref{smaps}. On either side of
a branch cut they are related by a permutation matrix as in
\eqref{eq:sbranchmaps}. 
(To trivialize the covering $\Sigma\to C$, we choose not only the
branch cuts connecting the two branch points, but also an additional two
branch cuts at $z=\infty$.) The bases of sections for each network
$\scrW^\pm$, as illustrated in Figure~\ref{AD2sections}, brings the connection $\nabla$
in an almost-diagonal gauge, so that it may be lifted to a $GL(1,\C)$
connection $\nablaab_\pm$ on the spectral cover $\Sigma$ by
identifying the two elements $(s_i,s_j)$ of the basis in each cell
with the sheets of $\Sigma$.
\end{example}
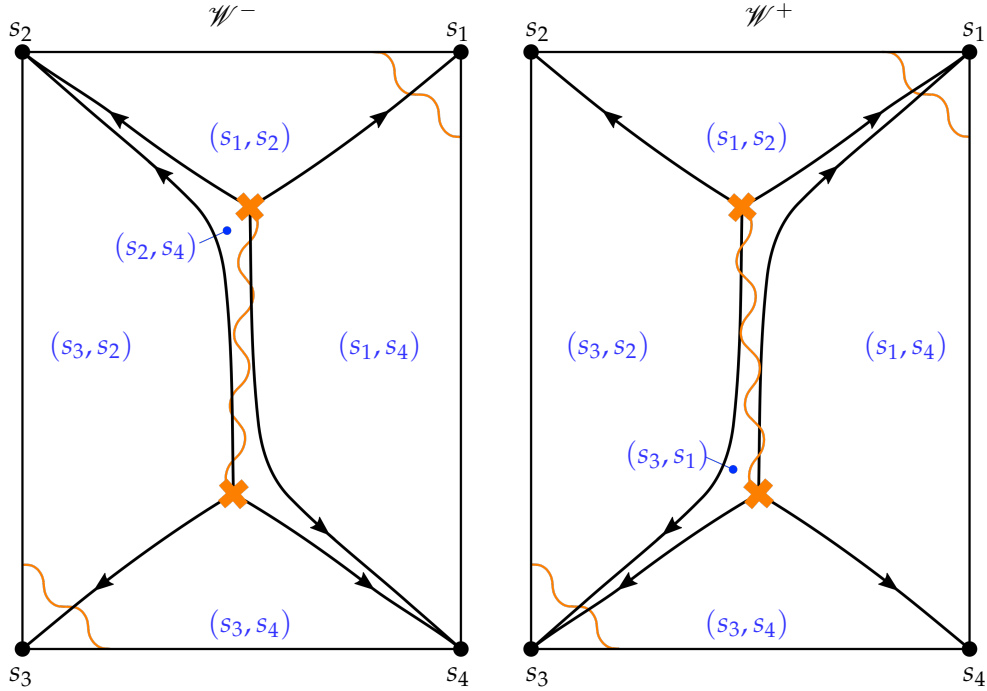
\begin{figure}[h!]
\centering
\small
\begin{overpic}
[width=0.80\textwidth]{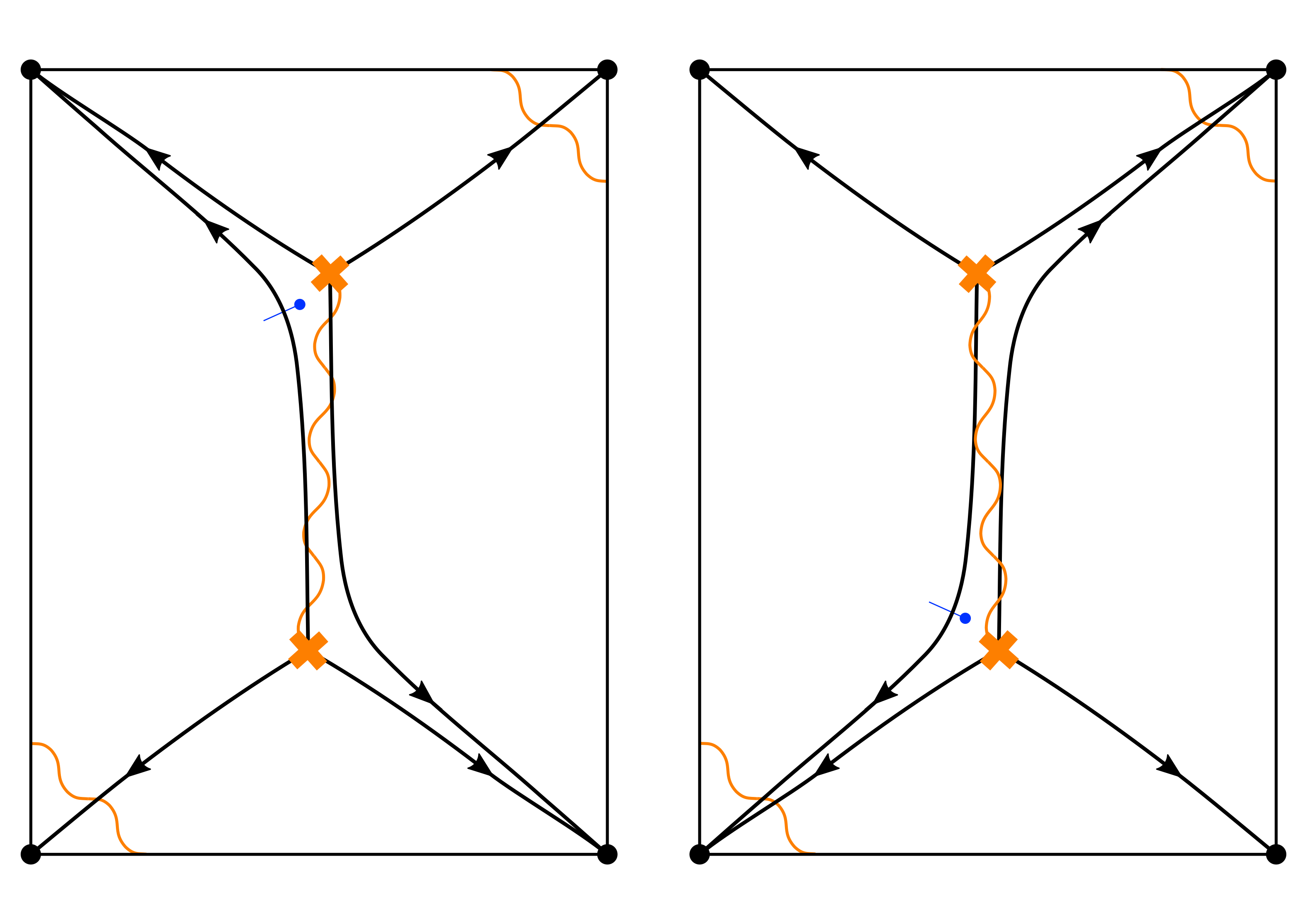}
\put(45,2){$s_4$}
\put(1,2){$s_3$}
\put(1,67){$s_2$}
\put(45,67){$s_1$}
\put(97,2){$s_4$}
\put(53,2){$s_3$}
\put(53,67){$s_2$}
\put(97,67){$s_1$}
\put(21,68){$\scrW^-$}
\put(75,68){$\scrW^+$}
\put(21,7){\textcolor{Blue}{$(s_3,s_4)$}}
\put(11.5,45){\textcolor{Blue}{$(s_2,s_4)$}}
\put(5,35){\textcolor{Blue}{$(s_3,s_2)$}}
\put(34,35){\textcolor{Blue}{$(s_1,s_4)$}}
\put(21,56){\textcolor{Blue}{$(s_1,s_2)$}}
\put(71,7){\textcolor{Blue}{$(s_3,s_4)$}}
\put(63,24){\textcolor{Blue}{$(s_3,s_1)$}}
\put(87,35){\textcolor{Blue}{$(s_1,s_4)$}}
\put(57,35){\textcolor{Blue}{$(s_3,s_2)$}}
\put(71,56){\textcolor{Blue}{$(s_1,s_2)$}}
\end{overpic}
\caption{\small Bases of sections $(s_i,s_j)$ in the cells of
  $C\setminus\scrW^{\pm}$ defining the abelianization of $\nabla$ for the two networks~$\scrW^{\pm}$.}
\label{AD2sections}
\normalsize
\end{figure}

\begin{example} \label{ex:pureab}
Let us now go back to Example~\ref{ex:purenetworks}, where $C=\C^\times$
and the spectral network $\scrW(p_2,\vartheta_{\rm c})$ is
defined by (see Figures~\ref{PureSU2Wjuggle} and~\ref{PureSU2resolutions})
\begin{align}
p_2(z) = \Big(\frac1{z^3}+\frac{9}{4\,z^2}+\frac1z\Big)\,\de
  z\otimes\de z \qquad \mbox{and} \qquad \vartheta_{\rm c}=\frac\pi2 \ .
\end{align}
Since $p_2$ has irregular singularities at $z=0$ and $z=\infty$,
we consider flat $SL(2,\C)$ connections $\nabla$ with a corresponding
irregular singularity at these points. Any such connection $\nabla$
may have a monodromy around the cylinder and is described by 
{Stokes theory} at $z=0$ and $z=\infty$. For this particular type of
singularity (the mildest case where $p_2$ has a pole of order
three), there is just one Stokes ray emitted from each puncture. 

To frame the connection $\nabla$ we choose $\nabla$-invariant line
bundles $\ell$ and $\ell'$ in the neighbourhood of $z=0$
and $z=\infty$, respectively, as well as an eigenline $\ell''$ of the
counterclockwise monodromy $\sf M$ of $\nabla$ around $z=0$. With this choice of framing data there is a
unique $\scrW^\pm$-abelianization of $\nabla$, for each of the two
resolutions $\scrW^\pm$ of $\scrW(p_2,\vartheta_{\rm c})$. The abelianization of $\nabla$ with respect to the resolution $\scrW^-$ is illustrated in
Figure~\ref{WminusSections}. 
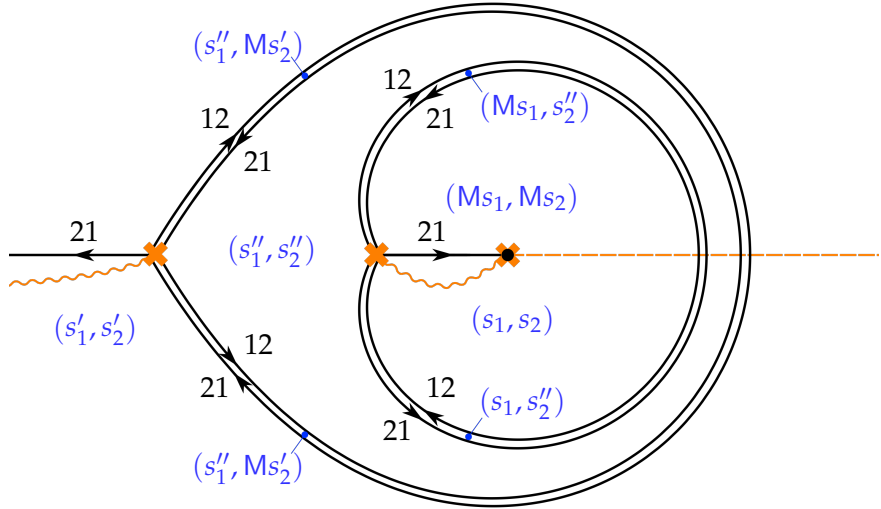
\begin{figure}[h!]
\centering
\small
\begin{overpic}
[width=0.70\textwidth]{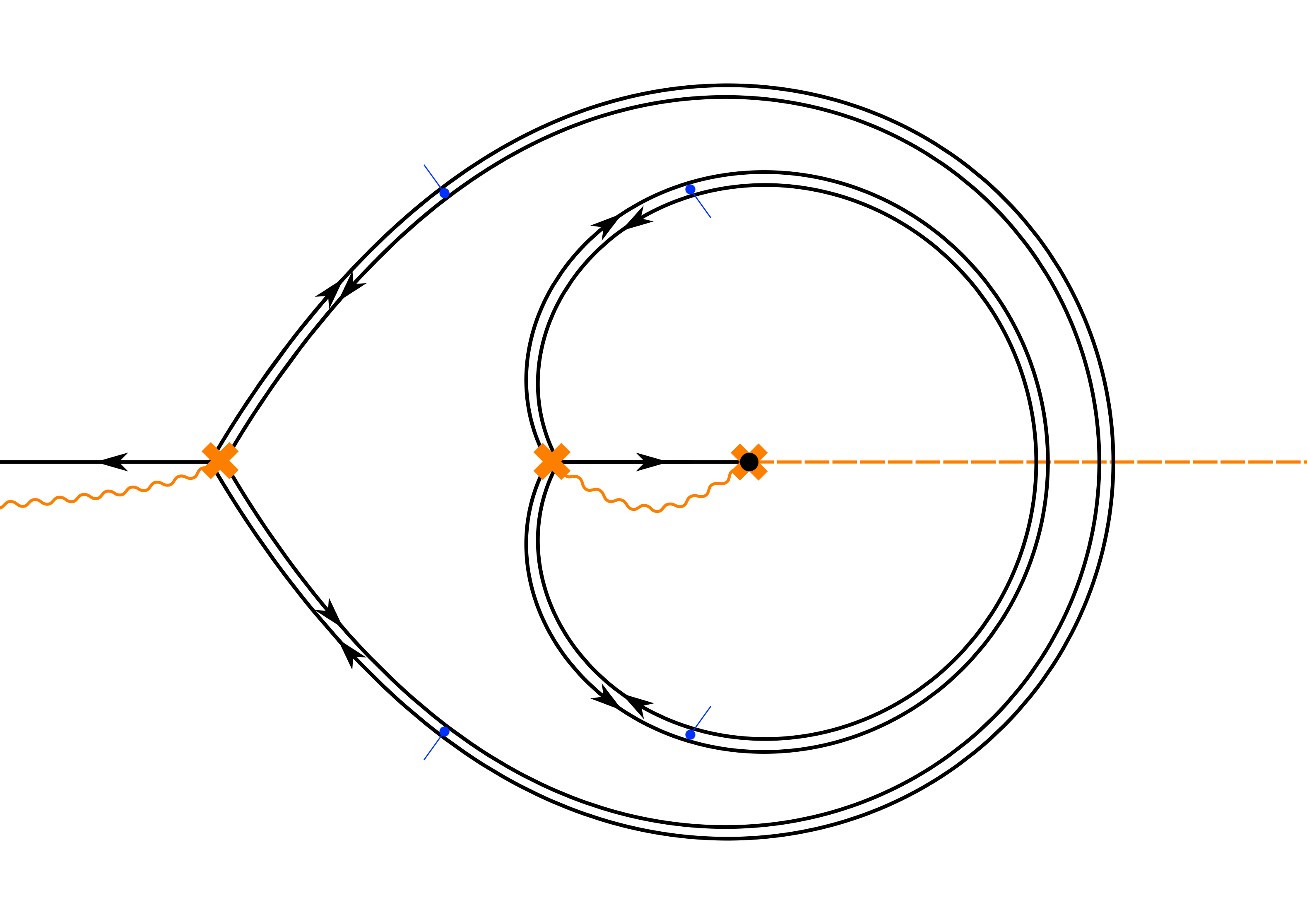}
\put(7,37){$21$}
\put(22,50){$12$}
\put(22,19){$21$}
\put(48,50){$21$}
\put(48,19){$12$}
\put(47,37){$21$}
\put(27,45){$21$}
\put(27,24){$12$}
\put(43,54.5){$12$}
\put(43,14){$21$}
\put(25,35){\textcolor{Blue}{$\left(s_1'',s_2''\right)$}}
\put(53,27){\textcolor{Blue}{$\left(s_1,s_2\right)$}}
\put(50,41){\textcolor{Blue}{$\left({\sf M}s_1,{\sf M}s_2\right)$}}
\put(5,26){\textcolor{Blue}{$\left(s_1',s_2'\right)$}}
\put(54,51.5){\textcolor{Blue}{$\left({\sf M}s_1,s_2''\right)$}}
\put(54,17.5){\textcolor{Blue}{$\left(s_1,s_2''\right)$}}
\put(21,58.5){\textcolor{Blue}{$\left(s_1'',{\sf M}s_2'\right)$}}
\put(21,10){\textcolor{Blue}{$\left(s_1'',{\sf M}s_2'\right)$}}
\end{overpic}
\caption{\small Bases of sections for the resolution $\scrW^-$. The dashed orange line is a monodromy cut for the monodromy around the cylinder. Note that ${\sf M}s_1=s_2$ and ${\sf M}s_1'=s_2'$.}
\label{WminusSections}
\normalsize
\end{figure}

This abelianization is characterized by the
bases of sections shown in Figure~\ref{WminusSections} (in agreement with~\cite[Section~4.3]{hollands2019exact}). In this figure $s_i$ and $s_i'$
are local bases of sections in the neighbourhood of the incoming trajectories
at $z=0$ and $z=\infty$, respectively, with $s_1\in\ell$ and
$s_1'\in\ell'$, and $s_i''$ are local sections in some
simply-connected domain of the intermediate annulus, with $s_2''\in\ell''$ (in the conventions of~\cite[Sections~5.2 and~5.3]{hollands2016spectral}). 

After possibly rescaling some of these sections, the change
of basis matrices can be brought in the desired triangular form. For instance, the matrix taking $(s_1,s_2)$ to $(s_1'',s_2'')$ has the form
\begin{align}
\bigg(\begin{matrix} * & * \\ 0 &
  * \end{matrix}\bigg)\,\bigg(\begin{matrix} * & 0 \\ * &
  * \end{matrix}\bigg)
\end{align}
                                                         relative to the basis $(s_1,s_2)$ in
                                                         resolution
                                                         $\scrW^+$, and
                                                         \begin{align}
                                                         \bigg(\begin{matrix}
                                                           * & 0 \\ * &
* \end{matrix}\bigg)\,\bigg(\begin{matrix} * & * \\ 0 &
* \end{matrix}\bigg)
\end{align}
in resolution $\scrW^-$.
\end{example}

It is not always that easy (or even possible) to find a suitable framing
on $\nabla$ which makes the abelianization process one-to-one. A nice
example of this is the circular higher Fenchel-Nielsen network of
Example~\ref{ex:higherFN}. Here it turns out that there are
generically 12 abelianizations which may be identified with the
singular fibers of an auxiliary rational elliptic
surface~\cite[Section~6]{hollands2019exact}.

\subsection{Spectral Coordinates}\label{sec:SpectralCoords}
\noindent
One of the applications of $\scrW$-abelianization is the
construction of a system of {Darboux coordinates} on the moduli
space $\scrM_{\rm flat}^{\scrW}(C,SL(K,\C))$ of $\scrW$-framed flat
$SL(K,\C)$ connections on
$C$. This moduli space is of importance for example in the study of
the {Hitchin system}, and it also shows up in a variety of
other problems in mathematical physics.

Given a basis of
one-cycles on $\Sigma$, we first define a coordinate system on the
moduli space $\scrM'_{\rm flat}(\Sigma,GL(1,\C))$ of equivariant
$GL(1,\C)$ connections on $\Sigma$ via the holonomies
\begin{equation}\label{chidef}
\cX_{\gamma}=\text{Hol}_{\gamma}\nablaab \ \in \ \mathbb{C}^{\times}
\ ,
\end{equation}
which depend only on the homology class $[\gamma]\in \Gamma = H_1(\overline{\Sigma},\Z)$
of a one-cycle $\gamma$. Through $\scrW$-abelianization they also give a system of
coordinates \smash{$\big\{\cX^\scrW_{\gamma}\}_{\gamma\in\Gamma}$} on $\scrM_{\rm flat}^{\scrW}(C,SL(K,\C))$, by first abelianizing $\nabla$ into $\nabla_\scrW^{\rm ab}$ and then
using (\ref{chidef}). We refer to these coordinates
 as {spectral coordinates}, since they are
defined in terms of {spectral data}, i.e. data on the spectral
curve~$\Sigma$. Note that this coordinate system depends \emph{only} on the
isotopy class of the network $\scrW$.

The moduli space $\scrM_{\rm flat}^{\scrW}(C,SL(K,\C))$ has a natural holomorphic
symplectic form given locally by the Atiyah-Bott-Goldman formula~\cite{Atiyah:1982fa,Goldman1984}
\begin{equation}\label{eq:symplecticform}
\Omega=\frac{1}{2}\,\int_C\, \Tr\big(\delta\cA\wedge\delta\cA\big) \ ,
\end{equation}
where $\cA$ is the connection one-form of a $\scrW$-framed flat $SL(K,\C)$ connection $\nabla$ on $C$.
Any spectral coordinate system $\{\cX_\gamma\}_{\gamma\in\Gamma}$ consists of (exponentiated) holomorphic Darboux coordinates with respect to this holomorphic symplectic form, as it brings the corresponding
holomorphic Poisson structure $\{\cdot,\cdot\}$ on $\scrM_{\rm flat}^{\scrW}(C,SL(K,\C))$ to the form
\begin{equation}
\{\cX_{\gamma},\cX_{\gamma'}\}=\langle\gamma,\gamma'\rangle\,
\cX_{\gamma}\,\cX_{\gamma'}=\langle\gamma,\gamma'\rangle\,
\cX_{\gamma+\gamma'} \ ,
\end{equation}
where $\langle\cdot,\cdot\rangle$ is the intersection pairing on $\Gamma$.
In examples, we may compute the coordinates $\cX_\gamma$ explicitly
through abelianization in terms of framing data~\cite{gaiotto2012spectral}.

\begin{example}
\label{eq:FGcoords}
Suppose $K=2$, and consider a Fock-Goncharov network $\scrW$. Given a
$\scrW$-framed $SL(2,\C)$ connection $\nabla$ and a
one-cycle $\gamma$ on $\Sigma$, we would like to compute the holonomy $\cX_\gamma$
of the corresponding $GL(1,\C)$ connection $\nablaab$
along $\gamma$. The local geometry of a generic one-cycle $\gamma$ in
a Fock-Goncharov network is illustrated in Figure~\ref{AbelCycle}.
\begin{figure}[h!]
\centering
\small
\begin{overpic}
[width=0.80\textwidth]{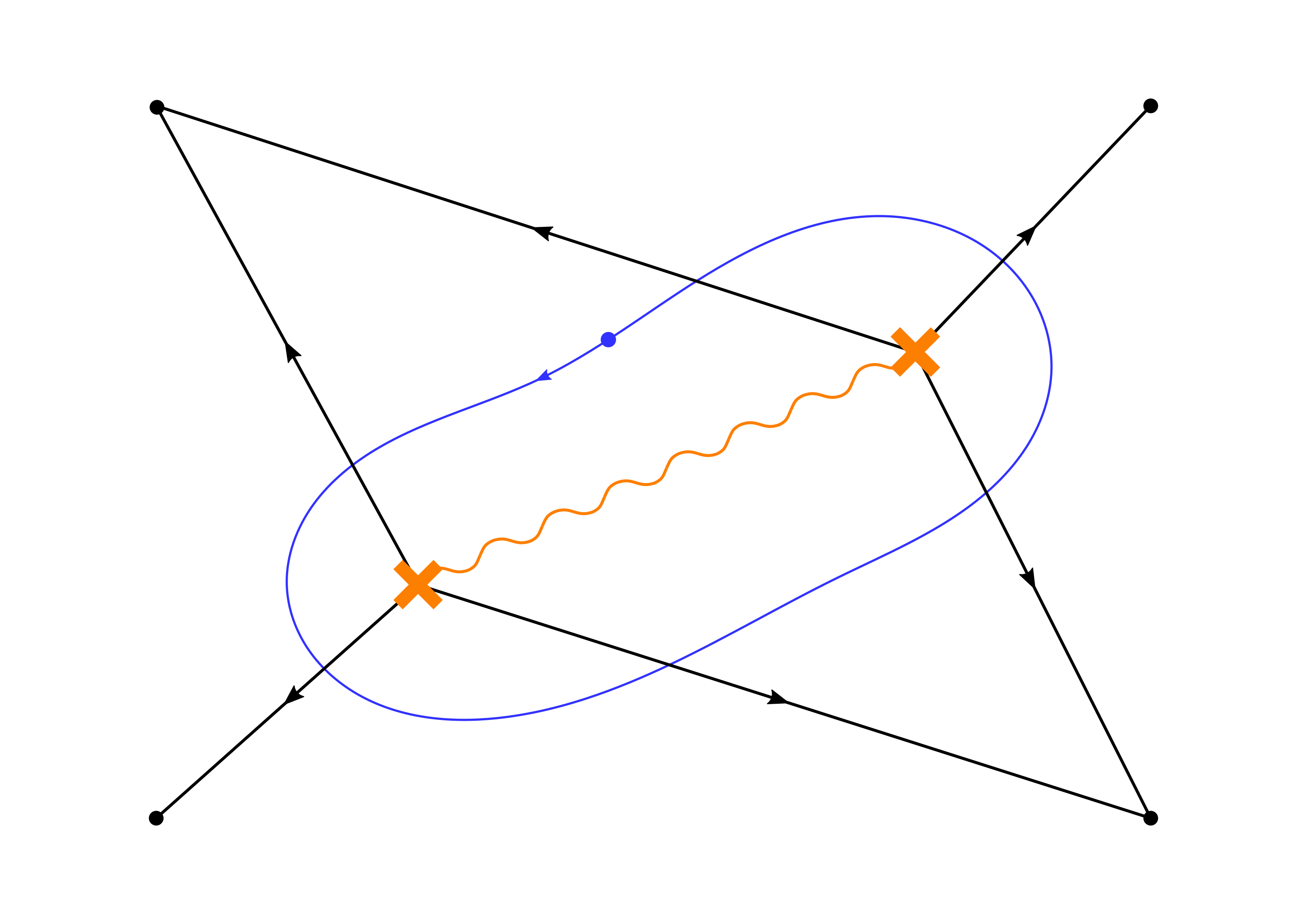}
\put(8,62){$z_l$}
\put(7,6){$z_m$}
\put(89,6){$z_n$}
\put(89,62){$z_k$}
\put(15,46){$21$}
\put(33,57){$21$}
\put(64,11){$21$}
\put(82,22){$21$}
\put(17,10){$12$}
\put(80,58){$12$}
\put(29,46){\textcolor{Blue}{$(s_l,s_n)$}}
\put(69,17){\textcolor{Blue}{$(s_n,s_l)$}}
\put(11,26){\textcolor{Blue}{$(s_l,s_m)$}}
\put(36,9){\textcolor{Blue}{$(s_n,s_m)$}}
\put(81,30){\textcolor{Blue}{$(s_n,s_k)$}}
\put(51,57){\textcolor{Blue}{$(s_l,s_k)$}}
\end{overpic}
\caption{\small A local Fock-Goncharov network with one-cycle $\gamma$ indicated in blue, and the data needed to compute its holonomy.}
\label{AbelCycle}
\normalsize
\end{figure}
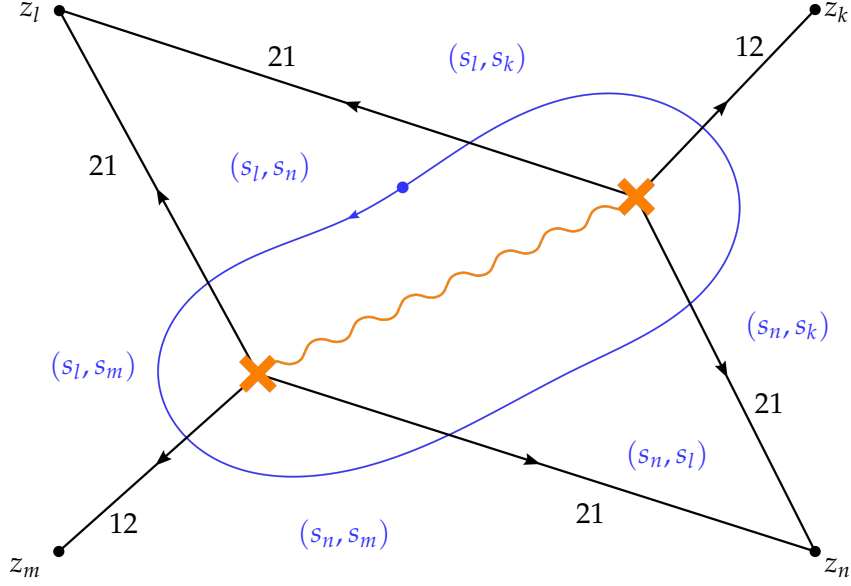

Before starting
the computation, let us recall that crossing a trajectory of type $12$ maps the basis from $(s_i,s_j)$ to 
\begin{equation}
s_j\longmapsto s_j \qquad \mbox{and} \qquad s_i\longmapsto
s_i+c_{ij}\, s_j \ .
\end{equation}
Suppose that the basis of sections on the other side of the trajectory
is $(s_k,s_j)$. Then $s_i'=s_i+c_{ij}\,s_j$ must be proportional to
the already chosen basis section $s_k$:
\begin{align}
s_i' = \lambda\, s_k \ .
\end{align}
Taking the exterior product on both sides with $s_j$, we find
\begin{align}
\lambda=\frac{s_i\wedge s_j}{s_k\wedge s_j} \ .
\end{align}
We may thus alternatively write the mapping of $s_i$ as
\begin{equation}\label{smap}
s_i\longmapsto s_i' = \frac{s_i\wedge s_j}{s_k\wedge s_j} \, s_k \ .
\end{equation}
In a similar manner, a trajectory of type $21$ leaves the first section in
the basis $(s_i,s_j)$ unchanged and maps the second section according
to (\ref{smap}). It is useful to note that the trajectory going into a
puncture $z_i$ leaves invariant the section $s_i$ associated to it.

Let us now begin the computation of the holonomy of $\nablaab$ along
$\gamma$. Start at the marked point on the first sheet with section
$s_l$. Note that parallel transport by $\nablaab$ along $\gamma$ will
multiply the section $s_l$ by this holonomy:
\begin{align}
s_l\longmapsto \cX_\gamma \, s_l \ .
\end{align}
We can compute $\cX_\gamma$ by following the path $\gamma$ and using
the rules \eqref{smap} when crossing a trajectory.
The first trajectory we cross goes into $z_l$ and thus does not change $s_l$. We then cross the trajectory going into $z_m$ and find
\begin{equation}
s_l\longmapsto \frac{s_l\wedge s_m}{s_n\wedge s_m} \, s_n \ .
\end{equation}
This is left unchanged as we cross the third and fourth trajectories, as both of these go into $z_n$. The next trajectory goes into $z_k$ and implies
\begin{equation}
s_n\longmapsto\frac{s_n\wedge s_k}{s_l\wedge s_k} \, s_l \ ,
\end{equation}
which is again left invariant when we cross the last trajectory and close the cycle.
Hence in the end, by following the path $\gamma$ we arrive at the
section $s_l$ multiplied by the cross-ratio
\begin{equation}
\cX_{\gamma}=\frac{s_n\wedge s_k}{s_l\wedge s_k} \, \frac{s_l\wedge
  s_m}{s_n\wedge s_m} \ .
\end{equation}
This is the spectral coordinate that we set out to
compute. It is indeed the familiar formula for a Fock-Goncharov
coordinate on the moduli space of framed flat $SL(2,\C)$-connections on $C$,
attached to the ideal triangulation $\scrT$ of $C$ dual to the
Fock-Goncharov network
$\scrW$~\cite{gaiotto2013wall,Gaiotto:2012db}. These coordinates and
their generalizations to higher rank are for instance useful for studying the quantization of the moduli spaces $\scrM_{\rm flat}^{\scrW}(C,SL(K,\C))$.
\end{example}

\begin{example}\label{ex:AD2spectralcoordinates}
  Going back to Example~\ref{ex:AD2abelianisation}, we first need to
  address a subtlety. Flat $SL(2,\C)$ connections on this geometry may
  experience a monodromy around the irregular singularity at infinity,
  which is usually fixed. This implies that the moduli space of such
  connections is actually zero-dimensional. However, we may consider a
  larger moduli space (of complex dimension two) by considering flat
  $SL(2,\C)$ connections with any monodromy around infinity, and where we
  instead fix the gauge at infinity. This latter requirement means
  choosing a trivialization of the bundle $E$ at infinity.

  This extended moduli space can be parameterized by two spectral
  coordinates $\cX_A$ and $\cX_B$ corresponding to the two one-cycles
  shown in Figure~\ref{AD2Cycles}. 
	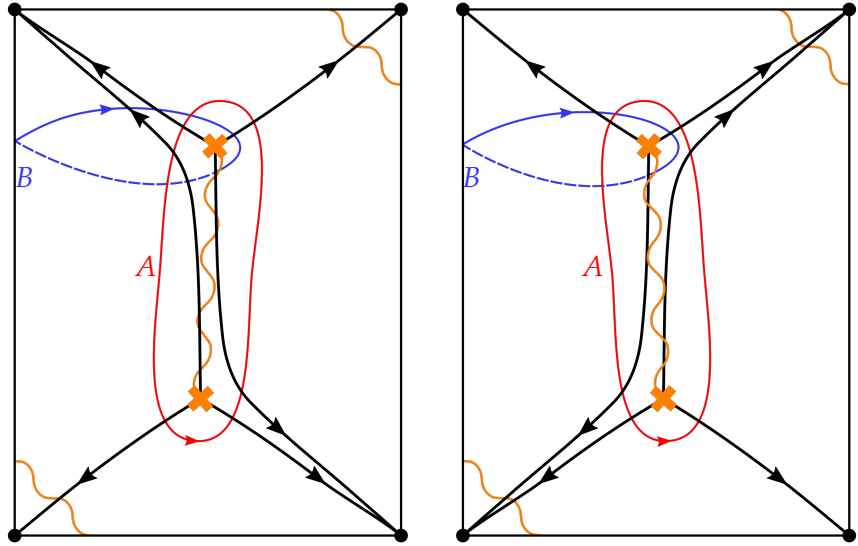
\begin{figure}[h!]
\centering
\small
\begin{overpic}
[width=0.80\textwidth]{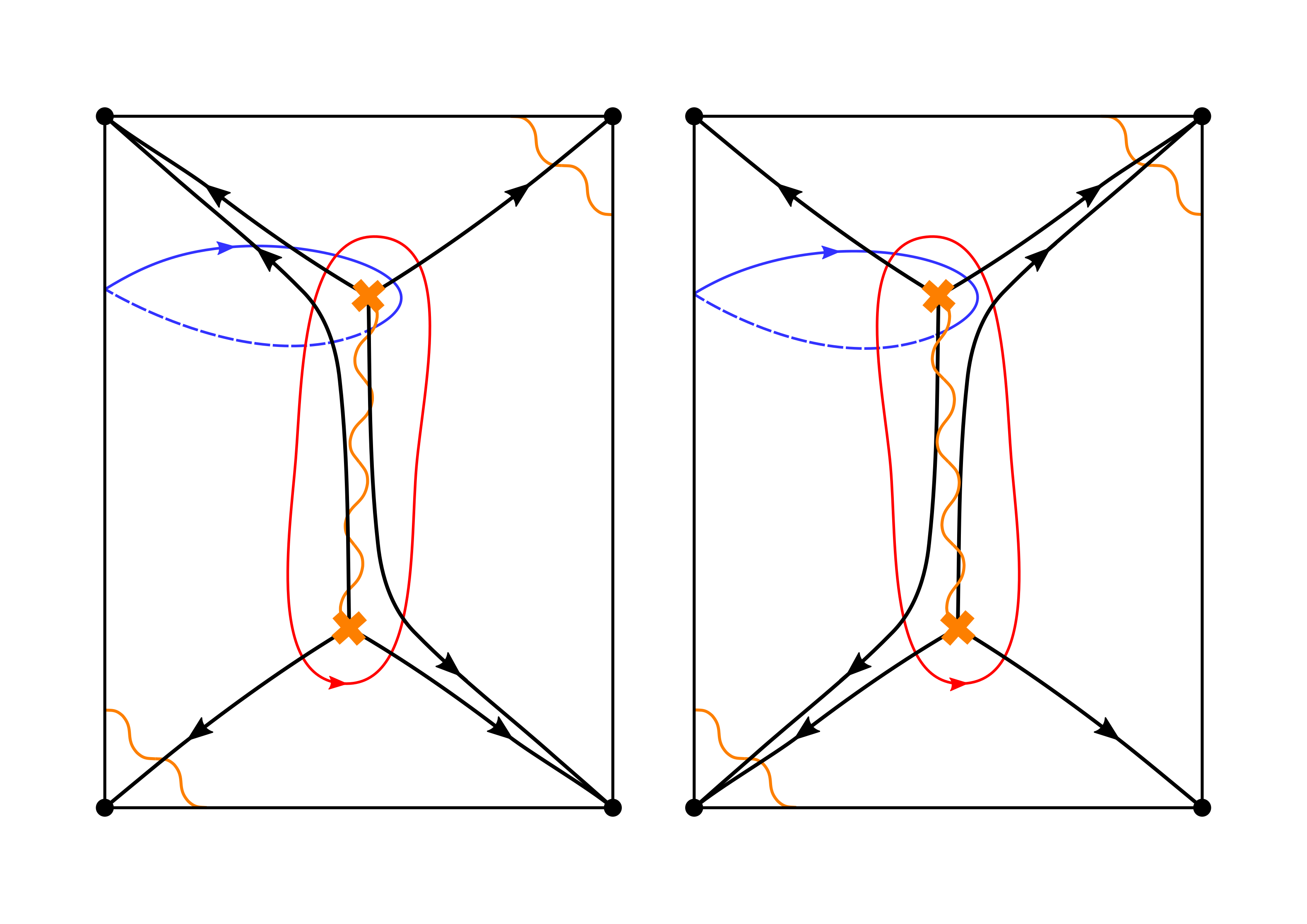}
\put(8,44){\textcolor{Blue}{$B$}}
\put(53,44){\textcolor{Blue}{$B$}}
\put(20,35){\textcolor{Red}{$A$}}
\put(65,35){\textcolor{Red}{$A$}}
\end{overpic}
\caption{\small Choice of $A$-cycle and $B$-cycle on $\Sigma$ for $\scrW^{\pm}$. The $B$-cycle is non-compact and requires the gauge at infinity to be fixed.}
\label{AD2Cycles}
\normalsize
\end{figure}
	Note that the $B$-cycle begins and ends on a
  different sheet of the spectral covering $\Sigma$, and $\cX_B$ is
  only a holonomy invariant because we fixed the gauge at
  infinity. The coordinates $\cX_A$ and $\cX_B$ can be computed
  similarly to Example~\ref{eq:FGcoords}. We find that $\cX_A$ is
  given by the cross-ratio
\begin{equation}
\cX_ A= \frac{s_1\wedge s_2}{s_3\wedge s_2} \, \frac{s_3\wedge s_4}{s_1\wedge s_4}  \ ,
\end{equation}
for both resolutions $\scrW^+$ and $\scrW^-$, while $\cX_B$ can be
brought in the form
\begin{align}
  \cX_B^+=\frac{s_1\wedge s_4}{s_2\wedge s_4} \qquad \mbox{and}
  \qquad \cX_B^-=-\frac{s_1\wedge s_3}{s_1\wedge s_2}
\end{align}
for the networks $\scrW^+$ and $\scrW^-$, respectively. The gauge
fixing at infinity means that the sections $s_3$ and $s_4$ are
completely fixed (a $GL(1,\C)$ gauge transformation would multiply
them by a constant), so that $\cX_B^\pm$ are indeed
invariants. It may seem like there are various other choices
  to be made for the $B$-cycle, but it actually turns out that all of
  them result in a spectral coordinate that is equal to $\cX_B$ up to
  multiplication by a power of $\cX_A$. This follows easily after
  realizing that the exterior products $s_1\wedge s_2$, $s_1\wedge s_4$, $s_3\wedge s_4$ and $s_3\wedge s_2$ are equal to
  one another, up to the monodromy $\cX_A$. This is because $\nabla$ is an $SL(2,\C)$ connection and
  hence the exterior products $s_i\wedge s_j$ are
  invariant after crossing a trajectory.\footnote{One should exercise
    caution here, since these relations only hold locally and do not
    account for possible monodromies in non-simply connected regions.}
  \end{example}

  \begin{example}\label{ex:FNpureab}
Consider the Fenchel-Nielsen network from Example~\ref{ex:pureab}, with the choice of $A$-cycle and $B$-cycle on $\Sigma$ as
shown in Figure~\ref{SU2CyclesV2}. 
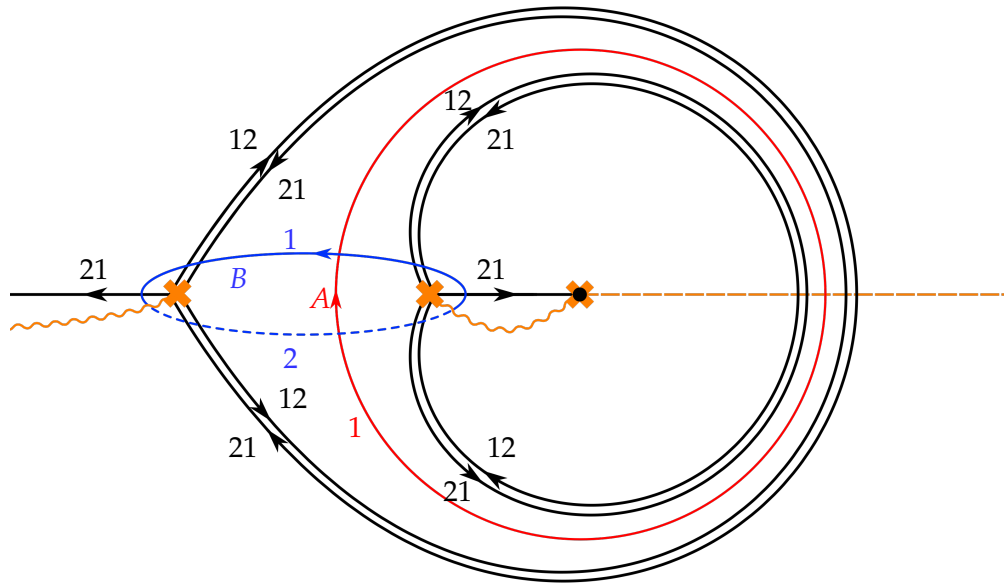
\begin{figure}[h!]
\centering
\small
\begin{overpic}
[width=0.80\textwidth]{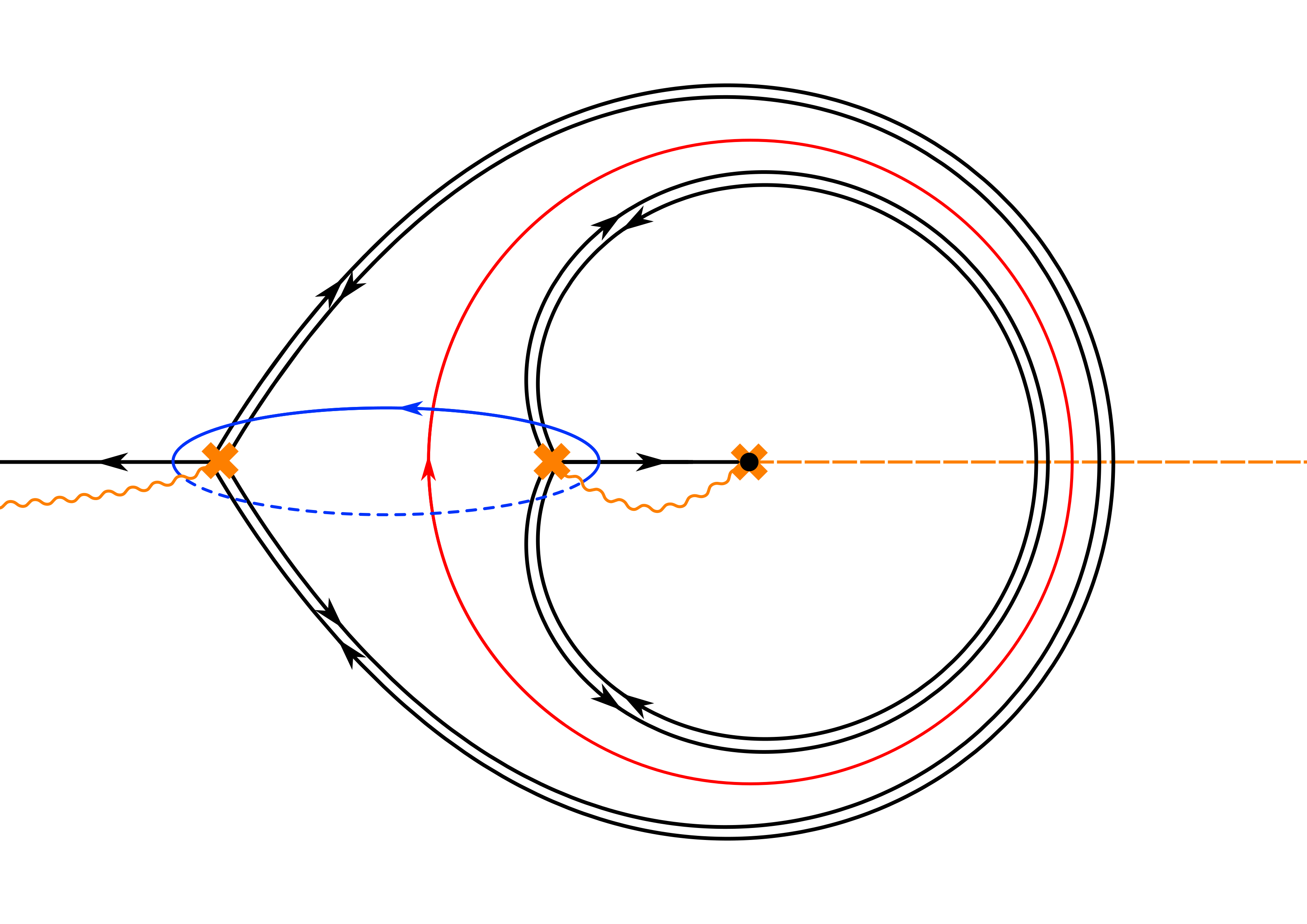}
\put(7,37){$21$}
\put(22,50){$12$}
\put(22,19){$21$}
\put(48,50){$21$}
\put(48,19){$12$}
\put(47,37){$21$}
\put(27,45){$21$}
\put(27,24){$12$}
\put(43.6,54){$12$}
\put(43.6,14.6){$21$}
\put(22,36){\textcolor{Blue}{$B$}}
\put(30,34){\textcolor{Red}{$A$}}
\put(27.5,40){\textcolor{Blue}{$1$}}
\put(27.5,28){\textcolor{Blue}{$2$}}
\put(34,21){\textcolor{Red}{$1$}}
\end{overpic}
\caption{\small Choice of $A$-cycle and $B$-cycle for $\scrW^-$.}
\label{SU2CyclesV2}
\normalsize
\end{figure}
The spectral coordinate $\cX_A$ is simply an eigenvalue of the monodromy
matrix $\mathsf{M}$ of $\nabla$, since the  one-cycle $A$ on $\Sigma$ does
not cross any trajectories. (More precisely, it is the eigenvalue
corresponding to the eigenvector $s_2''$.) The coordinate $\cX_B$, on
the other hand, is given by the cross-ratio
\begin{align}
\cX_B^- = \cX_A^2 \, \frac{(s_1\wedge s_2^{\prime\prime})^2\,(s_1'\wedge
  s_1^{\prime\prime})^2}{(s_1\wedge s_2)\,(s_1'\wedge
  s_2^{\prime})\,(s_1''\wedge s_2'')^2}
\end{align}
in resolution $\scrW^-$, and by
\begin{align}
\cX_B^+ = \cX_A^{-2} \, \frac{(s_1\wedge s_2)\,(s_1'\wedge s_2')\,(s_1''\wedge s_2'')^2}{(s_1\wedge s_1'')^2\,(s_1'\wedge s_2'')^2}
\end{align}
in resolution $\scrW^+$,
where the sections $s_i$, $s_i'$ and $s_i''$ are all defined in
Example~\ref{ex:pureab} (see Figure~\ref{WminusSections}). Similarly to the previous examples, this follows by computing the parallel transport of $\nablaab$ along the one-cycle $B$ on $\Sigma$ and using the
rules \eqref{smap} when crossing a trajectory. 

To compute $\cX_B^-$, consider the setup in Figure~\ref{WminusSections}, where we have introduced a mondromy cut (the dashed orange line) to take care of the monodromy that the local sections $s_i$, $s_i'$ and $s_i''$ experience under parallel transport around the $A$-cycle. Using the identities $\mathsf{M}s_1=s_2$, $\mathsf{M}s_1'=s_2'$, $\mathsf{M}s_1''=\mu^{-1} \, s_1''$ and $\mathsf{M}s_2''=\mu\,s_2''$ with $\mu=\cX_A$, as well as the fact that $\mathsf{M}s\wedge\mathsf{M}s'=s\wedge s'$, we indeed find
\begin{align}
\cX_B^- &= \frac{s_2'\wedge s_1''}{s_2''\wedge s_1''} \, \frac{s_2''\wedge s_1}{s_2\wedge s_1} \, \frac{s_2\wedge s_2''}{s_1''\wedge s_2''} \, \frac{s_1''\wedge\mathsf{M}s_2'}{s_2'\wedge\mathsf{M}s_2'} = \mu^2 \, \frac{(s_1\wedge s_2^{\prime\prime})^2\,(s_1'\wedge
  s_1^{\prime\prime})^2}{(s_1\wedge s_2)\,(s_1'\wedge
  s_2^{\prime})\,(s_1''\wedge s_2'')^2} \ .
\end{align}
An analogous computation of $\cX_B^+$ considers instead Figure~\ref{WminusSections} in the opposite resolution, that is, with the saddle trajectories interchanged, and correspondingly modified bases of sections ``in between'' the saddle trajectories. This yields
\begin{align}
\cX_B^+ &= \frac{s_2'\wedge s_1'}{s_2''\wedge s_1'} \, \frac{s_2''\wedge s_1''}{s_2\wedge s_1''} \, \frac{s_2\wedge\mathsf{M}s_2}{s_1''\wedge\mathsf{M}s_2} \, \frac{s_1''\wedge s_2''}{s_2'\wedge s_2''} = \mu^{-2} \, \frac{(s_1\wedge s_2)\,(s_1'\wedge s_2')\,(s_1''\wedge s_2'')^2}{(s_1\wedge s_1'')^2\,(s_1'\wedge s_2'')^2} \ .
\end{align}
Following the averaging prescription of~\cite{hollands2018higher,hollands2019exact}, we may thus associate the average $B$-cycle coordinate
\begin{align}
\cX_B = \sqrt{\cX_B^- \, \cX_B^+} = \frac{s_1'\wedge s_1''}{s_1\wedge s_1''} \, \frac{s_1\wedge s_2''}{s_1'\wedge s_2''}
\end{align}
to the Fenchel-Nielsen network $\scrW$ at the critical phase (in particular, this agrees with the computation in~\cite[Appendix~A]{hollands2019exact}).
\end{example}

\begin{example}\label{ex:FNcoordinates}
A construction analogous to Example~\ref{ex:FNpureab} for $K=2$ in the case of a Fenchel-Nielson
network on a surface $C$ with only regular punctures yields a
complexified version of the well-known Fenchel-Nielsen
coordinates. Indeed, one finds that there are two spectral coordinates
$\cX_A$ and $\cX_B$ associated to each pants cycle in the pants
decomposition of $C$, dual to the Fenchel-Nielsen network
$\scrW$. Here $\cX_A$ is an eigenvalue of the monodromy of $\nabla$
along the pants cycle, while $\cX_B$ is characterized by a certain
property under what is called the `twist flow' on the moduli space of
suitably framed flat $SL ( 2 ,\C)$ connections on $C$ (details are
found in~\cite[Section~8.4]{hollands2016spectral}). These are
exponentiated and complexified versions of the original length and
twist coordinates introduced in the context of hyperbolic geometry,
which for instance played an important role in Kontsevich's
proof~\cite{Kontsevich:1992ti} of Witten's conjecture on the
intersection numbers of the moduli space of
curves~\cite{Witten:1990hr}.

The spectral coordinates obtained in Example~\ref{ex:FNpureab} are
examples of Fenchel-Nielsen type coordinates on a Riemann surface with
irregular punctures. Higher rank generalizations of Fenchel-Nielsen length-twist 
coordinates have been obtained through abelianization
in~\cite{hollands2018higher},\footnote{Another
  proposal for $SL(3,\C)$ Darboux coordinates on the four-punctured
  sphere is
  found in~\cite{Jeong:2018qpc}.} while higher rank Fenchel-Nielsen type coordinates on the three-punctured sphere have been introduced and studied in~\cite{hollands2019exact} by abelianization with respect to the circular Fenchel-Nielsen network in Figure~\ref{E6Networks}.\footnote{Other accounts and examples of abelianization may be found in e.g.~\cite{Coman:2018uwk,Nikolaev:2019img,Yan:2020kkb}.}
\end{example}

\subsection{Nonabelianization}\label{sec:nonabelianization}
\noindent
It is an interesting exercise to express the monodromies of a flat
$SL(K,\C)$ connection $\nabla$ in terms of the data of the $GL(1,\C)$
connection $\nablaab$. This is called \emph{nonabelianization}~\cite{gaiotto2012spectral}. It enables one to find the monodromy representation of $\nabla$ in terms of spectral coordinates $\cX_\gamma$.

\begin{example} \label{ex:purenonab}
Returning to Example~\ref{ex:pureab}, let us describe the
corresponding nonabelianization. For this, we suppose instead that $\nablaab_\pm$
is the resulting $GL(1,\C)$ connection on $\Sigma$. Then we may
construct the $SL(2,\C)$ monodromies of $\nabla$ in terms of the
parallel transport of $\nablaab_\pm$ as follows. Consider the
{path groupoid} depicted in Figure~\ref{pathgroupoid}. 
\begin{figure}[h!]
\centering
\small
\begin{overpic}
[width=0.80\textwidth]{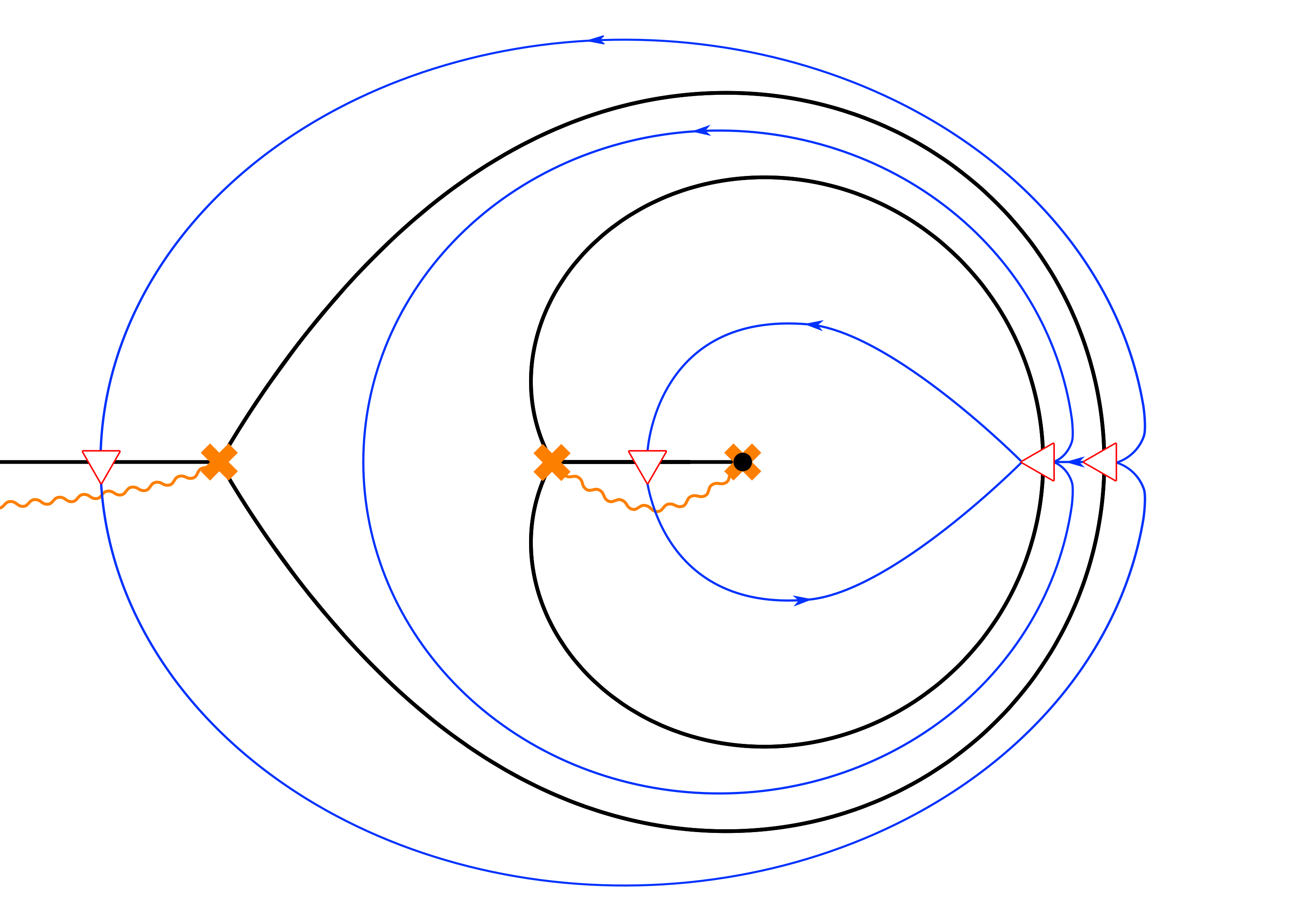}
\put(9,37){\textcolor{Red}{$\sf S_1$}}
\put(46,37){\textcolor{Red}{$\sf S_4$}}
\put(74,35){\textcolor{Red}{$\sf S_3$}}
\put(88,34){\textcolor{Red}{$\sf S_2$}}
\put(45,65){\textcolor{Blue}{$\varrho_1$}}
\put(45,5){\textcolor{Blue}{$\varrho_2$}}
\put(37,52){\textcolor{Blue}{$\varrho_6$}}
\put(59,47){\textcolor{Blue}{$\varrho_4$}}
\put(59,22){\textcolor{Blue}{$\varrho_5$}}
\put(81.9,33){\textcolor{Blue}{$\varrho_3$}}
\end{overpic}
\caption{\small A choice of paths $\varrho_i$ on $C=\mathbb{C}^\times$ for the network $\scrW$. Each path begins and ends at a trajectory of the network on an arrow which is labeled by a matrix ${\sf S}_k$.}
\label{pathgroupoid}
\normalsize
\end{figure}
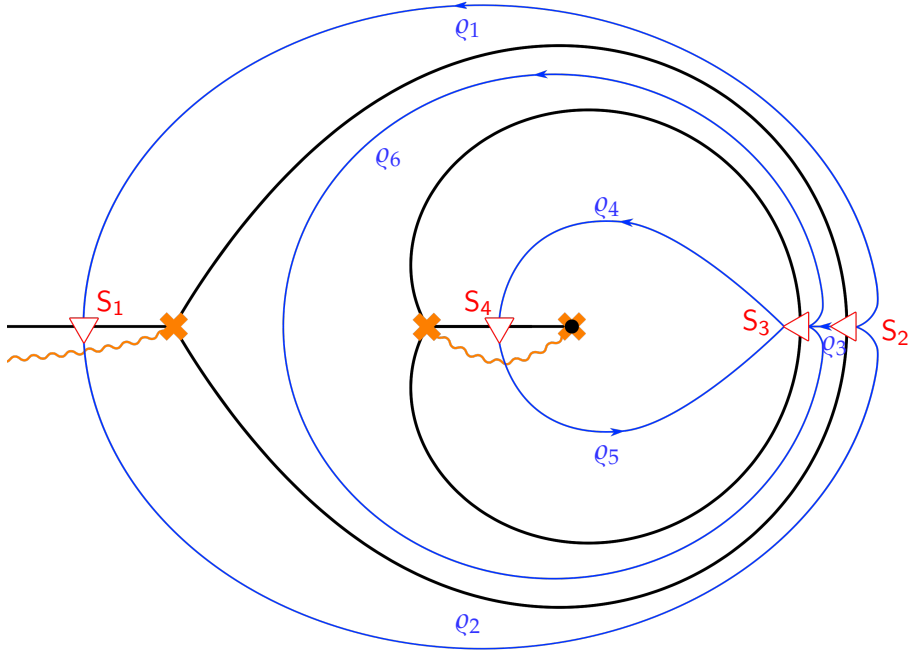
Associate a diagonal
$SL(2,\C)$ matrix ${\sf D}_i$ to each path $\varrho_i$ that does not cross a
branch cut, and an off-diagonal $SL(2,\C)$ matrix $\tilde{\sf D}_i$ if
the path $\varrho_i$ crosses a branch cut. We also associate a unipotent
matrix ${\sf S}_k$ with non-zero entry at position $ji$ to each
trajectory of type $ij$.

Let the entries of the matrices
\begin{align}
  {\sf D}_i=\bigg(\begin{matrix} d_i & 0 \\ 0 &
    d_i^{-1}\end{matrix}\bigg) \qquad \mbox{and} \qquad \tilde{\sf
                                                D}_i=\bigg(\begin{matrix}
                                                0 & -\tilde d_i \\
                                                \tilde d_i^{-1} &
                                                0 \end{matrix}\bigg)
\end{align}
encode the parallel transport of the $GL(1,\C)$ connection
$\nablaab_\pm$ along lifts of the paths $\varrho_i$ to $\Sigma$. That is, we
choose the non-zero elements $d_i$ and $\tilde d_i$ of the matrices
${\sf D}_i$ and $\tilde{\sf D}_i$ as
\begin{align}
d_1 = \frac{g_1}{g_2} \ , \quad \tilde d_2 = g_1\,g_2\, \cX_B \ ,
      \quad d_3 = \frac{g_3}{g_2} \ , \quad
  d_4 = \frac{g_4}{g_3} \ , \quad \tilde d_5 = g_3\,g_4 \ , \quad d_6
        = \frac1{\cX_A} \ ,
\end{align}
for some complex numbers $g_1$, $g_2$, $g_3$, $g_4$, where $\cX_A$
and $\cX_B$ are the spectral coordinates computed in Example~\ref{ex:FNpureab}. 

The unipotent transformations are of the form
\begin{align}
{\sf S}_1 &= \bigg(\begin{matrix} 1&f_1 \\ 0&1\end{matrix}\bigg) \qquad
                                           ,
                                             \qquad {\sf S}_2 =
                                             \bigg(\begin{matrix} 1&0
                                               \\
                                               f_2&1\end{matrix}\bigg)\,\bigg(\begin{matrix}
                                               1&f_3 \\
                                               0&1\end{matrix}\bigg) \
                                                  , \nonumber \\[4pt]
  {\sf S}_3&= \bigg(\begin{matrix} 1&0 \\
    f_4&1 \end{matrix}\bigg)\,\bigg(\begin{matrix} 1&f_5\\
    0&1\end{matrix}\bigg) \qquad , \qquad {\sf S}_4 = \bigg(\begin{matrix}
    1&f_6 \\ 0&1 \end{matrix}\bigg) \ , \label{eq:smiley1}
\end{align}
or
\begin{align}
{\sf S}_1 &= \bigg(\begin{matrix} 1&\tilde f_1 \\
  0&1\end{matrix}\bigg) \qquad ,
                                             \qquad {\sf S}_2 =
                                             \bigg(\begin{matrix}
                                               1&\tilde f_2 \\
                                               0&1\end{matrix}\bigg)\,\bigg(\begin{matrix} 1&0
                                               \\
                                               \tilde f_3&1\end{matrix}\bigg) \
                                                  , \nonumber \\[4pt]
  {\sf S}_3 &= \bigg(\begin{matrix} 1&\tilde f_4\\
    0&1\end{matrix}\bigg)\,\bigg(\begin{matrix} 1&0 \\
    \tilde f_5&1 \end{matrix}\bigg) \qquad , \qquad {\sf S}_4 = \bigg(\begin{matrix}
    1&\tilde f_6 \\ 0&1 \end{matrix}\bigg) \ , \label{eq:smiley2}
\end{align}
for the resolution $\scrW^+$ or $\scrW^-$, respectively. Solving for the
off-diagonal elements $f_k$ and $\tilde f_k$, by requiring that the
$SL(2,\C)$ monodromy around the branch points on $C$ is equal to~$\mathbb{1}$, yields
\begin{align}
f_1 &= -\frac{\cX_B}{\cX_A}\,\big(1+\cX_A^2\big)\,g_1^2 = \tilde f_1 \
      , \nonumber \\[4pt]
  f_2 &= \frac{\cX_A}{\cX_B\,g_2^2} \ \frac1{1-\cX_A^2} \ , \qquad
        \tilde f_2 = -\frac1{\cX_A\,\cX_B\,g_2^2} \ , \nonumber
  \\[4pt]
  f_3 &= -\frac{\cX_A}{\cX_B}\,g_2^2 \ , \qquad \tilde f_3 =
        \frac{\cX_A\,\cX_B\,g_2^2}{1-\cX_A^2} \ , \nonumber \\[4pt]
  f_4 &= \frac1{\cX_A\,g_3^2} \ , \qquad \tilde f_4 =
        \frac{g_3^2}{\cX_A} \ , \nonumber \\[4pt]
  f_5 &= -\frac{\cX_A\,g_3^2}{1-\cX_A^2} \ , \qquad \tilde f_5 =
        -\frac{\cX_A}{g_3^2\,(1-\cX_A^2)} \ , \nonumber \\[4pt]
  f_6 &= -\frac{g_4^2}{\cX_A}\,\big(1+\cX_A^2\big) = \tilde f_6 \ . 
\end{align}
Then the monodromies of the $SL(2,\C)$ connection $\nabla$ along any
path on $C$, written as a concatenation of the paths $\varrho_i$, are given
by multiplying the corresponding matrices ${\sf D}_i$, $\tilde{\sf
  D}_i$ and ${\sf S}_k$. Notice that although there are various
other parameters $g_i$ around (corresponding to an abelian gauge
choice), the monodromy invariants of $\nabla$ can be expressed
entirely in terms of the spectral coordinates $\cX_A$ and $\cX_B$.
  
A final point worth mentioning here is that the off-diagonal
elements $f_k$ of the Stokes matrices ${\sf S}_k$ have an
interpretation in terms of the $GL(1,\C)$ connection $\nablaab_\pm$ as
well: They represent the parallel transport of $\nablaab_\pm$ along
``detour paths'' on $\Sigma$ that follow the corresponding trajectory
back to the branch point they are emitted from. Multiple terms in
$f_k$ correspond to multiple trajectories in the network. For example,
each term in the power series expansion
\begin{align}
f_2 = \frac{\cX_A}{\cX_B\,g_2^2}\,\big(1+\cX_A^2+\cX_A^4+\cdots\big) 
\end{align}
corresponds to the parallel transport of $\nablaab_\pm$ along a detour
path which encircles the $A$-cycle on $\Sigma$ multiple times.
\end{example}

A more thorough description of nonabelianization (and its relation to
abelianization) can be found in~\cite{hollands2016spectral}, where nonabelianization is shown to be one-to-one for the Fock-Goncharov and Fenchel-Nielsen type networks in rank $K=2$. However, this is not always the case, see for instance the discussion in~\cite{hollands2019exact} on nonabelianization for the higher rank Fenchel-Nielsen networks on the three-punctured sphere, illustrated in Figure~\ref{E6Networks}. Nonabelianization is extended to other Lie groups in~\cite{Ionita:2021tqn}.

\clearpage
\newpage
\section{\tops{Theories of Class $\cS$ and their BPS States}{Theories of Class S and their BPS States}}\label{sec:classS}
\noindent
In this section we introduce class $\cS$ theories, which are a rich
subclass of $\N=2$ supersymmetric field theories in four
dimensions. We start with a brief introduction to 
four-dimensional $\N=2$ supersymmetric field theories. In
Section~\ref{sec:SWTheory} we go back to the seminal work of Seiberg
and Witten~\cite{Seiberg:1994rs,seiberg1994monopoles}, introducing
Seiberg-Witten geometry;\footnote{However, we do not review the most
physically interesting idea of Seiberg-Witten theory, which is the
relation to monopole condensation and confinement.} here we follow the
reviews~\cite{Lerche:1997sm,Dijkgraaf:1997ip}, see also~\cite{Martone:2020hvy} for a more recent account.\footnote{We further
  recommend the lectures of Witten for an introduction to
  Seiberg-Witten theory, which are available at {\tt
    https://www.youtube.com/watch?v=EC1SvnjYWsA} and {\tt
    https://www.youtube.com/watch?v=9nPU3WNhH-0}.} In
Section~\ref{ClassSTheories} we
specialize to a class of four-dimensional $\N=2$ theories 
known as class~$\cS$ theories, which was first
discovered by Gaiotto~\cite{Gaiotto:2009we}; for more extensive
reviews see
e.g.~\cite{tachikawa2013n,moore2012lecture,teschner2016new,Pestun:2016zxk}. Finally,
in Section~\ref{4dBPS} we make contact with the spectral networks
introduced in Section~\ref{WKBSNs}. We explain how the BPS spectra of
class $\cS$ theories may be obtained from spectral networks,
particularly for the example of Seiberg-Witten theory; more details about
this can be found in~\cite{gaiotto2012spectral}.

\subsection{Seiberg-Witten Geometry}\label{sec:SWTheory}
\noindent
A four-dimensional $\N=2$ theory is a supersymmetric field theory
on $\R^{3,1}$ which is invariant under two supercharges in the
fundamental representation of $Spin(3,1)\cong SL(2,\mathbb{C})$, or
equivalently under eight real supercharges. One usually writes them as
$Q^I_{\alpha}$ and $\bar{Q}^I_{\dot{\alpha}}$, where $I=1,2$ counts
the supercharges (they are called {R-symmetry} indices: the
$SU(2)_R\subset SL(2,\C)$ R-symmetry rotates them), and $\alpha,\dot{\alpha}=1,2$ are spinor indices.
There are two different particle multiplets in an $\N=2$ gauge
theory. The {vector multiplet} consists of a gauge field
$A_{\mu}$, two spinors $\lambda^I_\alpha$ and
$\bar\lambda_{\dot\alpha}^I$, and a complex scalar $\phi$. All of them
transform in the adjoint
representation of the gauge group $G$ (or more precisely the gauge algebra~$\fg$). The other multiplet is the
{hypermultiplet}. It consists of a pair of complex scalars $q^I$, and
two spinors $\psi_\alpha^I$ and $\bar\psi_{\dot\alpha}^I$. These may
transform in any representation of the gauge group $G$ (or more
precisely $\fg$).\footnote{More precisely, $q^1$ and $q^2$ live
in different representations of $G$ which are conjugate to each
other, and similarly for $\psi_\alpha^I$ and
$\bar\psi_{\dot\alpha}^I$.} The hypermultiplet representation can
sometimes be reduced to half of its degrees of freedom, which is then
called a {half-hypermultiplet}. 

We start by studying the pure $SU(2)$ gauge theory which just consists
of a single vector multiplet in the adjoint representation of $\fg=\mathfrak{s}\mathfrak{u}(2)$. The Lagrangian includes a potential term for the scalar field $\phi$ which is given by
\begin{equation}
V(\phi)=\Tr\big[\phi,\phi^{\dagger}\big]^2 \ .
\end{equation}
This potential vanishes if $\phi$ is restricted to the complexified
maximal torus of $SU(2)$, i.e.
\begin{equation}
\phi=a\,\sigma_3 \ ,
\end{equation}
where $a\in\C$ and $\sigma_3$ generates the maximal torus of
$SU(2)$. The potential $V(\phi)$ thus has flat
directions or {moduli}. Physically distinct vacua are parametrized by the gauge invariant Casimir element
\begin{equation}
u(a)=\Tr\phi^2 \ ,
\end{equation}
so that the classical {moduli space of vacua} is $\scrB_{\rm
  c}\cong\mathbb{P}^1$ (if we add a point at infinity). The space
$\scrB_{\rm c}$ is also called the (classical) {Coulomb
  branch}. Indeed, if $u\neq0$ then all fields will acquire masses by
the Higgs mechanism, except for the component of the vector multiplet
in the direction of $\phi$. After integrating out the massive fields
(after all, we are interested in the infrared physics), we are left
with an effective abelian theory. This is obviously not possible at
$u=0$ and $u=\infty$, and we therefore regard these two points as
singularities of~$\scrB_{\rm c}$.

The effective abelian theory is characterized by a {prepotential}
$\cF_0(a)$. The complexified gauge coupling
\begin{equation}
\tau(a)=\frac{\theta}{\pi}+\frac{8\pi \,\I}{g_{\textrm{\tiny YM}}^2}
\end{equation}
is given as its second derivative
\begin{equation}
\tau(a)=\frac{\partial^2\cF_0(a)}{\partial a^2} \ .
\end{equation}
Furthermore, the pair
\begin{equation}
\Omega = \bigg(\begin{matrix} a_{\textrm{\tiny D}} \\
  a \end{matrix}\bigg) \ ,
  \end{equation}
  where
\begin{equation}\label{aD}
a_{\textrm{\tiny D}}=\frac{\partial\cF_0(a)}{\partial a} \ ,
\end{equation}
is known to define a holomorphic section of an $Sp(2,\Z)\cong
SL(2,\Z)$ vector bundle over the Coulomb branch $\scrB_{\rm c}$.
Seiberg and Witten have shown us how to find an explicit description
of the quantum moduli space $\scrB=\scrB_{\rm q}$, i.e. an explicit description of this
$SL(2,\Z)$ bundle~\cite{Seiberg:1994rs}.

The starting point for the Seiberg-Witten solution is the behaviour of
$\cF_0(a)$ around $u=\infty$, where the leading contribution can be
computed in perturbation theory. The coupling constant has an
expansion
\begin{equation}
\tau(a) = \tau_0 + \frac{\I}{\pi}\,\log\Big(\frac a{\Lambda^2}\Big) +
\sum_{n=1}^\infty\, c_n \, \Big(\frac\Lambda a\Big)^{4n} \ ,
\end{equation}
where the first term $\tau_0$ is the classical contribution, the
second term is the one-loop contribution (this is the only
perturbative correction), and the remaining terms are instanton
corrections. The ultraviolet cutoff $\Lambda$ is exponentially small
from the classical point of view, so that quantum mechanically the
single classical singularity there splits in two. We will say more about this expansion in
Section~\ref{sec:NRS}.

What is important here is that the logarithmic contributions determine
the local monodromy of the $SL(2,\Z)$ bundle around $u=\infty$. If we
follow a loop around $u=\infty$ we find the transformation
\begin{equation}
\bigg(\begin{matrix} a_{\textrm{\tiny D}} \\ a \end{matrix}\bigg)
\longmapsto \bigg(\begin{matrix} -a_{\textrm{\tiny D}}+2\,a \\
  -a \end{matrix}\bigg) = {\sf M}_\infty \bigg(\begin{matrix}
  a_{\textrm{\tiny D}} \\ a \end{matrix} \bigg) \qquad \mbox{with}
\quad {\sf M}_\infty = \bigg(\begin{matrix} -1 & 2 \\ 0 &
  -1 \end{matrix}\bigg) \ .
\end{equation}
The simplest solution would be to extend the $SL(2,\Z)$ vector bundle
over $\mathbb{P}^1$ by having the opposite monodromy around the origin
$u=0$. This however turns out to violate the positivity constraint
${\rm Im}(\tau) = 8\pi /g_{\textrm{\tiny YM}}^2>0$. The
Seiberg-Witten solution assumes the minimal amount of a total of three
solutions, say at $u=\infty$ and $u=\pm\,\Lambda^2$, with monodromy
matrices generating a subgroup $\Gamma\subset
SL(2,\Z)$. The quantum moduli space is then given by $\scrB =
\mathbb{H}/\Gamma$, where $\mathbb{H}$ is the upper complex
half-plane.

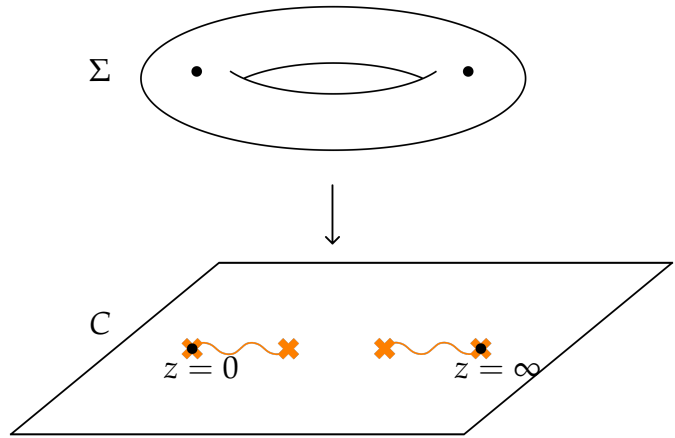
\begin{figure}[h!]
\centering
\begin{overpic}
[width=0.60\textwidth]{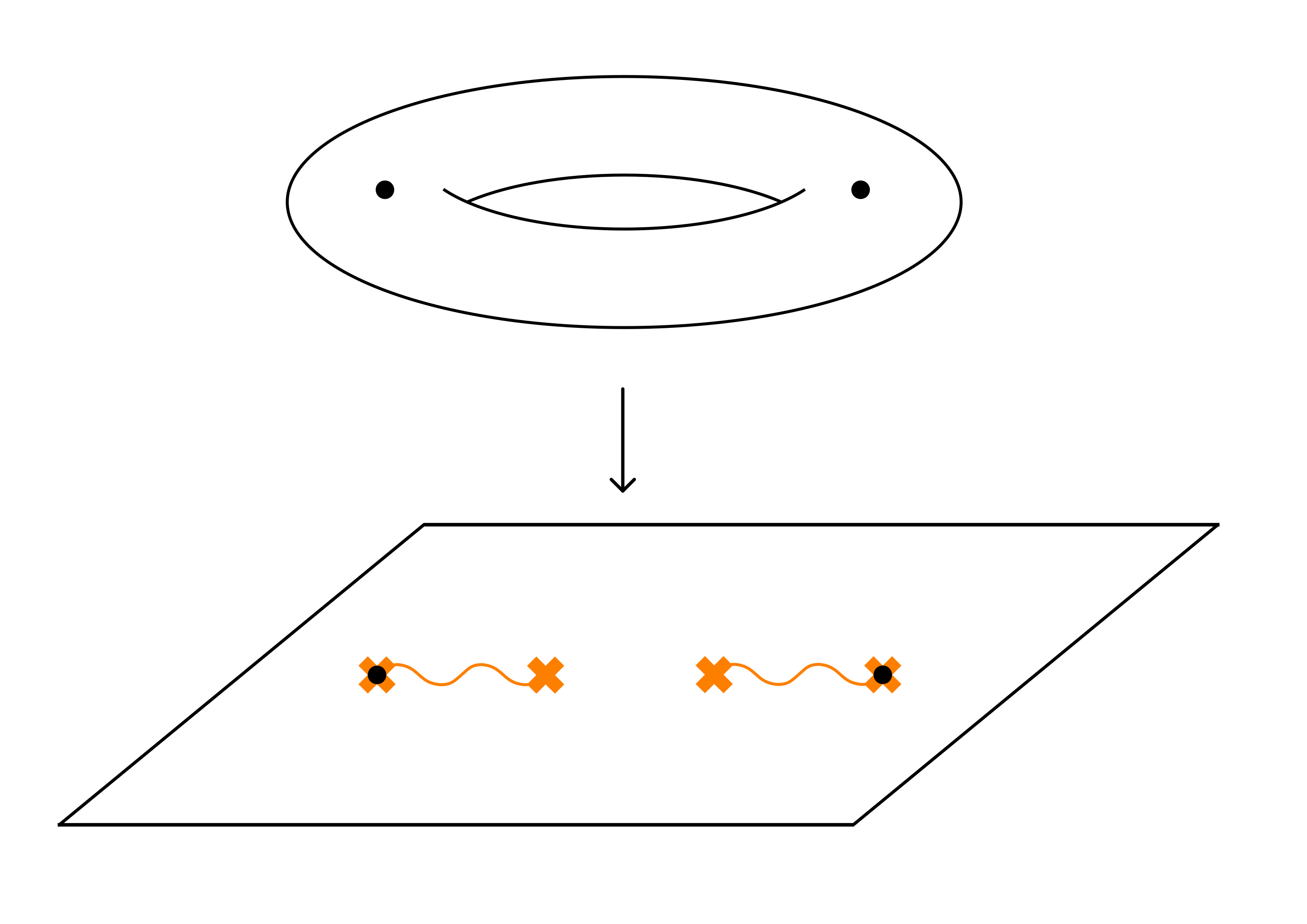}
\put(15,55){$\Sigma$}
\put(15,21){$C$}
\put(25,15){$z=0$}
\put(64,15){$z=\infty$}
\end{overpic}
\caption{\small Seiberg-Witten geometry for the pure $SU(2)$ theory. The ultraviolet curve $C$ is a sphere with irregular singularities (of order three) at $z=0$ and $z=\infty$, while the Seiberg-Witten curve $\Sigma$ is topologically a torus with two double punctures.}
\label{SU2SW}
\end{figure}

This moduli space also happens to parametrize elliptic
curves of the form
\begin{equation}\label{eq:pureSU2SWcurve}
\Sigma\colon \quad w^2= \frac{\Lambda^2}{z^3} - \frac{2\,u}{z^2} + \frac{\Lambda^2}z \ ,
\end{equation}
now known as {Seiberg-Witten curves},\footnote{This is a different formulation of $\Sigma$ compared to the original
formulation given by Seiberg and Witten. It is the more
recent class $\cS$ theory formulation~\cite{Gaiotto:2009we}, which is needed to apply
spectral network techniques.} where the first
homology of $\overline{\Sigma}$ may be identified with the charge lattice
$\Gamma$ of the $SL(2,\Z)$ bundle:
\begin{equation} 
\Gamma = H_1(\overline{\Sigma},\Z) \ .
\end{equation}
Note that we have already encountered the family of curves $\Sigma$ (with $\Lambda=1$ and $u=-\frac98$) in Section~\ref{WKBSNs}, see Examples~\ref{ex:purenetworks}, \ref{ex:pureab}, \ref{ex:FNpureab} and~\ref{ex:purenonab}, as a branched covering $\Sigma\to C$ defined by the quadratic differential $p_2(z)$ from \eqref{eq:pureSU2p2}.
From the Riemann-Hurwitz formula it follows that $\Sigma$ is topologically a torus with two double punctures, see Figure~\ref{SU2SW}.

\begin{figure}[h!]
\small
\centering
\begin{overpic}
[width=0.80\textwidth]{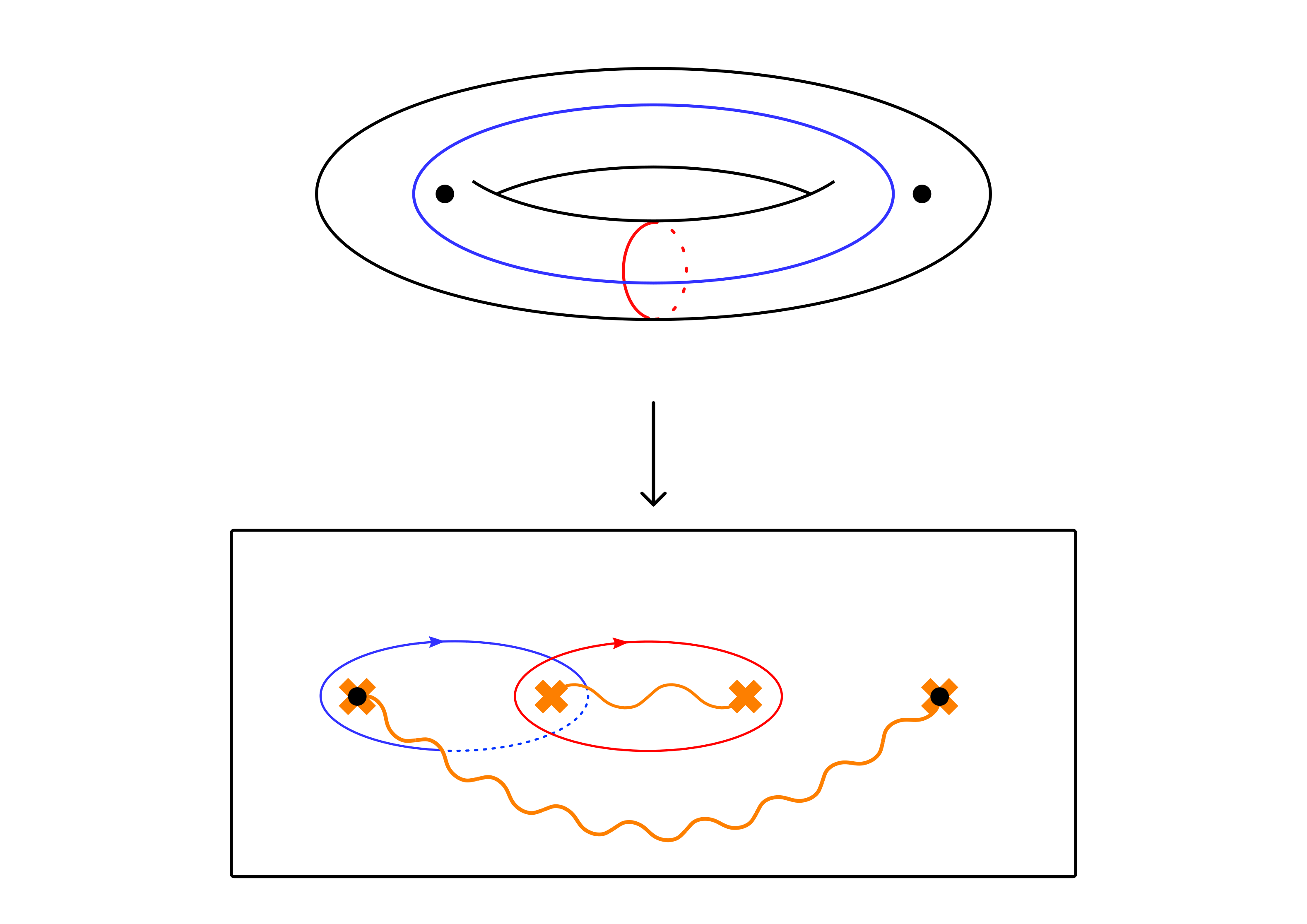}
\put(32,22.5){\textcolor{Blue}{$B$}}
\put(48,22.5){\textcolor{Red}{$A$}}
\put(45,51){\textcolor{Red}{$A$}}
\put(28.5,55){\textcolor{Blue}{$B$}}
\put(24,13){\textcolor{Orange}{$0$}}
\put(40,12){\textcolor{Orange}{$-\Lambda$}}
\put(55,12){\textcolor{Orange}{$+\Lambda$}}
\put(73,15){\textcolor{Orange}{$\infty$}}
\end{overpic}
\caption{\small Canonical basis $(A,B)$ of one-cycles on $\Sigma$
  viewed as a two-sheeted branched cover of a base curve $C$.}
\label{SymplecticBasis}
\normalsize
\end{figure}
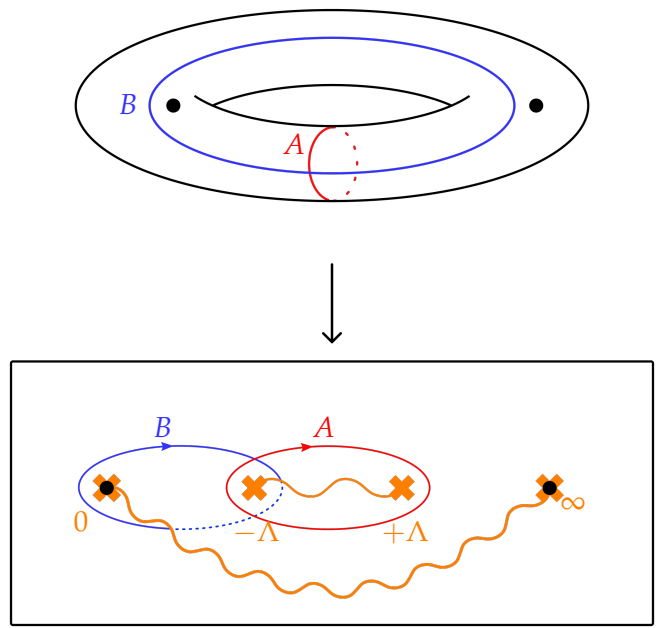

Seiberg and Witten discovered that this geometry is crucial for
understanding the infrared properties of the pure $SU(2)$ gauge
theory. Indeed, one can show that the moduli $a$ and $a_{\textrm{\tiny
    D}}$ are given by period integrals
\begin{equation}\label{eq:pureperiods}
a=\oint_{ A}\,\lambda \qquad \mbox{and} \qquad a_{\textrm{\tiny
    D}}=\oint_{ B}\,\lambda \ , 
\end{equation}
where $(A,B)$ is a canonical basis of $H_1(\Sigma,\Z)$ (see Figure~\ref{SymplecticBasis}) and $\lambda$
is the unique one-form on $\Sigma$ given by
\begin{equation}
\lambda = w \, \de z \ ,
\end{equation}
which is called the {Seiberg-Witten differential}.
The singularities at $u=\infty$ and $u=\pm\,\Lambda^2$ on $\scrB$ are then the points in the moduli space where the Seiberg-Witten
curve $\Sigma$ degenerates, and some combination of one-cycles
vanishes. As we will see later in Section~\ref{4dBPS}, one can read
off the BPS spectrum of the field theory from the Seiberg-Witten
curve. The singularities correspond to moduli where certain BPS states
become massless: $W^\pm$-bosons become massless at $u=\infty$
(restoring the $SU(2)$ symmetry), whereas certain dyons become
massless at $u=\pm\,\Lambda^2$.

The Seiberg-Witten solution was based on the premise that the abelian
gauge field was the only massless field in the theory. Thus we expect
the formalism to break down at $u=\infty$ and $u=\pm\,\Lambda^2$,
where there are additional massless BPS particles (that could not be
integrated out). Reinstating these massless states removes
these singularities. The singularities on $\scrB$ are
therefore not physical.

Following the works of Seiberg and Witten, similar techniques have
been used to find Seiberg-Witten curves for many other Lagrangian
$\N=2$ theories. More importantly, it was realized that Seiberg-Witten
curves are not just auxiliary mathematical objects which are helpful
in the study of $\N=2$ gauge theories, but they can actually be
identified as physical objects appearing in various string theory
constructions of $\N=2$ theories, see
e.g.~\cite{Klemm:1996bj,Banks:1996nj,Witten:1997sc}. Perhaps the most
insightful construction is obtained by considering M5-branes in the
M-theory background $\R^{3,1}\times X\times \R^3$, with an M5-brane
wrapping $\R^{3,1}\times\Sigma\times\{0\}$, where $\Sigma\subset
X$ is the Seiberg-Witten curve embedded in an ambient four-dimensional
space $X$ and $\{0\}\subset\R^3$ is the origin of $\R^3$. This realizes
the corresponding $\N=2$ theory as the worldvolume theory on the
M5-brane. 

\subsection{\tops{Class $\cS$ Geometry}{Class S Geometry}}\label{ClassSTheories}
\noindent
So far we have emphasized the feature that the low energy properties
of \emph{any} $\N=2$ field theory are encoded in a geometrical object,
the Seiberg-Witten curve $\Sigma$, together with the Seiberg-Witten
differential $\lambda$. From now on we will specialize to a subclass of
$\N=2$ theories for which the Seiberg-Witten curve $\Sigma$ comes with
some additional structure. Namely, $\Sigma$ is a (possibly branched)
cover of another (possibly punctured) Riemann surface $C$, where
$\Sigma\subset T^*C$ and $\lambda$ is the tautological one-form on the
cotangent bundle $T^*C$ restricted to $\Sigma$. In contrast to
$\Sigma$, the {ultraviolet curve} $C$, together with some
additional data at its punctures, encodes the microscopic
properties of the $\N=2$ field theory, and it defines the $\N=2$
theory uniquely. These $\N=2$ theories are
called {class $\cS$} theories~\cite{Gaiotto:2009we}. Note that
  this is the geometric setup of Section~\ref{WKBSNs}, where we
  considered $K$-fold coverings $\Sigma\to C$ defined by a collection of
  higher differentials on $C$. Here the integer $K-1$
  corresponds to the rank of the gauge algebra $\fg$ of the $\N=2$
  theory, so that $K=2$ sheeted coverings describe $SU(2)$ gauge
  theories. 

The complex structure moduli of $C$ correspond to complexified gauge
couplings of the $\N=2$ theory, which is often a quiver gauge theory
consisting of various gauge theories coupled to one another using
matter fields. The
additional data needed at the higher rank punctures is determined by
the type of singularities of the differentials. Regular singularities
are characterized by Lie
algebra homomorphisms $\varrho_i:\mathfrak{su}(2)\to\fg$. Roughly
speaking, they encode
the {flavour symmetries} of the $\N=2$ theory, which are the global
symmetries acting on the matter fields of the $\N=2$ theory, with
group $G^{\varrho_i}$ given by the 
centralizer of $\varrho_i$ in the gauge group $G$. We will take the
gauge algebra to be $\fg=\mathfrak{su}(K)$, and we
denote the corresponding $\N=2$ theory of class $\cS$ by ${\sf T}_K[C,\cD]$, where
$\cD$ represents the additional data at the punctures,
which in this case correspond to Young diagrams with $K$
boxes. The theory ${\sf T}_K[C,\cD]$ may be realized physically as the
worldvolume theory of $K$ M5-branes wrapped on $\R^{3,1}\times
C\times\{0\}$ in the M-theory background $\R^{3,1}\times
T^*C\times\R^3$, or equivalently as a (twisted) compactification of
the six-dimensional $(2,0)$-theory $\fX[K]$ on the Riemann surface
$C$. 

\subsubsection*{Class $\cS$ Theories of Type $\boldsymbol{SU(2)}$}
\begin{figure}[h!]
\centering
\begin{overpic}
[width=0.80\textwidth]{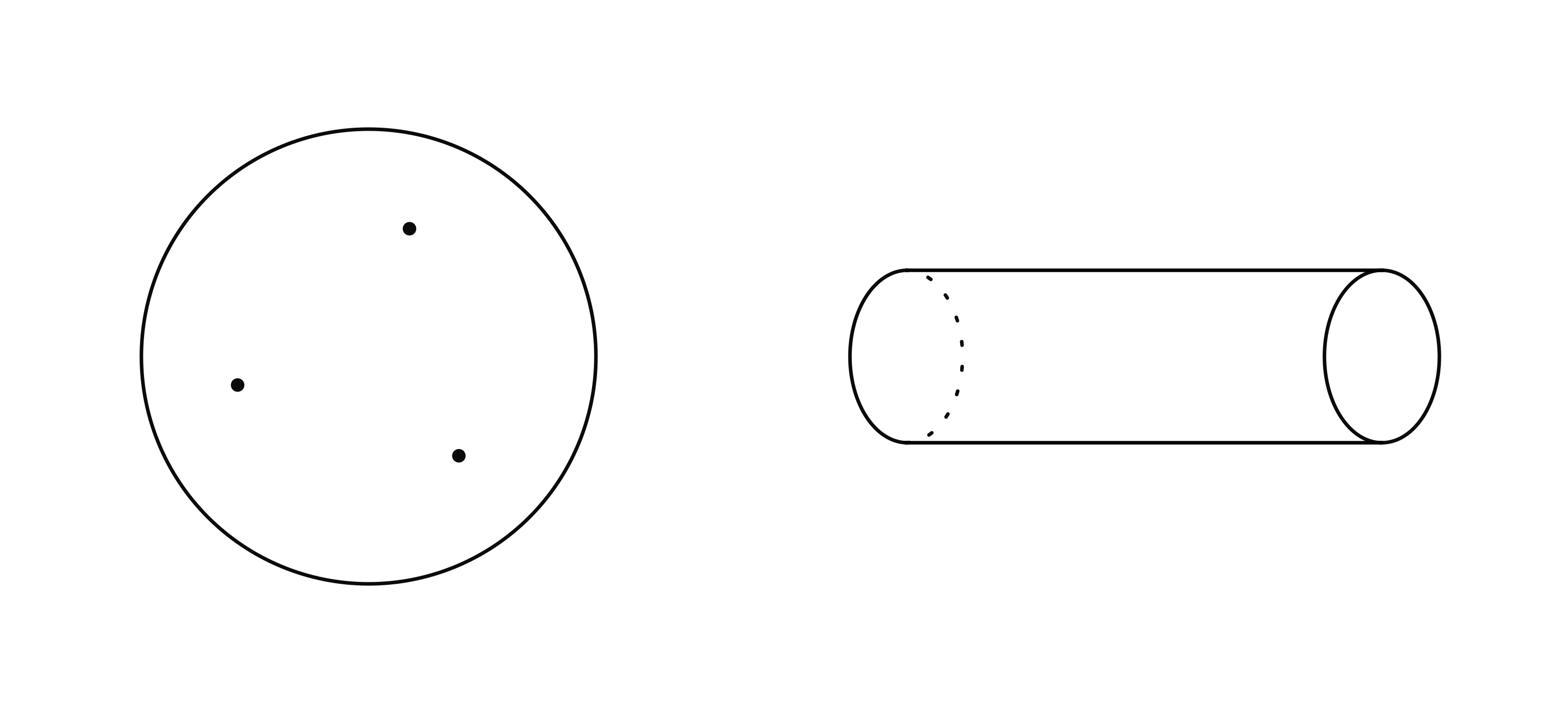}
\end{overpic}
\caption{\small The three punctured sphere and the cylinder which constitute the building blocks for $SU(2)$ class $\cS$ theories.}
\label{bblockssutwo}
\end{figure}

\noindent
Conformal $SU(2)$ theories of class $\cS$ can be constructed out of
just two elementary building blocks (see Figure~\ref{bblockssutwo}):
\begin{itemize}
\item the three-punctured sphere with an associated $SU(2)$ flavour
  symmetry group at each puncture (this is only possible for $K=2$ as
  there is only a single Young diagram with two boxes at each puncture);
  and
\item the cylinder with complex structure parameter
  \begin{equation}
\tau_{\textrm{\tiny
    UV}}=\frac{\theta}{2\pi}+\I\,\frac{4\pi}{g_{\textrm{\tiny YM}}^2}
\ ,
\end{equation}
where $g_{\textrm{\tiny YM}}$ is real and $\theta$ is periodic with
period $2\pi$.
\end{itemize}
The cylinder corresponds to an $\N=2$ vector multiplet with gauge
group $SU(2)$ and complexified gauge coupling $\tau_{\textrm{\tiny
    UV}}$, while the three-punctured sphere corresponds to a
half-hypermultiplet in the trifundamental representation of
$SU(2)_a\times SU(2)_b\times SU(2)_c$ (see
\cite{Hollands:2011zc,tachikawa2015review} for further details about 
this half-hypermultiplet as well as a Lagrangian description).
All theories that are constructed using these building blocks have a
Lagrangian description and may be checked to be conformal (if all
masses are set to zero). This means in particular that the
complexified coupling $\tau_{\textrm{\tiny
    UV}}$ is a dimensionless parameter. (However, in contrast to
$\N=4$ supersymmetric Yang-Mills theory we should differentiate here
between the (exactly marginal) microscopic coupling $\tau_{\textrm{\tiny
    UV}}$ and the infrared coupling $\tau$.)

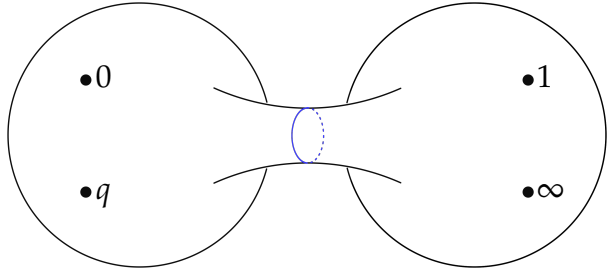
\begin{figure}[h!]
\centering
\begin{overpic}
[width=0.50\textwidth]{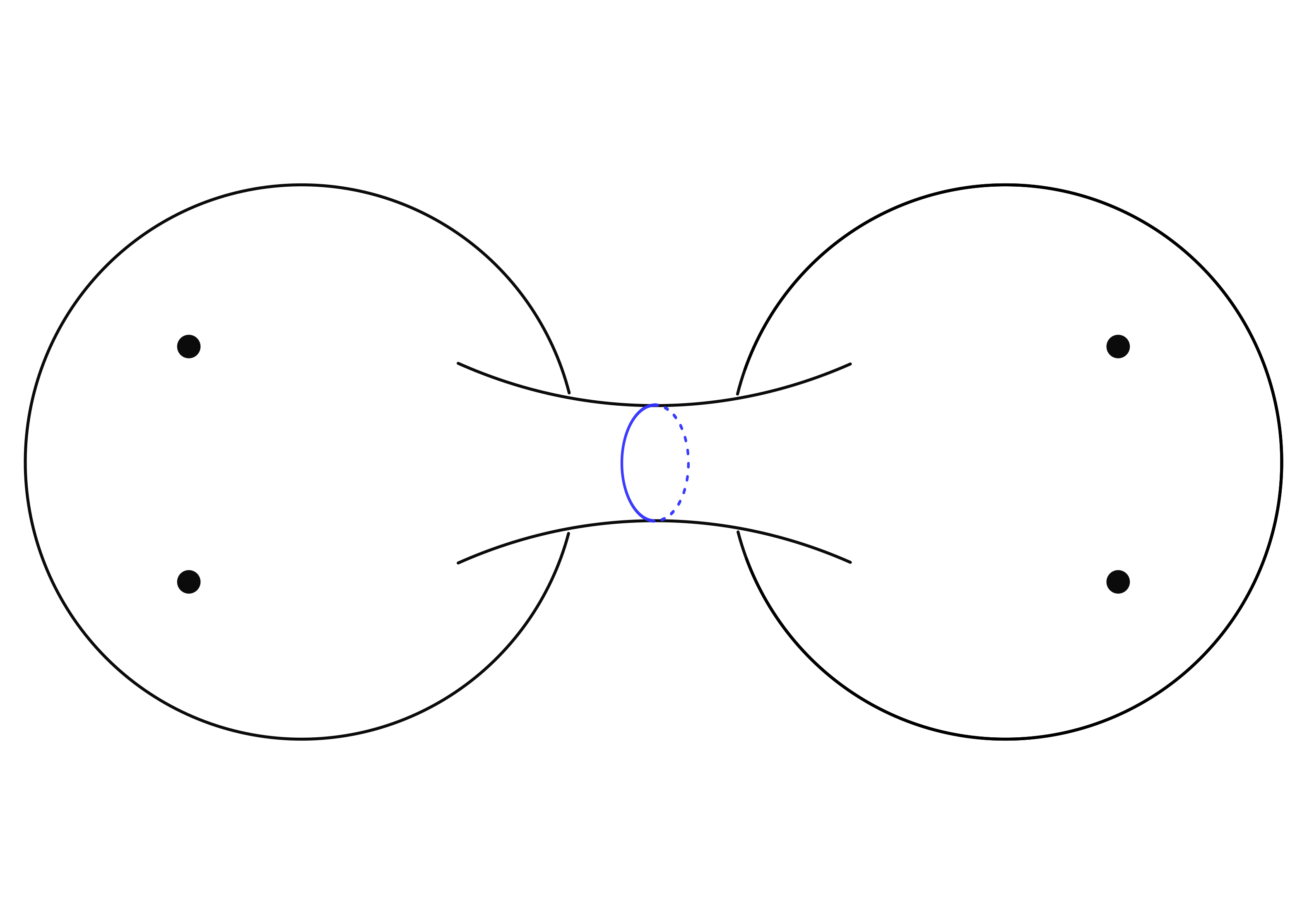}
\put(16,43){$0$}
\put(16,25){$q$}
\put(87,43){$1$}
\put(87,25){$\infty$}
\end{overpic}
\caption{\small Plumbing fixture construction of the four-punctured sphere.}
\label{SpherePoP}
\end{figure}

For example, we may construct the $SU(2)$ gauge theory coupled to four
additional fundamental hypermultiplets by gluing two three-punctured spheres using
the plumbing fixture method. For this, pick a puncture on each sphere,
with the first puncture at $z=0$ in a local coordinate $z$ and the
second puncture at $w=0$ in a local coordinate $w$. Then make the
identification $z\,w = q$, where
\begin{align}
q=\E^{\,2\pi
  \,\I\,\tau_{\textrm{\tiny UV}}}
\end{align}
is the exponentiated gauge
coupling (see Figure~\ref{SpherePoP}).
Indeed, on the gauge theory side this means
coupling the two corresponding $SU(2)$ flavour symmetries to a
dynamical $SU(2)$ gauge field with complexified coupling
$\tau_{\textrm{\tiny UV}}$. The four leftover $SU(2)$ flavour
symmetries combine to give the enhanced $SO(8)$ flavour symmetry group
of the $SU(2)$ gauge theory coupled to four hypermultiplets. Note
that, in general, the Lagrangian for a theory with four
hypermultiplets in a complex representation of the gauge group is
invariant under an $SU(4)$ flavour symmetry rotating the four
hypermultiplets. However, the case of $SU(2)$ gauge group is special,
since its fundamental representation is pseudoreal. This implies that
the flavour symmetry is in fact enhanced to an $SO(8)$ flavour
symmetry which rotates the eight half-hypermultiplets. We
can embed the flavour symmetries $SU(2)_i$ acting on the individual
hypermultiplets into this flavour symmetry group:
\begin{equation}
SO(8)\supset SU(2)_a\times SU(2)_b\times SU(2)_c\times SU(2)_d \ .
\end{equation}

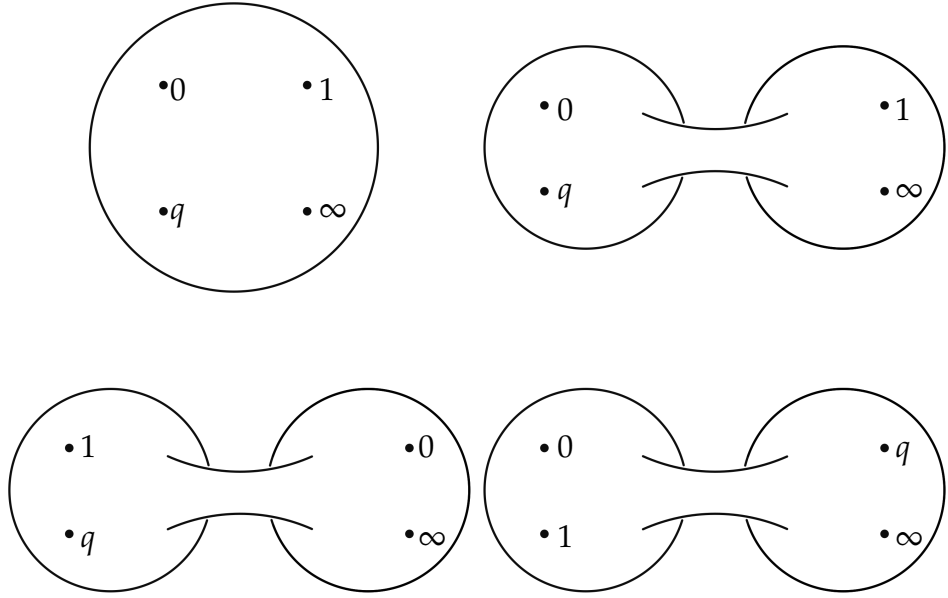
\begin{figure}[h!]
\centering
\begin{overpic}
[width=0.80\textwidth]{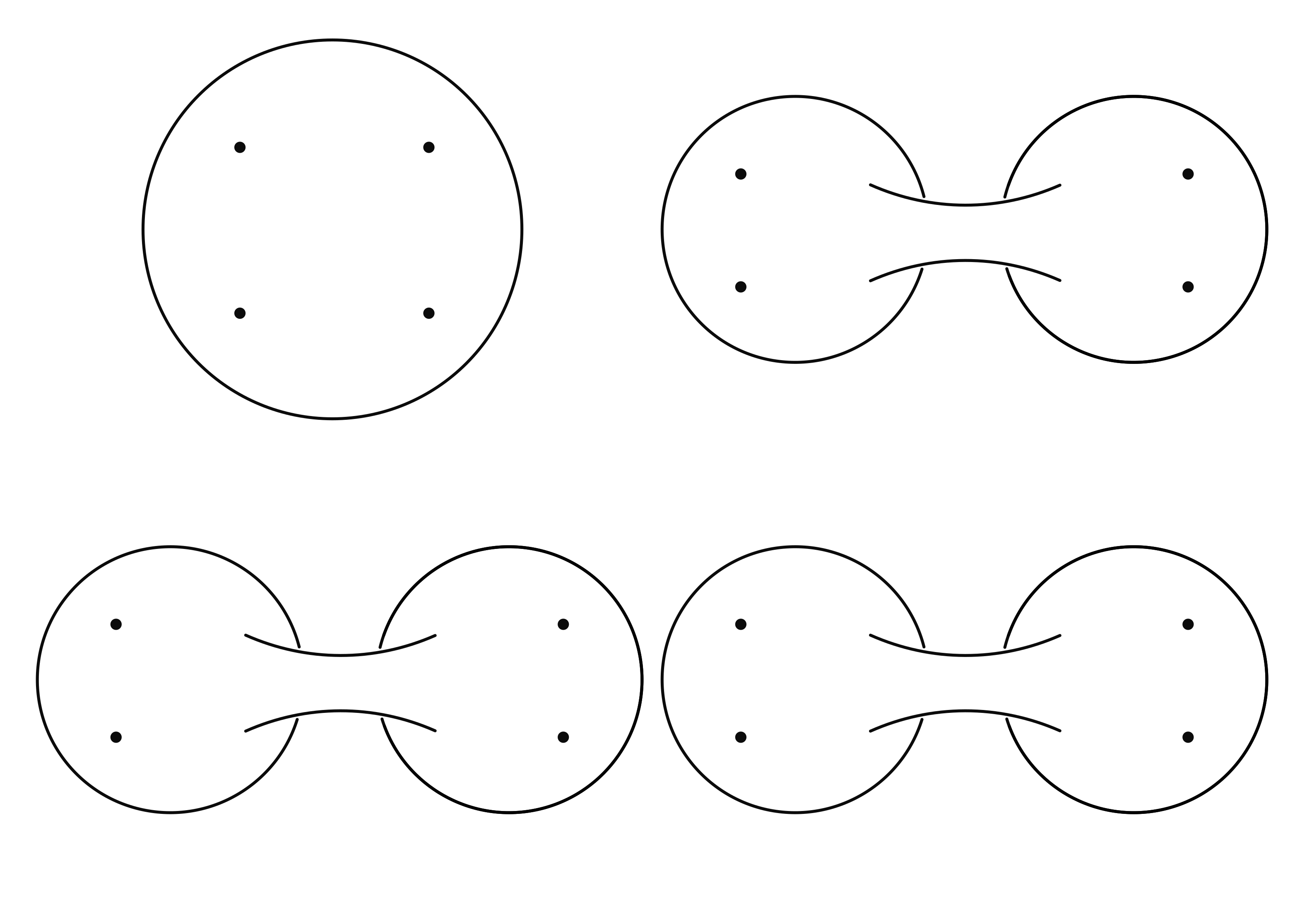}
\put(19,58){$0$}
\put(19,46){$q$}
\put(34,58){$1$}
\put(34,46){$\infty$}
\put(58,56){$0$}
\put(58,48){$q$}
\put(92,56){$1$}
\put(92,48){$\infty$}
\put(58,22){$0$}
\put(58,13){$1$}
\put(92,22){$q$}
\put(92,13){$\infty$}
\put(10,22){$1$}
\put(10,13){$q$}
\put(44,22){$0$}
\put(44,13){$\infty$}
\end{overpic}
\caption{\small The four-punctured sphere and its limits when $q\rightarrow0,1,\infty$.}
\label{SphereLimits}
\end{figure}

Changing the complex structure parameter of the four-punctured sphere
changes the complexified gauge coupling $\tau_{\textrm{\tiny
    UV}}$. But we can ask about the effect of a
$PSL(2,\Z$) action on $\tau_{\textrm{\tiny UV}}$ given by
\begin{equation}
\tau_{\textrm{\tiny UV}}\longmapsto\frac{a\,\tau_{\textrm{\tiny
      UV}}+b}{c\,\tau_{\textrm{\tiny UV}}+d} \qquad \mbox{for} \quad
\bigg(\begin{matrix} a & b \\ c & d \end{matrix}\bigg) \in SL(2,\Z) \ .
\end{equation}
This action leaves the complex structure of the four-punctured sphere
invariant and just permutes the punctures. It suggests that when the
$SU(2)$ gauge theory with $N_{\rm f}=4$ flavour hypermultiplets is
strongly coupled, that is when $q\to1$ or $q\to\infty$, we can
equivalently describe it as a weakly coupled theory with $q\to0$ and
permuted flavour symmetry groups (see Figure~\ref{SphereLimits}). 
This is indeed the S-duality found in the $SU(2)$ gauge theory coupled
to four hypermultiplets by Argyres and
Seiberg~\cite{Argyres:2007cn}. In~\cite{Gaiotto:2009we} this S-duality
was generalized to any theory of class $\cS$: any theory ${\sf
  T}_K[C,\cD]$ is invariant under a generalized S-duality that is
realized geometrically as the mapping class group acting on the
ultraviolet curve $C$. This is a main reason for naming them theories
of class $\cS$.

Let us finally relate this discussion to Section~\ref{sec:SWTheory} by
noting that any asymptotically free theory of class $\cS$ may be
realized in this framework by sending some masses to infinity. For
instance, the pure $SU(2)$ gauge theory is obtained by sending all four
masses $m_i$ of the hypermultiplets in the $N_{\rm
  f}=4$ $SU(2)$ theory to infinity, leaving $q=\E^{\,2\pi\,\I\,\tau_{\textrm{\tiny
    UV}}}=\Lambda^4/m_1\,m_2\,m_3\,m_4$ finite, where $\Lambda$ is the
ultraviolet scale of the pure $SU(2)$ theory. The resulting
ultraviolet curve $C$ is a sphere with two irregular punctures. The corresponding Seiberg-Witten geometry is described by the quadratic differential
\begin{align}
p_2(z) = \Big(\frac{\Lambda^2}{z^3}-\frac{2\,u}{z^2}+\frac{\Lambda^2}z\Big)\, \de
  z\otimes\de z \ ,
\end{align}
as anticipated from \eqref{eq:pureSU2SWcurve}.

\subsubsection*{Class $\cS$ Theories of Type $\boldsymbol{SU(K)}$}
\noindent
As for spectral networks, the story changes quite drastically for
higher rank gauge groups $SU(K)$ with $K>2$. Conformal $SU(K)$
theories of class $\cS$ can again be built out of three-punctured
spheres and cylinders which are now labeled by Young diagrams with
$K$ boxes. However, the corresponding $\N=2$ field theories are
often ``non-Lagrangian'', meaning that they are {intrinsically}
stongly coupled quantum field theories which do not admit a weakly
coupled description that can be studied using perturbation
theory.\footnote{They can be investigated physically as infrared
  limits of $\N=1$ gauge theories, where supersymmetry becomes
  enhanced at the fixed points, see
  e.g.~\cite{Gadde:2015xta,Maruyoshi:2016tqk}.}

\begin{figure}[h!]
\centering
\begin{overpic}
[width=0.80\textwidth]{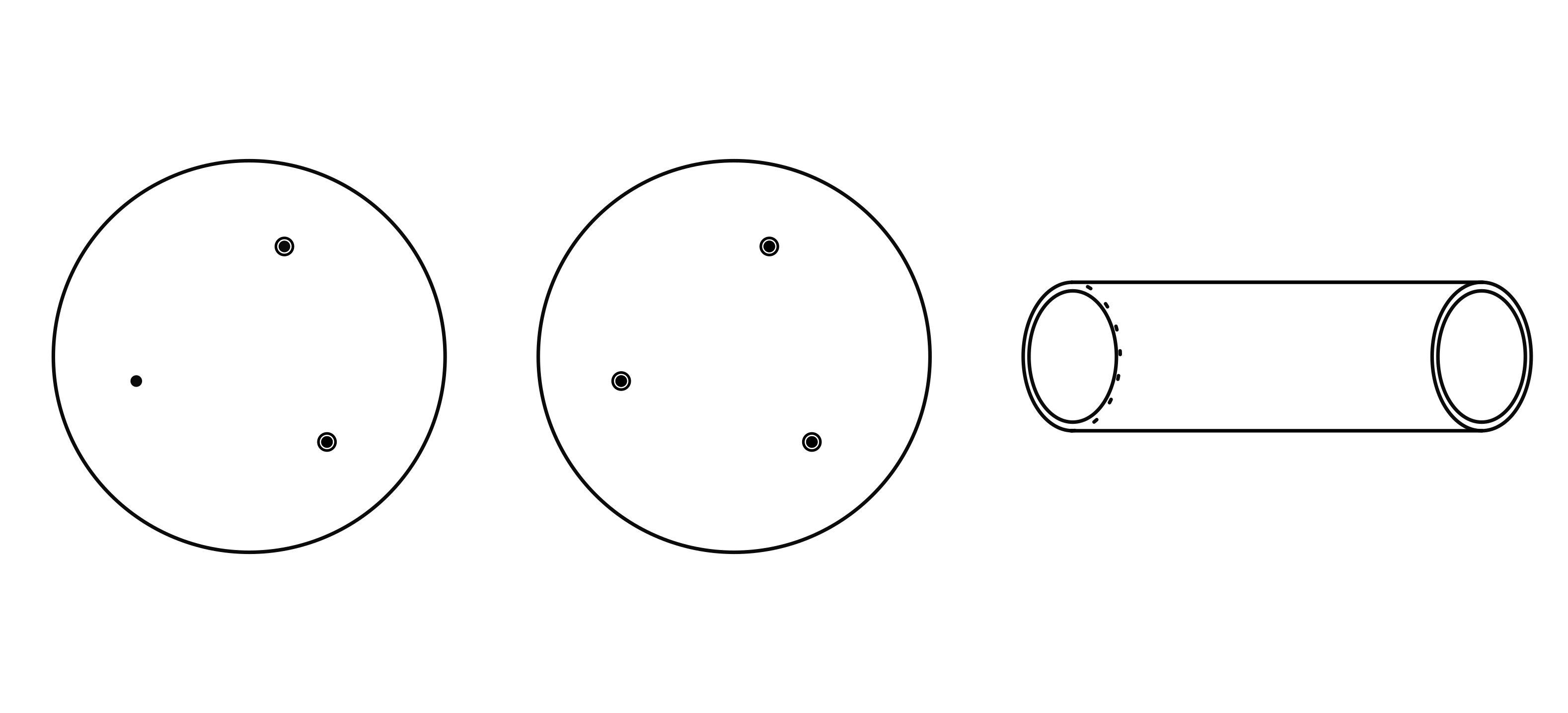}
\put(15.5,5){$\mathbf{A}$}
\put(47,5){$\mathbf{B}$}
\put(80,5){$\mathbf{C}$}
\end{overpic}
\caption{\small $SU(3)$ class $\cS$ theories are built from: $\mathbf{A}$ the sphere with two maximal and one minimal puncture; $\mathbf{B}$ the sphere with three maximal punctures; $\mathbf{C}$ the cylinder with two maximal punctures.}
\label{bblockssuthree}
\end{figure}

As an example, let us consider $K=3$. Class $\cS$ theories of type
$SU(3)$ can be constructed out of three elementary building blocks
(see Figure~\ref{bblockssuthree}):
\begin{itemize}
\item the three-punctured sphere with one ``minimal'' and two
  ``maximal'' punctures;
\item the three-punctured sphere with three ``maximal'' punctures;
    and
   \item the cylinder with complex structure parameter
     $\tau_{\textrm{\tiny UV}}$ labeled by two ``maximal'' punctures. 
   \end{itemize}
Here a {maximal puncture} refers to a puncture labeled by a Young
diagram consisting of one row with three boxes, together with mass
parameters $(m_1,m_2,-m_1,-m_2)$ labeling the residues of the
Seiberg-Witten differential $\lambda$ at the (regular) puncture; 
the associated flavour symmetry group is $SU(3)$ since the multiplicity of
column heights is three. A {minimal
  puncture} refers to a puncture labeled
by a Young diagram consisting of one column with two boxes and one
column with one box, together with mass parameters $(m,m,-2m)$; the
associated flavour symmetry group is $U(1)=S(U(1)\times
U(1))$ since the multiplicities of column heights are
$(1,1)$. Similarly to the rank one case $K=2$, the cylinder
corresponds to an $\N=2$ vector multiplet with gauge group $SU(3)$ and
complexified coupling $\tau_{\textrm{\tiny UV}}$, while the
three-punctured sphere with two maximal and one minimal puncture now
corresponds to a free hypermultiplet in the bifundamental
representation of $SU(3)_a\times SU(3)_c$.

\begin{figure}[h!]
\centering
\begin{overpic}
[width=0.70\textwidth]{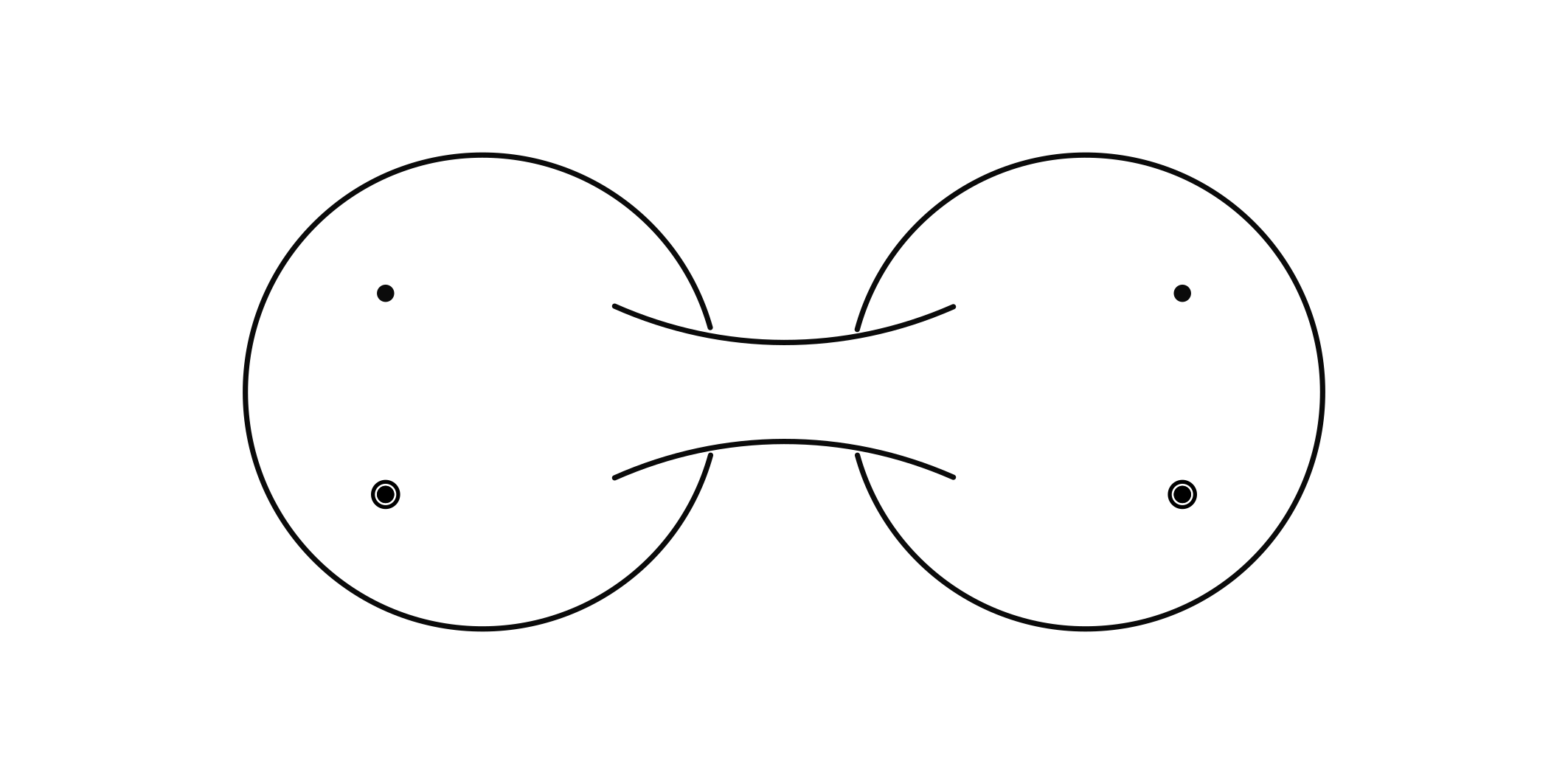}
\put(25.5,30){$a$}
\put(25.5,17){$b$}
\put(76.5,30){$d$}
\put(76.5,17){$c$}
\end{overpic}
\caption{\small The ultraviolet curve for the superconformal $SU(3)$ theory as $q\rightarrow0$.}
\label{suthreepopab1}
\end{figure}

\begin{figure}[h!]
\centering
\begin{overpic}
[width=0.70\textwidth]{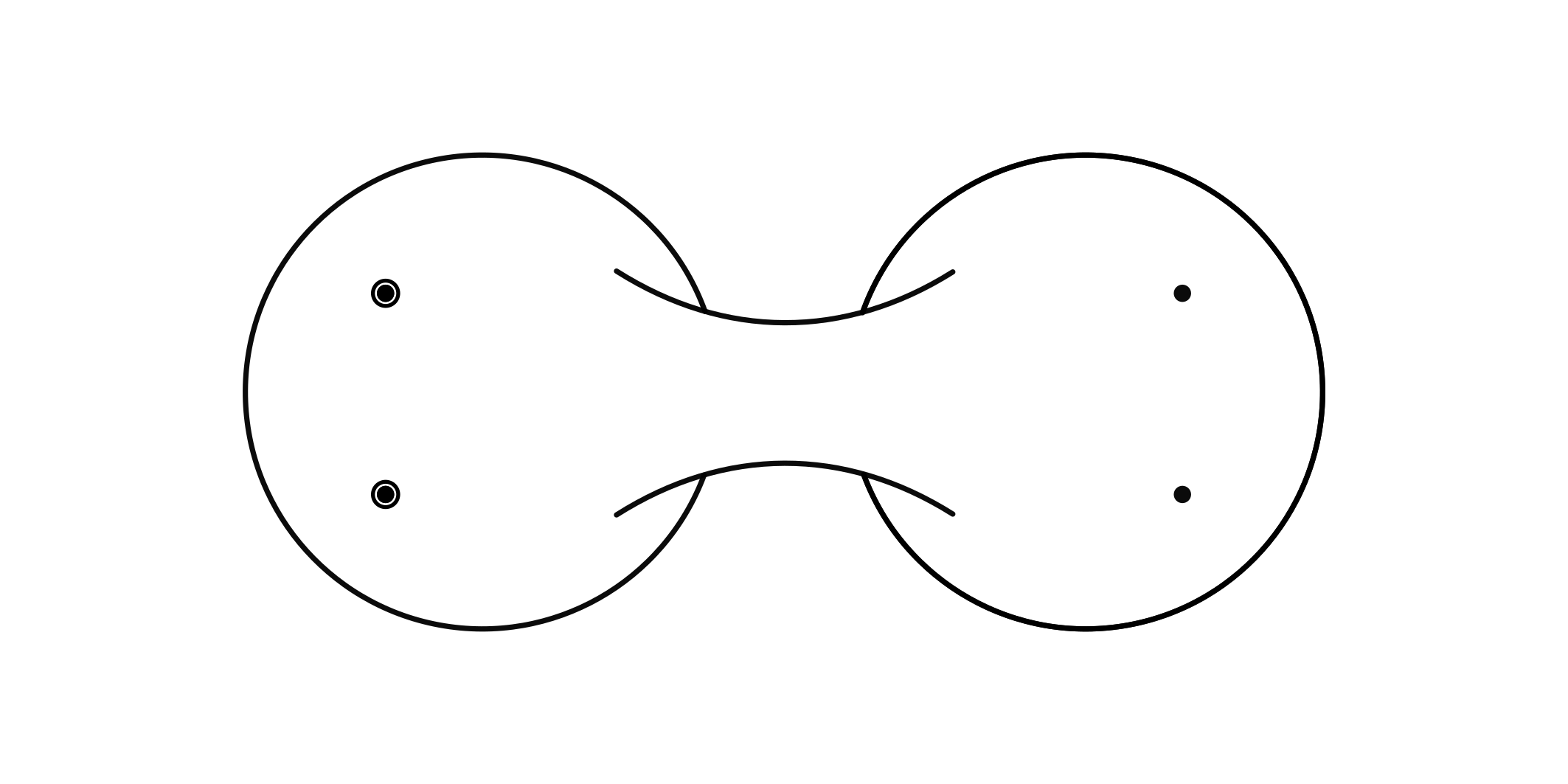}
\put(25.5,30){$b$}
\put(25.5,17){$c$}
\put(76.5,30){$a$}
\put(76.5,17){$d$}
\end{overpic}
\caption{\small The ultraviolet curve for the superconformal $SU(3)$ theory as $q\rightarrow1$.}
\label{suthreepopab2}
\end{figure}

We may then construct the superconformal $SU(3)$ gauge theory coupled
to six hypermultiplets (this is the $SU(3)$ analogue of the
superconformal $N_{\rm f}=4$ $SU(2)$ theory) by gluing two such
three-punctured spheres using the plumbing fixture method; indeed, we
can check that the $U(6)$ flavour symmetry rotating the six
hypermultiplets decomposes into
\begin{equation}
U(6) \supset U(3)\times U(3) = \big(SU(3)_a\times U(1)_b\big) \times
\big(SU(3)_c\times U(1)_d\big) \ .
\end{equation}
This is illustrated in Figure~\ref{suthreepopab1}. Suppose the punctures are positioned at
$z=0,1,q,\infty$, respectively, with $q$ close to $0$ in this weakly
coupled description of the $N_{\rm f}=6$ $SU(3)$ theory.
In the strongly coupled limit $q\to1$, the ultraviolet curve can instead be
described by the (degenerated) four-punctured sphere, as illustrated
in Figure~\ref{suthreepopab2} (to be precise, here we apply a mapping class group
action to permute the punctures). This decomposes into a three-punctured sphere with three
maximal punctures, and another three-punctured sphere with one maximal
and two minimal punctures.

Gaiotto interpreted this geometry
in~\cite{Gaiotto:2009we} as the S-dual description of the strongly
coupled $N_{\rm f}=6$ $SU(3)$ theory. It has been argued by Argyres and
Seiberg~\cite{Argyres:2007cn} that this S-dual description is given by
an intrinsically strongly coupled interacting superconformal field
theory, the $E_6$ Minahan-Nemeschansky theory~\cite{minahan1996n},
coupled to a weakly coupled $SU(2)$ theory.
In particular, the field theory description of the three-punctured
sphere with three maximal punctures is given by the ``non-Lagrangian''
$E_6$ Minahan-Nemeschansky
theory~\cite{Gaiotto:2009we}. Microscopically, its flavour symmetry
group is given by $SU(3)_a\times SU(3)_b\times SU(3)_c$, which in the low energy
limit is enhanced to $E_6$. A classification of all the building
blocks (or ``tinkertoys'') for theories of class $\cS$ has been given
in e.g.~\cite{Chacaltana:2010ks}.

To summarise, there is a large class of quantum field theories ${\sf
  T}_K[C,\cD]$, including most known Lagrangian\footnote{$\N=2$
  complete BPS quiver gauge theories were classified in~\cite{Cecotti:2011rv}; there
    are $11$ exceptional theories that are not of class $\cS$.} as
  well as new non-Lagrangian $\N=2$ theories, whose
  {microscopic} data is encoded in a (possibly punctured) Riemann
  surface $C$. These theories may be built out of elementary blocks
  corresponding to three-punctured spheres and cylinders. The
  (quantum) Coulomb branch $\scrB$ of infrared vacua of a theory ${\sf
    T}_K[C,\cD]$ is parameterized by $K-1$-tuples of meromorphic
  $k$-differentials $(p_2,\dots,p_K)$ on $C$, with $2\leq
  k\leq K$, whose poles at the punctures are characterized by the
  singularity data $\cD$; just as in Section~\ref{sec:SpectralHigherRank},
  regular punctures correspond to poles of order $k$, and wild
  punctures have higher order singularities. Each tuple
  $(p_2,\dots,p_K)$ defines a spectral curve
  \begin{align}\label{eq:SWcurve}
\Sigma\colon \quad w^K - \sum_{k=2}^K \, p_k\, w^{K-k} = 0
  \end{align}
  in the cotangent bundle $T^*C$, with local coordinates $(z,w)$ where
  $z\in C$. The curve $\Sigma$ is the
  Seiberg-Witten curve for the theory ${\sf T}_K[C,\cD]$ at the vacuum
  corresponding to $(p_2,\dots,p_K)$. The Seiberg-Witten
  differential is simply the tautological one-form $\lambda = w\,\de
  z$ on $T^*C$ restricted to $\Sigma$. Explicit descriptions of
  Seiberg-Witten curves and differentials for the theories
  discussed above may be found in e.g.~\cite{hollands2018higher}. 

\subsection{BPS States from Spectral Networks}\label{4dBPS}
\noindent
In Section~\ref{ClassSTheories} we introduced $\N=2$ theories ${\sf
  T}_K[C,\cD]$ of class $\cS$ and found that their moduli spaces of Coulomb vacua are
parametrized by tuples $u = (p_2,\dots,p_K)$ of
$k$-differentials on $C$, each defining a $K$-fold (possibly branched)
cover $\Sigma\to C$. We also know from Section~\ref{WKBSNs} that,
together with a phase $\vartheta \in \R/2\pi\,\Z$, this data allows us
to construct a spectral network $\scrW(u,\vartheta)$ on $C$. One
may wonder at this stage whether this spectral network encodes any
physical data of the theory ${\sf T}_K[C,\cD]$ in the vacuum
$u$. The answer is affirmative: The spectral network
$\scrW(u,\vartheta)$ encodes in a beautiful way information
about various types of BPS
states~\cite{gaiotto2013wall,Gaiotto:2010be,gaiotto2012spectral,Gaiotto:2012db}. In
the following we will explain how spectral networks can be used to
find the spectrum of {BPS particles} in the theory ${\sf T}_K[C,\cD]$.

Any theory with extended supersymmetry has special
representations, called {BPS states}, which are annihilated by
certain linear combinations of supercharges. These BPS states have
masses $M$ which saturate the {BPS bound}
\begin{align}
M\geq|Z| \ ,
\end{align}
where $Z$ is the central charge of the supersymmetry algebra. They are
an important tool in understanding the theory and its dualities.

Any $\N=2$ field theory has half-BPS states, which are annihilated by
half of the supercharges and form small representations of dimension
two (rather than four). In the low energy limit, when the $\N=2$
theory is described by an abelian gauge theory with prepotential
$\cF_0(a)$, the central charge $Z$ takes the form
\begin{align}
Z = q_i\,a^i + p^i\,\frac{\partial\cF_0(a)}{\partial a^i} \ .
\end{align}
The pair of integer vectors $(\vec q,\vec p)$ labels the electric charges $\vec
q$ and the magnetic charges $\vec p$ of the BPS state. The central
charge can be expressed
in terms of the Seiberg-Witten curve $\Sigma$ through the period integrals
\begin{align}
a^i = \oint_{A^i} \, \lambda \qquad \mbox{and} \qquad a_{\textrm{\tiny
  D},i} = \frac{\partial\cF_0(a)}{\partial a^i} = \oint_{B_i} \, \lambda
  \ ,
\end{align}
where $(A^i,B_i)$ generate a symplectic basis\footnote{A {symplectic
    basis} is a homology basis for which the only non-zero intersection
  pairings are given by $\langle A^i, B_j\rangle=\delta^{i}_{j}$.} of the first homology
group $H_1(\overline{\Sigma},\Z)$ and $\lambda$ is the Seiberg-Witten
differential. (This is a generalization of the formula
\eqref{eq:pureperiods} for the pure $SU(2)$ gauge theory discussed in
Section~\ref{sec:SWTheory}.)

A four-dimensional BPS particle is a one-particle state in $\R^{3,1}$
that preserves half of the supersymmetry. Its electromagnetic charge
$(\vec q,\vec p)$ determines a one-cycle
\begin{align}
\gamma = q_i\,A^i+p^i\,B_i
\end{align}
on $\Sigma$. Its central charge is then given by
\begin{align}
  Z_\gamma = \oint_\gamma \, \lambda
\end{align}
and its mass by $M_\gamma = |Z_\gamma|$.

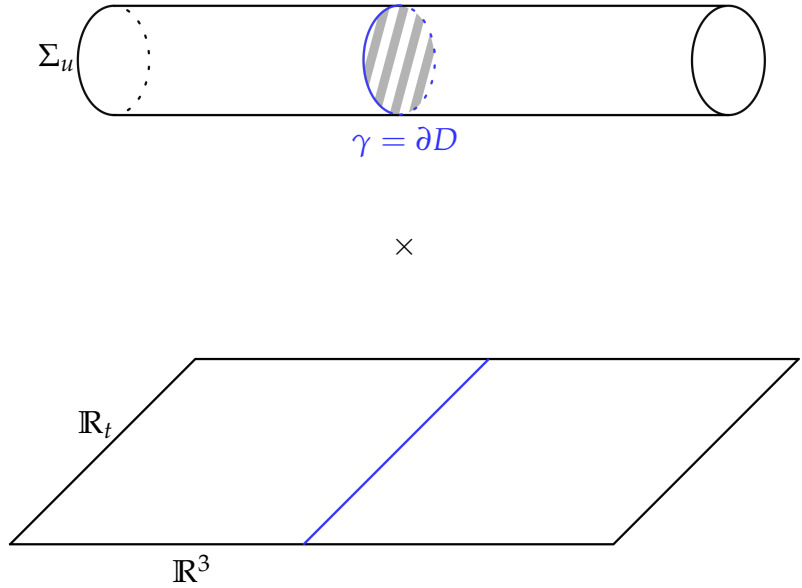
\begin{figure}[h!]
\centering
\begin{overpic}
[width=0.80\textwidth]{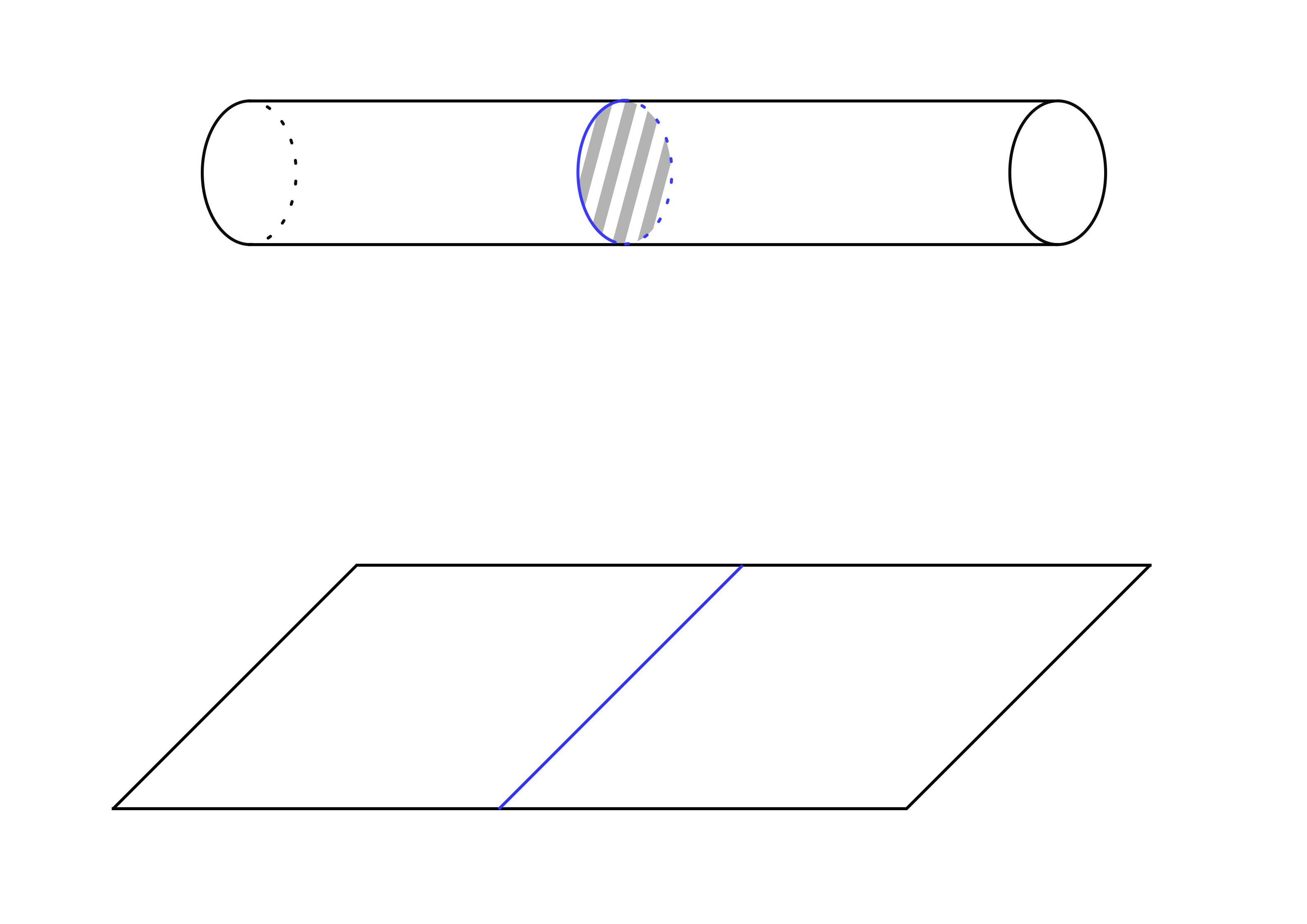}
\put(25,5){$\mathbb{R}^3$}
\put(15.5,20){$\mathbb{R}_t$}
\put(11.5,57){$\Sigma_{ u}$}
\put(43,48.5){\textcolor{Blue}{$\gamma=\partial D$}}
\put(47,38){$\times$}
\end{overpic}
\caption{\small A local picture of an M5-brane wrapping the product of
  the Seiberg-Witten curve $\Sigma$ with
  $\mathbb{R}^{3,1}=\mathbb{R}^3\times \mathbb{R}_t$, with an M2-brane
  wrapping the product of a two-cycle $D$ with the wordline $\R_t$ of a BPS particle in $\mathbb{R}^{3,1}$.}
\label{M2Wrap}
\end{figure}

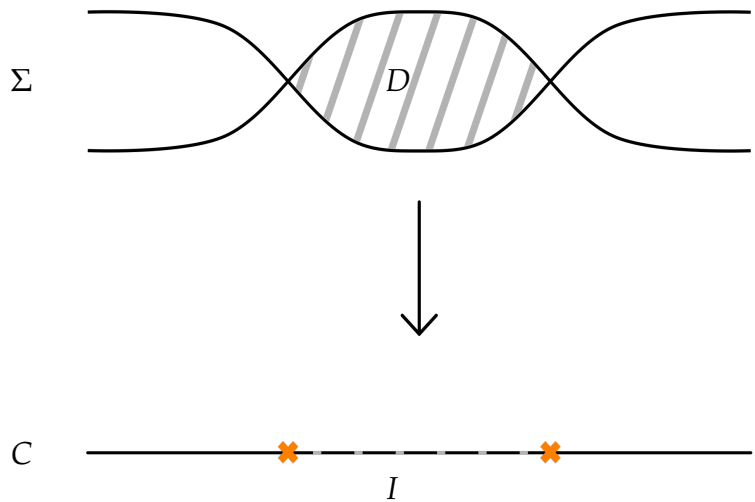
\begin{figure}[h!]
\centering
\begin{overpic}
[width=0.70\textwidth]{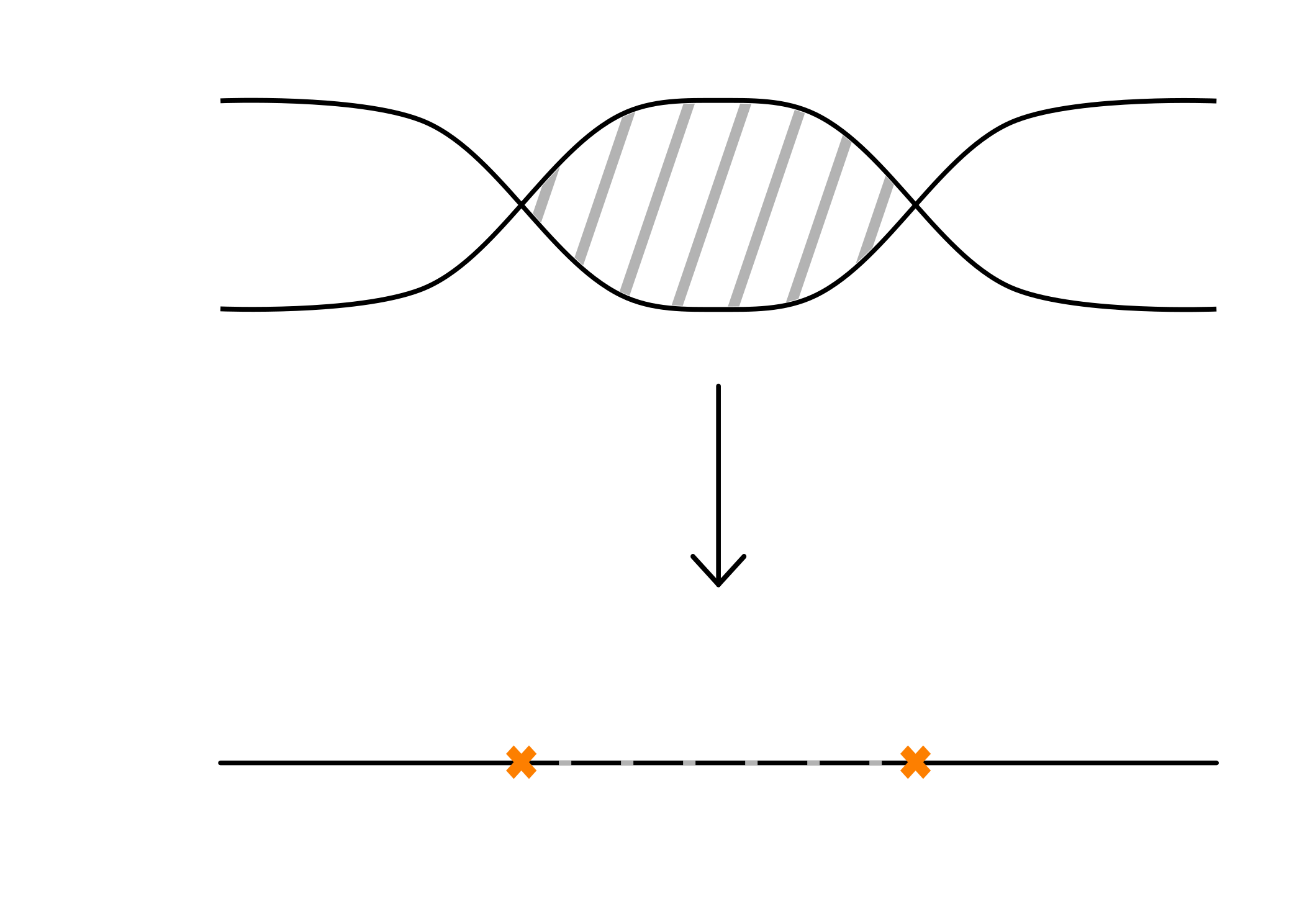}
\put(8,11){$C$}
\put(8,54){$\Sigma$}
\put(51,54){$D$}
\put(51,7){$I$}
\end{overpic}
\caption{\small A disc $D$ bounded by $\Sigma$ which projects down to an interval $I$ on $C$.}
\label{M2Disc}
\end{figure}

Such a BPS particle in four dimensions can be regarded as a BPS string in six
dimensions~\cite{Klemm:1996bj}. It can be embedded in
M-theory as an M2-brane wrapping
the product of a compact two-cycle $D$ in $T^*C$, which ends on
$\Sigma$, with its worldline in $\R^{3,1}$ (see Figure~\ref{M2Wrap}).
Its
electromagnetic charge $\gamma$ is encoded in the boundary $[\partial
D]=\gamma$, while the BPS condition translates into a calibration
condition on the two-cycle $D$: $D$ should be a holomorphic Lagrangian
submanifold of $T^*C$ with boundary $[\partial D]=\gamma$. In the
simplest case, $D$ is topologically a disc, and its projection to the
ultraviolet curve $C$ is an interval $I$ bounded on either side by a
branch point of type $(ij)$ of the covering $\Sigma\to C$ (see
Figure~\ref{M2Disc}). 
Let $\lambda_i$ be the restriction of the Seiberg-Witten
differential to the $i$-th sheet; the tension of an $(ij)$-string is
given by $\lambda_i-\lambda_j$. Then the central charge of the
corresponding BPS particle is given by
\begin{align}
Z_{ij} =\oint_\gamma\,\lambda = \int_I\, (\lambda_i-\lambda_j) \ ,
\end{align}
while its mass is
\begin{equation}
M_{ij}=\int_I\,|\lambda_i-\lambda_j| \ .
\end{equation}
The BPS condition $M_{ij}=|Z_{ij}|$ is thus satisfied if and only if the
absolute value of the integral of the one-form $\lambda_i-\lambda_j$ is the same as the integral of its
absolute value, which is true if and only if
\begin{equation}\label{eq:BPSM}
\E^{-\I\,\vartheta}\, (\lambda_i-\lambda_j)(v) \ \in \ \mathbb{R}^\times
\end{equation}
for any tangent vector $v$ along the projection of $\gamma$ to $C$,
where $\vartheta={\rm arg}(Z_{ij})$. (This also specifies which
supercharges are preserved by this BPS state.)
Note that \eqref{eq:BPSM} is exactly the condition for a trajectory of type
$(ij)$. In other words, we can
identify such a BPS particle of charge $\gamma$ with a saddle of
the spectral network
$\scrW(u,\vartheta)$ on $C$ that
lifts to $\gamma$ on~$\Sigma$.

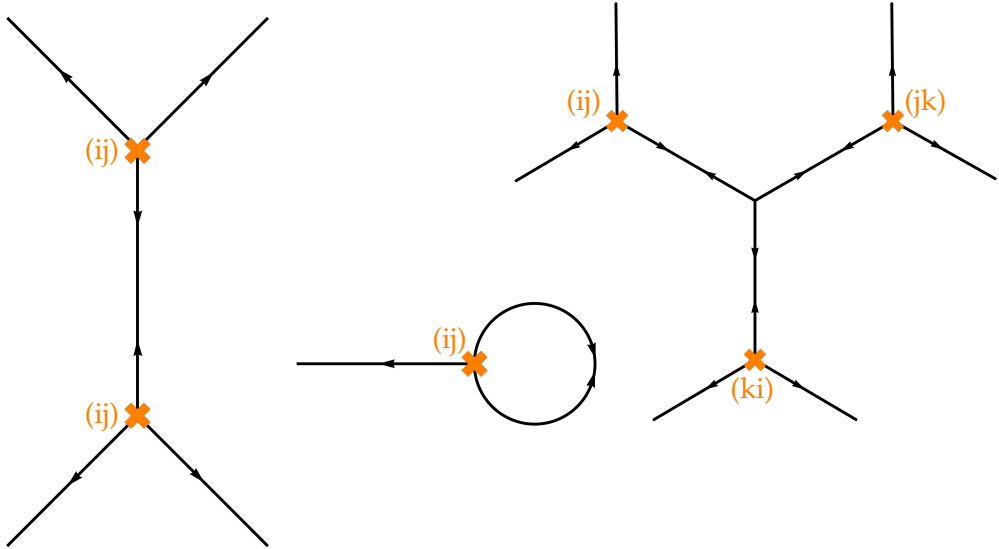
\begin{figure}[h!]
\small
\centering
\begin{overpic}
[width=0.8\textwidth]{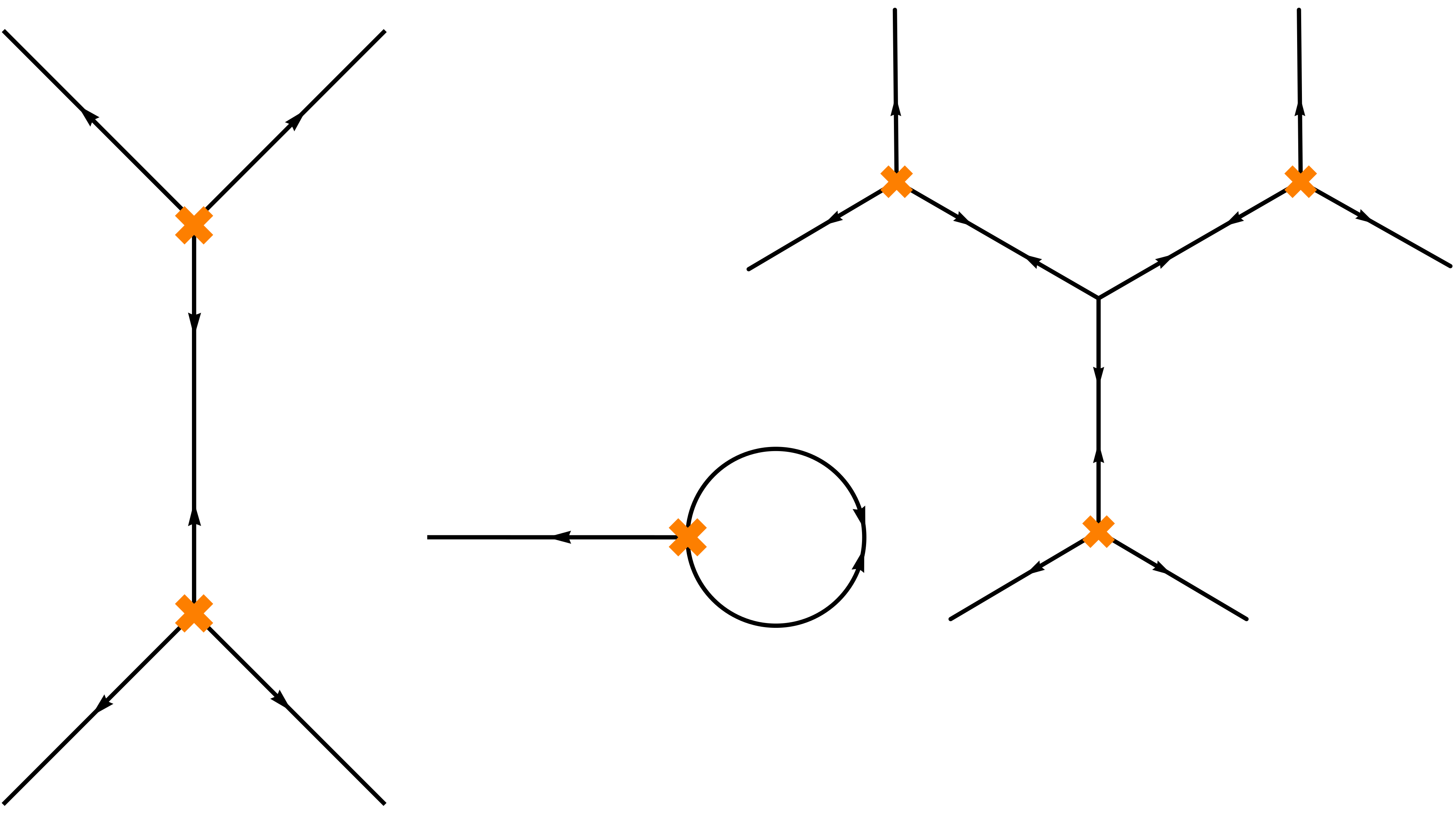}
\put(8,41){\textcolor{Orange}{(ij)}}
\put(8,14.5){\textcolor{Orange}{(ij)}}
\put(43,22){\textcolor{Orange}{(ij)}}
\put(56.5,46){\textcolor{Orange}{(ij)}}
\put(90.5,46){\textcolor{Orange}{(jk)}}
\put(73,17){\textcolor{Orange}{(ki)}}
\end{overpic}
\caption{\small Three examples of finite webs. For class $\cS$
  theories with $SU(2)$ gauge group,
  the first two webs are the only possibilities, but for
  $SU(3)$ and higher rank gauge groups, more complicated webs can form, such as the third web shown. We mark the branch points with pairs of indices $ij$ that correspond to the sheets $\Sigma_i$, $\Sigma_j$ that come together there, as well as the trajectories that emerge from them, which are of type $(ij)$ or $(ji)$.}
\label{fig:finiteweb}
\normalsize
\end{figure}

More generally, any BPS particle in the vacuum $u\in\scrB$ with
central charge $Z=\E^{\,\I\,\vartheta}\,M$ is encoded in the spectral
network $\scrW(u,\vartheta)$ as a {finite web}~\cite{gaiotto2012spectral}, a web
consisting of a finite number of saddles that have both ends on
branch points or junctions (see Figure~\ref{fig:finiteweb} for some examples). 
If we vary
$\vartheta$ and systematically scan for such finite webs, we uncover
the spectrum of BPS particles of the class $\cS$ theory ${\sf
  T}_K[C,\cD]$ in the vacuum $u$. The rays in the complex $Z$-plane for which there exists a BPS particle in the vacuum $u$ are sometimes called `BPS rays'. If we vary the point $u$ on the Coulomb branch, the central charges $Z_\gamma$ change and thus the BPS rays move in the $Z$-plane.
  
There might be real codimension one walls in the Coulomb branch $\scrB$ where the phases of some of the BPS particles coincide. Such walls are called `walls of marginal stability'. These are the only places where BPS bound states can form or decay. The walls of marginal stability thus divide the Coulomb branch into different regions where there may be a different spectrum of BPS states.

\begin{example}\label{ex:AD2BPS}
Recall Example~\ref{ex:CAD2} with $C=\C$ and the quadratic differential
\begin{equation}\label{AD2Sigma}
p_2(z)=(z^2+m)\, \de z \otimes\de z \ .
\end{equation}
The corresponding four-dimensional $\N=2$ field theory is the simplest
example of an {Argyres-Douglas theory}~\cite{Argyres:1995jj}, which is
the theory of a single hypermultiplet of mass $m$. It is sometimes
called the AD$_2$ theory~\cite{gaiotto2013wall}.

The Coulomb
branch $\scrB$ of the AD$_2$ theory is a single point, corresponding
to the differential $p_2(z)$ with fixed parameter $m$ (the mass
of the hypermultiplet), so this theory contains a
unique vacuum. Its Seiberg-Witten geometry is illustrated in Figure~\ref{AD2SWCurve}; the
charge lattice $\Gamma=H_1(\Sigma,\Z)\cong\Z$ is generated by the
one-cycle $\gamma$.
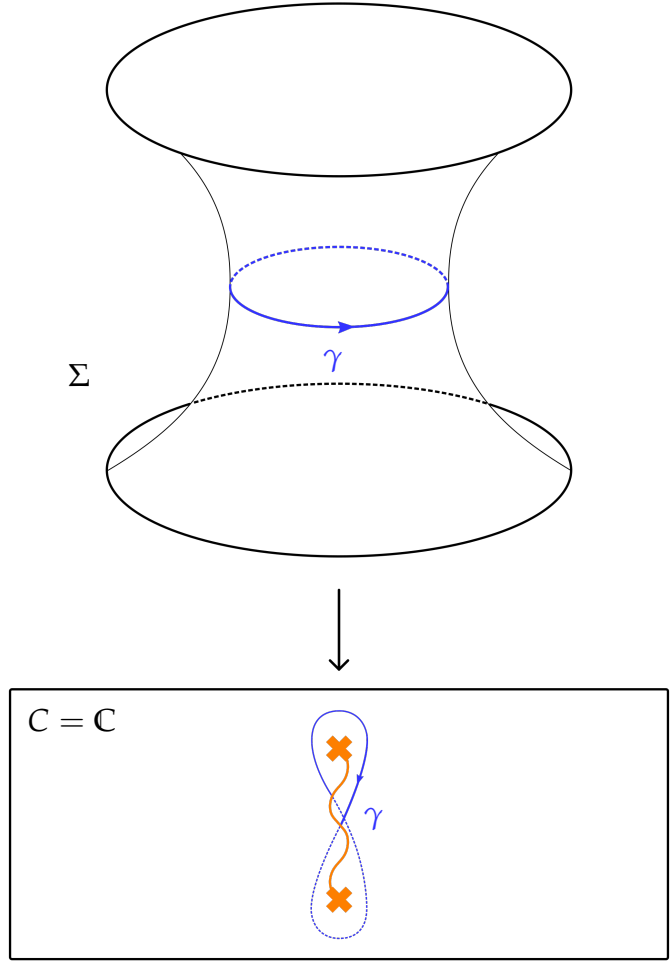
\begin{figure}[h!]
\centering
\begin{overpic}
[width=0.60\textwidth]{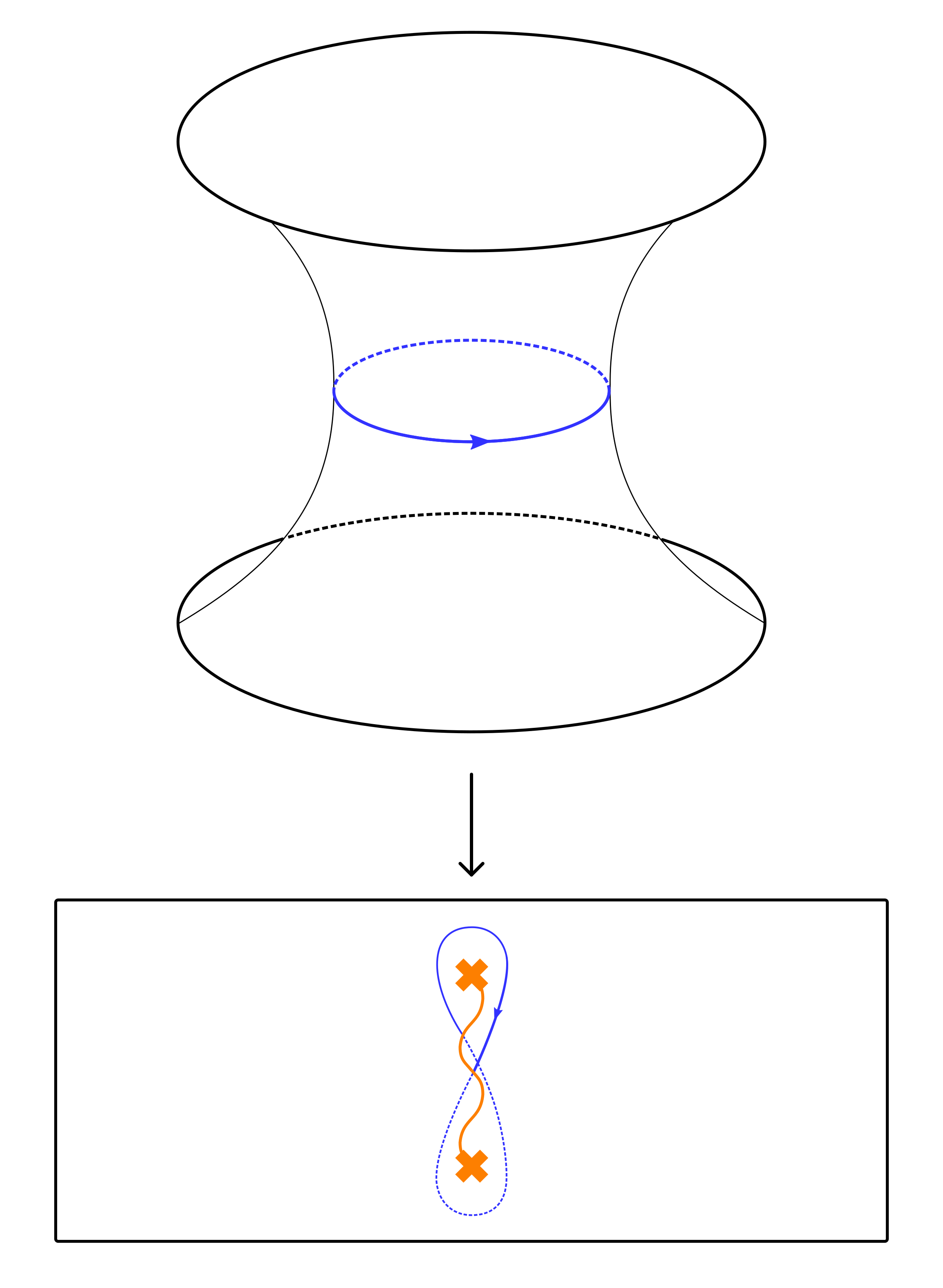}
\put(10,60){$\Sigma$}
\put(35,62){\textcolor{Blue}{$\gamma$}}
\put(39,17){\textcolor{Blue}{$\gamma$}}
\put(6,26){$C=\mathbb{C}$}
\end{overpic}
\caption{\small Seiberg-Witten geometry of the AD$_2$ theory.}
\label{AD2SWCurve}
\end{figure}
The spectral networks for this theory were discussed in
Example~\ref{ex:CAD2}: if we vary $\vartheta$ from $0$ to $\pi$ a
saddle appears at $\vartheta=\vartheta_{\rm c}=\pi/2$, see Figure~\ref{AD2jump}. 
The corresponding BPS particle is of course the BPS
hypermultiplet of mass $m$. When $m=0$ the two branch points collide, the one-cycle $\gamma$ collapses, and the hypermultiplet becomes massless.
\end{example}

\begin{example}\label{ex:PureSU2BPS}
Let us now determine the BPS spectrum of the pure $SU(2)$ gauge
theory, which we introduced in
Section~\ref{sec:SWTheory}, using spectral networks. The
Seiberg-Witten geometry is determined by the quadratic differential
\begin{equation}
p_2(z)=\Big(\frac{\Lambda^2}{z^3} - \frac{2\,u}{z^2} +
\frac{\Lambda^2}{z}\Big)\, \de z \otimes\de z \ ,
\end{equation}
where $\Lambda$ is the ultraviolet scale and $u$ parametrizes the Coulomb branch $\scrB$.
This differential has irregular (or wild) singularities (of order three) at
$z=0$ and $z=\infty$, and two zeroes at
\begin{equation}
z_{\pm}=-\frac{u}{\Lambda^2}\pm\sqrt{\Big(\frac{u}{\Lambda^2}\Big)^2-1}
\ .
\end{equation}
The Seiberg-Witten curve $\Sigma$ is defined as $w^2=p_2(z)$. As we saw in Section~\ref{sec:SWTheory}, this is topologically a torus with two double punctures, see 
Figure~\ref{SU2SW}. 

Note that there are two special points on the Coulomb branch: at
$u=\pm\,\Lambda^2$ both zeroes of $p_2(z)$ collide. These are
the two quantum singularities of $\scrB$, where the Seiberg-Witten
curve $\Sigma$ degenerates and some one-cycle $\gamma$ vanishes (see Figure~\ref{SWmoduli}). 
\begin{figure}[h!]
\centering
\begin{overpic}
[width=0.80\textwidth]{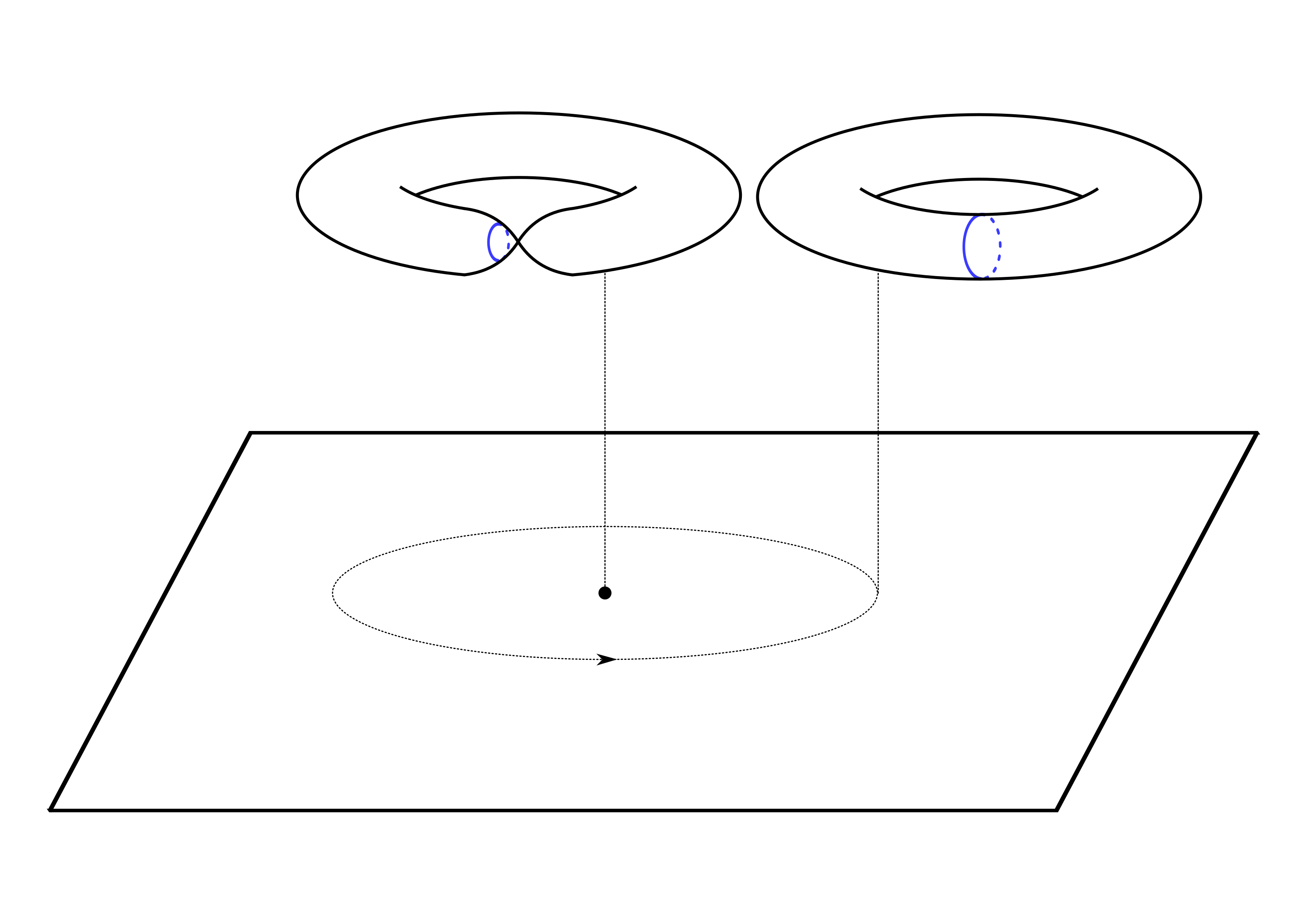}
\put(5,23){$\scrB$}
\put(86,47){$\Sigma_u$}
\put(74,47){\textcolor{Blue}{$\gamma_{\pm}$}}
\put(37,47){\textcolor{Blue}{$\gamma_{\pm}$}}
\end{overpic}
\caption{\small At the singularities $u=\pm\,\Lambda^2$ the Seiberg-Witten curve $\Sigma_u$ degenerates because a one-cycle $\gamma_{\pm}$ vanishes. This leads to an $SL(2,\mathbb{Z})$ monodromy in the charge lattice $\Gamma$ around these two points.}
\label{SWmoduli}
\end{figure}
We
now know that this signals the presence of massless BPS particles of
electromagnetic charge $\gamma$.

Indeed, this can be verified by examining the spectral networks
$\scrW(u,\vartheta)$ close to the singularities. It actually
turns out that the $u$-plane is divided into two regions in which the
networks $\scrW(u,\vartheta)$ behave very differently. These
two regions are known as the strong coupling region (for small $u$) and
the weak coupling region (for large $u$). They are divided by a {wall of marginal stability}, which is defined by the
equation ${\rm Im}(a_{\textrm{\tiny D}}/a)=0$ and can be approximated
by an ellipse passing through the singularities (see Figure~\ref{CouplingRegions}). 
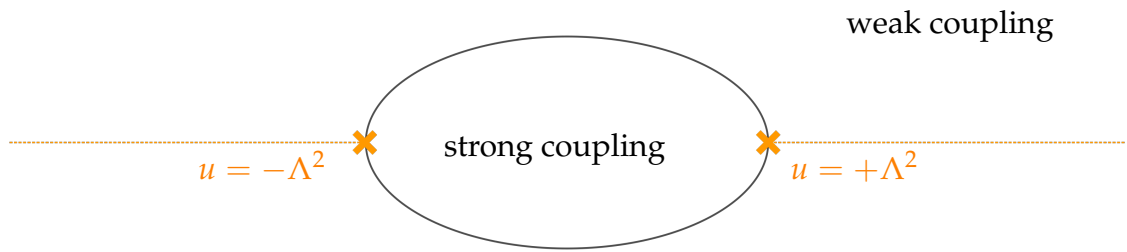
\begin{figure}[h!]
\centering
\begin{overpic}
[width=0.90\textwidth]{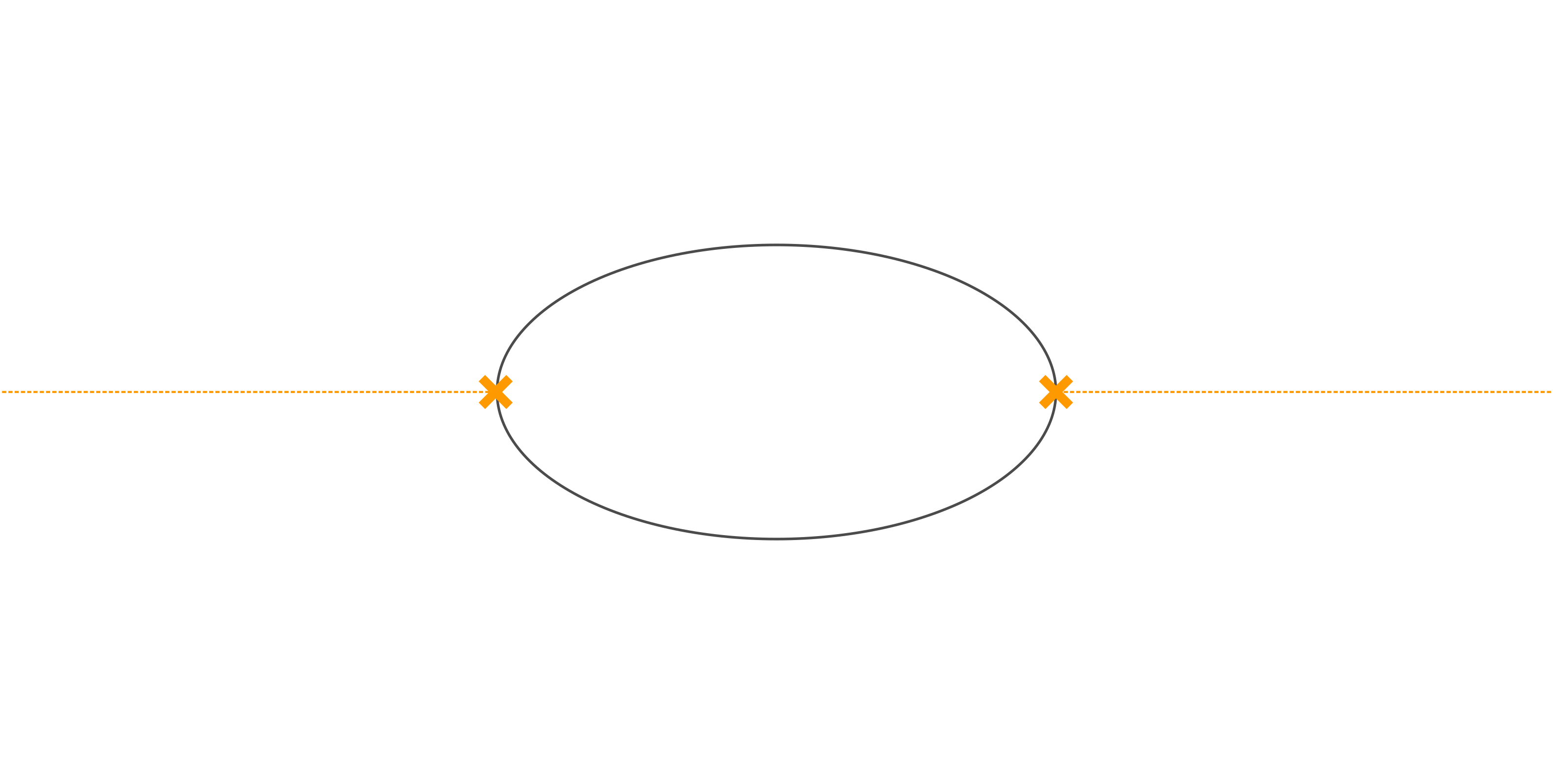}
\put(17,22){\textcolor{Orange}{$u=-\Lambda^2$}}
\put(70,22){\textcolor{Orange}{$u=+\Lambda^2$}}
\put(39,24){strong coupling}
\put(75,35){weak coupling}
\end{overpic}
\vspace{-2cm}
\caption{\small Coulomb branch, or $u$-plane, for the pure $SU(2)$
  theory. The wall of marginal stability, shown in black, separates a
  strongly coupled region near $u=0$ from a weakly coupled region near
  $u=\infty$. Since there is an $SL(2,\Z)$ monodromy around
  the singularities at $u=\pm\,\Lambda^2$, we have chosen the cut,
  shown in orange, to trivialize the corresponding local system.}
\label{CouplingRegions}
\end{figure}
We
choose cuts for the $SL(2,\Z)$ monodromy as in
Figure~\ref{CouplingRegions}, and trivialize the corresponding local
system by choosing a basis of the charge lattice $\Gamma$ as in Figure~\ref{SU2Cycles}.
\begin{figure}[h!]
\centering
\begin{overpic}
[width=0.80\textwidth]{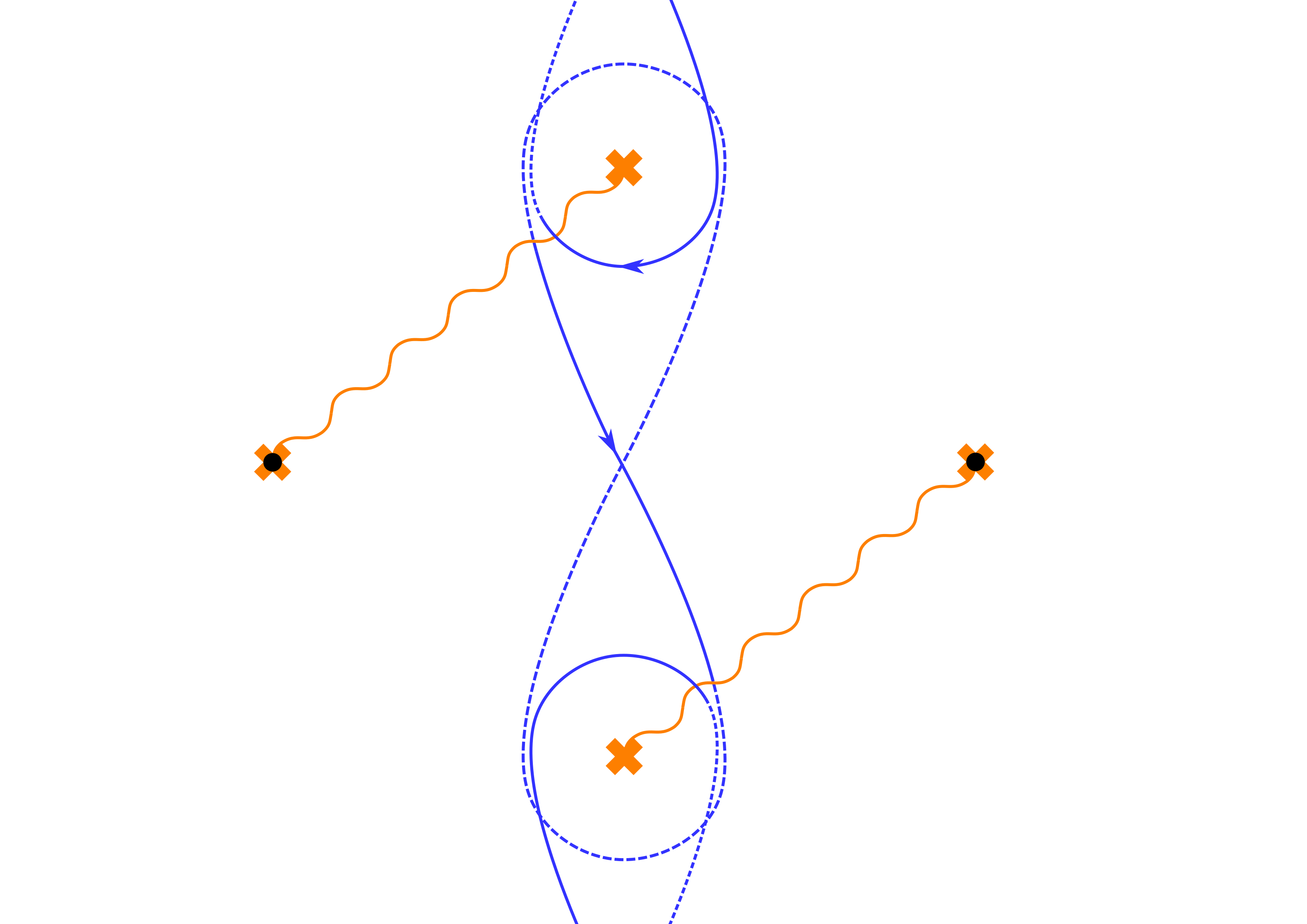}
\put(52,30){\textcolor{Blue}{$\gamma_+$}}
\put(54,66){\textcolor{Blue}{$\gamma_-$}}
\end{overpic}
\caption{\small Choice of a basis of one-cycles $\gamma_{\pm}$ of
  $\Gamma$ drawn on the ultraviolet curve $C$, where we mapped $z=\infty$ to finite distance.}
\label{SU2Cycles}
\end{figure}

Let us consider the family of spectral networks $\scrW(u,\vartheta)$
for varying phase $\vartheta$ at a point $u$ in the strongly coupled
region first. In Figure~\ref{SU2Flips} we illustrate the topology changes that occur
as $\vartheta$ is varied from $-\frac\pi4$ to $\frac{3\pi}4$ at $u=0$. (The picture
would be similar at any other point in the strongly coupled region.)
\begin{figure}[h!]
\centering
\begin{overpic}
[width=0.80\textwidth]{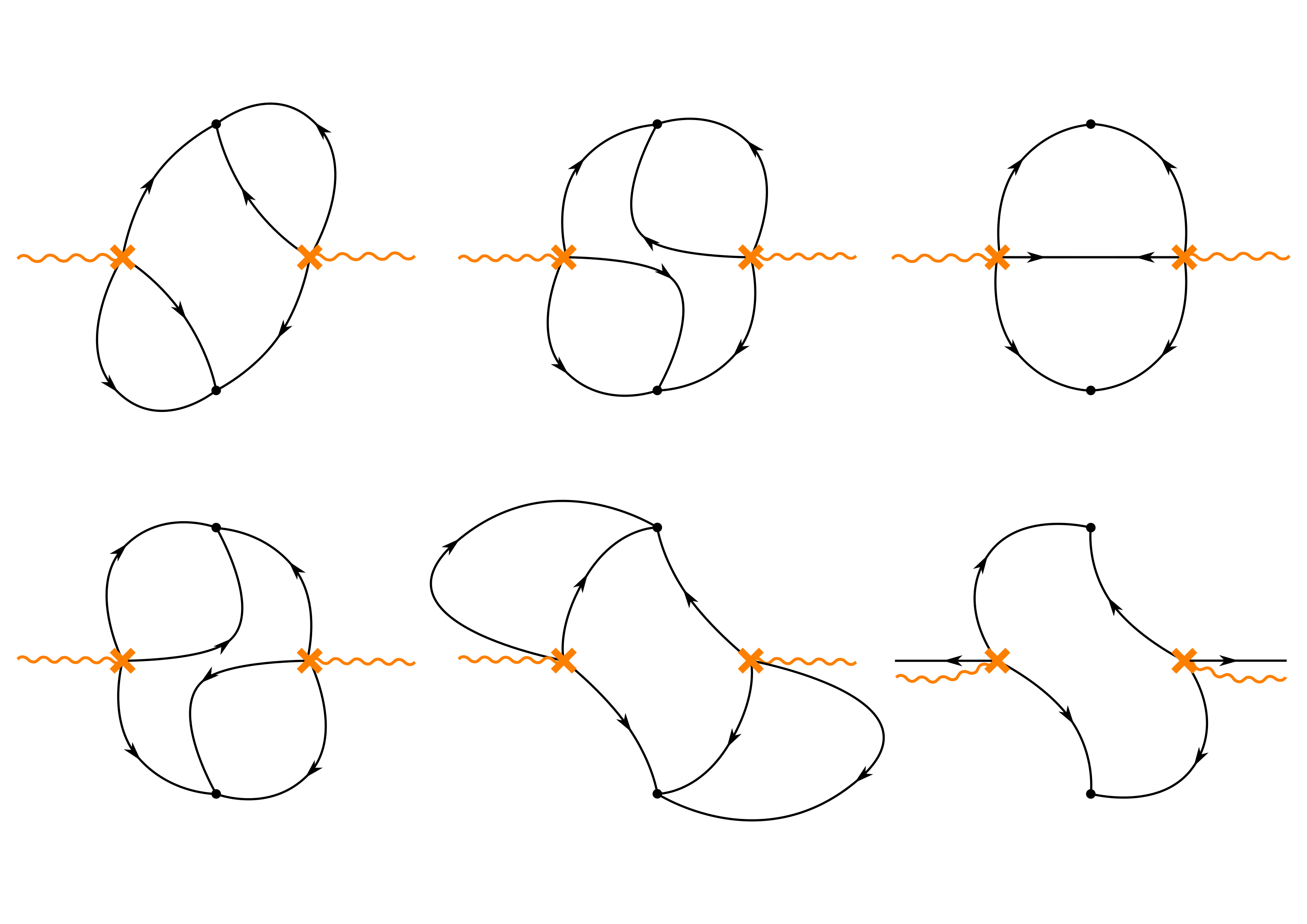}
\end{overpic}
\caption{\small The spectral networks for the pure $SU(2)$ theory as $\vartheta$ is varied over two flips. The networks are rotated through $-90^\circ$.}
\label{SU2Flips}
\end{figure}
In this family we encounter two saddles, at $\vartheta=0$ and
$\vartheta=\frac\pi2$. The first saddle corresponds to a hypermultiplet of
electromagnetic charge $\gamma_+$ and the second saddle to a
hypermultiplet of electromagnetic charge $\gamma_-$. If we continue
varying $\vartheta$ from $\frac{3\pi}4$ to $\frac{7\pi}4$, we come across two saddles
corresponding to their antiparticles. The BPS spectrum in the strongly
coupled region thus consists of two hypermultiplets, traditionally
called the {monopole} and the {dyon}, and their charge
conjugates.

The story in the weakly coupled region is slightly more
complicated. Fortunately we have already studied the relevant networks
in Example~\ref{ex:purenetworks}. Topology changes occur when
$\vartheta$ is tuned to a critical phase $\vartheta_{\rm c}$ (as well
as $\pi+\vartheta_{\rm c}$), see Figure~\ref{PureSU2Wjuggle}.
As $\vartheta$ increases
to $\vartheta_{\rm c}$ we encounter an infinite number of flips, which
correspond to an infinite tower of hypermultiplets with charges
$(n+1)\,\gamma_++n\,\gamma_-$ for $n\in\Z_{\geq0}$. The juggle at
$\vartheta=\vartheta_{\rm c}$ encodes a vector multiplet of charge
$\gamma_++\gamma_-$, and the infinite tower of flips as $\vartheta$
decreases to $\vartheta_{\rm c}$ correspond to another infinite number
of hypermultiplets with charges $n\,\gamma_++(n+1)\,\gamma_-$. At the
phase $\vartheta=\pi+\vartheta_{\rm c}$ we find all of their charge
conjugates.

Evidently, the spectra on either side of the wall of marginal
stability are not the same, and some BPS states have decayed. This may
be captured by the formula
\begin{align}
K_{\gamma_+}\,K_{\gamma_-} =
  K_{\gamma_-}\,K_{\gamma_++2\gamma_-}\,K_{2\gamma_++3\gamma_-} \cdots
  K_{\gamma_++\gamma_-}^2 \cdots
  K_{3\gamma_++2\gamma_-}\,K_{2\gamma_++\gamma_-}\,K_{\gamma_+} \ , 
\end{align}
where $K_\gamma$ is the transformation
\begin{align}
K_\gamma\, : \, \cX_{\gamma'}\longmapsto
  \cX_{\gamma'}\,\big(1-(-1)^{\langle\gamma',\gamma\rangle} \,
  \cX_\gamma \big)^{\langle\gamma',\gamma\rangle}
\end{align}
of the spectral coordinates $\cX_{\gamma'}$ under a flip~\cite[Section~7.6]{gaiotto2013wall}. This is the
famous Kontsevich-Soibelman wall-crossing
formula~\cite{Kontsevich:2008fj}. For a good introduction to
wall-crossing in $\N=2$ field theories, we
recommend~\cite{moorePiTPlecture}.
\end{example}

By now there is rich body of work on BPS states and wall-crossing in
$\N=2$ theories, from many points of view. Of special note is the BPS
quiver method, which is particularly powerful when there is a region
(or chamber) in the Coulomb branch with a finite BPS spectrum. This led to the
determination of the BPS spectra of all ``complete'' $\N=2$
theories~\cite{Alim:2011ae}.

The beautifully geometric spectral network method can be applied to
any $\N=2$ theory of class $\cS$ at any point $u\in\scrB$ of its
Coulomb branch. It turns out to be especially valuable for conformal
$\N=2$ theories with a single infinite BPS chamber. Examples include
the intrinsically strongly coupled $E_6$ and higher rank Minahan-Nemeschansky theories discussed in
Section~\ref{ClassSTheories}, whose spectra are determined
in~\cite{Hollands:2016kgm,Hao:2019ryd}, where higher rank
Fenchel-Nielsen networks play a crucial role. In these examples one
finds not only saddles corresponding to
hypermultiplets or closed loops corresponding to vector multiplets,
but also more complicated topologies arise, such as the finite webs of
Figure~\ref{fig:finiteweb}. The method of spectral networks can also
be extended to compute the spin content of BPS states, with the spin
identified with the writhe of paths on the Seiberg-Witten
curve $\Sigma$~\cite{Galakhov:2014xba}, which is related to a quantization of
the moduli space of flat abelian connections on $\Sigma$.
Further developments and applications of
the spectral network method can be found in
e.g.~\cite{galakhov2013wild,Longhi:2016rjt,Longhi:2016wtv},
while the connection between the techniques based on spectral networks
and BPS quivers is discussed in~\cite{Gabella:2017hpz}.

\clearpage
\newpage
\section{Twisted Superpotentials, Hitchin Systems and Opers}\label{sec:NRS}
\noindent
In this section we discuss a certain physical
quantity, called the effective twisted superpotential \smash{$\Weff$} of an $\N=2$
field theory ${\sf T}_K[C,\cD]$ of class $\cS$ in the
$\frac12\Omega$-background; this is introduced in Section~\ref{sec:Weff}. Interestingly, in~\cite{nekrasov2011darboux} it was conjectured that \smash{$\Weff$} has a concrete
geometric meaning, which may be extended to any $\N=2$ theory of class
$\cS$, in the Hitchin moduli space. This conjecture will be formulated in Section~\ref{sec:newNRS}, while in the remainder of this section we explain the various geometric ingredients: the Hitchin integrable
system in Section~\ref{sec:Hitchin}, and its Lagrangian subspace of oper connections in Section~\ref{sec:opers}. We give an explicit description of the relevant oper connections in our main
examples, and relate the abelianization and spectral coordinates
associated to these connections with exact WKB methods in
Section~\ref{sec:WKB}. 

\subsection{Effective Twisted Superpotentials}\label{sec:Weff}
\noindent
For any $\N=2$ theory ${\sf T}$, the low energy dynamics on its
Coulomb branch $\scrB$ can be described in terms of its holomorphic
prepotential $\cF_0(a;m;q)$. This is a multi-valued analytic function
depending on the Coulomb moduli $a$, the
mass parameters $m$, and the exponentiated ultraviolet gauge couplings
$q$. We have seen that in terms of
Seiberg-Witten geometry it can be obtained from period integrals 
\begin{equation}
a_i=\oint_{ A^i}\,\lambda \qquad \mbox{and} \qquad \ a^i_{\textrm{\tiny D}}=\oint_{ B_i}\, \lambda
\end{equation}
on
the Seiberg-Witten curve $(\Sigma,\lambda)$ via the relation
\begin{equation}
a^i_{\textrm{\tiny D}}=\frac{\partial\cF_0(a;m;q)}{\partial a_i} \ .
\end{equation}

Nekrasov computed $\cF_0$ from first principles for $\N=2$ theories $\sf T$ with a Lagrangian formulation (that is, in terms of gauge fields possibly coupled to matter fields)~\cite{nekrasov2003seiberg} by
considering a deformation of the theory $\sf T$ labeled by two complex
parameters $\epsilon_1$ and $\epsilon_2$. This deformation is called
the $\Omega$-background,\footnote{The precise construction of this background (starting from six dimensions) is given for instance in~\cite[Section~1.4]{nekrasov2011darboux}.} where the two parameters $\epsilon_1$ and
$\epsilon_2$, both with dimensions of mass, correspond to two
isometries rotating two-planes in $\R^4$ according to the splitting
\begin{equation}
\mathbb{R}^4=\mathbb{R}^2_{\epsilon_1}\times\mathbb{R}^2_{\epsilon_2}
\ .
\end{equation}
The low energy dynamics of the resulting theory ${\sf
  T}_{\epsilon_1,\epsilon_2}$ is described by a prepotential
$\cF(a;m;q;\epsilon_1,\epsilon_2)$ which is a deformation of
$\cF_0(a;m;q)$ in the sense that its limit, as
$\epsilon_1$ and $\epsilon_2$ are sent to zero, is the prepotential $\cF_0(a;m;q)$. More precisely, in an expansion in
$\epsilon_1,\epsilon_2$ near zero, the function
$\epsilon_1\,\epsilon_2\,\cF(a;m;q;\epsilon_1,\epsilon_2)$ is analytic and we write
\begin{equation}
\cF(a;m;q;\epsilon_1,\epsilon_2)=\frac{1}{\epsilon_1\,\epsilon_2}\,\cF_0(a;m;q)+\text{terms regular in $\epsilon_1,\epsilon_2$}
\ .
\end{equation}

The gauge theory partition function
$Z^{\rm Nek}(a;m;q;\epsilon_1,\epsilon_2)=\exp \cF(a;m;q;\epsilon_1,\epsilon_2)$
of the theory ${\sf T}_{\epsilon_1,\epsilon_2}$ is sometimes
called the Nekrasov partition function. After twisting the theory with
a Donaldson twist, the Nekrasov partition function may be computed as an equivariant integral over the
moduli space of instantons on $\R^4$. The resulting prepotential $\cF$ can be decomposed into a
classical term plus contributions from one-loop and instanton effects as
\begin{equation}
\cF=\cF_{\text{cl}}+\cF_{1\text{-loop}}+\cF_{\text{inst}} \ .
\end{equation}
The classical contribution is
\begin{equation}
\cF_{\text{cl}}=\frac{1}{\epsilon_1\,\epsilon_2}\,\log q \, \sum_{i=1}^r\, a_i^2 \ ,
\end{equation}
where $r$ is the rank of the gauge algebra $\fg$.
The one-loop contribution is independent of $q$ and may be computed as a product of determinants
of differential operators. The result depends on the particle multiplets involved;
for example, in the case of a hypermultiplet of mass $m$ one finds
\begin{equation}
\cF_{1\text{-loop}}=\frac{1}{\epsilon_1\,\epsilon_2}\,\Big(\frac{1}{2}\,m^2\log
m-\frac{3}{4}\,m^2\Big)+ \text{terms regular in $\epsilon_1,\epsilon_2$} \ .
\end{equation}
Lastly, the instanton contribution may be written as a sum over Young diagrams (for $\fg=\mathfrak{su}(K)$), and it has a power series expansion in the
exponentiated gauge couplings $q=\E^{\,2\pi
  \,\I\,\tau_{\textrm{\tiny UV}}}$ of the form
\begin{equation}
\cF_{\text{inst}}=\sum_{k=1}^{\infty}\, c_k\, q^k \ .
\end{equation}
More about Nekrasov partition functions may be found
in~\cite{tachikawa2016review,
  nekrasov2010quantization,nekrasov2012seiberg,Jaewon:2012wsa,Szabo:2015wua}.

In the following we will be interested in the special case where
$\epsilon_2=0$ while $\epsilon_1=\epsilon$ is kept
finite, which is also known as the {Nekrasov-Shatashvili limit} or the
{$\frac12\Omega$-background}. The
resulting theory ${\sf T}_{\epsilon}={\sf T}_{\epsilon,0}$ 
preserves a two-dimensional $\N=(2,2)$ supersymmetry in the
$\R^2_{34}$-plane.\footnote{\label{fnote:4supercharges} This can be shown by decomposing the
  supercharges of the original theory $\sf T$, which transform in the
  representation $({\boldsymbol 2}^+\oplus{\boldsymbol
    2}^-,{\boldsymbol 2})$ of
  $\mathfrak{so}(4,\C)\times\mathfrak{sl}(2,\C)_R$, under the subalgebra
    $\mathfrak{gl}(1,\C)\times\mathfrak{gl}(1,\C)\times\mathfrak{gl}(1,\C)_R$
    of complexified rotations in the $12$-plane and the $34$-plane,
    and R-symmetry rotations in the $12$-plane; this decomposes into
    eight weight spaces with weights $(\pm\,1,\pm\,1,\pm\,1)$. Four of the resulting
    charges are invariant under the generator
    $\epsilon\,(J^{12}+J^{12}_R)$ of the $\frac12\Omega$-background,
    where $J^{12}$ is a generator of rotations in $\R^2_{12}$ and
    $J^{12}_R$ is an R-symmetry generator, and these generate
    $\N=(2,2)$ supersymmetry in the $34$-plane.} 
    The physics of ${\sf T}_\epsilon$ has been discussed
extensively in the work of Nekrasov and Shatashvili, see in
particular~\cite{nekrasov2010quantization}. It is proposed that in the infrared limit, at energies $E\ll|\epsilon|$, the
theory ${\sf T}_{\epsilon}$ is described by $r$ abelian vector
multiplets, coupled to an {effective twisted superpotential}
$\Weff(\boldsymbol{\Sigma};m;q;\epsilon)$ built from the twisted
chiral superfields $\boldsymbol\Sigma$ in the abelian vector
multiplets. Moreover, if one restricts
$\Weff(\boldsymbol{\Sigma};m;q;\epsilon)$ to the lowest components
$\sigma_i=a_i$ of $\boldsymbol\Sigma$, it was proposed that
\begin{equation}
\Weff(a;m;q;\epsilon)=\lim_{\epsilon_2\rightarrow0}\,\epsilon_2\,\cF(a;m;q;\epsilon_1=\epsilon,\epsilon_2) \ .
\end{equation}
In particular, the theory ${\sf T}_{\epsilon}$ has a discrete set of
vacua determined as solutions to the quantization condition
\begin{equation}
\exp\Big(\frac{\partial\Weff(a;m;q;\epsilon)}{\partial a_i}\Big)=1 \ .
\end{equation}

\begin{example}
The effective twisted superpotential $\Weff$ of the simplest
Argyres-Douglas theory, the AD$_2$ theory from
Example~\ref{ex:AD2BPS}, in the $\frac12\Omega$-background has only a
one-loop contribution from the free hypermultiplet of mass
$m$. Hence\footnote{\label{fn:Weff1loop} The one-loop contribution
  \smash{$\exp\widetilde{\cW}^{\textrm{eff}}_{\textrm{1-loop}}$} may be computed as a
  product of determinants of differential operators. There is a
  certain freedom in its definition due to ambiguities in the regularization
  of divergences, which implies that it is only determined up to a
  phase~\cite{Pestun:2007rz,Vartanov:2013ima}. For a distinguished
  choice of phase, \smash{$\exp\Weff$} can be identified with the square root
  of the product of two Liouville three-point functions in the
  Nekrasov-Shatashvili (or $c\to\infty$)
  limit. The one-loop contributions that we use are computed in this
  ``Liouville scheme''.}
\begin{align}\label{eq:AD2Weff}
\Weff(m;\epsilon) =
  \frac\epsilon2\,\Upsilon\Big(\frac12+\frac{m}{2\,\epsilon}\Big) 
\end{align}
where
\begin{align}
\Upsilon(x) = \int_{\frac12}^x\, \log \frac{\Gamma(x')}{\Gamma(1-x')}\,
  \de x' \ . 
\end{align}
\end{example}

\begin{example}
The effective twisted superpotential $\Weff$ of the pure $SU(2)$
theory, discussed in Section~\ref{sec:SWTheory}, in the
$\frac12\Omega$-background has a classical, a one-loop and an
instanton contribution\footnote{To be precise,
the Nekrasov partition function computes the instanton contributions
to the $U(2)$ gauge theory. It is possible to extract the ``spurious''
$U(1)$ contribution, which does not depend on the Coulomb parameter
$a$, by comparing it to a dual Liouville conformal
block~\cite{Alday:2009aq}, or alternatively by computing the $Sp(1)$
Nekrasov partition function. The latter is
related to the former by a change of ultraviolet regularization
scheme~\cite{Hollands:2010xa}.} as a series expansion in powers of the
ultraviolet scale parameter $\Lambda$ (which replaces the instanton
coupling $q$ in asymptotically free theories). Explicitly
\begin{align}\label{eq:pureWeff}
\widetilde{\cW}^{\rm eff}_{\rm cl}(a;\Lambda;\epsilon) &=
                                                         \frac{a^2}{\epsilon}\,\log\Big(\frac\Lambda\epsilon\Big)
                                                         \ , \nonumber
  \\[4pt]
  \widetilde{\cW}^{\rm eff}_{\textrm{1-loop}}(a;\epsilon) &=
                                                              - \frac\epsilon2\,\Upsilon\Big(-\frac
                                                            a\epsilon\Big)
                                                               -
                                                               \frac\epsilon2\,\Upsilon\Big(\frac
                                                            a\epsilon\Big)
                                                               \ ,
                                                               \nonumber
  \\[4pt]
  \widetilde{\cW}^{\rm eff}_{\rm inst}(a;\Lambda;\epsilon) &=
                                         \frac2{\epsilon\,(a^2-\epsilon^2)}
                                         \, \Lambda^4 + \frac{5\,a^2 + 7\,\epsilon^2}{\epsilon\,(a^2-4\,\epsilon^2)\,(a^2-\epsilon^2)^3} \, \Lambda^8 + O(\Lambda^{12}) \ .
\end{align}
As shown by~\cite{Mironov:2009uv}, the parameter $a$ in \eqref{eq:pureWeff} is not proportional to the  classical period \smash{$\Pi_A^{(0)} = \oint_A\,\lambda$}, but instead receives $\epsilon$-corrections. As we will see later, the $\epsilon$-expansion for $a$ may be obtained by computing a quantum period $\Pi_A(\epsilon)$ which is defined in \eqref{eq:quantumperiod}, while the quantum period $\Pi_B(\epsilon)$ computes the $\epsilon$-expansion for \smash{$\Weff$}.\footnote{More precisely, if we define the $\epsilon$-expansion for $a$ by $\frac{\pi\,\I}\epsilon\,a = \Pi_A(\epsilon)$, then the $\epsilon$-expansion for \smash{$\Weff(a;\Lambda;\epsilon)$} is obtained from the equation \smash{$\Pi_B(\epsilon) = 2 \, \partial_a\Weff(a;\Lambda;\epsilon)$}.}
\end{example}

Although the Nekrasov
partition function can only be
computed from first principles for $\N=2$ field theories with a
Lagrangian description, one expects that a similar object can be
defined for any $\N=2$ theory and, in particular, for any $\N=2$
theory of class $\cS$.
For attempts to
compute this object from either string theory, five dimensions or
conformal field theory, see
e.g.~\cite{Huang:2011qx,Aganagic:2013tta,Aganagic:2014oia,Bao:2013pwa,Mitev:2014isa,Mitev:2014jza,Coman:2017qgv}. 

\begin{example}
The effective twisted superpotential $\Weff$ of the intrinsically strongly coupled ``non-Lagrangian''
$E_6$ Minahan-Nemeschansky theory, discussed in
Section~\ref{ClassSTheories}, in the $\frac12\Omega$-background is (to
our knowledge) so far unknown in explicit form. 
We will return to this open problem in 
Section~\ref{ex:recipeE6} below.
\end{example}

\subsection{Hitchin Systems}
\label{sec:Hitchin}
\noindent
Let us now introduce the Hitchin moduli space~\cite{Hitchin:1986vp}. We start by explaining how
four-dimensional $\N=2$ field theories are related to integrable
systems (more details can be found in the review~\cite{neitzke2013hitchin}). Of particular
relevance is the notion of a (classical) algebraic integrable
system. This is a triple $(\scrI,\Omega,\scrB)$, where
$(\scrI,\Omega)$ is a complex $2r$-dimensional holomorphic symplectic
manifold and $\scrB$ is a complex $r$-dimensional manifold, together
with a holomorphic fibration $\pi:\scrI\to\scrB$ whose generic fibers
$\scrI_u=\pi^{-1}(u)$ are polarized abelian varieties of complex
dimension $r$.

Recall that the low energy description of any $\N=2$ theory is encoded in
the auxiliary Seiberg-Witten geometry $(\Sigma,\lambda)$.
Soon after the discoveries of Seiberg and Witten, it was realized that
this description can be formulated mathematically as an
algebraic integrable system of dimension $r$ given by the rank of the
gauge algebra~\cite{Donagi:1995cf}. Its base manifold $\scrB$ is the
Coulomb branch of the $\N=2$ theory and its fibers are the complex
tori $\C^r/\Gamma_u$, where $\Gamma_u=H_1(\overline{\Sigma}_u,\Z)$ is the charge
lattice at the vacuum $u$, polarized by the choice of $A$-cycles and
$B$-cycles on $\overline{\Sigma}_u$ (the electric-magnetic splitting). 
Even more concretely, it was found for instance that the pure
$SU(K)$ gauge theory corresponds to the periodic $A_{K-1}$ Toda
chain~\cite{Gorsky:1995zq}.

However, so far only the base $\scrB$ of the algebraic integrable
system has a description in terms of the fields of the $\N=2$
theory. The complete picture emerges if we compactify the
four-dimensional $\N=2$ theory on a circe $S^1$ of radius
$R$~\cite{Seiberg:1996nz}. At low energies $E\ll R^{-1}$, the
resulting three-dimensional field theory ${\sf T}[R]$ can be described
in terms of $r$ complex scalars and $r$ abelian gauge fields in three
dimensions, together with $2r$ periodic real scalars. The latter are
the holonomies of the $r$ abelian and $r$ dualized abelian gauge
fields in four dimensions around the compactified direction $S^1$. The
complex scalars parameterize a sigma-model into $\scrB$, and we can
think of the $2r$ periodic real scalars as giving a map into a complex
$r$-dimensional torus. Taking into account electric-magnetic
duality transformations, we may conclude that ${\sf T}[R]$ is a
three-dimensional sigma-model whose target $\scrM$ is diffeomorphic
to the total space $\scrI$ of the algebraic integrable system.

Invariance under three-dimensional $\N=4$
supersymmetry demands that the target space $\scrM$ of the sigma-model
is then a {hyperk\"ahler} manifold.\footnote{By compactifying a
  four-dimensional $\N=2$ theory to three dimensions, the eight real
  supercharges that make up the $\N=2$ supersymmetry algebra in four
  dimensions are rearranged into an $\N=4$ supersymmetry algebra in three
  dimensions, as the respective spin group changes from
  $SL(2,\mathbb{C})$ to $SL(2,\mathbb{R})$. This is easiest to see by
  taking the compactified direction to be the $x_2$-axis. With the
  standard parametrization of $SL(2,\mathbb{C})$ matrices in terms of
  coordinates of $\R^{3,1}$, we then get real instead of complex
  matrices.} This is a \kahler manifold which has a triple of complex
structures $(I, J, K)$ with respect to each of which its metric $g$ is
\kahler. The complex structures $(I,J,K)$ satisfy the quaternion
algebra
\begin{align}
  I^2=J^2=K^2=-1=I\,J\,K \ .
\end{align}
With the \kahler forms $\omega_I$, $\omega_J$ and $\omega_K$ of
$\scrM$ one can construct holomorphic symplectic forms $\Omega_I$,
$\Omega_J$ and $\Omega_K$ given by 
\begin{equation}
\Omega_I(\cdot,\cdot)=g\big((J+\I\,K)\,\cdot,\cdot\big)=\omega_J(\cdot,\cdot)+\I\,\omega_K(\cdot,\cdot) \ ,
\end{equation}
and its cyclic permutations in $(I,J,K)$. One of these complex
structures, say $I$,  is distinguished since the target space
$\scrM^I$, regarded as a holomorphic symplectic manifold with respect
to this complex structure, is biholomorphic to the integrable system
$(\scrI,\Omega)$. For later use  we note that, given a \hk structure, we can use the twistor prescription to construct a family of complex structures $J_{\boldsymbol{r}}$ by combining
\begin{equation}
J_{\boldsymbol{r}}=r_1\, I+r_2\, J+ r_3\, K \qquad \mbox{for} \quad \boldsymbol{r}\in S^2\hookrightarrow\mathbb{R}^3 \ .
\end{equation}
It is customary to identify $S^2\!\cong\!\PP^1$ and parametrize it by
complex numbers $\zeta$. The \hk manifold $\scrM$ is \kahler with
respect to the family $J_{\zeta}$ and we can define a corresponding
family of holomorphic symplectic forms $\Omega_{\zeta}$ parameterized
by $\zeta\in\PP^1$.

For any $\N=2$ theory ${\sf T}={\sf T}_K[C,\cD]$ of class $\cS$, the
resulting three-dimensional theory ${\sf T}[R]$ is a three-dimensional
sigma-model into the moduli space $\scrM_{\textrm{\tiny H}}(R)$ of solutions to the
{Hitchin equations} on the Riemann surface
$C$~\cite{Cherkis:2000cj,gaiotto2013wall}. This is the moduli space of
solutions $(A,\varphi)$ to the equations 
\begin{align}
F_A+R^2\,[\varphi,\overline{\varphi}]=0 \ ,\nonumber\\[4pt]
\overline{\partial}_A\varphi=0 \ ,\nonumber\\[4pt]
\partial_A\overline{\varphi}=0 \ ,
\end{align}
where $A$ is a connection on an $SU(K)$-bundle $E\rightarrow C$ with
curvature $F_A$, and $\varphi$ is a
{Higgs field} which is a holomorphic one-form
$\varphi\in\Omega^{1,0}(C,\text{End}\,E)$ with suitable boundary
conditions at the punctures of $C$.\footnote{A rigorous discussion of
  these boundary conditions for Hitchin's equations and the
  construction of their \hk moduli spaces has been given in~\cite{Konno,Nakajima}
  for regular singularities, and in~\cite{Biquard} for irregular
  singularities. For our purposes the discussion of boundary
  conditions on flat $SL(K,\C)$ connections in Section~\ref{WKBSNs}
  (and in Section~\ref{sec:WKB} below) is
  sufficient.} The holomorphic structure is defined by the twisted
Dolbeault differential $\overline{\partial}_A$, and $\overline{\varphi}$ denotes the
Hermitean conjugate of $\varphi$.

Indeed, the $\N=2$ theory $\sf T$
can be obtained from the six-dimensional $(2,0)$-theory $\fX[K]$ by
twisted compactification on $C$. If we reverse the order of
compactifications, by first compactifying on $S^1$, the resulting
theory in the infrared limit is the five-dimensional supersymmetric
Yang-Mills theory with gauge algebra
$\mathfrak{su}(K)$. Compactifying this theory on $C$, with a partial
twist on $C$ that changes the internal field into a complex one-form
$\varphi$, gives an effective description of the three-dimensional
theory ${\sf T}[R]$ as an $\N=4$ sigma-model into the moduli space of
vacuum configurations of five-dimensional supersymmetric Yang-Mills
theory on $C\times\mathbb{R}^{2,1}$ that are translation invariant in the non-compact directions. 
These are precisely the solutions to the {Hitchin equations}. The
Hitchin equations can similarly be obtained by dimensional reduction of the self-dual
Yang-Mills equations in four dimensions to equations in two
dimensions~\cite{Harvey:1995tg,Bershadsky:1995vm}.

The Hitchin moduli space $\scrM_{\textrm{\tiny H}}(R)$ is well-known to be a \hk
manifold, with a family of complex structures $J_\zeta$ for
$\zeta\in\PP^1$. Its complex geometry depends crucially on whether
$\zeta\in\{0,\infty\}$ or $\zeta\in\C^\times$.

Let us start with $\zeta=0$, when $(\scrM_{\textrm{\tiny H}},J_0)$ is
biholomorphic to the moduli space $\Mdol(C)$ of stable Higgs bundles
$(V,\varphi,\overline{\partial}_A)$ on $C$. A {Higgs bundle} is a
holomorphic vector bundle $(V,\overline{\partial}_A)$ of rank $K$ on $C$ together
with a Higgs field $\varphi$, which is a holomorphic one-form on $C$
taking values in the endomorphism bundle ${\rm End}\,V$. 
The moduli space $\Mdol(C)$ has the structure of an algebraic
integrable system, where the projection $\pi:\Mdol(C)\rightarrow\scrB$
is defined by taking the characteristic polynomial of $\varphi$. Any
point $u\in\scrB$ thus defines an algebraic curve $\Sigma\subset T^*C$
as
\begin{equation}\label{Hitfib}
\det(w-\varphi)=w^K - \sum_{k=2}^K\,p_k\,w^{K-k}=0 \ ,
\end{equation}
which is the same as the spectral curve \eqref{eq:SWcurve}. The
Coulomb branch $\scrB$ of any $\N=2$ field theory of class $\cS$ may
therefore be identified with the space of $K-1$-tuples
$(p_2,\dots,p_K)$ of meromorphic $k$-differentials on
$C$, as asserted in Section~\ref{ClassSTheories}. The differentials
$p_k$ may be thought of as a maximal set of algebraically
independent commuting Hamiltonians of the integrable
system. Similarly, at $\zeta=\infty$ the Hitchin moduli space is
biholomorphic to $\overline{\Mdol(C)}$.

For any $\zeta\in\C^\times$, it is useful to consider the
complex-valued connection 
\begin{equation}\label{eq:flatfamily}
\cA=\frac{R}{\zeta}\,\varphi+A+R\,\zeta\, \overline{\varphi} \ .
\end{equation}
Since the Hitchin equations are equivalent to the flatness of $\cA$, the Hitchin moduli
space $(\scrM_{\textrm{\tiny H}},J_\zeta)$ for $\zeta\in\C^\times$ may be
identified with de~Rham
moduli space $\scrM_{\rm flat}(C,SL(K,\C))$ of flat $SL(K,\C)$
connections on $C$, with natural holomorphic symplectic form $\Omega_\zeta$ given by \eqref{eq:symplecticform}. 
These equivalences are made precise by the
non-abelian Hodge correspondence, which gives a diffeomorphism between
$\Mdol(C)$ and $\scrM_{\rm
  flat}(C,SL(K,\C))$~\cite{Donaldson1983,Hitchin:1986vp,Simpson1992}. 

The different descriptions of $(\scrM_{\textrm{\tiny H}}, J_\zeta)$ for
$\zeta\in\C$ may be combined into a new notion known as a flat
{$\lambda$-connection} (see e.g.~\cite{Simpson}). This is an
object $\nabla_\lambda$
that interpolates between a Higgs bundle and a flat
connection. Instead of the usual Leibniz rule
\begin{align}
\nabla(f\, s) = \de f\otimes s + f\,\nabla s
\end{align}
obeyed by a connection $\nabla$ acting on the
multiplication of a section $s$ by a function $f$, it obeys the more
general Leibniz rule 
\begin{align}
\nabla_\lambda(f\, s) = \lambda\,\de f\otimes s + f\, \nabla_\lambda s
\end{align}
for $\lambda\in\C$. In the present case, multiplying
\eqref{eq:flatfamily} by $\epsilon=\zeta/R$ gives a flat
$\epsilon$-connection, which interpolates between a Higgs field
$\varphi$ for $\epsilon=0$ and a flat connection for
$\epsilon=1$.\footnote{What we refer to here as a flat
  $\lambda$-connection is usually called a holomorphic
  $\lambda$-connection. However, any flat connection $\nabla$ is
  holomorphic in the complex structure induced by its $(0,1)$ part. In
particular, the flat $\epsilon$-connection $\epsilon\,\cA$ corresponds
to the holomorphic $\epsilon$-connection
$D_\epsilon=\epsilon\,\partial_A+\varphi$ in the complex structure
$\overline{\partial}=\overline{\partial}_A +
  R^2\,\epsilon\,\overline{\varphi}$.}

\subsection{Opers}\label{sec:opers}
\noindent
The moduli space of flat $\epsilon$-connections $\scrM_{\rm
  flat}^\epsilon(C,SL(K,\C))$ has, for $\epsilon\neq0$, a
distinguished holomorphic Lagrangian submanifold \smash{$\scrL^{\rm
  oper}_\epsilon\subset\scrM_{\rm
  flat}^\epsilon(C,SL(K,\C))$}, which is known as the
{brane of $\epsilon$-opers}~\cite{BD2005}. They feature
mathematically in the geometric Langlands
program~\cite{Frenkel:2005pa} and its gauge theory
interpretation~\cite{kapustin2006electric}. They also play an
important role in the non-abelian Hodge
correspondence~\cite{dumitrescu2016opers,dumitrescu2016journey},
following a conjecture by Gaiotto~\cite{gaiotto2014opers}. This
conjecture states that in the ``conformal limit'' $R,\zeta\to0$ with
$\zeta/R=\epsilon$ fixed, the $\epsilon$-connection $\epsilon\,\cA$
reduces to an $\epsilon$-oper.

Since $\epsilon$-opers are also a fundamental ingredient in the NRS
conjecture, let us now give a concrete characterization of them in the
simplest case of flat $SL(2,\C)$ connections. In this case
$\epsilon$-opers are also known as Schr\"odinger operators.

An $SL(2,\C)$ $\epsilon$-oper is defined locally by a second order
ordinary differential equation 
\begin{equation}\label{operdef}
{\rm D}_\epsilon \psi(z) = \epsilon^2\,\psi''(z)+q_2(z)\,\psi(z)=0 \ ,
\end{equation}
where $\psi$ is a $(-\frac12)$-differential on $C$. That is, the operator
${\rm D}_\epsilon$ is a bundle map ${\rm D}_\epsilon:K_C^{-1/2}\to
K_C^{3/2}$, where $K_C$ is the canonical line bundle on $C$. This
means that ${\rm D}_\epsilon$ is not automatically well-defined as a
global object on $C$. Let us consider what happens to the differential
equation under a holomorphic change of coordinates.

Under an arbitrary holomorphic coordinate change $z\mapsto z(w)$, the
$(-\frac12)$-differential $\psi$ transforms into
\begin{equation}\label{halfdiff}
\tilde{\psi}(w)=\psi(z(w))\Big(\frac{\de z}{\de w}\Big)^{-1/2} \ .
\end{equation}
This implies that
\begin{equation}
\epsilon^2\,\tilde{\psi}''(w)=z'(w)^{3/2}\,\Big(\epsilon^2\,\psi''(z(w))-\frac{\epsilon^2}{2}\,\{\!\!\{w,z\}\!\!\}\,\psi(z(w))\Big) \ ,
\end{equation}
where the bracket $\{\!\!\{\cdot,\cdot\}\!\!\}$ denotes the {Schwarzian derivative}
\begin{equation}
\{\!\!\{w,z\}\!\!\}=\frac{w'''(z)}{w'(z)}-\frac{3}{2}\,\Big(\frac{w''(z)}{w'(z)}\Big)^2 = -\frac{\{\!\!\{z,w\}\!\!\}}{z'(w)^2} \ .
\end{equation}
Inserting these transformation laws into (\ref{operdef}), we find that
the differential equation changes into
\begin{equation}
z'(w)^{3/2}\,\Big(\epsilon^2\,\psi''(z(w))-\frac{\epsilon^2}{2}\,\{\!\!\{w,z\}\!\!\}\,\psi(z(w))+z'(w)^{-2}\,\tilde{q}_2(w)\,\psi(z(w))\Big)=0
\ , 
\end{equation}
if $q_2(z)$ would simply transform into $\tilde q_2(w)$.

To ensure
that the oper ${\rm D}_\epsilon$ remains invariant under arbitrary
holomorphic coordinate transformations, we must require that $q_2(z)$
transforms as 
\begin{equation}\label{eq:t2ztransform}
q_2(z)\longmapsto\tilde{q}_2(w)=z'(w)^2\, q_2(z(w))+\frac{\epsilon^2}{2}\,\{\!\!\{z,w\}\!\!\} \ .
\end{equation}
In other words, the coefficient function $q_2$ should transform as
what is called a {projective connection} on $C$.

Another possibility would be to restrict to a coordinate atlas for
which the transition functions are M\"obius transformations
\begin{equation}
z\longmapsto\frac{a\,z+b}{c\,z+d} \qquad \mbox{with} \quad \begin{pmatrix} a & b \\ c & d\end{pmatrix}\in SL(2,\C) \ .
\end{equation}
Under such a transformation, the Schwarzian derivative vanishes,
$\{\!\!\{w,z\}\!\!\}=0$. Thus if we
restrict to such an atlas, i.e. we fix a {projective structure}
on $C$, the oper ${\rm D}_\epsilon$ is globally defined again, and
$q_2(z)$ simply transforms as a quadratic differential.

The oper ${\rm D}_\epsilon$ becomes the spectral curve
\begin{align}
  \Sigma: \quad w^2-p_2(z) = 0 
\end{align}
in the classical limit $\epsilon\to0$ where the projective connection
$q_2$ becomes a quadratic differential $-p_2$, when we replace
$\epsilon\,\partial_z$ with the momentum $w$. It is therefore 
known as a {quantum curve}. The opers ${\rm D}_\epsilon$ may also
be thought of as the quantum Hamiltonians of the Hitchin integrable
system (see e.g.~\cite{FeiginFrenkel}).

\begin{example} \label{ex:operharmonic}
Let $C=\C$ with an irregular singularity at $z=\infty$ (a pole of
$p_2$ of order six, i.e. of type $L=4$),
as in Examples~\ref{ex:CAD2}, \ref{ex:AD2abelianisation},
\ref{ex:AD2spectralcoordinates} and~\ref{ex:AD2BPS}. The brane
of $\epsilon$-opers $\scrL^{\rm oper}_\epsilon$ on $C$ consists of a
single point, given by the Schr\"odinger equation
\begin{equation}
{\rm D}_{\epsilon} \psi(z) = \epsilon^2\, \psi''(z) - \big(z^2+m\big)\,\psi(z) = 0 \ .
\end{equation}
The oper ${\rm D}_\epsilon$ reduces in the classical limit
$\epsilon\to0$ to the spectral curve
\begin{equation}
\Sigma\colon \quad w^2 = z^2 + m \ .
\end{equation}
Fixing the $SL(2,\C)$ gauge symmetry at infinity, as we did in
Example~\ref{ex:AD2spectralcoordinates}, turns $m$ into a parameter on
a complex one-dimensional brane of opers. In the dual Argyres-Douglas
theory, this corresponds to gauging the $SU(2)$ flavour symmetry of
the hypermultiplet.
\end{example}

\begin{example}\label{ex:operpure}
Let $C=\C^\times$ with irregular singularities of type $L=1$ at $z=0$
and $z=\infty$, as in Examples~\ref{ex:purenetworks}, \ref{ex:pureab},
\ref{ex:FNpureab}, \ref{ex:purenonab} and~\ref{ex:PureSU2BPS}. In this case there is a
complex one-dimensional family of opers parametrized by the equation
\begin{align}
{\rm D}_\epsilon\psi(z) = \epsilon^2\,\psi''(z) -
  \Big(\frac{\Lambda^2}{z^3}-\frac{2\,u+\epsilon^2/4}{z^2} +
  \frac{\Lambda^2}z\Big)\,\psi(z) = 0 \ .
\end{align}
This equation is equivalent to the Mathieu differential equation with parameters $2\,u$ and $\Lambda^2/2$
(after the coordinate transformation $z=\E^{\,\I\,x}$ and the
redefinition given by $\psi(z)=(\I\,z)^{1/2}\,\tilde\psi(x)$).\footnote{The parametrization of the term $2\,u+\epsilon^2/4$ is chosen for convenience, it is not relevant in the following.} In the classical
limit $\epsilon\to0$, the oper ${\rm D}_\epsilon$ reduces to the
spectral curve
\begin{align}
\Sigma: \quad w^2 = \frac{\Lambda^2}{z^3} - \frac{2\,u}{z^2} + \frac{\Lambda^2}z \ .
\end{align}
\end{example}

\begin{example}
Let $C=\PP^1_{0,1,\infty}$ with three regular punctures at
$z=0,1,\infty$, which corresponds to the pair of pants building block
that we discussed in Section \ref{ClassSTheories}. For a fixed
choice of residues $\pm\,m_l/2$, the three-punctured sphere admits the
unique quadratic differential
\begin{equation}
p_2(z) =
\frac{m^2_0}{4\,z^2}+\frac{m^2_1}{4\,(z-1)^2}+\frac{m^2_{\infty}-m^2_0-m^2_1}{4\,z\,(z-1)}
\end{equation}
with at most second order poles at all punctures. Similarly, $C$
admits a unique $SL(2,\C)$ $\epsilon$-oper given by
\begin{equation}\label{hypergeometric}
{\rm D}_{\epsilon}\psi(z)=\epsilon^2\, \psi''(z)+\Big(\frac{\delta_0}{z^2}+\frac{\delta_1}{(z-1)^2}+\frac{\delta_{\infty}-\delta_0-\delta_1}{z\,(z-1)}\Big)\,\psi(z)=0 \ ,
\end{equation}
where
\begin{equation}
\delta_l=\frac{\epsilon^2-m_l^2}{4} \ .
\end{equation}
This oper is equivalent (after a simple and standard transformation)
to the classical Gauss hypergeometric differential equation.
\end{example}

We may turn the equation (\ref{operdef}) into the first order
differential equation 
\begin{equation}\label{noperdef}
\nabla^{\rm oper}_\epsilon \Psi(z)=\epsilon\,\frac{\de \Psi(z)}{\de z}\,\de z+A_z\,\de z\, \Psi(z)=0
\end{equation}
where the $1$-jet $\Psi$ and connection coefficient $A_z$ are given by
\begin{equation}\label{operconnection}
\Psi(z)=\begin{pmatrix} -\epsilon\,\psi'(z)\\\psi(z)\end{pmatrix} \qquad \text{and} \qquad A_z=\begin{pmatrix}0 & -q_2(z)\\ 1 & 0\end{pmatrix} \ .
\end{equation}
This defines the oper ${\rm D}_\epsilon$ locally as a flat $SL(2,\C)$
$\epsilon$-connection $\nabla_\epsilon^{\rm oper}$. Under a
holomorphic change of coordinates $z\mapsto z(w)$, the $1$-jet $\Psi$
transforms to
\begin{align}
\tilde\Psi(w) = {\sf G}^{-1}(w)\,\Psi(z(w)) = \bigg(\begin{matrix}
  g(w)^{-1} & -\epsilon\,g'(w) \\ 0 & g(w) \end{matrix}\bigg)
                                      \Psi(z(w)) \ ,
\end{align}
where $g(w)=(z'(w))^{-1/2}$, which obeys
\begin{equation}
\frac{\de\tilde{\Psi}(w)}{\de w}+\tilde{A}_w\, \tilde{\Psi}(w)=0
\end{equation}
with the new connection again of the form
\begin{equation}
\tilde A = \tilde{A}_w\,\de w= {\sf G}^{-1}\,\de{\sf G} + {\sf
  G}^{-1}\,A\,{\sf G} = \begin{pmatrix}0 & -\tilde
  q_2(w) \\ 1 &
  0\end{pmatrix}\, \de w \ .
\end{equation}
Hence the $SL(2,\C)$ $\epsilon$-oper defined locally by
(\ref{operdef}) is equivalent to the flat $SL(2,\C)$
$\epsilon$-connection $\nabla^{\rm oper}_\epsilon$ defined locally by
\eqref{noperdef}.

In a global description, the form of the transition functions ${\sf
  G}$ imply that $\nabla_\epsilon^{\rm oper}$ is a connection on the
rank two vector bundle $V_\epsilon$ defined as the unique
extension~\cite{Gunning}
\begin{align}
0\longrightarrow K_C^{1/2}\longrightarrow V_\epsilon\longrightarrow
  K_C^{-1/2}\longrightarrow 0 \ .
\end{align}
Rescaling the extension class gives an isomorphic bundle, so 
the bundles $V_\epsilon$ for all $\epsilon\in\C^\times$ are in fact
isomorphic. In the classical limit $\epsilon\to0$, the transition
functions ${\sf G}$ become diagonal, so that $(V_\epsilon,\nabla^{\rm
  oper}_\epsilon)$ degenerates to a Higgs bundle $(V_0,\varphi)$ with
rank two vector bundle
$V_0=K_C^{1/2}\oplus K_C^{-1/2}$ and Higgs field
\begin{align}
\varphi = \bigg(\begin{matrix} 0 & -p_2 \\ 1 & 0\end{matrix}\bigg) \ .
\end{align}
That is, the brane of
$\epsilon$-opers $\scrL^{\rm oper}_\epsilon$ is a quantization of the
Hitchin section $s_0:\scrB\to(\scrM_{\textrm{\tiny H}}, J_0)$, which is 
defined precisely by sending the quadratic differential $p_2$ to the Higgs
bundle $(V_0,\varphi)$ as above and embeds the Coulomb branch $\scrB$
in the integrable system $(\scrM_{\textrm{\tiny H}}, J_0)$.

This whole story generalizes rather straightforwardly to higher rank
gauge groups. For example, an $SL(K,\C)$ $\epsilon$-oper on $C$ corresponds to a choice
 of projective structure on $C$ together with $K-1$ meromorphic
 $k$-differentials $p=(p_2,\dots,p_K)$, which is equivalent to a flat
 $SL(K,\C)$ $\epsilon$-connection defined locally by a linear ordinary
 differential equation of order $K$. This is spelled out in detail for
 $K=3$ in~\cite{hollands2018higher}, where it is also shown how to
 obtain the brane of $\epsilon$-opers for a surface $C$ with
 non-maximal punctures.

\begin{example}\label{ex:K3oper}
Let $K=3$ and $C=\PP^1_{1,\omega,\omega^2}$ with three maximal singularities at
the third roots of unity \smash{$z=\omega^l$}, with $l=0,1,2$ and
\smash{$\omega=\E^{\,2\pi\,\I/3}$}, corresponding to the $E_6$
Minahan-Nemeschansky theory described in
Section~\ref{ClassSTheories}. The complex one-dimensional brane of
$\epsilon$-opers for this geometry is parametrized by the differential
equation
\begin{align}
{\rm D}_\epsilon\psi(z) = \epsilon^3\,\psi'''(z) +
  \epsilon\,q_2(z)\,\psi'(z) + q_3(z)\,\psi(z)=0 \ ,
\end{align}
with
\begin{align}
q_2(z) = \frac{9\,z}{(z^3-1)^2} \qquad \mbox{and} \qquad q_3(z) =
  \frac u{(z^3-1)^2} + \frac\epsilon2 \, q_2'(z) \ .
\end{align}
This is called the $T_3$-equation in~\cite{hollands2019exact}. In the
classical limit $\epsilon\to0$, this brane of $\epsilon$-opers reduces
to the Coulomb branch of the (massless) $E_6$ Minahan-Nemeschansky
theory. 
\end{example}

Curiously, in all examples that we have studied, opers corresponding
to building blocks of Lagrangian $\N=2$ field theories are defined by
well-known differential equations whose solutions are given by special
functions, whereas opers corresponding to building blocks of
non-Lagrangian $\N=2$ theories are rather unfamiliar and are not
similarly characterized.

\subsection{Exact WKB Analysis}\label{sec:WKB}
\noindent
Recall that we constructed Darboux coordinates on
the moduli space $\scrM_{\rm flat}(C,SL(K,\C))$ of flat $SL(K,\C)$
connections $\nabla$ on a (punctured) Riemann surface $C$ in Section~\ref{sec:SpectralCoords}. This is
easily generalized to the moduli space $\scrM_{\rm
  flat}^\epsilon(C,SL(K,\C))$ of flat $\epsilon$-connections
$\nabla_\epsilon$ (where $\epsilon\in\C^\times$), as multiplying
$\nabla_\epsilon$ by $\epsilon^{-1}$ gives an ordinary flat
connection on $C$. Since any $\epsilon$-oper
${\rm D}_\epsilon$ is in particular a flat $SL(K,\C)$ $\epsilon$-connection
$\nabla^{\rm oper}_\epsilon$, it is natural to wonder how to
characterize these spectral coordinates on the brane of
$\epsilon$-opers.

Fix an $\epsilon$-oper ${\rm D}_\epsilon$. Consider the WKB spectral network $\scrW(u,\vartheta)$ defined
by $K-1$ meromorphic $k$-differentials $u=(p_2,\dots,p_K)$,
obtained in the classical limit $\epsilon\to0$ from $\nabla^{\rm
  oper}_\epsilon$, and choose the phase $\vartheta={\rm
  arg}\,\epsilon$. Then there is a distinguished
$\scrW(u,\vartheta)$-abelianization which is determined by the
{exact WKB method}, see~\cite{hollands2019exact}
and~\cite[Section~11]{hollands2018higher}. 

Let us explain this point in more detail. The exact WKB
method is a scheme for studying the monodromy (or bound states, or more
generally Stokes data) of ordinary linear differential equations;
see~\cite{Aoki:1993ra,iwaki2014exact,Takeibook} for some nice
reviews.
Here we focus on the case $K=2$
to simplify the notation. In this case the
exact WKB method starts with the set-up at the beginning of
Section~\ref{sec:opers}, that is, a holomorphic Schr\"odinger equation
\begin{align}\label{eq:Schrodinger}
{\rm D}_\epsilon\psi(z) = \epsilon^2\,\psi''(z) +
  q_2(z;\epsilon)\,\psi(z) = 0
\end{align}
on a (punctured) Riemann surface $C$, where $\psi(z)$ is a section of
$K_C^{-1/2}$, and we have made the dependence of $q_2$ on $\epsilon$
explicit. Central to the method are exact \emph{local} solutions to
\eqref{eq:Schrodinger} of the form
\begin{align}
\psi(z) = \exp\Big(\frac1\epsilon \, \int_{z_0}^z\, S(z';\epsilon) \, \de z'\Big)
  \ ,
\end{align}
where $z$ is a local coordinate in a contractible open subset
$U\subset C$ and $z_0\in U$. This implies that $S(z;\epsilon)$ obeys the
Riccati equation
\begin{align}\label{eq:Riccati}
\epsilon\,\partial_zS + S^2 + q_2 = 0 \ .
\end{align}

The first step in constructing $S$ is to develop a formal series
solution $S^{\rm for}$ in powers of $\epsilon$. The leading order
in $\epsilon$ of the differential equation \eqref{eq:Riccati} is
\begin{align}
w^2 + p_2 = 0 \,
\end{align}
which defines a branched double covering $\Sigma$ over $C$. On sheet
$i=1,2$ of $\Sigma$, the higher order expansion of $S_i^{\rm for}$
is then uniquely determined by \eqref{eq:Riccati} and takes the form
\begin{align}
S_i^{\rm for} (z;\epsilon) = w_i + \sum_{n=1}^\infty\, \epsilon^n
  \, S_n(z) \ .
\end{align}
This series is not convergent in general. However, using a technique called {Borel
  resummation} it can be given an analytic meaning.

\begin{example}
  The logarithm of the gamma-function has the asymptotic expansion
  \begin{align}
\log \Gamma\Big(\frac m\epsilon\Big) = \frac m\epsilon \log\frac
    m\epsilon - \frac m\epsilon-\frac12\log\frac m{2\pi\,\epsilon} +
    \sum_{g=1}^\infty \, \frac{B_{2g}}{2g\,(2g-1)} \,
    \Big(\frac\epsilon m\Big)^{2g-1}
  \end{align}
when $\epsilon\to0$, where $B_{2g}$ are the Bernoulli numbers. This is
a divergent series, since $B_{2g}$ grows factorially as $(2g)!$. The
Borel transform of the series
\begin{align}\label{eq:fx}
f(t) = \sum_{g=1}^\infty\, \frac{B_{2g}}{2g\,(2g-1)} \, t^{2g-1}
\end{align}
is defined as
\begin{align}
\cB[f](s) = \sum_{g=1}^\infty\, \frac{B_{2g}}{(2g-1)\,(2g)!}\,
  s^{2g-1} \ .
\end{align}
The Borel sum of $f$ in the direction $\vartheta$ is then
\begin{align}
\cS_\vartheta f(t) &= \int_0^{\infty\,\E^{\,\I\,\vartheta}} \, \de s \ \E^{-s}
                     \, \cB[f](t\,s) \nonumber \\[4pt]
  &= \int_0^{\infty\,\E^{\,\I\,\vartheta}} \, \de s \ 
  \E^{-s} \ \sum_{g=1}^\infty \, \frac{B_{2g}}{(2g-1)\,(2g)!}\,
  (t\,s)^{2g-1}  \ .
\label{eq:cSfx}\end{align}
Since $\cB[f](s)$ has no singularity along the real $s$-axis we can
choose $\vartheta=0$. The integral over $s$ then reproduces the series
\eqref{eq:fx} since
\begin{align}
\int_0^\infty\, \de s \ \E^{-s} \, s^{2g-1} = (2g-1)! \ .
\end{align}

Instead, consider the integral representation of the Bernoulli numbers
\begin{align}
B_{2g} = 4g\,(-1)^{g-1} \, \int_0^\infty \, \de z \ 
  \frac{z^{2g-1}}{\E^{\,2\pi\,z}-1} 
\end{align}
for $g\geq1$. Substituting this into \eqref{eq:cSfx} yields
\begin{align}
\cS_0f(t) &= 2 \, \int_0^\infty \, \de s \ \E^{-s} \ \sum_{g=1}^\infty
  \, \frac{(-1)^{g-1}}{(2g-1)\,(2g-1)!} \ \int_0^\infty \, \de z \
            \frac{(t\,s\,z)^{2g-1}}{\E^{\,2\pi\,z}-1} \nonumber \\[4pt]
  &= 2 \, \int_0^\infty \, \frac{\de s}t \ \E^{-s/t} \ \sum_{g=1}^\infty
  \, \frac{(-1)^{g-1}}{(2g-1)\,(2g-1)!} \ \int_0^\infty \, \de z \
    \frac{(s\,z)^{2g-1}}{\E^{\,2\pi\,z}-1}  \nonumber \\[4pt]
  &= 2 \, \int_0^\infty \, \de s \ \E^{-s/t} \ \sum_{g=1}^\infty \,
    \frac{(-1)^{g-1}\,s^{2g-2}}{(2g-1)!} \ \int_0^\infty \, \de z \
    \frac{z^{2g-1}}{\E^{\,2\pi\,z}-1} \ , 
\end{align}
where in the last equality we integrated by parts over $s$. 
Now using
\begin{align}
\frac1{\E^{\,s}-1} -\frac1s +\frac12 = 2 \, \sum_{g=1}^\infty \,
  (-1)^{g-1} \, \frac{s^{2g-1}}{(2g-1)!} \ \int_0^\infty \, \de z \
  \frac{z^{2g-1}}{\E^{\,2\pi\,z}-1} 
\end{align}
we obtain
\begin{align}
\cS_0f(t) = \int_0^\infty \, \frac{\de s}{s} \ \E^{-s/t} \,
  \Big(\frac1{\E^{\,s}-1} - \frac1s + \frac12\Big) \ .
\end{align}
The final formula
\begin{align}
\cS_0\big[\log\Gamma(t)\big] = t\log t -t - \frac12\log\frac t{2\pi} +
  \cS_0f\Big(\frac1t\Big)
\end{align}
is known as {Binet's first formula} for the logarithm of the
gamma-function. In this example the Borel sum is thus exact.
\end{example}

Going back to the Riccati equation \eqref{eq:Riccati}, it is believed
that the solution $S_i^{\rm for}(z;\epsilon)$, while not 
being convergent in general, \emph{is} Borel summable in the direction
$\vartheta={\rm arg}(\epsilon)$~\cite{KoikeSchafke,Nikolaev}. More
precisely, the Borel sum $S_i^\vartheta$ gives an analytic solution of
the Riccati equation away from the trajectories of the WKB
spectral network $\scrW(u,\vartheta)$. Recall that these
trajectories are defined by the condition
\begin{align}
\E^{-\I\,\vartheta} \, \sqrt{p_2(v)} \ \in \ \R^\times
\end{align}
for any tangent vector $v$ to the trajectory; the spectral network
$\scrW(u,\vartheta)$ is known as the {Stokes graph} in exact
WKB analysis. Furthermore, $S_i^\vartheta$ has the expansion
$S_i^{\rm for}(z;\epsilon)$ when $\epsilon\to0$ while remaining in
the closed half-plane with ${\rm
  Re}(\E^{-\I\,\vartheta}\,\epsilon)\geq0$.

The Borel sum $S_i^\vartheta$ can be integrated to give an exact
solution
\begin{align}
\psi_i^\vartheta(z) = \exp\Big(\frac1\epsilon\,\int_{z_0}^z\,
  S_i^\vartheta(z') \, \de z' \Big)
\end{align}
of the holomorphic Schr\"odinger equation
\eqref{eq:Schrodinger}. Suppose that we have two sets of solutions
$\psi_{i}^\vartheta$ and $\tilde\psi_{i}^\vartheta$ in two
neighbouring cells of $C\setminus\scrW(u,\vartheta)$ divided by a
trajectory of type $ij$. Then $\psi_i^\vartheta$ and
$\tilde\psi_i^\vartheta$ are related by the {connection formulas}
\begin{align}\label{eq:Stokesmult}
\psi_i^\vartheta\longmapsto \tilde\psi_i^\vartheta =
  \psi_i^\vartheta+c_{ij}\, \psi_j^\vartheta \qquad \mbox{and} \qquad
  \psi_j^\vartheta\longmapsto\tilde\psi_j^\vartheta = \psi_j^\vartheta
  \ ,
\end{align}
where the functions $c_{ij}$ are known as {Stokes multipliers}.
Note that these connection formulas are precisely of the form of the
unipotent transformations from \eqref{smaps}. Hence the exact WKB
solutions $\psi_i^\vartheta$ provide a basis of sections which
abelianize the flat $SL(2,\C)$ $\epsilon$-connection $\nabla^{\rm
  oper}_\epsilon$. The corresponding almost flat $GL(1,\C)$
$\epsilon$-connection $\nabla_\epsilon^{\rm ab}$ has the explicit form
\begin{align}
\big(\epsilon\,\partial_z - S_i^\vartheta\big)\psi_i^\vartheta = 0 \ .
\end{align}

\begin{example}\label{ex:AD2WKB}
Perhaps the simplest example of the exact WKB method is provided by
the (actual) Schr\"odinger equation for the complex harmonic oscillator
\begin{align}\label{eq:AD2Deps}
{\rm D}_\epsilon\psi(z) = \epsilon^2\,\psi''(z) - \big(z^2+m\big)\,\psi(z)
  = 0
\end{align}
from Example~\ref{ex:operharmonic}~\cite{Aoki:1993ra}. Note that the differential equation \eqref{eq:AD2Deps} is invariant under $\epsilon\mapsto-\epsilon$ and $z\mapsto-z$. We fix ${\rm
  arg}(\epsilon)$. If we take $\vartheta={\rm
  arg}(\epsilon)=\frac\pi2$ and $m>0$, then the corresponding spectral
network $\scrW=\scrW(u,\vartheta)$ is shown in
Figure~\ref{AD2jump}.\footnote{Changing ${\rm arg}(\epsilon)$ rotates
  the trajectories at infinity; changing $\epsilon$ and $m$, while
  leaving ${\rm arg}(m/\epsilon)$ invariant, rotates the entire
  network.} Let us choose these values for convenience. The
formal series solutions to \eqref{eq:AD2Deps} are of the form
\begin{align}
\psi_1^{\rm for}(z) &= \E^{-t^2/4} \, t^\mu \, \sum_{n=0}^\infty \,
                      (-1)^n \,
                      \frac{\mu\,(\mu-1)\cdots(\mu-2n+1)}{n!
                      \, 2^n \, t^{2n}} \ , \nonumber \\[4pt]
  \psi_2^{\rm for}(z) &= \E^{\,t^2/4} \, t^{-\mu-1} \,
                        \sum_{n=0}^\infty \,
                        \frac{(\mu+1)\,(\mu+2)\cdots(\mu+2n)}{n!
                        \, 2^n \, t^{2n}} \ ,
\end{align}
where
\begin{align}
t=\sqrt{\frac2\epsilon} \, z \qquad \mbox{and} \qquad \mu=-\frac
  m{2\,\epsilon}-\frac12 \ ,
\end{align}
and we assume $\mu\notin\mathbb{Z}$. The formal series
$\psi_1^{\rm for}(z)$ decreases fastest along the line
$t=\sqrt{2/\epsilon} \, z \in \R$, while the formal series
$\psi_2^{\rm for}(z)$ decreases fastest along the line
$t=\sqrt{2/\epsilon} \, z \in \,\I\,\R$. These are the trajectories
going to infinity at angles $\frac\pi4$ and $\frac{3\pi}4$ in
Figure~\ref{AD2jump}; see also
Figure~\ref{Stokesrays}.
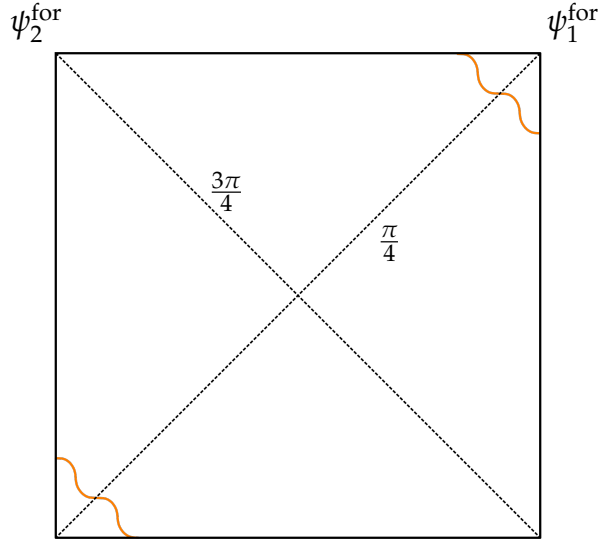
\begin{figure}[h!]
\centering
\begin{overpic}
[width=0.80\textwidth]{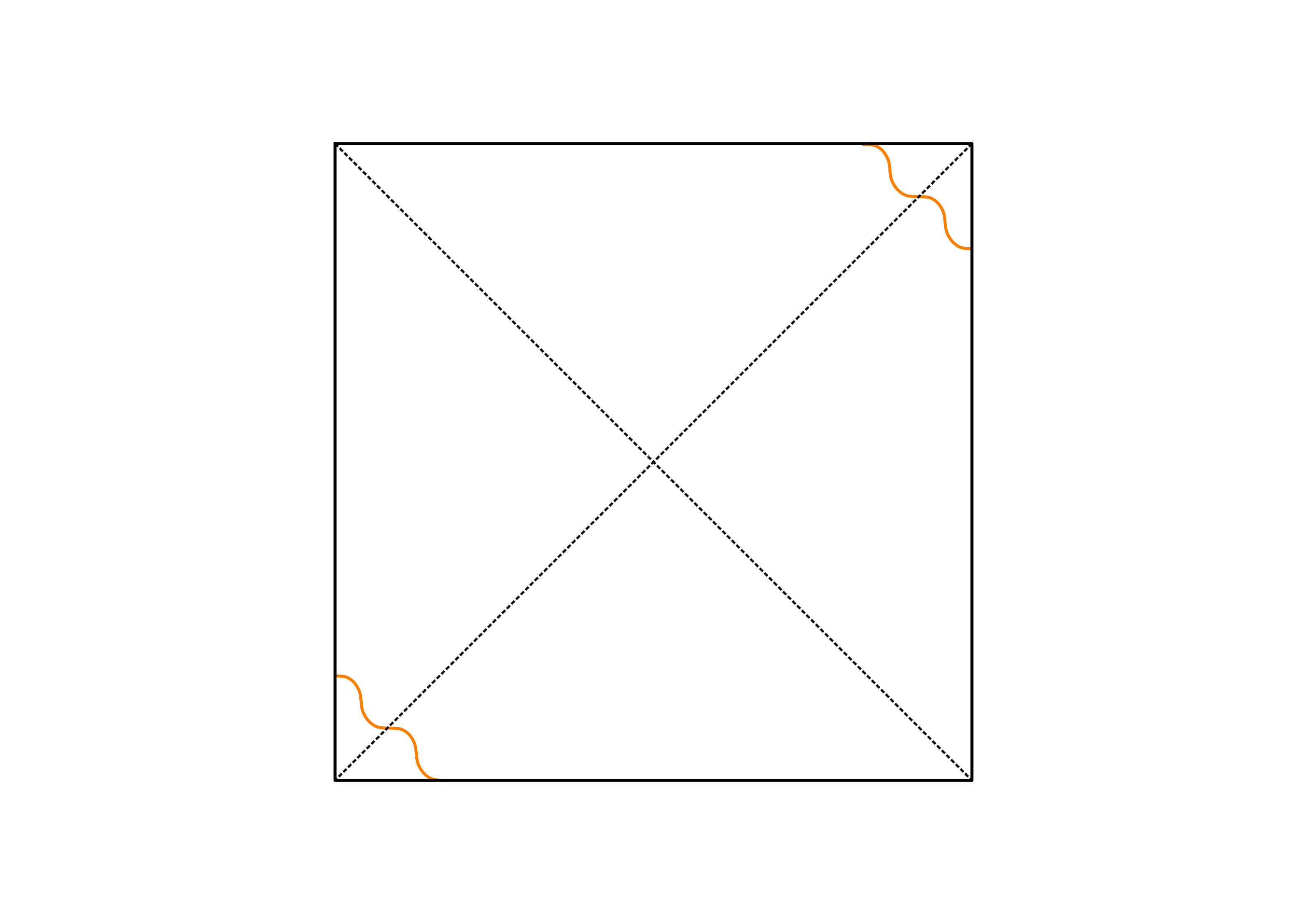}
\put(21,62){$\psi_2^{\rm for}$}
\put(75,62){$\psi_1^{\rm for}$}
\put(58,40){$\frac{\pi}{4}$}
\put(41,45){$\frac{3\pi}{4}$}
\end{overpic}
\caption{\small Stokes rays for the Schr\"odinger equation \eqref{eq:AD2Deps} with arg$(\epsilon)=\pi/2$.}
\label{Stokesrays}
\end{figure}

Let us therefore consider the Borel sums of $\psi_1^{\rm for}(z)$ and
$\psi_2^{\rm for}(z)$. The Borel transform of the series
\begin{align}
f(t) = t^{\mu} \, \sum_{n=0}^\infty \, (-1)^n \,
                      \frac{\mu\,(\mu-1)\cdots(\mu-2n+1)}{n!
                      \, 2^n \, t^{2n}}
\end{align}
is given by
\begin{align}
\cB[f](s) &= \sum_{n=0}^\infty \, (-1)^n \,
                      \frac{\mu\,(\mu-1)\cdots(\mu-2n+1)}{n!
                      \, 2^n \, \Gamma(-\mu+2n)} \,
            s^{2n-\mu-1} \nonumber \\[4pt]
  &= \sum_{n=0}^\infty \, \frac{(-1)^n}{n! \, 2^n \,
    \Gamma(-\mu)} \, s^{2n-\mu-1} \nonumber \\[4pt]
  &=
  \frac{s^{-\mu-1}}{\Gamma(-\mu)} \, \E^{-s^2/2} \
    , \label{eq:cBpsi1} 
\end{align}
where the weight $\Gamma(-\mu+2n)$ in \eqref{eq:cBpsi1} is
determined by the prefactor $t^\mu$ in $\psi_1^{\rm for}(z)$. The
Borel sum of $f$ for $t>0$ is thus
\begin{align}
\cS_0f(t) = \frac1{\Gamma(-\mu)} \, \int_0^\infty \, \de s \
  \E^{-t\,s-\frac{s^2}2} \, s^{-\mu-1} \ .
\end{align}
The holomorphic function
\begin{align}
\psi_1(z) = \E^{-t^2/4} \, \cS_0f(t) =
  D_\mu\Big(\sqrt{\frac2\epsilon} \, z\Big)
\end{align}
is an integral representation of the {Weber parabolic cylinder
  function}. This is indeed an exact solution of the Schr\"odinger
equation \eqref{eq:AD2Deps}. The Borel sum $\psi_1(z)$ can be
analytically continued to ${\rm arg}(z)\in
[-\frac\pi4,\frac{3\pi}4]$. Similarly, for $\I\,t=\sqrt{2/\epsilon} \,
\I \, z<0$ the Borel sum $\psi_2(z)$ of the formal series $\psi_2^{\rm
  for}(z)$ coincides with the Weber parabolic cylinder function
\begin{align}
\psi_2(z) = \E^{-\frac{\pi\,\I}2\,(\mu+1)} \,
  D_{-\mu-1}\Big(-\sqrt{\frac2\epsilon} \, \I \, z\Big) \ ,
\end{align}
and it can be analytically continued to ${\rm
  arg}(z)\in[\frac\pi4,\frac{5\pi}4]$.

For $t=\sqrt{2/\epsilon} \, z<0$ the Borel sum of $\psi_1^{\rm
  for}(z)$ coincides with
\begin{align}
\psi_3(z) = \E^{\,\pi\,\I\,\mu} \,
  D_\mu\Big(-\sqrt{\frac2\epsilon} \, z\Big) \ ,
\end{align}
and it can be analytically continued to ${\rm
  arg}(z)\in[\frac{3\pi}4,\frac{7\pi}4]$. For $\I\,t=\sqrt{2/\epsilon}
\, \I \, z>0$ the Borel sum of $\psi_2^{\rm for}(z)$ is equal to
\begin{align}
\psi_4(z) = \E^{-\frac{\pi\,\I}2\,(3\,\mu-1)} \,
  D_{-\mu-1}\Big(\sqrt{\frac2\epsilon} \, \I \, z\Big) \ ,
\end{align}
which can be analytically continued to ${\rm arg}(z) \in
[-\frac{3\pi}4,\frac\pi4]$. We thus find four Stokes sectors at
infinity, with bases of exact solutions as shown in Figure~\ref{Localsections}.
\begin{figure}[h!]
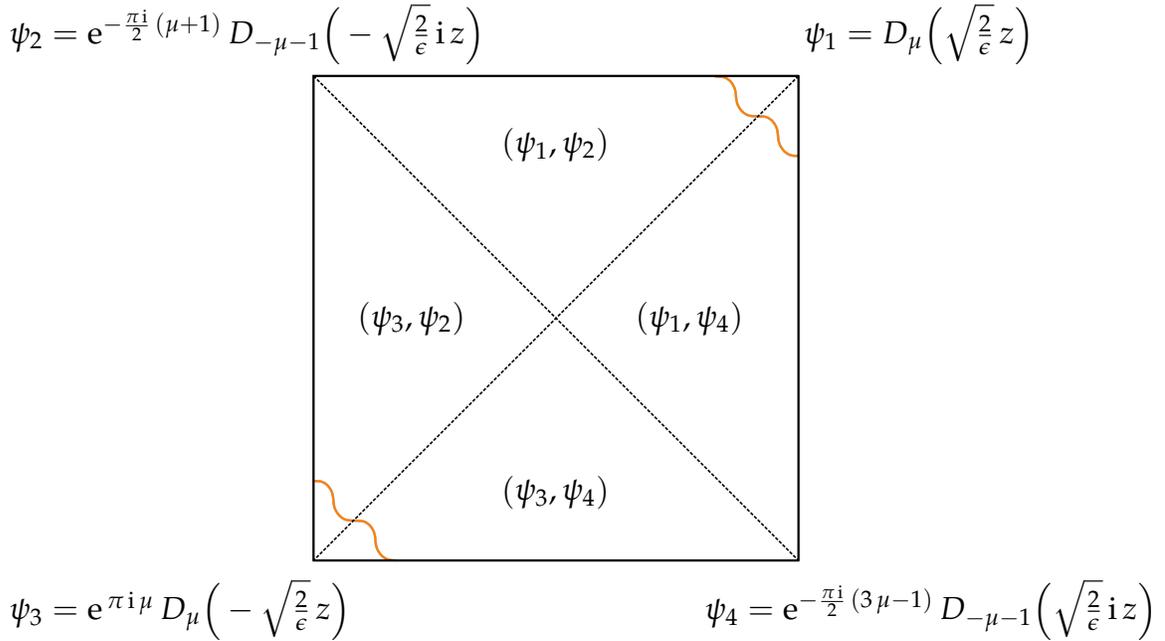

\centering
\begin{overpic}
[width=0.80\textwidth]{Stokes_Rays.png}
\put(75,63){$\psi_1=D_{\mu}\Big(\sqrt{\frac2\epsilon}  \, z\Big)$}
\put(-5,63){$\psi_2=\E^{-\frac{\pi\,\I}2\,(\mu+1)} \,D_{-\mu-1}\Big(-\sqrt{\frac2\epsilon} \, \I \, z\Big)$}
\put(-5,5){$\psi_3=\E^{\,\pi\,\I\,\mu}\,D_\mu\Big(-\sqrt{\frac2\epsilon}\,z\Big)$}
\put(65,5){$\psi_4=\E^{-\frac{\pi\,\I}2\,(3\,\mu-1)}\,D_{-\mu-1}\Big(\sqrt{\frac2\epsilon} \, \I \, z\Big)$}
\put(44.5,17){$\left(\psi_3,\psi_4\right)$}
\put(44.5,52){$\left(\psi_1,\psi_2\right)$}
\put(30,34.5){$\left(\psi_3,\psi_2\right)$}
\put(58,34.5){$\left(\psi_1,\psi_4\right)$}
\end{overpic}
\caption{\small Local basis of solutions at infinity for the Schr\"odinger equation \eqref{eq:AD2Deps} with arg$(\epsilon)=\pi/2$.}
\label{Localsections}
\end{figure}

The exact solutions $\psi_l(z)$ and $\psi_{l+2}(z)$ are related across
the Stokes ray labelled by $\psi_{l+1}(z)$ by the connection formula
\begin{align}
\psi_l = \psi_{l+2} + c_{l,l+2} \, \psi_{l+1} \ ,
\end{align}
where the Stokes multipliers
\begin{align}
c_{l,l+2} =
  \frac{\text{Wr}(\psi_l,\psi_{l+2})}{\text{Wr}(\psi_{l+1},\psi_{l+2})} 
\end{align}
may be obtained as a ratio of {Wronskians} of $\psi_l(z)$ and
$\psi_{l+2}(z)$, and of $\psi_{l+1}(z)$ and $\psi_{l+2}(z)$. This
yields
\begin{align}
\psi_1(z) &= \psi_3(z) + \frac{\sqrt{2\pi} \,
  \E^{\,\pi\,\I\,(\mu+1)}}{\Gamma(-\mu)} \, \psi_2(z) \ ,
  \nonumber \\[4pt]
  \psi_2(z) &= \psi_4(z) - \frac{\sqrt{2\pi} \,
              \I \, \E^{-2\,\pi\,\I\,\mu}}{\Gamma(\mu+1)} \, \psi_3(z) \
              , \nonumber \\[4pt]
  \psi_3(z) &= \E^{\,2\pi\,\I\,\mu} \, \psi_1(z) -
              \frac{\sqrt{2\pi} \, \E^{\,\pi\,\I\,(3\,\mu+1)}}{\Gamma(-\mu)} \,
              \psi_4(z) \ , \nonumber \\[4pt]
  \psi_4(z) &= \E^{-2\pi\,\I\,\mu} \, \psi_2(z) +
              \frac{\sqrt{2\pi} \, \I \, \E^{-2\pi\,\I\,\mu}}{\Gamma(\mu+1)} \,
              \psi_1(z) \ , \label{eq:psiconnection}
\end{align}
where we used $\psi_5(z)=\E^{\,2\pi\,\I\,\mu} \, \psi_1(z)$ and
$\psi_6(z)=\E^{-2\pi\,\I\,\mu} \, \psi_2(z)$. These formulas
indeed agree with the classical connection
formulas for parabolic cylinder functions
(see~\cite[Chapter~12]{DLMF})
\begin{align}
D_\mu(t) &= \E^{\mp\,\pi\,\I\,\mu} \, D_\mu(-t) +
               \frac{\sqrt{2\pi}}{\Gamma(-\mu)} \,
               \E^{\mp\,\frac{\pi\,\I}2\,(\mu+1)} \,
               D_{-\mu-1}(\pm\,\I\,t) \ , \nonumber \\[4pt]
  D_{-\mu-1}(\pm\,\I\,t) &=
                                                                   -\E^{\mp\,\pi\,\I\,\mu}
                                                                   \,
                                                                   D_{-\mu-1}(\mp\,\I\,t)
                                                                   +
                                                                   \frac{\sqrt{2\pi}}{\Gamma(\mu+1)}
                                                                   \,
                               \E^{\mp\,\frac{\pi\,\I}2\,\mu} \, 
                                                                   D_\mu(t)
                                                                   \ .
\end{align}
Since the connection formulas \eqref{eq:psiconnection} are of the form of the unipotent
transformations \eqref{eq:Stokesmult} or \eqref{smaps}, the sections
\begin{align}
s_l(z)=\begin{pmatrix} -\epsilon\,\psi_l'(z) \\ \psi_l(z) \end{pmatrix}
\end{align}
determine a distinguished $\scrW$-abelianization of the Schr\"odinger oper
\eqref{eq:AD2Deps}. In particular, the exact WKB method has provided
us with the framing data $s_l$.
\end{example}

Now that we have found a distinguished abelianization for the flat
$\epsilon$-connection $\nabla_\epsilon^{\rm oper}$, for ${\rm arg}(\epsilon)=\vartheta$, with respect to
the WKB spectral network $\scrW(u,\vartheta)$, we can consider the
spectral coordinates associated to this abelianization. Recall that
for a one-cycle $\gamma$ on $\Sigma$, the corresponding spectral
coordinate is defined by
\begin{align}
\cX^{\scrW(u,\vartheta)}_\gamma(\nabla_\epsilon^{\rm oper}) = {\rm Hol}_\gamma
  \nabla_\epsilon^{\rm ab} \ .
\end{align}
In exact WKB analysis, the logarithm of this spectral coordinate, for ${\rm arg}(\epsilon)=\vartheta$, is known as the Voros symbol $V_\gamma(\epsilon)$. It follows from the discussion above that the spectral coordinate \smash{$\cX^{\scrW(u,\vartheta)}_\gamma(\nabla_\epsilon^{\rm oper})$}, now with ${\rm arg}(\epsilon)$ a free parameter, has good
asymptotics when $\epsilon\to0$ in the half-plane with ${\rm
  Re}(\E^{-\I\,\vartheta}\,\epsilon)\ge0$. It is given by
  $\exp\big(\Pi_\gamma(\epsilon)\big)$, where
  \begin{align}\label{eq:quantumperiod}
\Pi_\gamma(\epsilon) = \frac1\epsilon\,\oint_\gamma\, S_i^{\rm
    for}(z;\epsilon) \, \de z
  \end{align}
is called a {quantum period}.\footnote{This is also the conformal
limit of the $\zeta\to0$ asymptotics of $\cX_\gamma(\cA)$, where $\cA$
is the flat connection from
\eqref{eq:flatfamily}~\cite{gaiotto2013wall}.}

We should point out one subtlety. The exact WKB analysis, just like abelianization, is not defined at a critical phase $\vartheta_{\rm c}$ when the corresponding network $\scrW(u,\vartheta_{\rm c})$ develops one or more saddle trajectories. Again, in parallel to our approach with abelianization, one may apply the exact WKB method to the resolutions $\scrW(u,\vartheta_{\rm c}^\pm)$ of the critical network. In exact WKB analysis this is called lateral Borel resummation, while the average Voros symbol $V_\gamma^{\rm c} = \frac12\,\big(V_\gamma^+ + V_\gamma^-\big)$ is said to be obtained using median summation.

Suppose we keep the vacuum $u$ fixed while varying the phase $\vartheta$. As long as we do not cross any BPS ray (a critical phase where the spectral network $\scrW(u,\vartheta)$ develops a saddle trajectory), the spectral coordinate \smash{$\cX^{\scrW(u,\vartheta)}_\gamma(\nabla_\epsilon^{\rm oper})$} or Voros symbol $V_\gamma(\epsilon)$, with ${\rm arg}(\epsilon)=\vartheta$, is an analytic function of $\epsilon$.

However, the spectral coordinate \smash{$\cX^{\scrW(u,\vartheta)}_\gamma(\nabla_\epsilon^{\rm oper})$} or Voros symbol $V_\gamma(\epsilon)$, with ${\rm arg}(\epsilon)=\vartheta$, will have a jump discontinuity across a BPS ray. From the perspective of abelianization this is because the topology of the spectral network changes, while in the exact WKB method this occurs because of the need to move a contour of integration across a singularity in the Borel plane. This jump will generally be of the form of a Kontsevich-Soibelman cluster transformation associated to the BPS states supported by the BPS wall.

Hence the spectral coordinates \smash{$\cX^{\scrW(u,\vartheta)}_\gamma(\nabla_\epsilon^{\rm oper})$} or Voros symbols $V_\gamma(\epsilon)$, with ${\rm arg}(\epsilon)=\vartheta$, are piecewise analytic functions on the parameter space. Yet, for a fixed phase $\vartheta$, these coordinates may be analytically continued to analytic functions on the whole $\epsilon$-plane. The resulting analytically continued functions agree with the spectral coordinates \smash{$\cX_\gamma^\scrW$} obtained by the abelianization method, evaluated at the flat $\epsilon$-connection $\nabla_\epsilon^{\rm oper}$, where we fix the isotopy class $\scrW$ of the spectral network $\scrW(u,\vartheta)$.

For further background, details and examples we refer
to~\cite{hollands2019exact}
and~\cite[Section~11]{hollands2018higher}. A particularly interesting
feature is that the spectral coordinates
\smash{$\cX^{\scrW(u,\vartheta)}_\gamma(\nabla_\epsilon^{\rm oper})$} or Voros periods $V_\gamma(\epsilon)$, for ${\rm arg}(\epsilon)=\vartheta$, may also be characterized as
solutions to integral equations in the $\epsilon$-plane, called
{TBA equations}. The general form of these equations has been
formulated in~\cite{gaiotto2014opers}, and derived (from reasonable
analytic assumptions) in~\cite{Ito:2018eon}, while more specifics and
examples are found
in~\cite{hollands2019exact,Grassi:2019coc,Grassi:2021wpw} where the various methods for computing quantum periods are compared numerically. Similar equations in the context of
topological string theory appear in the work of
  Bridgeland~\cite{bridgeland2019riemann}.

\begin{example}\label{ex:WKBE6}
Consider the one-parameter family of $\epsilon$-opers
$\nabla_\epsilon^{\rm oper}(u)$ from
Example~\ref{ex:K3oper} with $u\in\R$ and ${\rm
  arg}\,\epsilon=0=\vartheta$. For these choices the spectral
network $\scrW(u,\vartheta)$ is isotopic to the circular Fenchel-Nielsen
network from Figure~\ref{E6Networks}. We introduce one-cycles $A$, $B$ and $C$ with $C=-A-B$ on the spectral cover $\Sigma$ as shown in Figure~\ref{E6Cycles}. 
\begin{figure}[h!]
\centering
\begin{overpic}
[width=0.80\textwidth]{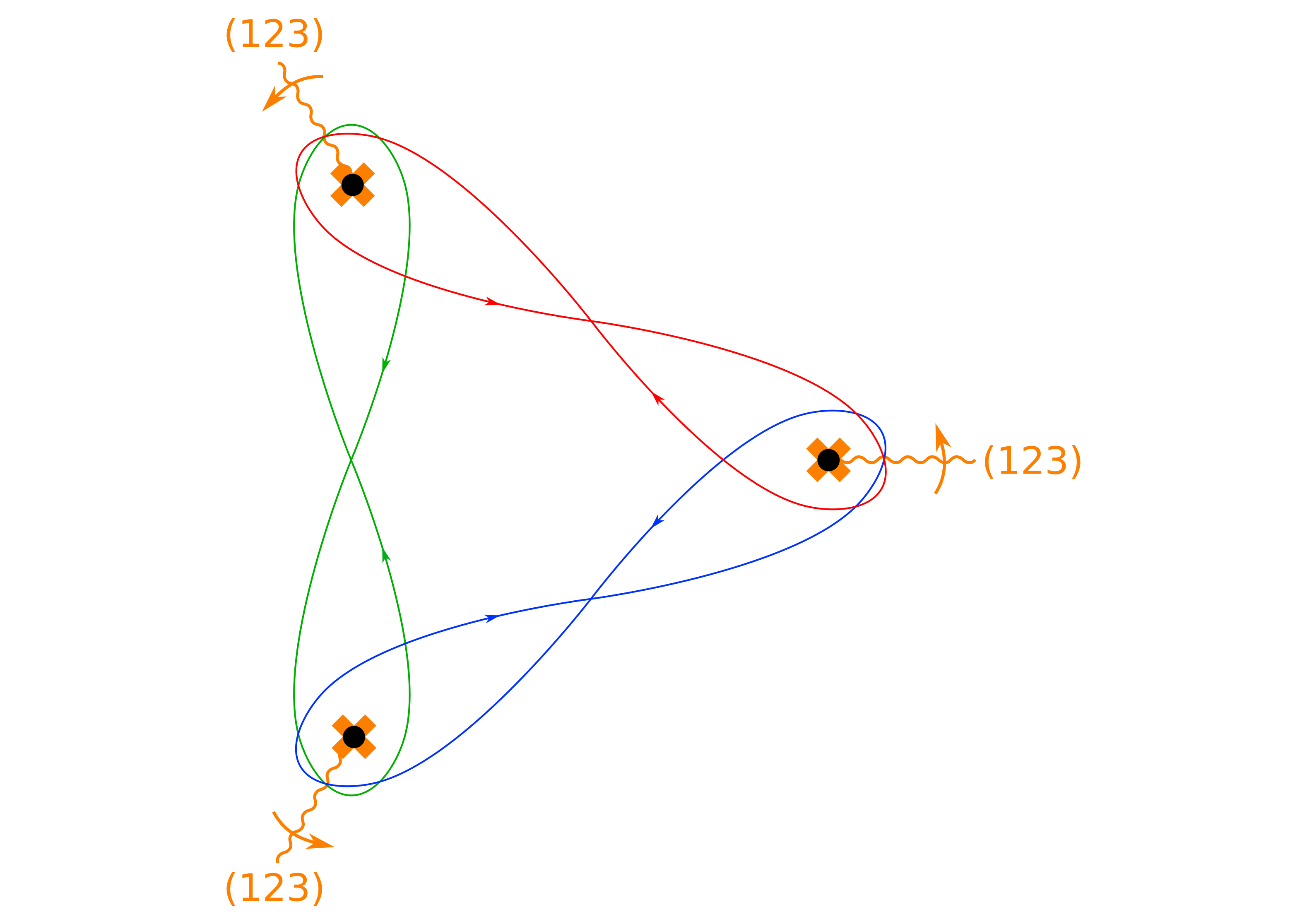}
\put(50,47){\textcolor{Red}{$A$}}
\put(42,16){\textcolor{Blue}{$B$}}
\put(22,39){\textcolor{Green}{$C$}}
\end{overpic}
\caption{\small Three one-cycles $A$, $B$ and $C=-A-B$ on $\Sigma$.}
\label{E6Cycles}
\end{figure}

The spectral coordinates $\cX_A$ and $\cX_B$ (for the distinguished WKB
abelianization), evaluated at the family of $\epsilon$-opers $\nabla_\epsilon^{\rm
  oper}(u)$, have an asymptotic expansion determined by the Riccati equation, as $\epsilon\to0$ in the
half-plane with ${\rm Re}(\E^{-\I\,\vartheta}\,\epsilon)\ge0$. This has been verified numerically in~\cite[Section~6.7]{hollands2019exact}. The Riccati equation in this case is found by making the ansatz
\begin{align}
\psi(z;\epsilon) = \exp\bigg(\sum_{n=-1}^\infty \, \epsilon^n \ \int_{0}^z \, S_n(z') \, \de z'\bigg)
\end{align}
for the solutions of the $T_3$-equation
\begin{align}
\epsilon^3 \, \frac{\partial^3\psi(z)}{\partial z^3} + \epsilon \, q_2(z) \, \frac{\partial\psi(z)}{\partial z} + q_3(z) \, \psi(z) = 0 \ ,
\end{align}
with
\begin{align}
q_2(z) = \frac{9\,z}{(z^3-1)^2} \qquad \mbox{and} \qquad q_3(z) = \frac{u}{(z^3-1)^2} + \frac\epsilon2 \, q_2'(z) \ .
\end{align}
This yields
\begin{align}
S_{-1}^{(i)} = -\frac{\omega^i\, u^{1/3}}{(z^3-1)^{2/3}} \ , \quad S_0^{(i)} = \frac{2\,z^2}{z^3-1} \ , \quad S_1^{(i)} =\frac{\omega^{-i} \, u^{-1/3} \, z}{3\,(z^3-1)^{4/3}} \ , \quad \dots
\end{align}
for $i=0,1,2$, where we fixed the labelling of the sheets ${}^{(i)}$ by choosing $S_{-1}^{(i)} = -\omega^i\,u^{1/3}$ at $z=0$, with \smash{$\omega=\E^{\,2\pi\,\I/3}$}. The spectral coordinate $\cX_A$ then has an asymptotic expansion
\begin{align}
\cX_A(\epsilon) \sim \exp\big(\Pi_A(\epsilon)\big) &= \exp\bigg(\sum_{n=-1}^\infty \, \epsilon^n \ \oint_A \, S_n(z') \, \de z'\bigg) \nonumber \\[4pt]
&= \exp\bigg(\sum_{n=-1}^\infty \, \epsilon^n \, \Big(\int_1^\omega \, S_n^{(2)}(z') \, \de z' + \int_\omega^1 \, S_n^{(3)}(z') \, \de z'\Big)\bigg) \ .
\end{align}
With $t=u^{1/3}/\epsilon$, we find
\begin{align}\label{eq:E6PiAt}
\Pi_A(t) &=
  -\frac{\Gamma(\frac13)\,\Gamma(\frac16)}{2^{2/3}\,\sqrt\pi} \, t -
  \frac{3\,\sqrt\pi\,\Gamma(\frac53)}{2^{1/3}\,\Gamma(\frac16)} \,
  t^{-1} +
  \frac{11\,\Gamma(\frac13)\,\Gamma(\frac76)}{270\cdot2^{2/3}\,\sqrt\pi}
                  \, t^{-5} \nonumber \\
  & \quad \, -
  \frac{83\,\sqrt\pi\,\Gamma(\frac53)}{2268\cdot2^{1/3}\,\Gamma(\frac76)}
  \, t^{-7} +
    \frac{44857\,\Gamma(\frac13)\,\Gamma(\frac76)}{96228\cdot2^{2/3}\,\sqrt\pi}
    \, t^{-11} + \cdots \ .
\end{align}
Similarly, the spectral coordinates $\cX_B$ and $\cX_C$ have asymptotic expansions determined by the quantum periods $\Pi_C(\omega^2\,t) = \Pi_B(\omega\,t) = 
\Pi_A(t)$.

A challenge for the interested reader is to try to find the
Borel sum $\cS_\vartheta\cX_A(t)$ of the quantum period $\Pi_A(t)$ in the
direction $\vartheta$. In~\cite[Section~6.7]{hollands2019exact} it is explained
how to obtain $\cX_A(t)$ numerically, by both solving the
abelianization problem for the $T_3$-equation as well as by writing
down and approximating suitable integral equations. An exact answer is not
presently known (and might require the definition of a new sort of
``special function'').
\end{example}

\clearpage
\newpage
\section{The Nekrasov-Rosly-Shatashvili Correspondence}\label{sec:newNRS}
\noindent
In this section we formulate a recipe for computing the effective twisted superpotential \smash{$\Weff$} of an $\N=2$
field theory ${\sf T}_K[C,\cD]$ of class $\cS$ in the
$\frac12\Omega$-background, geometrically in terms of spectral
coordinates and opers. This recipe is based on a proposal of
Nekrasov, Rosly and Shatashvili (NRS), who conjectured that \smash{$\Weff$} is
essentially the difference between the generating functions of two holomorphic
Lagrangian subspaces of the associated moduli space $\scrM_{\rm
  flat}(C,SL(K,\C))$ of flat $SL(K,\C)$ connections on $C$, in the
appropriate Darboux coordinates on $\scrM_{\rm
  flat}(C,SL(K,\C))$~\cite{nekrasov2011darboux}; this conjecture is formulated in Section~\ref{sec:conjecture}. Evidence for this
conjecture has so far been given for the conformal $SU(2)$ and $SU(3)$
theories coupled to four and six
hypermultiplets~\cite{nekrasov2011darboux,Vartanov:2013ima,hollands2018higher,Jeong:2018qpc}.

Here we will consider two new examples, the rank one
Argyres-Douglas theory encountered in Example~\ref{ex:AD2BPS} and the
pure $SU(2)$ theory  introduced in Section~\ref{sec:SWTheory}. The
novel ingredients involved are the new Fenchel-Nielsen type
coordinates found in Section~\ref{WKBSNs}, see
Examples~\ref{ex:AD2spectralcoordinates}, \ref{ex:FNpureab} and~\ref{ex:FNcoordinates}, respectively.
In Section~\ref{sec:HOex} we formulate the
recipe for 
computing \smash{$\Weff$} geometrically. This has been tested in~\cite{hollands2018higher} for the conformal $SU(2)$ and $SU(3)$ examples. Here we test it for the rank one Argyres-Douglas theory and the weakly coupled pure $SU(2)$ theory in Sections~\ref{ex:recipeAD2} and \ref{ex:recipepure}, whereas in Section~\ref{ex:recipeE6} we use it to propose a new expansion of
\smash{$\Weff$} for the $E_6$ Minahan-Nemeschansky theory discussed in Section~\ref{ClassSTheories}. In Section~\ref{sec:QFT} we give more background
information and a string theory derivation of  the NRS
proposal, mostly following the treatment of Nekrasov and Witten in~\cite{Nekrasov:2010ka}, while in Section~\ref{sec:QFT3d} we discuss the conjecture with more general boundary conditions. 

\subsection{The NRS Conjecture}\label{sec:conjecture}
\noindent
In Section~\ref{sec:NRS} we have encountered the parameter $\epsilon$
in two different contexts. Firstly, as deformation parameter of the
$\frac12\Omega$-background in Section~\ref{sec:Weff}, and secondly as
a quantization parameter $\epsilon=\zeta/R$ in the Hitchin moduli
space $\scrM_{\textrm{\tiny H}}$ in Sections~\ref{sec:Hitchin}--\ref{sec:WKB}. In~\cite{nekrasov2010quantization} these two
instances of $\epsilon$ are related. It is proposed that the low
energy physics of the $\N=2$ theory ${\sf T}_K[C,\cD]$ of class $\cS$
in the $\frac12\Omega$-background $\R_\epsilon^2\times\R^2$ is
described by quantization of the Hitchin integrable system
$(\scrM_{\textrm{\tiny H}}(C,SU(K)),J_0)$ (or more generally $(\scrM_{\textrm{\tiny
  H}}(C,{}^\lang G),J_0)$ where ${}^\lang G$ is the Langlands dual of the
gauge group $G$). 

In particular, Nekrasov and Shatashvili argue
that the generators of the `twisted ring' of the effective
two-dimensional $\N=(2,2)$ theory ${\sf T}_\epsilon$ may be identified
with the quantum Hamiltonians, and that its supersymmetric vacua may
be identified with the `Bethe states' of the quantum integrable
system. The latter claim implies that the effective twisted
superpotential \smash{$\Weff$} of the $\N=(2,2)$ theory ${\sf T}_\epsilon$
corresponds to the {Yang-Yang function} $Y$  of the quantum
Hitchin system. That is,
\begin{align}
\Weff(a) = Y(\nu) \ ,
\end{align}
where the vacuum expectation values $\sigma_i=a_i$ of the scalars in
the twisted chiral superfields $\boldsymbol\Sigma$ are identified with
the spectral parameters (or rapidities) $\nu_i$.

This proposal was made more tangible in~\cite{nekrasov2011darboux}. It was conjectured that the
effective twisted superpotential \smash{$\Weff$} corresponds to the difference
of two generating functions of holomorphic Lagrangian submanifolds of the
moduli space
$\scrM_{\rm flat}^\epsilon(C,SL(K,\C))$ in appropriate Darboux coordinates, and
that the supersymmetric vacua may be identified with the intersection
points of these holomorphic Lagrangian submanifolds. Let us explain this in a few more words.

Given any system of holomorphic Darboux
coordinates $(x,y)$ on $\scrM^\epsilon_{\rm
  flat}(C,SL(K,\C))$,
\begin{align}
\{x_i,y^j\}=\delta_{i}^j \ , 
\end{align}
we can define a generating function $\cW$ of any holomorphic Lagrangian
submanifold of the moduli space $\scrM_{\rm
  flat}^\epsilon(C,SL(K,\C))$ in this coordinate chart through the equation
\begin{equation}
y^i=\frac{\partial\cW(x)}{\partial x_i} \ ,
\end{equation}
which uniquely determines it up to an $x$-independent
function.

As can be anticipated from Section~\ref{sec:NRS}, one holomorphic Lagrangian submanifold is the brane of $\epsilon$-opers $\scrL_\epsilon^{\rm oper}$. As explained in Section~\ref{sec:opers}, this submanifold may be characterized as the quantization of the Coulomb branch $\scrB$ with quantization parameter $\epsilon$. The second holomorphic Lagrangian submanifold
corresponds to a boundary condition in the $\N=(2,2)$ theory ${\sf
  T}_\epsilon$ at infinity in $\R^2_{\epsilon}$. It might be surprising that such a boundary condition is needed, given that the original computation of the four-dimensional Nekrasov partition function did not require one. However, as already observed in~\cite{nekrasov2010quantization}, further emphasized in~\cite{Nekrasov:2010ka}, and as we will expand on later, the two-dimensional perspective reveals that the definition of ${\sf T}_\epsilon$ requires more data than just the choice of a four-dimensional theory $\sf T$ and a parameter $\epsilon$. 

In~\cite{nekrasov2011darboux} the appropriate Darboux
coordinates are described in the example of the conformal $SU(2)$ theory
coupled to four hypermultiplets (which is generalized to the conformal
$SU(K)$ theory coupled to $2K$ hypermultiplets
in~\cite{Jeong:2018qpc}). In this case the appropriate Darboux
coordinates are basically the complex
Fenchel-Nielsen coordinates described in
Example~\ref{ex:FNcoordinates}.\footnote{These are called Darboux
  coordinates rather than complex Fenchel-Nielsen coordinates
  in~\cite{nekrasov2011darboux} because the shift ambiguity in the
  Fenchel-Nielsen twist coordinate is fixed in the NRS Darboux
  coordinates.} We will outline the string theory derivation of the NRS conjecture (based on~\cite{Nekrasov:2010ka}) in
Section~\ref{sec:QFT}. A gauge theory proof of the conjecture, in the
conformal $SU(2)$ and $SU(3)$ examples, was given
in~\cite{Jeong:2018qpc}, whereas a conformal field theory perspective
on the conjecture was offered in~\cite{Vartanov:2013ima}.

\subsection{A Geometric Recipe}\label{sec:HOex}
\noindent
We shall now make some additional remarks, with the goal of turning the NRS conjecture into a concrete recipe for computing the effective twisted
superpotential \smash{$\Weff$} geometrically for any $\N=2$ theory ${\sf T} = {\sf
  T}_K[C,\cD]$ of class $\cS$. Recall from Section~\ref{sec:Weff} that the effective twisted superpotential \smash{$\Weff(a;\epsilon)$} may be obtained as the Nekrasov-Shatashvili limit
\begin{align}
\Weff(a;\epsilon) = \lim_{\epsilon_2\to0} \, \epsilon_2 \, \log Z^{\rm Nek}(a;\epsilon_1=\epsilon,\epsilon_2)
\end{align}
of the Nekrasov partition function $Z^{\rm Nek}(a;\epsilon_1,\epsilon_2)$. The latter partition function may be computed from first principles for any $\N=2$ theory $\sf T$ with a Lagrangian formulation (that is, in terms of gauge fields possibly coupled to matter fields).

However, recall that an $\N=2$ theory $\sf T$ may have distinct Lagrangian descriptions in different regions of its moduli space (this is the generalized S-duality discussed in Section~\ref{ClassSTheories}), or it may not even have a Lagrangian formulation at all (see as well Section~\ref{ClassSTheories}). The characterization of the corresponding two-dimensional theory ${\sf T}_\epsilon$ (the theory $\sf T$ in the $\frac12\Omega$-background), as well as the explicit expression for \smash{$\Weff(a;\epsilon)$}, therefore depends on more data than just the theory $\sf T$ and the parameter $\epsilon$. In particular, it matters in which region of the moduli space we are considering $\sf T$. For example, the standard Lagrangian description of the pure $SU(2)$ theory is only valid in the weak coupling region of its moduli space, and is distinct from the dual weakly coupled description near its strong coupling points. We thus obtain two distinct two-dimensional $\N=(2,2)$ theories
${\sf T}_\epsilon^{\rm w}$ and ${\sf T}_\epsilon^{\rm s}$ when we
place either description in the $\frac12\Omega$-background, each with its own effective
twisted superpotential \smash{$\Weff$}. 

This extra data may be encoded in a half-BPS boundary condition at infinity in $\R_{\epsilon}^2$. This is the boundary condition that defines the second Lagrangian submanifold in the NRS conjecture. A full classification and analysis of such boundary conditions is likely to be quite rich and has not been completed thus far (to our knowledge), however see~\cite{Nekrasov:2010ka,Dimofte:2013lba,Cecotti:2011iy} for various examples and related discussions. In particular, some simple choices have been studied in~\cite[Section~3.4]{Nekrasov:2010ka}. Especially relevant here is the type~I (or standard Neumann) boundary condition, which is characterized in a four-dimensional \emph{gauge} theory by Neumann boundary conditions for the components of the gauge fields parallel to the boundary, and Dirichlet boundary conditions for the component of the gauge field normal to the boundary. For an abelian gauge theory this would correspond to a Lagrangian submanifold in the Hitchin moduli space defined by the condition $\exp( \oint_B \, \lambda/\epsilon)=1$. For a non-abelian gauge theory, with a given choice of an electric-magnetic duality frame, this Lagrangian submanifold would be \emph{approximated} by the condition
\begin{align}
\exp\Big(\frac1\epsilon \, \oint_{B_i} \, \lambda\Big) = 1 \ ,
\end{align}
for all $i=1,\dots,\frac12\,h_1(\overline{\Sigma})$, near infinity on the Coulomb branch.

Here, expanding on~\cite{hollands2019exact}, we want to speculate more generally, for any $\N=2$ theory ${\sf T}={\sf T}_K[C,\cD]$ of class $\cS$, that the relevant class of half-BPS boundary conditions are labelled in the {infrared} by a distinguished spectral network $\scrW=\scrW(u,\vartheta_{\rm c})$ together with a polarization (a choice of $A$-cycles and $B$-cycles) of the Seiberg-Witten curve $\Sigma_u$. The spectral network $\scrW$, together with the electric-magnetic splitting, determine a choice of holomorphic exponentiated Darboux coordinates $\cX_\gamma$ on the moduli space of flat $\epsilon$-connections $\scrM_{\rm flat}^\epsilon(C,SL(K,\C))$ (with respect to the complex structure $J_\epsilon$). In terms of these Darboux coordinates, the second Lagrangian submanifold in the NRS conjecture is given \emph{exactly} by
\begin{align}
\cX_{B_i} = 1
\end{align}
for all $i=1,\dots,\frac12\,h_1(\overline{\Sigma})$.\footnote{This proposal is somewhat similar in flavour to the boundary conditions described in~\cite{Dimofte:2013lba,Cecotti:2011iy}.}

For example, any theory $\sf T$ with a weakly coupled Lagrangian description (such as the pure $SU(2)$ theory at weak coupling) comes with a canonical choice of $A$-cycles $A^i$ and $B$-cycles $B_i$ on the Seiberg-Witten curve
$\Sigma_u$: in a limit where an exponentiated gauge coupling $q_i$ tends to zero, $\exp(\oint_{A^i} \, \lambda)$ becomes exponentially small, whereas $\exp(\oint_{B_i} \, \lambda)$ becomes exponentially large. Furthermore, there is a critical phase
$\vartheta_{\rm c}$ in a neighbourhood of this weak coupling point for which the $A$-cycle period satisfies $\E^{-\I\,\vartheta_{\rm c}} \, \oint_{A^i} \, \lambda < 0$, and the corresponding spectral network
$\scrW^\FN = \scrW(u,\vartheta_{\rm c})$ is a Fenchel-Nielsen type network.\footnote{Whenever the description of the theory $\sf T$ includes non-zero mass parameters, the isotopy class of the Fenchel-Nielsen type network depends on the values of these parameters (or better, on the wall-crossing chamber these values are part of), and is thus not quite unique. We believe these differences play no significant role in the  following. See~\cite{Coman:2019eex} for a related discussion in five dimensions where these jumps are related to flop transitions.} The
closed trajectories of this network correspond to an infinite tower
of BPS particles, including the $W^\pm$-bosons, accumulating at this
phase. 

On the other hand, suppose that $\sf T$ is an asymptotically free theory near a strong coupling point with a dual weakly coupled description. For example, say $\sf T$ is the pure $SU(2)$ theory at strong coupling.\footnote{This example will be treated in detail in a forthcoming paper.} Then the distinguished spectral network $\scrW(u,\vartheta_{\rm c})$ would be a critical Fock-Goncharov type network, with a saddle signalling the presence of a massless dyon. In this case, the massless dyon determines the $A$-cycle in the polarization of the Seiberg-Witten curve $\Sigma_u$.\footnote{One might worry that there is no unique such critical Fock-Goncharov type network nor choice of $B$-cycle, but  these ambiguities just illustrate the $SL(2,\Z)$ monodromy around the strong coupling point.}

Finally, suppose that $\sf T$ is an intrinsically strongly coupled theory. For example, say $\sf T$ is the $E_6$ Minahan-Nemeschansky theory. In this case we propose that the different boundary conditions are labelled by the higher rank Fenchel-Nielsen type spectral networks $\scrW(u,\vartheta_{\rm c})$ where the critical phase $\vartheta_{\rm c}$ satisfies $\E^{-\I\,\vartheta_{\rm c}}\,\oint_{\gamma}\,\lambda<0$ for a one-cycle $\gamma$ on the Seiberg-Witten curve $\Sigma_u$. This one-cycle also determines the polarization of $\Sigma_u$.

With all this in mind, we now turn the NRS conjecture into a concrete geometric recipe for computing the effective twisted superpotential \smash{$\Weff$} for any $\N=2$ theory ${\sf T}={\sf T}_K[C,\cD]$ of class $\cS$ in the $\frac12\Omega$-background, with respect to a chosen half-BPS boundary condition at infinity in the $\R^2_{\epsilon}$ plane, that is characterized by a spectral network $\scrW$ as well as a polarization of the Seiberg-Witten curve $\Sigma_u$. Following~\cite{hollands2018higher}, this recipe reads:
\begin{enumerate}
\item Write down the brane of $\epsilon$-opers $\scrL_\epsilon^{\rm oper}$
  associated to the theory $\sf T$ (using the explanations of
  Section~\ref{sec:opers}).
\item Compute the ``length'' and ``twist'' coordinates
  \begin{align}\label{eq:lengthtwist}
    x_i=\frac\epsilon{\pi\,\I} \log (-\cX_{A^i}) \qquad \mbox{and} \qquad
    y^i=\frac1{2\,\epsilon} \log \cX_{B_i}
  \end{align}
  associated to the spectral network
  $\scrW$ (following the instructions from
  Section~\ref{WKBSNs}). 
\item Evaluate the coordinates $(x_i,y^i)$ on the brane of
  $\epsilon$-opers $\scrL_\epsilon^{\rm oper}$, with respect to the WKB-framing
  (as discussed in Section~\ref{sec:WKB}).\footnote{Recall that there may be multiple $\scrW$-abelianizations, corresponding to distinct $\scrW$-framings.}
\item Finally, extract $\Weff$ through
  \begin{align}
y^i = \frac1\epsilon \, \frac{\partial\Weff(x;\epsilon)}{\partial x_i} \ .
  \end{align} 
\end{enumerate}
This recipe computes the effective twisted superpotential $\Weff$ exactly in the parameter $\epsilon$. As explained in Section~\ref{sec:WKB}, the spectral coordinates $\cX_\gamma$, evaluated on the brane of $\epsilon$-opers, may be interpreted as non-perturbative completions of the quantum periods \smash{$\Pi_\gamma(\epsilon) = 
\frac1\epsilon \, \big( \oint_\gamma \, \lambda + O(\epsilon) \big)$}.\footnote{More precisely, $\log\cX_\gamma(\nabla_\epsilon^{\rm oper})$ is an analytic continuation of the Voros symbol $V_\gamma(\epsilon)$ computed at ${\rm arg}(\epsilon) = \vartheta$. If $\vartheta=\vartheta_{\rm c}$ is a critical phase, it is an analytic continuation of the average of the Voros symbols computed at ${\rm arg}(\epsilon) = \vartheta_{\rm c}^\pm$.} The $\epsilon$-expansion for \smash{$\Weff$} may then be computed from the equations \smash{$\Pi_{B_i}(\epsilon)=2\,\partial_{x_i}\Weff(x;\epsilon)$}, where for $x_i$ we substitute the quantum period $\epsilon\,\Pi_{A^i}(\epsilon)/\pi\,\I$.

For theories $\sf T$ with a weakly coupled Lagrangian formulation, this recipe computes the well-known Nekrasov-Shatashvili limit of the Nekrasov partition function with respect to the distinguished Fenchel-Nielsen type network $\scrW^\FN$ described above. The coordinate $x_i$ then corresponds to a complexified Fenchel-Nielsen length parameter, which we will denote in the following by $-a_{\FN,i}$ since it may be obtained by Borel resummation of the quantum period $\Pi_{A^i}(\epsilon)$ with respect to the Fenchel-Nielsen type network $\scrW^\FN$.
This was already tested in~\cite{hollands2018higher} for the weakly coupled conformal $SU(2)$ and $SU(3)$ examples. Let us take a look at some new illustrative examples here, starting with the very simplest model of a theory of class $\cS$. 

\subsection{AD$_2$ Theory}\label{ex:recipeAD2}
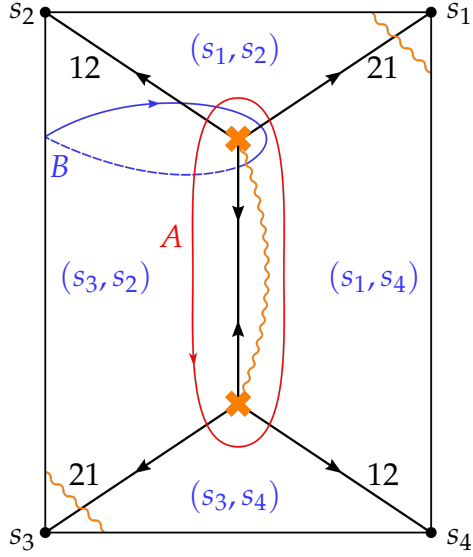
\begin{figure}[h!]
\centering
\begin{overpic}
[width=0.80\textwidth]{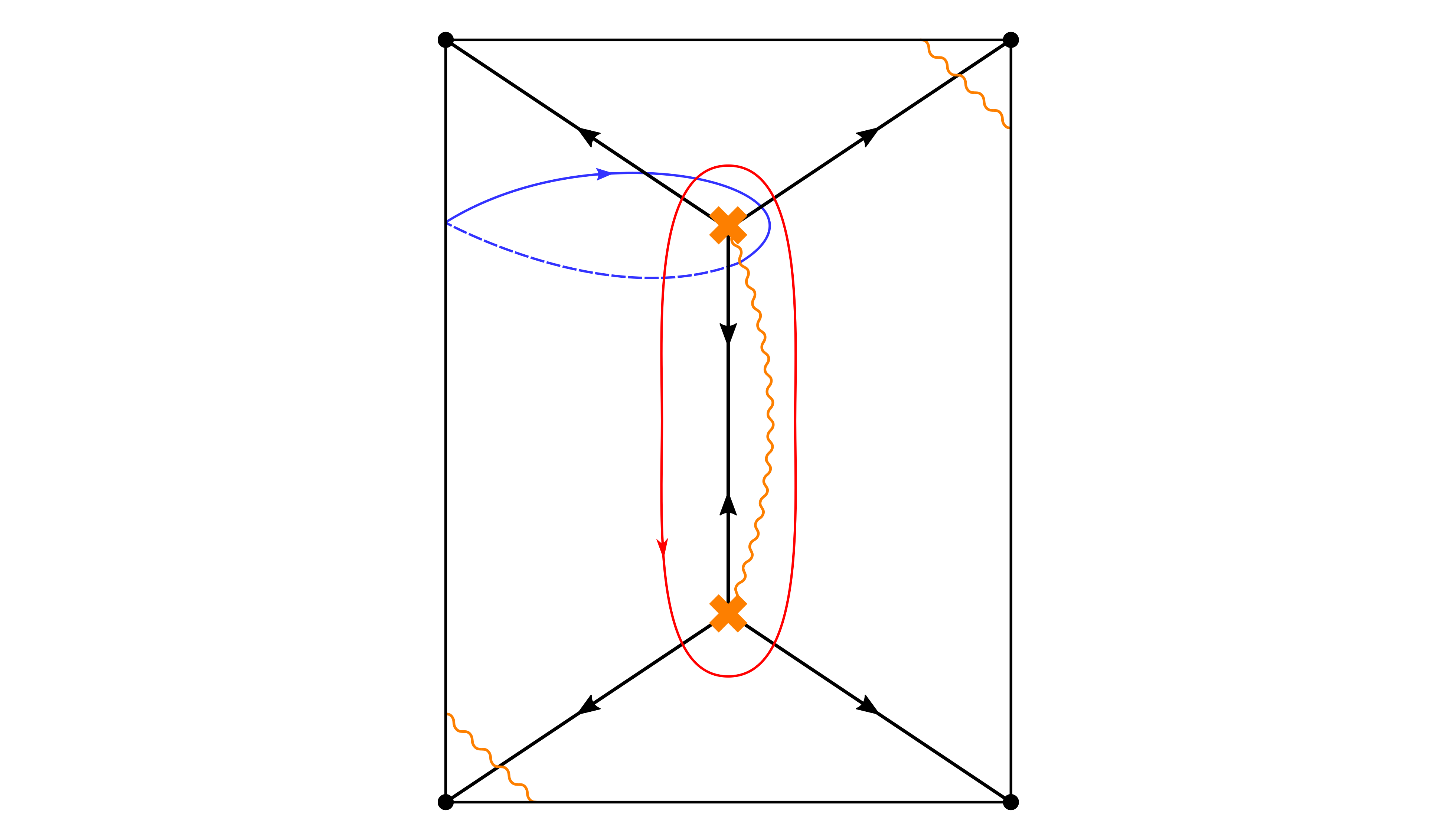}
\put(71,54){$s_1$}
\put(27,54){$s_2$}
\put(27,1){$s_3$}
\put(71,1){$s_4$}
\put(33,48){$12$}
\put(63,48){$21$}
\put(33,7){$21$}
\put(63,7){$12$}
\put(45,50){\textcolor{Blue}{$(s_1,s_2)$}}
\put(45,5){\textcolor{Blue}{$(s_3,s_4)$}}
\put(32,27){\textcolor{Blue}{$(s_3,s_2)$}}
\put(59,27){\textcolor{Blue}{$(s_1,s_4)$}}
\put(42,31){\textcolor{Red}{$A$}}
\put(31,38){\textcolor{Blue}{$B$}}
\end{overpic}
\caption{\small Critical network $\scrW$ for the AD$_2$ theory together with a choice of $A$-cycle and $B$-cycle as well as framing data $s_1,s_2,s_3,s_4$.}
\label{AD2FNCycles}
\end{figure}

\noindent
Let us compute $\Weff$ for the Argyres-Douglas theory AD$_2$ following
the geometric recipe. Recall that this is the four-dimensional $\N=2$
field theory of a single free hypermultiplet of mass $m$,
corresponding to the Seiberg-Witten curve
\begin{align}
\Sigma\colon \quad w^2=z^2+m \ ,
\end{align}
viewed as a branched covering $\Sigma\to C$ of degree two over
$C=\C$. This geometry is illustrated in
Figure~\ref{AD2SWCurve}. Although the AD$_2$ Lagrangian has merely a
free hypermultiplet contribution, we can introduce gauge degrees of
freedom by gauging the flavour symmetry of hypermultiplet. As we saw
earlier, this turns $m$ into a variable parameterizing a complex
one-dimensional Coulomb branch $\scrB$. The corresponding $A$-cycle
and $B$-cycle are illustrated in Figure~\ref{AD2FNCycles}, together with the relevant
critical  network $\scrW$ (which we already
encountered in Example~\ref{ex:CAD2}, together with its resolutions
$\scrW^\pm$ in Example~\ref{ex:AD2abelianisation}).

The associated complex two-dimensional moduli space $\scrM_{\rm
flat}^\epsilon(C,SL(2,\C))$ of flat $\scrW$-framed
$\epsilon$-connections $\nabla_\epsilon$ was described in
Example~\ref{ex:AD2abelianisation}, while the corresponding length and twist coordinates
\begin{align}
x=\frac\epsilon{\pi\,\I} \log(-\cX_A) \qquad \mbox{and} \qquad y^\pm=\frac1{2\,\epsilon}
  \log\cX_B^\pm 
\end{align}
were constructed in Example~\ref{ex:AD2spectralcoordinates}. Recall
that the framing of $\nabla_\epsilon$ (which is the same for the
network $\scrW$ and its resolutions $\scrW^\pm$) consists of
a local section $s_l$ at each of the marked points $z_l$ at infinity
(see Figure~\ref{AD2FNCycles}), and that the spectral coordinates
\begin{align}
\cX_A = \frac{s_1\wedge s_2}{s_3\wedge s_2} \, \frac{s_3\wedge s_4}{s_1\wedge s_4} \ , \quad \cX_B^+ = \frac{s_1\wedge
  s_4}{s_2\wedge s_4} \qquad \mbox{and} \qquad \cX_B^- =
  -\frac{s_1\wedge s_3}{s_1\wedge s_2}
\end{align}
are determined in terms of this framing data.

The brane of $\epsilon$-opers $\scrL_\epsilon^{\rm oper}$ was found
to be parameterized by the Schr\"odinger equation
\begin{align}
{\rm D}_\epsilon\psi(z) = \epsilon^2\,\psi''(z) - (z^2+m)\,\psi(z)
  = 0
\end{align}
in Example~\ref{ex:operharmonic}. As we saw in
Example~\ref{ex:AD2WKB}, when ${\rm arg}(\epsilon)=\pi/2$ and
$m>0$ the spectral network $\scrW$ matches with the Stokes
graph for the Schr\"odinger oper ${\rm D}_\epsilon$. In that case the
spectral coordinates $\cX_A$ and $\cX_B^\pm$ may be computed as Voros
symbols using exact WKB analysis.

Here we let $m$ and $\epsilon$ be \emph{free} variables,
and evaluate $\cX_A$ and $\cX_B^\pm$ on the complete family of
opers ${\rm D}_\epsilon$. We frame the flat $\epsilon$-oper connection
$\nabla_\epsilon^{\rm oper}$ by choosing
\begin{align}
s_l(z)=\begin{pmatrix} -\epsilon\,\psi_l'(z) \\ \psi_l(z) \end{pmatrix}
\end{align}
as in Example~\ref{ex:operharmonic}. That is, the abelianization of
$\nabla_\epsilon^{\rm oper}$ is described in terms of the parabolic
cylinder functions
\begin{align}
\psi_1(z) &=
            D_\mu\Big(\sqrt{\frac2\epsilon} \, z\Big) \ , \nonumber \\[4pt]
\psi_2(z) &= \E^{-\frac{\pi\,\I}2\,(\mu+1)} \,
  D_{-\mu-1}\Big(-\sqrt{\frac2\epsilon} \, \I \, z\Big) \ , \nonumber
  \\[4pt]
\psi_3(z) &= \E^{\,\pi\,\I\,\mu} \,
            D_\mu\Big(-\sqrt{\frac2\epsilon} \, z\Big) \ , \nonumber \\[4pt]
\psi_4(z) &= \E^{-\frac{\pi\,\I}2\,(3\,\mu-1)} \,
  D_{-\mu-1}\Big(\sqrt{\frac2\epsilon} \, \I \, z\Big) \ ,
\end{align}
with $\mu=-\frac m{2\,\epsilon}-\frac12$. This is the framing provided by
the exact WKB method when
$m>0$ and ${\rm arg}(\epsilon)=\frac\pi2$, but now analytically
continued for all choices of $m$ and $\epsilon$.
Since the exterior product of these sections is given by the Wronskian
\begin{align}
s_k\wedge s_l = \epsilon \, \text{Wr}(\psi_k,\psi_l) \ ,
\end{align}
we find
\begin{align}
\cX_A(\nabla_\epsilon^{\rm oper}) &= -\exp\Big(-\pi\, \I\, \frac{m}{\epsilon}\Big) \ , \nonumber
  \\[4pt]
\cX_B^+(\nabla_\epsilon^{\rm oper}) &= \frac{\I}{\sqrt{2\pi}} \,
          \Gamma\Big(\frac12 - \frac m{2\,\epsilon}\Big) \ , \nonumber
  \\[4pt]
\cX_B^-(\nabla_\epsilon^{\rm oper}) &= \frac{\I \, \sqrt{2\pi} \, \E^{-\pi\,\I\,\frac
          m{2\,\epsilon}}}{\Gamma\big(\frac12 + \frac m{2\,\epsilon}\big)} \ .
\end{align}

Finally, we now have all the ingredients we need to compute
$\Weff$. We define the length and twist coordinates $x$ and $y$ by averaging
over the coordinates in the two resolutions (similarly
to~\cite[Section~10]{hollands2018higher}) to get
\begin{align}
x &= \frac\epsilon{\pi\,\I} \log (-\cX_ A)  =  -m \ , \nonumber \\[4pt]
y &= \frac1{2\,\epsilon} \log \sqrt{\cX_ B^+\,\cX_B^-} = \frac{1}{4\,\epsilon} \log
    \frac{\Gamma\big(\frac{1}{2}+\frac{x}{2\,\epsilon}\big)}{\Gamma\big(\frac{1}{2}-\frac{x}{2\,\epsilon}\big)}
    \ ,
\label{eq:AD2xy}\end{align}
where we dropped a linear term $\frac1{4\,\epsilon^2}\,x$ from $y$.\footnote{That is,
  we fix the shift ambiguity in the twist coordinate by defining $y$
  as in \eqref{eq:AD2xy}.} The effective twisted superpotential
$\Weff$ is then obtained by integrating the relation
\begin{align}
y = \frac1\epsilon \, \frac{\partial\Weff(x;\epsilon)}{\partial x} \ .
\end{align}
Recall the special function from Section~\ref{sec:Weff} given by
\begin{align}
\Upsilon(x) = \int_{\frac12}^x\, \log \frac{\Gamma(x')}{\Gamma(1-x')}\, \de x' \ ,
\end{align}
with the properties
\begin{align}
\frac\partial{\partial x}\Upsilon(a+b\,x) = b \log
  \frac{\Gamma(a+b\,x)}{\Gamma(1-a-b\,x)} \qquad \mbox{and} \qquad
  \Upsilon(1-x) = \Upsilon(x) \ ,
\end{align}
for constants $a,b$.
It then follows that
\begin{equation}
\Weff(x;\epsilon) = \frac\epsilon2\, \Upsilon\Big( \frac{1}{2} + \frac{x}{2\,\epsilon} \Big) = \frac\epsilon2\, \Upsilon\Big( \frac{1}{2} + \frac{m}{2\,\epsilon} \Big)  \ ,
\end{equation}
up to an integration constant that is independent of $x$. This indeed agrees
with the effective twisted superpotential $\Weff(m;\epsilon)$ from
\eqref{eq:AD2Weff}.

\subsection{Pure $SU(2)$ Theory}\label{ex:recipepure}
\noindent
As our second example, we compute the superpotential $\Weff$ for the weakly coupled pure $SU(2)$ theory described in Section~\ref{sec:SWTheory}. Recall that the Seiberg-Witten curve corresponding to the pure $SU(2)$ theory is a ramified double covering $\Sigma\to C$ over the twice-punctured sphere, given by the equation
\begin{align}\label{eq:purecurverecipe}
w^2 = \frac{\Lambda^2}{z^3} - \frac{2\,u}{z^2} + \frac{\Lambda^2}z \ .
\end{align}
This geometry was illustrated in Figure~\ref{SymplecticBasis}. In \eqref{eq:purecurverecipe}, $u$ is a parameter on the complex one-dimensional Coulomb branch and $\Lambda$ is the ultraviolet scale. Since we consider the weakly coupled pure $SU(2)$ theory, we assume $|\Lambda^2/u|\ll 1$. 

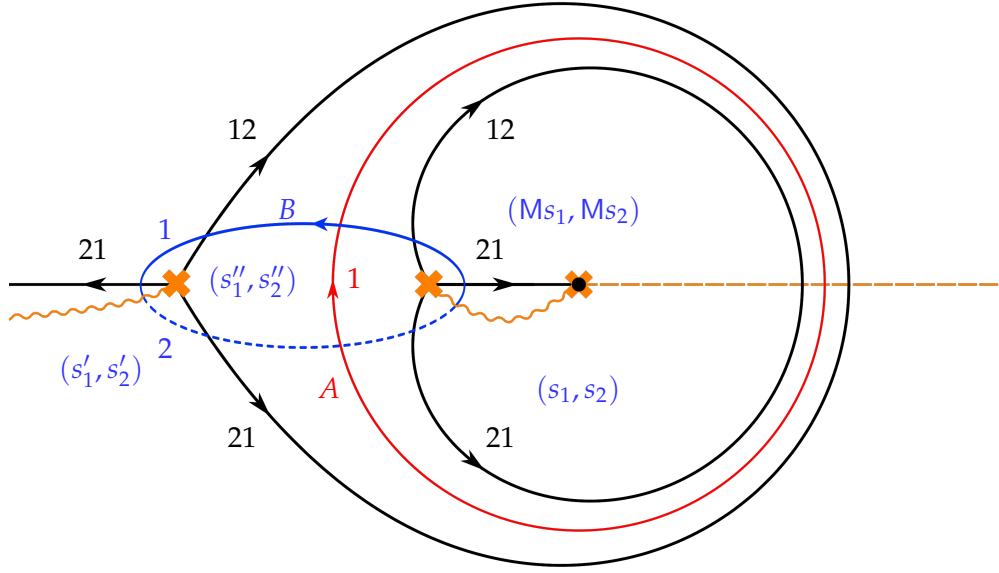
\begin{figure}[h!]
\centering
\small
\begin{overpic}
[width=0.80\textwidth]{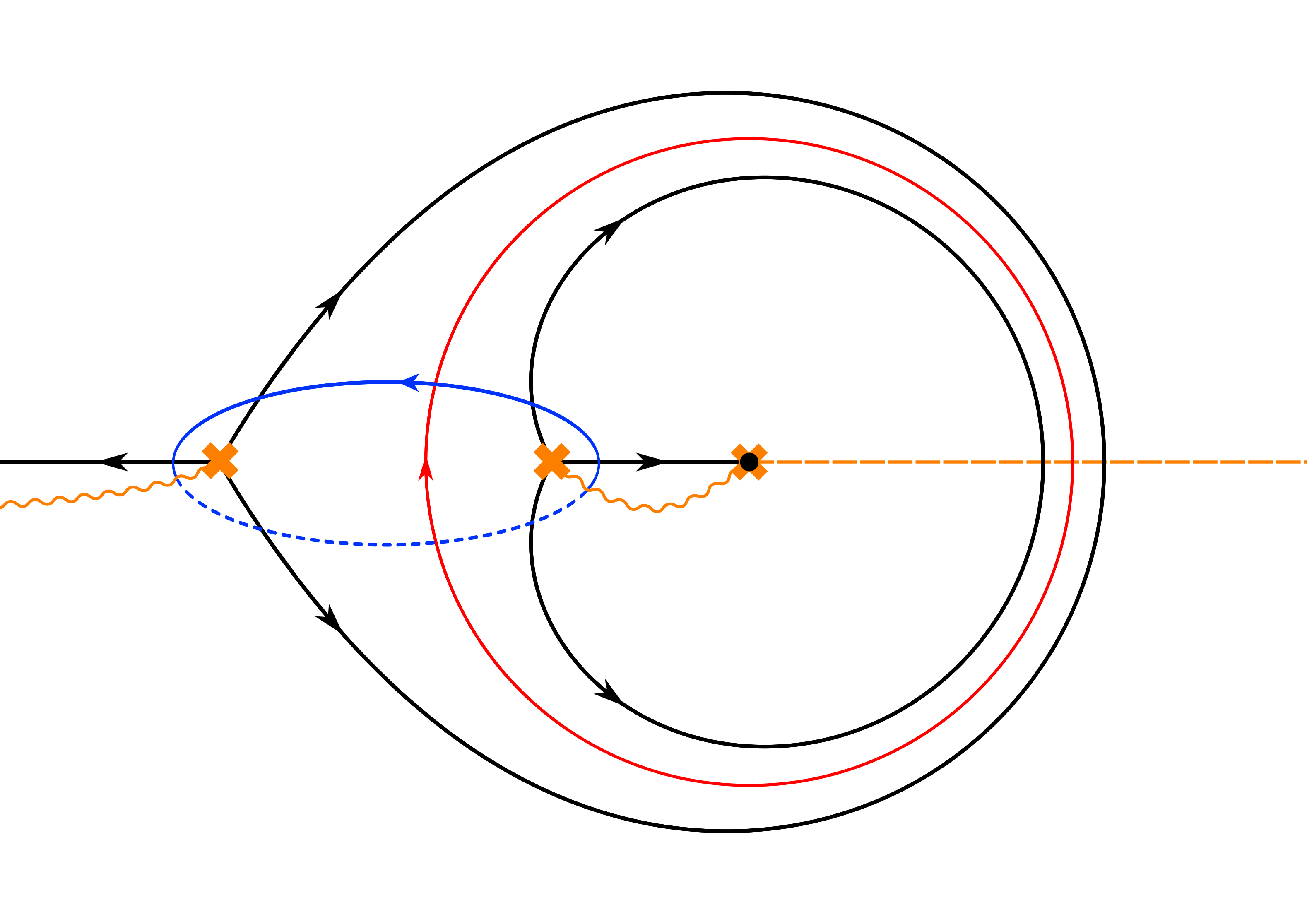}
\put(7,38){$21$}
\put(22,50){$12$}
\put(22,19){$21$}
\put(48,50){$12$}
\put(48,19){$21$}
\put(47,38){$21$}
\put(20,35){\textcolor{Blue}{$\left(s_1'',s_2''\right)$}}
\put(53,24){\textcolor{Blue}{$\left(s_1,s_2\right)$}}
\put(50,42){\textcolor{Blue}{$\left({\sf M}s_1,{\sf M}s_2\right)$}}
\put(5,26){\textcolor{Blue}{$\left(s_1',s_2'\right)$}}
\put(27,42){\textcolor{Blue}{$B$}}
\put(31,24){\textcolor{Red}{$A$}}
\put(34,35){\textcolor{Red}{$1$}}
\put(15,40){\textcolor{Blue}{$1$}}
\put(15,28){\textcolor{Blue}{$2$}}
\end{overpic}
\caption{\small Fenchel-Nielsen type network for the pure $SU(2)$ theory in the weak coupling regime together with a choice of $A$-cycle and $B$-cycle, drawn at $u=-\frac98$ and $\vartheta=\frac\pi2$.}
\label{PureSU2FNCycles}
\normalsize
\end{figure}

In Example~\ref{ex:FNpureab} we constructed the relevant Fenchel-Nielsen coordinates on the corresponding complex two-dimensional moduli space $\scrM_{\rm
  flat}^\epsilon(C,SL(2,\C))$ of framed flat $\epsilon$-connections with respect to the spectral network and its resolutions, see Figure~\ref{WminusSections}.\footnote{Although this network was drawn for specific values of $\Lambda$ and $u$, an isotopic network may be found at any point in the weak coupling region.} For convenience, we have drawn this network again in Figure~\ref{PureSU2FNCycles}. 
Recall that the framing data consists of a choice of local sections $s_1\in\ell$, $s_1'\in\ell'$ and $s_2''\in\ell''$, where $\ell$ and $\ell'$ are $\nabla$-invariant line bundles in a neighbourhood of $z=0$ and $z=\infty$, respectively, and $\ell''$ is an eigenline of the counterclockwise monodromy $\sf M$ of $\nabla$ (see Example~\ref{ex:pureab}). 
  
In terms of these choices, we found in Example~\ref{ex:FNpureab} that the spectral coordinate $\cX_A$ is the eigenvalue of $\sf M$ corresponding to the eigenvector $s_2''$, while the $B$-cycle coordinates are
\begin{align}\label{eq:XBrecipe}
\cX_B^- &= \cX_A^2 \, \frac{(s_1\wedge s_2^{\prime\prime})^2\,(s_1'\wedge
  s_1^{\prime\prime})^2}{(s_1\wedge s_2)\,(s_1'\wedge
  s_2^{\prime})\,(s_1''\wedge s_2'')^2} \ , \nonumber \\[4pt]
\cX_B^+ &= \cX_A^{-2} \, \frac{(s_1\wedge s_2)\,(s_1'\wedge s_2')\,(s_1''\wedge s_2'')^2}{(s_1\wedge s_1'')^2\,(s_1'\wedge s_2'')^2} \ , \nonumber \\[4pt]
\cX_B &= \sqrt{\cX_B^- \, \cX_B^+} = \frac{s_1'\wedge s_1''}{s_1\wedge s_1''} \, \frac{s_1\wedge s_2''}{s_1'\wedge s_2''} \ .
\end{align}
Bear in mind that the sections $s_1$, $s_1'$ and $s_2''$ should be evaluated at the same point $z\in C$.

The brane of $\epsilon$-opers \smash{$\scrL^{\rm
  oper}_\epsilon$} is parametrized by the Mathieu equation
\begin{align}\label{eq:recipeMathieu}
{\rm D}_\epsilon\psi(z) = \epsilon^2 \, \frac{\partial^2\psi(z)}{\partial z^2} - \Big(\frac{\Lambda^2}{z^3} - \frac{2\,u+\epsilon^2/4}{z^2} + \frac{\Lambda^2}{z}\Big)\,\psi(z) = 0 \ ,
\end{align}
for $|u|\gg\Lambda^2$, as found in Example~\ref{ex:operpure}. As in Section~\ref{ex:recipeAD2}, we let $\Lambda$, $u$ and $\epsilon$ be free variables with $|u/\Lambda^2|\gg1$, and evaluate $\cX_A$ and $\cX_B$ on the whole family of opers ${\rm D}_\epsilon$. The local sections $s(z)$ are related to the solutions $\psi(z)$ of \eqref{eq:recipeMathieu} as
\begin{align}\label{eq:spsirelation}
s(z) = \bigg(\begin{matrix} -\epsilon \, \frac{\partial\psi(z)}{\partial z} \\ \psi(z) \end{matrix} \bigg) \ .
\end{align}
We fix the framing data by letting $\psi_1(z,\Lambda)$ be the asymptotically small solution of \eqref{eq:recipeMathieu} at $z=0$ (when approaching $z=0$ along the negative real axis), while similarly $\psi_1'(z,\Lambda)$ is the asymptotically small solution at $z=\infty$ (when approaching $z=\infty$ along the negative real axis) and $\psi_2''(z,\Lambda)$ is the small eigenvector of the monodromy $\sf M$ of $\nabla_\epsilon^{\rm oper}$, all for $-u\gg\Lambda^2$ and ${\rm arg}(\epsilon)=\pi/2$. This is the framing provided by the exact WKB method for these choices of parameters, but now analytically continued to all $u,\epsilon\in\C$. 

Unlike in Section~\ref{ex:recipeAD2}, the functions $\psi_1(z,\Lambda)$, $\psi_1'(z,\Lambda)$ and $\psi_2''(z,\Lambda)$ are not known in exact form. But it is possible to find them as series expansions in $\Lambda^2$, while exact in $\epsilon$. For this, we follow the approach of~\cite[Section~9]{hollands2018higher}.

We start with the eigenfunctions $\psi_i''(z,\Lambda)$ of the monodromy $\sf M$. Note that the Mathieu equation \eqref{eq:recipeMathieu} is parametrized in terms of $\Lambda^2$. Since $\Lambda^2\ll|u|$, we expand
\begin{align}
u = \sum_{k=0}^\infty \, u_k \, \Lambda^{2k}
\end{align}
and
\begin{align}\label{eq:recipepsiexpand}
\psi_i''(z,\Lambda) = \sum_{k=0}^\infty \, f_{k,i}''(z) \, \Lambda^{2k} \ .
\end{align}
At order $\Lambda^0$, the Mathieu equation \eqref{eq:recipeMathieu} reads
\begin{align}
\epsilon^2 \, \frac{\partial^2f_0''(z)}{\partial z^2} + \frac{8\,u_0+\epsilon^2}{4\,z^2} \, f_0''(z) = 0 \ ,
\end{align}
with the two independent solutions
\begin{align}
f_{0,i}''(z) = \alpha_i \, z^{\frac12\pm\frac {a_\FN}{2\,\epsilon}}
\end{align}
if we choose $u_0= -a_\FN^2/8$. These solutions satisfy $f_{0,i}''(\E^{\,2\pi\,\I}\,z) = -\E^{\pm\,\pi\,\I\,a_\FN/\epsilon} \,f_{0,i}''(z)$; since we are looking for functions $\psi_i''(z,\Lambda)$ that diagonalize the monodromy around $z=0$ with eigenvalue $-\E^{\pm\,\pi\,\I\,a_\FN/\epsilon}$, this is the property we require for all $f_{k,i}''(z)$. Note that if we choose $a_\FN>0$, then \smash{$\psi_2''(z,\Lambda)$} is indeed the small eigenvector for $-u\gg\Lambda^2$ and ${\rm arg}(\epsilon)=\frac\pi2$.

This guides us to the ansatz
\begin{align}
f_{k,i}''(z) = \sum_{m=-k}^k \, c_{k,\pm}^{(m)} \, z^{\frac12\pm\frac {a_\FN}{2\,\epsilon} + m} \ .
\end{align}
This ansatz indeed solves the Mathieu equation \eqref{eq:recipeMathieu} at order $\Lambda^2$, which reads
\begin{align}
\epsilon^2 \, \frac{\partial^2f_1''(z)}{\partial z^2} + \frac{8\,u_0 + \epsilon^2}{4\,z^2} \, f_1''(z) - \Big(\frac{1}{z^3} - \frac{2\,u_1}{z^2} + \frac{1}z \Big) \, f_0''(z) = 0 \ ,
\end{align}
if we choose
\begin{align}
c_{1,\pm}^{(-1)} = \pm \, \frac1{\epsilon \, (a_\FN\pm\epsilon)} \ , \quad u_1=0 \qquad \mbox{and} \qquad c_{1,\pm}^{(1)} = \mp \, \frac1{\epsilon\,(a_\FN\mp\epsilon)} \ ,
\end{align}
while $c_{1,\pm}^{(0)}$ is unconstrained. The same strategy can be applied to go to any order in $\Lambda^2$. For instance, at order $\Lambda^4$ we find
\begin{align}
c_{2,\pm}^{(-1)} &= \frac1{2\,\epsilon^2\,(a_\FN\pm2\,\epsilon)\,(a_\FN\pm\epsilon)} \ , \quad
c_{2,\pm}^{(-1)} = \pm \, \frac{c_{1,\pm}^{(0)}}{\epsilon\,(a_\FN\pm\epsilon)} \ , \quad
u_2 = -\frac1{a_\FN^2-\epsilon^2} \ , \nonumber \\
c_{2,\pm}^{(1)} &= \mp \, \frac{c_{1,\pm}^{(0)}}{\epsilon\,(a_\FN\mp\epsilon)} \qquad \mbox{and} \qquad
c_{2,\pm}^{(2)} = -\frac1{2\,\epsilon^2\,(a_\FN\mp2\,\epsilon)\,(a_\FN\mp\epsilon)} \ , 
\end{align}
while $c_{2,\pm}^{(0)}$ is again undetermined, and so on.

Computing the asymptotically small solution $\psi_1(z,\Lambda)$ near $z=0$ is a little trickier. This requires explicit knowledge of the expansion
\begin{align}\label{eq:uexpansion}
u = -\frac{a_\FN^2}8 - \frac{\Lambda^4}{a_\FN^2-\epsilon^2} - \frac{5\,a_\FN^2+7\,\epsilon^2}{(a_\FN^2-4\,\epsilon^2)\,(a_\FN^2-\epsilon^2)^3} \, \Lambda^8 + O(\Lambda^{10})
\end{align}
which we just computed. We introduce a new coordinate $v=z/\Lambda^2$ which is of order~$1$ even if $z$ is very small. In this coordinate the Mathieu equation \eqref{eq:recipeMathieu} reads
\begin{align}\label{eq:recipeMathieuv}
\epsilon^2 \, \frac{\partial^2 g(v)}{\partial v^2} - \Big( \frac1{v^3} - \frac{8\,u+\epsilon^2}{4\,v^2} + \frac{\Lambda^4}v\Big) \, g(v) = 0 \ ,
\end{align}
with $g(v):=\psi(\Lambda^2\,v)$. As before, we expand $g(v)$, this time as a series in $\Lambda^4$:
\begin{align}
g(v) = \sum_{l=0}^\infty \, g_l(v) \, \Lambda^{4l} \ .
\end{align}

At order $\Lambda^0$, the rescaled Mathieu equation \eqref{eq:recipeMathieuv} reads
\begin{align}
\epsilon^2 \, \frac{\partial^2 g_0(v)}{\partial v^2} - \Big(\frac1{v^3} - \frac{8\,u_0+\epsilon^2}{4\,v^2}\Big) \, g_0(v) = 0 \ .
\end{align}
For the two independent solutions $g_{0,i}(v)$ we choose the Hankel functions
\begin{align}
g_{0,i}(v) = \sqrt{v} \ H^{([i+1])}\Big(\frac{a_\FN}\epsilon , \frac{2\,\I}{\epsilon\,\sqrt v}\Big) \ ,
\end{align}
where the index $[i+1]$ is understood modulo~$2$, because they have the correct asymptotics
\begin{align}\label{eq:recipeHankel}
H^{(i)}\Big(\frac{a_\FN}\epsilon , \frac{2\,\I}{\epsilon\,\sqrt v}\Big) \simeq \sqrt{-\I\,\epsilon} \ v^{1/4} \, \E^{\mp\,2/\epsilon\,\sqrt v}
\end{align}
when $v\to0$ with $-\pi<{\rm arg}(2\,\I/\epsilon\,\sqrt v\,)<2\pi$;\footnote{Note that the Hankel functions are multivalued functions, and so are really defined on the universal cover of $\C\setminus\{0\}$.} note that $H^{(1)}$ has exponent \smash{$-\frac{2}{\epsilon\,\sqrt v}<0$} when ${\rm arg}(\epsilon)=\frac\pi2$ and $v$ has negative real part as well as negative imaginary part. It is also useful to know~\cite[Equation~9.1.39]{Handbook}
\begin{align}
H^{(2)}(\alpha,\E^{-\pi\,\I}\,z) = -\E^{-\pi\,\I\,\alpha} \, H^{(1)}(\alpha,z) \ ,
\end{align}
which implies
\begin{align}
g_{0,1}(\E^{\,2\pi\,\I} \, v) = \E^{-\pi\,\I\,a_\FN/\epsilon} \, g_{0,2}(v) \ .
\end{align}
Since the solutions $\psi_1(z,\Lambda)$ and $\psi_2(z,\Lambda)$ near $z=0$ should be related as ${\sf M}\psi_1(z,\Lambda) = \psi_2(z,\Lambda)$, this equation implies that the leading order of the solution $\psi_i(z,\Lambda)$ needs to be
\begin{align}
\psi_i(z,\Lambda) = \E^{\pm\,\pi\,\I\,a_\FN/2\,\epsilon} \, g_{0,i}(v) + O(\Lambda^4) \ .
\end{align}

At order $\Lambda^4$, the equation \eqref{eq:recipeMathieuv} reads
\begin{align}\label{eq:recipeMathieuvorder4}
\epsilon^2 \, \frac{\partial^2g_1(v)}{\partial v^2} - \Big(\frac1{v^3} + \frac{a_\FN^2+\epsilon^2}{v^2}\Big) \, g_1(v) - \Big( \frac1v + \frac2{(a_\FN^2-\epsilon^2)\,v^2}\Big) \, g_0(v) = 0 \ .
\end{align}
Since we are looking for solutions $g(v)$ with the asymptotics \eqref{eq:recipeHankel} when $v\to0$, we propose the ansatz
\begin{align}
g_{1,i}(v) = \sum_{n=\pm\,1} \, d_{1,i}^{(n)} \, \sqrt v \ H^{([i+1])}\Big(\frac{a_\FN}\epsilon+2\,n,\frac{2\,\I}{\epsilon\,\sqrt v}\Big) \ .
\end{align}
This indeed solves \eqref{eq:recipeMathieuvorder4} when
\begin{align}
d_{1,i}^{(-1)} = \mp \, \frac1{a_\FN\,\epsilon\,(a_\FN\pm\epsilon)^2} \qquad \mbox{and} \qquad d_{1,i}^{(+1)} = \pm \, \frac1{a_\FN\,\epsilon\,(a_\FN\mp\epsilon)^2} \ .
\end{align}
We can extend this analysis to higher orders in $\Lambda^4$ by making the ansatz
\begin{align}
g_{l,i}(v) = \sum_{n=-l}^l \, d_{l,i}^{(n)} \, \sqrt v \ H^{([i+1])}\Big(\frac{a_\FN}\epsilon + 2\,n,\frac{2\,\I}{\epsilon\,\sqrt v}\Big) \ ,
\end{align}
and solving \eqref{eq:recipeMathieuv} at order $\Lambda^{4l}$ for $d_{l,i}^{(n)}$. 

We are not done yet though, as we are looking for solutions $\psi(z)$ of the original Mathieu equation \eqref{eq:recipeMathieu} with the correct asymptotics as $z\to0$, in a series expansion in $\Lambda^2$. Thus we need to analytically continue the functions $g_i(v)$ to $v\to\infty$, while keeping $z=\Lambda^2\,v$ of order~$1$. For this, we use the fact that the Hankel functions \smash{$H^{(i)}$} have the series expansions
\begin{align}
H^{(i)}\Big(\frac{a_\FN}\epsilon , \frac{2\,\I}{\epsilon\,\sqrt v}\Big) &= \mp \, \frac{\I}\pi \, \big(-\epsilon^2\,v\big)^{a_\FN/2\,\epsilon} \, \Gamma\Big(\frac{a_\FN}{\epsilon}\Big) \, \Big( 1 - \frac1{\epsilon\,(a_\FN-\epsilon)} \, v^{-1} + O(v^{-2}) \Big) \nonumber \\
& \quad \, \pm \frac{\I \, \E^{\mp\,\frac{\pi\,\I\,a_\FN}\epsilon} \, \big(-\epsilon^2\,v\big)^{-a_\FN/2\,\epsilon}}{\Gamma\big(\frac{a_\FN}\epsilon + 1\big) \sin\big(\frac{\pi\,a_\FN}\epsilon\big)} \, \Big(1 + \frac1{\epsilon\,(a_\FN + \epsilon)} \, v^{-1} + O(v^{-2}) \Big)
\end{align}
when $v\to\infty$. If we substitute $v=z/\Lambda^2$ in the corresponding series expansion in $v^{-1}$ for $g_i(v)$ and reorganize the resulting expression as a series in $\Lambda^2$, we find
\begin{align}
\psi_i(z,\Lambda) = \E^{\pm\,\pi\,\I\,a_\FN/2\,\epsilon} \, g_{i}(z/\Lambda^2) =  \sum_{k=0}^\infty \, \psi_{k,i}(z) \, \Lambda^{2k} \ ,
\end{align}
where $i\in\{1,2\}$; here $\psi_{0,i}(z)$ obtains a contribution only from the leading term in the series expansion of $g_{0,i}(v)$ when $v\to\infty$, while $\psi_{1,i}(z)$ gets a contribution from the next-to-leading term in the series expansion of $g_{0,i}(v)$ when $v\to\infty$ as  well as a contribution from the leading term in the series expansion of $g_{1,i}(v)$ when $v\to\infty$, and so on. Note that $\psi_1(z,\Lambda)$ is indeed asymptotically small near $z=0$ when ${\rm arg}(\epsilon) = \frac\pi2$ and $v$ has negative real part as well as negative imaginary part.

Since the resulting function $\psi_i(z,\Lambda)$ should be a solution of the original Mathieu equation \eqref{eq:recipeMathieu}, we should be able to express it as a linear combination of $\psi_1''(z,\Lambda)$ and $\psi_2''(z,\Lambda)$, the eigenfunctions of the monodromy $\sf M$ computed in \eqref{eq:recipepsiexpand}. Indeed we find
\begin{align}
\psi_i(z,\Lambda) = \frac{r_{i}}{\Lambda} \, \psi_2''(z,\Lambda) + \frac{s_{i}}{\Lambda} \, \psi_1''(z,\Lambda) \ ,
\end{align}
where
\begin{align}\label{eq:risi}
r_{i} &= \pm \, \frac\I{\pi} \, \Big(\frac{\I\,\Lambda}\epsilon\Big)^{\frac{a_\FN}\epsilon} \, \E^{\pm\,\frac{\pi\,\I\,a_\FN}{2\,\epsilon}} \, \Gamma\Big(-\frac{a_\FN}\epsilon\Big) \nonumber \\
& \hspace{1cm} \times \Big( 1 - \frac{1}{\epsilon^2\,(\epsilon-a_\FN)^2}\,\Lambda^4 + \frac{a_\FN^3+a_\FN^2\,\epsilon-29\,a_\FN\,\epsilon^2+43\,\epsilon^3}{4\,\epsilon^4\,(\epsilon+a_\FN)\,(2\,\epsilon-a_\FN)^2\,(\epsilon-a_\FN)^4} \, \Lambda^8 + O(\Lambda^{10}) \Big) \ , \nonumber \\[4pt]
s_{i} &= \pm \, \frac\I{\pi} \, \Big(\frac\epsilon{\I\,\Lambda}\Big)^{\frac{a_\FN}\epsilon} \, \E^{\mp\,\frac{\pi\,\I\,a_\FN}{2\,\epsilon}} \, \Gamma\Big(\frac{a_\FN}\epsilon\Big) \\
& \hspace{1cm} \times \Big(1-\frac{1}{\epsilon^2\,(\epsilon+a_\FN)^2}\,\Lambda^4 + \frac{-a_\FN^3+a_\FN^2\,\epsilon+29\,a_\FN\,\epsilon^2+43\,\epsilon^3}{4\,\epsilon^4\,(\epsilon-a_\FN)\,(2\,\epsilon+a_\FN)^2\,(\epsilon+a_\FN)^4} \, \Lambda^8 + O(\Lambda^{10}) \Big) \ , \nonumber
\end{align}
if we set the free parameters as $c_{1,\pm}^{(0)}=c_{3,\pm}^{(0)}=0$, $c_{2,\pm}^{(0)} = d_{2,\pm}^{(0)}$ and $c_{4,\pm}^{(0)}=d_{4,\pm}^{(0)}$. This required computing $g_i(v)$ up to order \smash{$\Lambda^{16}$}. As a check, note that
\begin{align}
{\sf M}\psi_1(z,\Lambda) &= - \frac{r_1\,\E^{-\pi\,\I\,a_\FN/\epsilon}}\Lambda \, \psi_2''(z,\Lambda) - \frac{s_1\,\E^{\,\pi\,\I\,a_\FN/\epsilon}}\Lambda \, \psi_1''(z,\Lambda) \nonumber \\[4pt]
&=  \frac{r_2}\Lambda \, \psi_2''(z,\Lambda) + \frac{s_2}\Lambda \, \psi_1''(z,\Lambda) \nonumber \\[4pt]
&= \psi_2(z,\Lambda) \ .
\end{align}

The same strategy at the other end of $C$, where $z\to\infty$ and it is useful to introduce a local coordinate $w=\Lambda^2\,z$, shows that
\begin{align}
\psi_i'(z,\Lambda) = \Lambda \, r_{[i+1]}\,\psi_1''(z,\Lambda) + \Lambda \, s_{[i+1]}\,\psi_2''(z,\Lambda) \ .
\end{align}
Indeed, we check that
\begin{align}
{\sf M}\psi_1'(z,\Lambda) &= -\Lambda \, r_2\,\E^{\,\pi\,\I\,a_\FN/\epsilon} \, \psi_1''(z,\Lambda) - \Lambda\,s_2\,\E^{-\pi\,\I\,a_\FN/\epsilon} \, \psi_2''(z,\Lambda) \nonumber \\[4pt]
&= \Lambda\,r_1\,\psi_1''(z,\Lambda) + \Lambda\,s_1\,\psi_2''(z,\Lambda) \nonumber \\[4pt]
&= \psi_2'(z,\Lambda) \ .
\end{align}

We finally have all the ingredients necessary to compute the spectral coordinates $\cX_A$ and $\cX_B$ at $\nabla_\epsilon^{\rm oper}$, and thus to extract the effective twisted superpotential $\Weff$ in a series expansion in $\Lambda$ for the pure $SU(2)$ theory at weak coupling. We find
\begin{align}\label{eq:xcXASU2}
\cX_A(\nabla_\epsilon^{\rm oper}) = -\exp\Big(-\pi\,\I\,\frac {a_\FN}\epsilon\Big) \ ,
\end{align}
while 
\begin{align}\label{eq:xcXBSU2}
\cX_B(\nabla_\epsilon^{\rm oper}) &= \frac{\textrm{Wr}(\psi_1',\psi_1'')}{\textrm{Wr}(\psi_1,\psi_1'')} \, \frac{\textrm{Wr}(\psi_1,\psi_2'')}{\textrm{Wr}(\psi_1',\psi_2'')} = \frac{s_1\,s_2}{r_1\,r_2} \ , \nonumber \\[4pt]
\cX_B^-(\nabla_\epsilon^{\rm oper}) &= \cX_A^2 \, \frac{s_1^2\,s_2^2}{(r_1\,s_2-r_2\,s_1)^2} \ , \nonumber \\[4pt]
\cX_B^+(\nabla_\epsilon^{\rm oper}) &= \cX_A^{-2} \, \frac{(r_1\,s_2-r_2\,s_1)^2}{r_1^2\,r_2^2}  \ ,
\end{align}
where we used $s\wedge\tilde s = \epsilon \, \text{Wr}(\psi,\tilde\psi)$.

From \eqref{eq:xcXASU2} we find
\begin{align}
x = \frac\epsilon{\pi\,\I} \log(-\cX_A) = -a_\FN \ .
\end{align}
Substituting \eqref{eq:risi} into \eqref{eq:xcXBSU2} and expanding to order $\Lambda^{8}$ we obtain
\begin{align}
\frac12 \log\cX_B &= -\frac{2\,a_\FN}\epsilon \log\Big(\frac\Lambda\epsilon\Big) - \frac12 \log\frac{\Gamma\big(-\frac {a_\FN}\epsilon\big)}{\Gamma\big(1+\frac {a_\FN}\epsilon\big)} \, \frac{\Gamma\big(1-\frac {a_\FN}\epsilon\big)}{\Gamma\big(\frac {a_\FN}\epsilon\big)} + \frac{4\,a_\FN\,\Lambda^4}{\epsilon\,(a_\FN^2-\epsilon^2)^2} \nonumber \\
& \quad \, + \frac{6\,(5\,a_\FN^5 - 2\,a_\FN^3\,\epsilon^2 - 37\,a_\FN\,\epsilon^4)}{\epsilon\,(a_\FN^2-4\,\epsilon^2)^2\,(a_\FN^2-\epsilon^2)^4} \, \Lambda^8 + O(\Lambda^{10}) \ ,
\end{align}
where we used
\begin{align}
\frac{\Gamma\big(-\frac {a_\FN}\epsilon\big)}{\Gamma\big(\frac {a_\FN}\epsilon\big)} = -\frac{\Gamma\big(1-\frac {a_\FN}\epsilon\big)}{\Gamma\big(1+\frac {a_\FN}\epsilon\big)} \ .
\end{align}

The effective twisted superpotential is now obtained by integrating $\frac12\log\cX_B$ with respect to $x=-a_\FN$. This gives
\begin{align}\label{eq:WeffpureSU2}
\Weff(a_\FN;\Lambda;\epsilon) &= - \int\,\frac12\log\cX_B \ \de a_\FN \nonumber \\[4pt]
&= \frac{a_\FN^2}\epsilon \log\Big(\frac\Lambda\epsilon\Big) - \frac\epsilon2\,\Upsilon\Big(-\frac {a_\FN}\epsilon\Big) - \frac\epsilon2\,\Upsilon\Big(\frac {a_\FN}\epsilon\Big) + \frac{2\,\Lambda^4}{\epsilon\,(a_\FN^2-\epsilon^2)} \nonumber \\
& \quad \, + \frac{5\,a_\FN^2 + 7\,\epsilon^2}{\epsilon\,(a_\FN^2-4\,\epsilon^2)\,(a_\FN^2-\epsilon^2)^3} \, \Lambda^8 + O(\Lambda^{10}) \ ,
\end{align}
which indeed matches the effective twisted superpotential from \eqref{eq:pureWeff} when we identify the monodromy parameter $a_\FN$ with the ($\epsilon$-corrected) Coulomb parameter $a$. As an additional check, we compute
\begin{align}
-\epsilon \, \frac\Lambda8 \, \frac{\partial\Weff}{\partial\Lambda} = - \frac{a_\FN^2}{8} - \frac{\Lambda^4}{a_\FN^2-\epsilon^2} - \frac{5\,a_\FN^2 + 7\,\epsilon^2}{(a_\FN^2-4\,\epsilon^2)\,(a_\FN^2-\epsilon^2)^3} \, \Lambda^8 + O(\Lambda^{10}) = u \ ,
\end{align}
which agrees with the expansion of $u$ in $\Lambda^2$ given in \eqref{eq:uexpansion}. This is known as the quantum Matone relation~\cite{Alday:2009fs,Drukker:2009id,Ashok:2015gfa}.

Finally, note that if we were to compute the effective twisted superpotential $\Weff$ with respect to the spectral coordinates $(\cX_A,\cX_B^\pm)$, the result would only differ in the one-loop contributions. Instead of
\begin{align}
\widetilde{\cW}^{\rm eff}_{\textrm{1-loop}}(a_\FN;\epsilon) &=
                                                               -\frac\epsilon2\,\Upsilon\Big(-\frac
                                                            {a_\FN}\epsilon\Big)
                                                               -
                                                               \frac\epsilon2\,\Upsilon\Big(\frac
                                                            {a_\FN}\epsilon\Big)
\end{align}
we would obtain
\begin{align}
\widetilde{\cW}^{{\rm eff},\pm}_{\textrm{1-loop}}(a_\FN;\epsilon) = \int\, \Big[\mp \log\Gamma\Big(\mp\,\frac {a_\FN}\epsilon\Big) \mp \log\Gamma\Big(1\mp\frac {a_\FN}\epsilon\Big)\Big] \, \de a_\FN \ ,
\end{align}
since 
\begin{align}
\cX^-_{B,\textrm{1-loop}} &= \frac1{\pi^2} \, \Gamma\Big(\frac{ a_\FN}\epsilon\Big)^2 \, \Gamma\Big(1+\frac {a_\FN}\epsilon\Big)^2 \ , \nonumber \\[4pt]
\cX^+_{B,\textrm{1-loop}} &= \frac{\pi^2}{\Gamma\big(-\frac {a_\FN}\epsilon\big)^2\,\Gamma\big(1-\frac {a_\FN}\epsilon\big)^2} \ .
\end{align}
Interestingly, these are the one-loop contributions to the Nekrasov partition function in the Nekrasov-Shatashvili limit, in the gauge theory or topological string scheme.\footnote{Recall that the one-loop contribution $\exp \widetilde{\cW}^{\rm eff}_{\textrm{1-loop}}$ is only determined up to a phase, see Footnote~\ref{fn:Weff1loop}.}

\subsection{$E_6$ Minahan-Nemeschansky Theory}\label{ex:recipeE6}
\noindent
The $E_6$ Minahan-Nemeschansky theory, introduced in Section~\ref{ClassSTheories}, is an example of an intrinsically strongly coupled $\N=2$ theory of class $\cS$. Its Seiberg-Witten curve, given by \eqref{eq:E6differentials}, is a ramified triple covering $\Sigma\to C$ over the three-punctured sphere with three maximal punctures (see Figure~\ref{bblockssuthree}~B). The three maximal punctures correspond to an $SU(3)\times SU(3)\times SU(3)$ flavour symmetry group in the ultraviolet, which is enhanced to $E_6$ upon flowing to the infrared. 

The $E_6$ theory is superconformal in the massless limit where the Seiberg-Witten curve is given by the equation
\begin{align}
w^3 + \frac u{(z^3-1)^2} = 0 \ ,
\end{align}
which in this case defines a triple unramified covering over the sphere with three maximal punctures at the third roots of unity $z=1,\omega,\omega^2$, where $\omega=\E^{\,2\pi\,\I/3}$. It has a one-dimensional Coulomb branch $\scrB$, parametrized by $u\in\C$. All points on the Coulomb branch are physically equivalent. In particular, there are no walls of marginal stability where BPS bound states can form or decay. The electromagnetic charge lattice $\Gamma$ has three distinguished one-cycles, called $A$, $B$ and $C$ in Example~\ref{ex:WKBE6} (see Figure~\ref{E6Cycles}). The periods $\oint_A \, \lambda = \omega \, \oint_B \, \lambda = \omega^2 \, \oint_C \,\lambda$ are all of the same order, hence there is no canonical choice of an electric-magnetic duality frame.

\begin{figure}[h!]
\centering
\begin{overpic}
[width=0.80\textwidth]{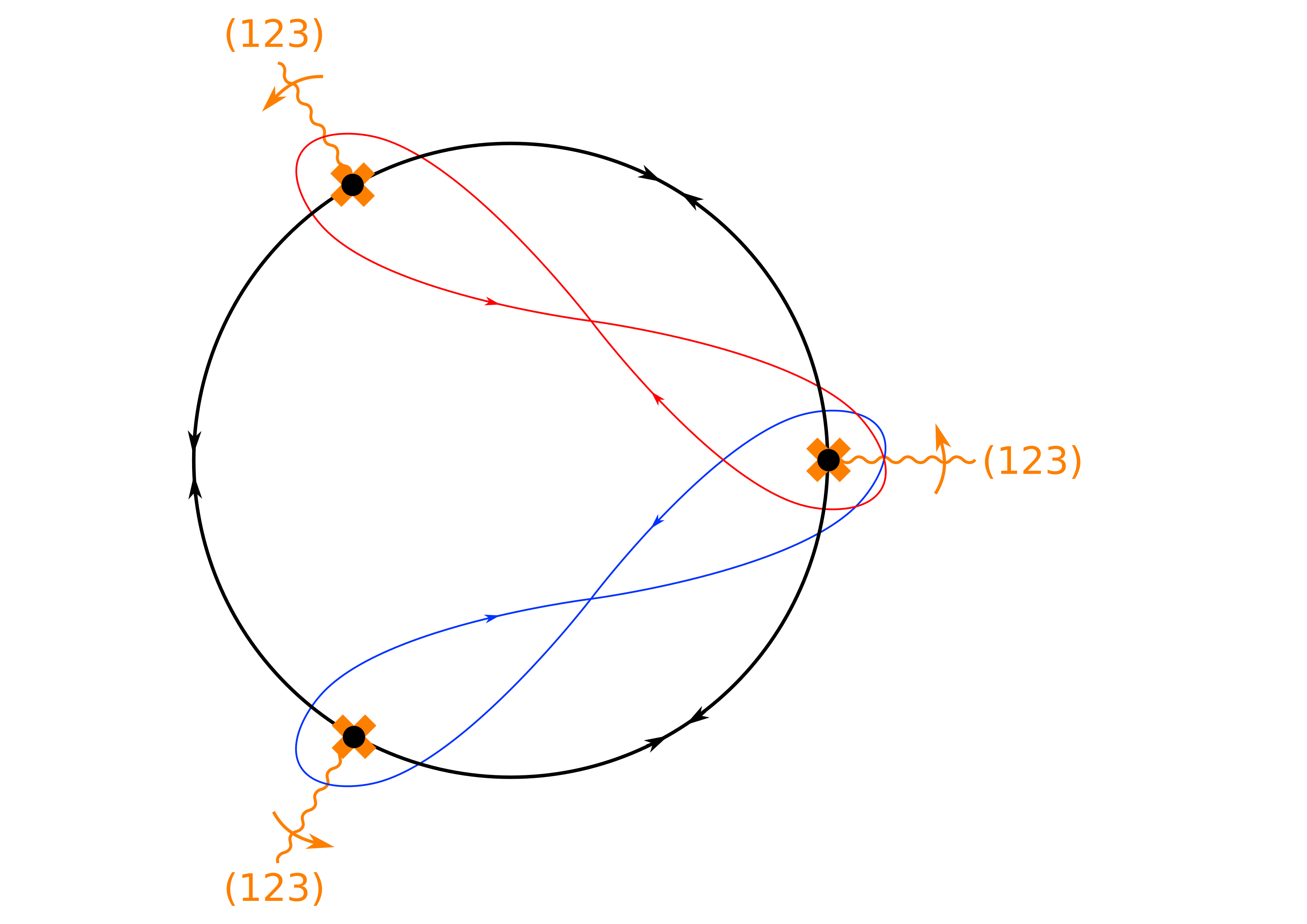}
\put(50,47){\textcolor{Red}{$A$}}
\put(35,44){\textcolor{Red}{$3$}}
\put(47,39){\textcolor{Red}{$2$}}
\put(42,16){\textcolor{Blue}{$B$}}
\put(35,24){\textcolor{Blue}{$1$}}
\put(47,31){\textcolor{Blue}{$2$}}
\put(55,55){$23$}
\put(47,59){$32$}
\put(55,13.5){$13$}
\put(47,10){$31$}
\put(11,37){$12$}
\put(11,32){$21$}
\end{overpic}
\caption{\small Higher rank Fenchel-Nielsen network for the $E_6$ Minahan-Nemeschansky theory together with a choice of $A$-cycle and $B$-cycle.}
\label{E6CyclesV2}
\end{figure}
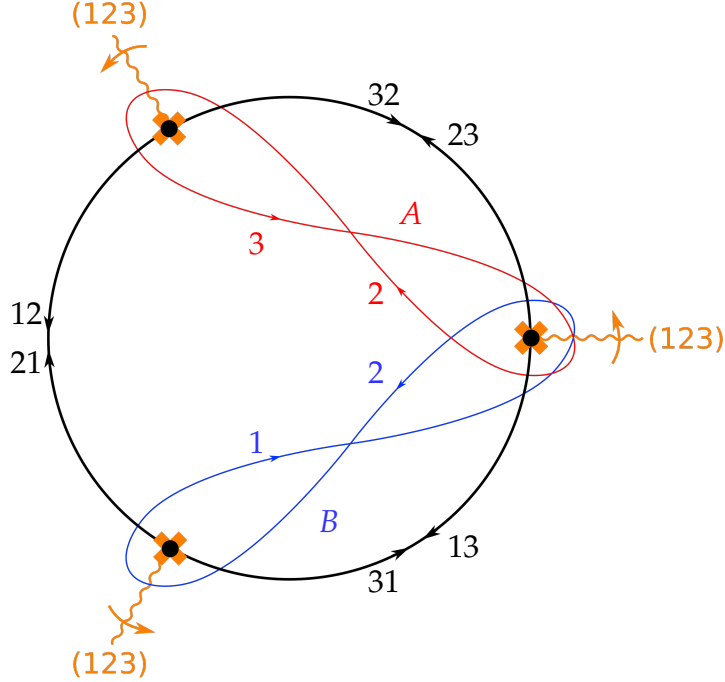

In Example~\ref{ex:higherFN} we noted that the topology of the spectral network $\scrW(u,\vartheta)$ only depends on the phase of the quantity $\E^{-3\,\I\,\vartheta} \, u$, and that the network seems to be ``wild'' at all phases $\vartheta$, except when 
\begin{align}
\E^{-\I\,\vartheta} \, \oint_\gamma \, \lambda \ \in \ \R
\end{align}
for some one-cycle $\gamma$ on $\Sigma$, in which case the network is of higher rank Fenchel-Nielsen type. The simplest topology appears when $\gamma$ is equal to either of the one-cycles $A$, $B$ or $C$. It is illustrated in Figure~\ref{E6Networks}~A for the choice $\gamma=A$. For convenience we have drawn this network again in Figure~\ref{E6CyclesV2}. (The networks corresponding to $\gamma=B,C$ are found simply by rotating this network through $120^\circ$.) 
The Fenchel-Nielsen coordinates $\cX_A$ and $\cX_B$ for the circular network $\scrW_{[1,0]}$ illustrated in Figure~\ref{E6Networks}~A have been obtained and studied in~\cite{hollands2019exact}, however in a form in which it is not easy to extract explicit expressions.

The effective twisted superpotential $\Weff$ for the $E_6$ theory is obtained by evaluating the coordinates $\cX_A$ and $\cX_B$ on the family of $T_3$-opers $\nabla_\epsilon^{\rm oper}(u)$ written down in Example~\ref{ex:K3oper}. 
Even though it is difficult to obtain exact expressions in $\epsilon$ for this example, we have learned in Example~\ref{ex:WKBE6} that the spectral coordinates $\cX_A$ and $\cX_B$ at $\nabla_\epsilon^{\rm oper}(u)$ have an asymptotic expansion, as $\epsilon\to0$ in the half-plane with ${\rm Re}(\E^{-\I\,\vartheta} \, \epsilon)\geq0$, determined by the Riccati equation. With the expansion \eqref{eq:E6PiAt}, and by defining \smash{$\Weff$} through
\begin{align}
x= \frac\epsilon{\pi\,\I} \log(-\cX_A) \qquad \mbox{and} \qquad \frac{\partial\Weff}{\partial x} = \frac12\,\log\cX_B \ ,
\end{align}
we find
\begin{align}
\Weff( x;\epsilon) &= -\frac{\pi^2}{4\,\epsilon} \, \E^{\,4\pi\,\I/3} \, x^2 + 6\pi\,\I\,\epsilon \log\frac x\epsilon + \frac{12\,\sqrt3\,\I\,\epsilon^3}{ x^2} - \frac{144\,\I\,\epsilon^5}{\pi\, x^4} \nonumber \\
& \quad \, + \frac{2\,\sqrt3\,\big(16800\pi^{13/2} + 23\cdot 2^{2/3} \,\Gamma(\frac13)^8 \, \Gamma(\frac76)^5\big)\,\epsilon^7}{35\pi^{17/2} \, x^6} + \cdots
\end{align}
in an expansion in $\epsilon$. Solving the challenge posed in Example~\ref{ex:WKBE6}, that is, finding the Borel sum $\cS_\vartheta\cX_A(t)$, would enable one to write down the exact effective twisted superpotential \smash{$\Weff$} in $\epsilon$ for the $E_6$ theory.

\subsection{Derivation from Quantum Field Theory}\label{sec:QFT}
\noindent
Now that we have explained how the geometric recipe works, let us take a closer look at the physics behind it. We start by sketching a derivation of the NRS correspondence from the point of view of quantum field theory, mostly following the treatments of~\cite{nekrasov2011darboux,Nekrasov:2010ka}.

Consider a theory $\sf T=\sf T_\fg[C,\cD]$ of class $\cS$ in the $\frac12\Omega$-background $\R^2_\epsilon\times S^1\times\R$. On the one hand, as we reviewed in Section~\ref{sec:Weff}, Nekrasov and Shatashvili argued in~\cite{nekrasov2010quantization} that the resulting theory ${\sf T}_\epsilon$ in the infrared limit may be described as a two-dimensional sigma-model (with worldsheet $S^1\times\R$) into the complexified maximal torus of the gauge group $G$ with the effective twisted superpotential \smash{$\Weff(a;\epsilon)$}.

On the other hand, as we will describe here, an alternative three-dimensional sigma-model argument shows that this superpotential is also equal to the difference of the generating functions of two holomorphic Lagrangian submanifolds of the moduli space $\scrM^\epsilon_{\rm flat}(C,{}^\lang G_\C)$ of flat ${}^\lang G_\C$ $\epsilon$-connections on the Riemann surface $C$, in suitable Darboux coordinates. Pivotal in this argument is the observation in~\cite{Nekrasov:2010ka} that the precise metric on \smash{$\R_\epsilon^2$} is not essential, as long as it is $U(1)$-invariant, so that we may equivalently replace it with a ``cigar'' metric of the form
\begin{equation}
\de s^2=\de r^2+f(r)\,\de\theta^2 \ ,
\end{equation}
with $0\leq r<\infty$ and $0\leq\theta<2\pi$, where $f(r)\simeq r^2$ for $r\rightarrow 0$ and $f(r)\rightarrow \rho^2$ as $r\rightarrow\infty$. We write $D$ for $\R^2$ endowed with such a cigar metric, and $D_R$ for $D$ restricted to $r\leq R$, with $R$ larger than the radius $r_0$ where the circle parametrized by $\theta$ reaches its asymptotic radius $\rho$. 

Since we take $R$ to be very large, $D_R$ looks globally like $I\times \widetilde{S}{\,}^1$, with the interval $I$ parametrized by $0\leq r\leq R$ and the circle \smash{$\widetilde{S}{\,}^1$} by $\theta$. We then turn on a $\frac12\Omega$-deformation with parameter $\epsilon$ corresponding to the $U(1)$ isometry of $D_R$ generated by \smash{$\frac\partial{\partial\theta}$}. We denote the resulting theory, that is, the theory $\sf T$ in the $\frac12\Omega$-background \smash{$D_{R,\epsilon}\times\R^2$}, by the same symbol ${\sf T}_\epsilon$. 

We now compactify the theory ${\sf T}_\epsilon$ over the fiber $\widetilde{S}\,^1$ of the cigar $D_{R,\epsilon}$. As demonstrated by~\cite{Nekrasov:2010ka}, the $\frac12\Omega$-deformation can be undone away from the tip of the cigar, where $r=0$, in exchange for a field redefinition and setting the asymptotic radius to $\rho=|\epsilon|^{-1}$. Then, away from the tip of the cigar, compactification of the $\frac12\Omega$-deformed theory ${\sf T}_\epsilon$ gives the same result as compactification of the {undeformed} theory, but now with field and coupling redefinitions.

Following the arguments of Section~\ref{sec:Hitchin}, we may thus describe the theory ${\sf T}_\epsilon$ at low energies $E\ll|\epsilon|$, at least away from the tip of the cigar, as a three-dimensional $\N{=}4$ sigma-model, whose target space is the hyperk\"ahler Hitchin moduli space $\scrM_{\textrm{\tiny H}}(C,G)$. In~\cite{Nekrasov:2010ka} it is then argued that the complete result, after compactifying the theory ${\sf T}_\epsilon$ over the fiber $\widetilde{S}\,^1$ of the cigar and then T-dualizing along either circle $\widetilde{S}\,^1$ or $S^1$,\footnote{Recall from Section~\ref{sec:Hitchin} that to arrive at the conventional description of this sigma-model in terms of hypermultiplets (rather than linear multiplets), it is necessary to dualize the abelian gauge fields in order to obtain enough periodic scalars. Alternatively, one may perform T-duality on either fiber $S^1$ or~\smash{$\widetilde{S}\,^1$}.} is a conventional three-dimensional $\N=4$ sigma-model with worldvolume $I\times S^1\times \R$ into either $\scrM_{\textrm{\tiny H}}(C,G)$ or $\scrM_{\textrm{\tiny H}}(C,{}^\lang G)$ (the Hitchin moduli space for the Langlands dual gauge group ${}^\lang G$). This sigma-model comes with two branes: a brane $\scrL'$ at $r=0$ which arises purely from the geometry at the tip of the cigar, and a brane $\scrL$ at $r=R$ which descends from a choice of half-BPS boundary condition in the four-dimensional theory. 

If we choose to T-dualize along the circle $\widetilde{S}\,^1$, the brane $\scrL'$ at $r=0$ is the space-filling (or canonical) coisotropic brane from~\cite{kapustin2006electric}. On the other hand, T-dualizing along the circle $S^1$ results in its three-dimensional mirror brane, the brane of $\epsilon$-opers $\scrL_\epsilon^{\rm oper}$ (also described in~\cite{kapustin2006electric} and introduced in Section~\ref{sec:opers} of these notes). The latter is obviously the interesting choice for us.

As proposed in Section~\ref{sec:HOex}, we assume that the brane $\scrL=\scrL_{\scrW,\varPi}$ at $r=R$ is labelled by the distinguished spectral network $\scrW=\scrW(u,\vartheta_{\rm c})$ together with a polarization $\varPi=\{A^i,B_i\}$ of the Seiberg-Witten curve $\Sigma_u$. That is, it corresponds to the holomorphic Lagrangian submanifold of the Hitchin moduli space defined by the equations
\begin{align}
\cX^\scrW_{B_i} = 1
\end{align}
in the system of holomorphic exponentiated Darboux coordinates $\big\{\cX^\scrW_{A^i},\cX^\scrW_{B_i}\big\}$ obtained by abelianization with respect to~$\scrW$. The brane of $\epsilon$-opers \smash{$\scrL_\epsilon^{\rm oper}$} is instead defined by its generating function $\cW^{\rm oper}(\epsilon)$ in these coordinates.\footnote{\label{fn:preserve4} In order to preserve four supersymmetries in the three-dimensional sigma-model, it is necessary to choose ${\rm arg}(\epsilon) = \vartheta$. However, we can analytically continue all of our results, and in particular the generating function $\cW^{\rm oper}(\epsilon)$, to any complex $\epsilon$. Indeed, recall from Section~\ref{sec:WKB} that the spectral coordinates \smash{$\cX_\gamma^\scrW$}, evaluated on the family of flat $\epsilon$-connections $\nabla_\epsilon^{\rm oper}$, are equal to the Voros periods $\exp V_\gamma(\epsilon)$ when ${\rm arg}(\epsilon) = \vartheta$, but then analytically continued to all $\epsilon\in\C^\times$.}

This implies that, when the interval $I$ is contracted to a point, the three-dimensional $\N=4$ sigma-model with target space $\scrM_{\textrm{\tiny H}}(C,{}^\lang G)$ may be equivalently described as a two-dimensional $\N=(2,2)$ sigma-model with target space $(\C^\times)^r$. This sigma-model is then characterized by a twisted superpotential\footnote{A similar statement holds in topologically twisted sigma-models, for example in the reduction of the two-dimensional A-model with worldsheet $I\times\R$ to quantum mechanics on $\R$, where this twisted superpotential is the Morse function generating the Fukaya category, or in the three-dimensional Rozansky-Witten model, which is explained in~\cite{Kapustin:2008sc}. These arguments are generalized to three-dimensional $\N=4$ gauge theories in~\cite{Bullimore:2016nji}.}
\begin{align}
\widetilde{\cW}{}^\sigma=\epsilon\,\cW^{\rm oper}(\epsilon) \ .
\end{align}

Hence, comparing the two arguments, we conclude that the effective twisted superpotential \smash{$\Weff$} characterizing the original $\N=2$ field theory ${\sf T}_\epsilon$ may be identified with the generating function $\cW^{\rm oper}(\epsilon)$ for the brane of $\epsilon$-opers $\scrL_\epsilon^{\rm oper}$. The vacua of the theory ${\sf T}_\epsilon$ correspond to the points of intersection $\scrL_\epsilon^{\rm oper}\cap \scrL$.

Another check of the NRS correspondence would be to compare vacuum expectation values of half-BPS line defects in the theory ${\sf T}_\epsilon$. Suppose we consider such line defects wrapping the fiber \smash{$\widetilde{S}\,^1$} of the cigar geometry $D_{R,\epsilon}$ while being situated at a point of $\R^2$. In the ultraviolet these line defects are labelled by a loop or network on $C$ as well as a phase $\zeta=\E^{\,\I\,\vartheta}$ (see e.g.~\cite{Drukker:2009tz,Gaiotto:2010be}). Their vacuum expectation values in an infrared vacuum $u$ have natural expansions
\begin{align}
\langle L_\zeta\rangle_u = \sum_{\gamma\in\Gamma_u} \, \underline{\overline\varOmega}(\gamma,L_\zeta;u) \ \cX^{\scrW(u,\vartheta)}_\gamma \ ,
\end{align}
in terms of the spectral coordinates $\cX_\gamma^{\scrW(u,\vartheta)}$ obtained by abelianization with respect to the spectral network $\scrW(u,\vartheta)$. Here \smash{$\underline{\overline\varOmega}(\gamma,L_\zeta;u)$} are indices which count what are called the framed BPS states of charge $\gamma$ supported by the line defect $L_\zeta$. 

In order to preserve four supersymmetries in the theory ${\sf T}_\epsilon$, the phase $\zeta$ of the line defect $L_\zeta$ has to be chosen consistently with the boundary conditions $\scrL_\epsilon^{\rm oper}$ and $\scrL_{\scrW,\varPi}$ (see Footnote~\ref{fn:preserve4}). The vacuum expectation value of the resulting half-BPS line defect $L_\epsilon$ has an expansion
\begin{align}
\langle L_\epsilon\rangle_u = \sum_{\gamma\in\Gamma_u} \, \underline{\overline\varOmega}(\gamma,L_\epsilon;u) \ \cX^{\scrW}_\gamma \ ,
\end{align}
in terms of the Darboux coordinates $\cX_\gamma^{\scrW}$ corresponding to the distinguished spectral network $\scrW$. That is, for a weakly coupled Lagrangian theory ${\sf T}_\epsilon$, vacuum expectation values of half-BPS line defects have a natural expansion in terms of complex Fenchel-Nielsen coordinates.

The NRS correspondence would thus be verified if we could perform an analogous computation in the four-dimensional field theory, for example by computing the vacuum expectation value $\langle L_\epsilon\rangle_u$ using supersymmetric localization in the $\frac12\Omega$-background. Although similar localization computations for weakly coupled Lagrangian theories indeed yield the correct expansion in terms of complex Fenchel-Nielsen coordinates, as far as we are aware the relevant calculation in the $\frac12\Omega$-background has yet to be done.\footnote{The supersymmetric localization calculations available~\cite{Ito:2011ea,Okuda:2014fja} indeed express the vacuum expectation values $\langle L_\epsilon\rangle_u$ in terms of complex Fenchel-Nielsen coordinates for weakly coupled Lagrangian field theories ${\sf T}_\epsilon$ and might appear very suggestive. See also~\cite{Brennan:2019hzm} for a similar investigation. However, beware that these supersymmetric localization computations are made in the $\Omega$-background denoted by $S^1\times_b\R^3$. This is an $\Omega$-background with a pair of parameters $(\epsilon_1,\epsilon_2)$ associated to the $U(1)$-isometry of $S^1$ and the $U(1)$-isometry of $\R^2\subset\R^3$, which is \emph{only} sypersymmetric if the ratio $b={\epsilon_1}/{\epsilon_2}$ is fixed. To perform an analogous calculation for the $\frac12\Omega$-background $\R_\epsilon^2\times\R^2$, one would need to consider the four supercharges preserved by the latter background (see Footnote~\ref{fnote:4supercharges}).}

\subsection{A More General Effective Twisted Superpotential}\label{sec:QFT3d}
\noindent
In Section~\ref{sec:HOex} we argued that the Lagrangian submanifold $\scrL_{\scrW,\varPi}$ corresponds to the type~I (or standard Neumann) boundary condition of~\cite{Nekrasov:2010ka}, at least when ${\sf T}_\epsilon$ is a weakly coupled gauge theory. A more general boundary condition may be obtained by coupling the four-dimensional theory ${\sf T}_\epsilon$ to three-dimensional degrees of freedom on its boundary. In~\cite{Dimofte:2011ju,Dimofte:2013lba} it is argued that the three-dimensional $\N=2$ theories ${\sf T}_K[M]$ of class~$\RR$ form a natural class of such boundary conditions. These superconformal field theories have ultraviolet Lagrangian descriptions as abelian Chern-Simons-matter theories, possibly deformed by superpotential terms that may contain monopole operators~\cite{Dimofte:2011ju,Dimofte:2011py,Dimofte:2013iv}.

Similarly to the $\N=2$ theories ${\sf T}_K[C]$ of class $\cS$ (see Section~\ref{ClassSTheories}), these three-dimen{-}sional theories may be obtained from a (twisted) compactification of the six-dimensional $(2,0)$-theory $\fX[K]$ on a three-manifold $M$. The three-manifolds $M$ that label the three-dimensional theories ${\sf T}_K[M]$ of class $\RR$ are assembled by gluing topological tetrahedra. This decomposition into tetrahedra determines an ideal triangulation $\scrT$ of their boundaries $\partial M$. The corresponding three-dimensional superconformal field theory depends on the triangulation $\scrT$ as well as a polarization $\varPi$ of the boundary,\footnote{More precisely, $\varPi$ is a polarization for an open subset of the moduli space $\scrM_{\rm flat}(\partial M,SL(K,\C))$ of framed flat $SL(K,\C)$ connections on $\partial M$.} and is therefore often denoted as ${\sf T}_K[M,\scrT,\varPi]$. Although the discussion which follows can be generalized rather straightforwardly to any $K$ (with the help of~\cite{Dimofte:2013iv,Gaiotto:2012db}), we will often restrict ourselves here to $K=2$ for simplicity.

For $K=2$, we call an edge of the triangulation $\scrT$ electric or magnetic if the corresponding one-cycle $\gamma\in H_1(\overline{\Sigma},\Z)$ is an $A$-cycle or a $B$-cycle, respectively. Each electric edge corresponds to a manifest $U(1)$ flavour symmetry in the three-dimensional superconformal field theory. For every such edge $E$ there is a chiral operator $\cO_E$ which transforms with charge $+1$ under the corresponding $U(1)$ flavour symmetry. If $E$ instead has a non-trivial magnetic charge, then an operator $\cO_E$ exists in the presence of a magnetic monopole background.  

Given a four-dimensional $\N=2$ theory ${\sf T}_K[C]$ in an infrared vacuum $u$, labelled by a (possibly) punctured Riemann surface $C$, we constructed the ``empty'' boundary condition $\scrL_{\scrW,\varPi}$ labelled by a distinguished spectral network $\scrW=\scrW(u,\vartheta_{\rm c})$ on $C$ together with a polarization $\varPi$ of the Seiberg-Witten curve $\Sigma_u$, which is given by the equations \smash{$\cX^\scrW_{B_i}=1$}. It should now be clear that we may generalize this boundary condition to a ``full'' boundary condition $\scrL_{M,\scrW',\varPi'}$ labelled by a three-manifold $M$ (assembled as above by gluing tetrahedra) with boundary $\partial M = C\sqcup C'$, whose boundary triangulation $\scrT\sqcup\scrT'$ is dual to the pair of spectral networks $\scrW\sqcup\scrW'$ on $C\sqcup C'$, and with boundary polarizations $\varPi$ on $C$ and $\varPi'$ on $C'$. The new boundary condition $\scrL_{M,\scrW',\varPi'}$ is defined by the equations
\begin{align}
\cX^{\scrW'}_{B'_i} = 1
\end{align}
in the system of exponentiated Darboux coordinates \smash{$\big\{\cX^{\scrW'}_{A'{}^i},\cX^{\scrW'}_{B'_i}\big\}$} obtained by abelianization with respect to the spectral network \smash{$\scrW'$}. 

In the field theory description when $K=2$, the abelian gauge fields of the four-dimensional theory ${\sf T}_2[C]$ in the vacuum $u$ can be used to gauge the flavour symmetries of the three-dimensional theory ${\sf T}_2[M]$ corresponding to the edges $E$ of the triangulation $\scrT$, whereas the electric BPS hypermultiplets of ${\sf T}_2[C]$ may be coupled to the corresponding chiral operators $\cO_E$ of ${\sf T}_2[M]$. This eliminates the dependence of the resulting four-dimensional theory on $\scrT$ and $\varPi$. 
A similar argument can be made for arbitrary $K$.

The simplest new boundary conditions are those that leave the vacuum $u$ invariant. 
Let us examine two types of examples. First, consider a boundary condition $\scrL_{M,\scrW,\varPi'}$ where only the polarization $\varPi\to\varPi'$ of the Seiberg-Witten curve $\Sigma_u$ is modified. This change corresponds to an electric-magnetic duality transformation ${\sf g}\in Sp(2r,\Z)$ in the infrared vacuum $u$, which is implemented by a generalized Legendre transform $\cL_{\sf g}$ at the level of the effective twisted superpotential, associated to the linear symplectomorphism \smash{$\big({}^x_y\big)\mapsto \big({}^{x'}_{y'}\big)={\sf g}\big({}^{x}_{y}\big)$} (see e.g.~\cite{Tul1977}). For the elementary canonical transformation \smash{$\big({}^{x'}_{y'}\big)={\sf g}_0\big({}^{x}_{y}\big)=\big({}^{-y}_{ \ \, x}\big)$}, this Legendre transform has the familiar classical expression
\begin{align}
\cL_{{\sf g}_0}\,:\,\Weff(x;\epsilon)\longmapsto \widetilde{\cW}{}'^{\,\rm eff}(x';\epsilon) = \epsilon\,x'\cdot x + \Weff(x;\epsilon) \ .
\end{align}

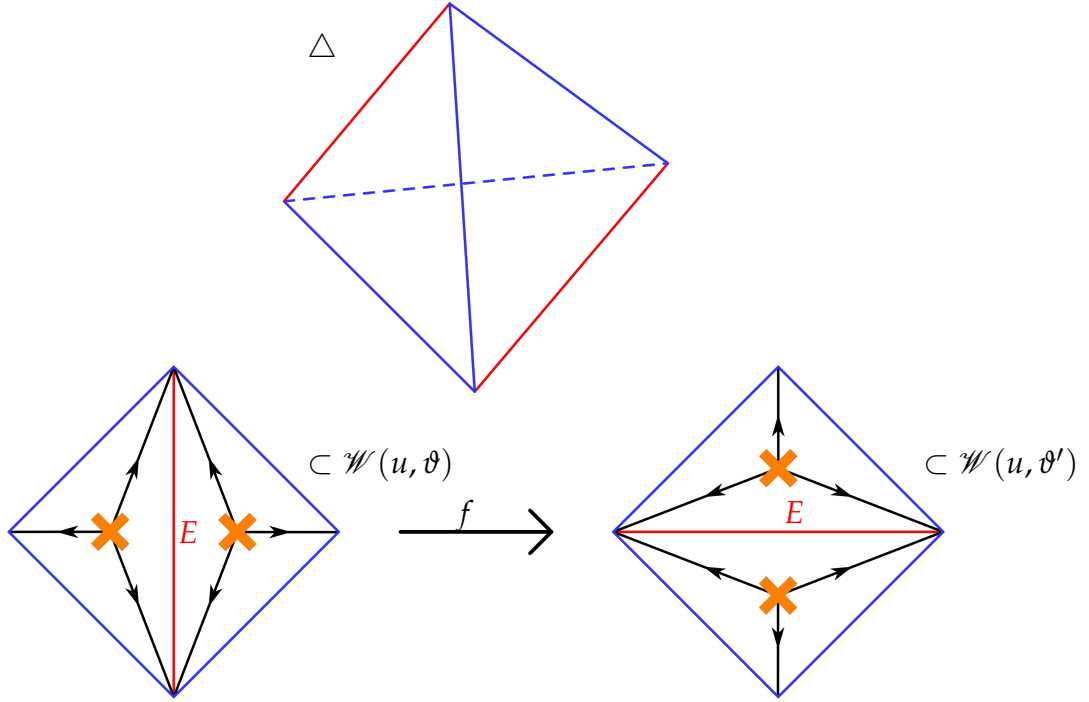
\begin{figure}[h!]
\centering
\begin{overpic}
[width=0.80\textwidth]{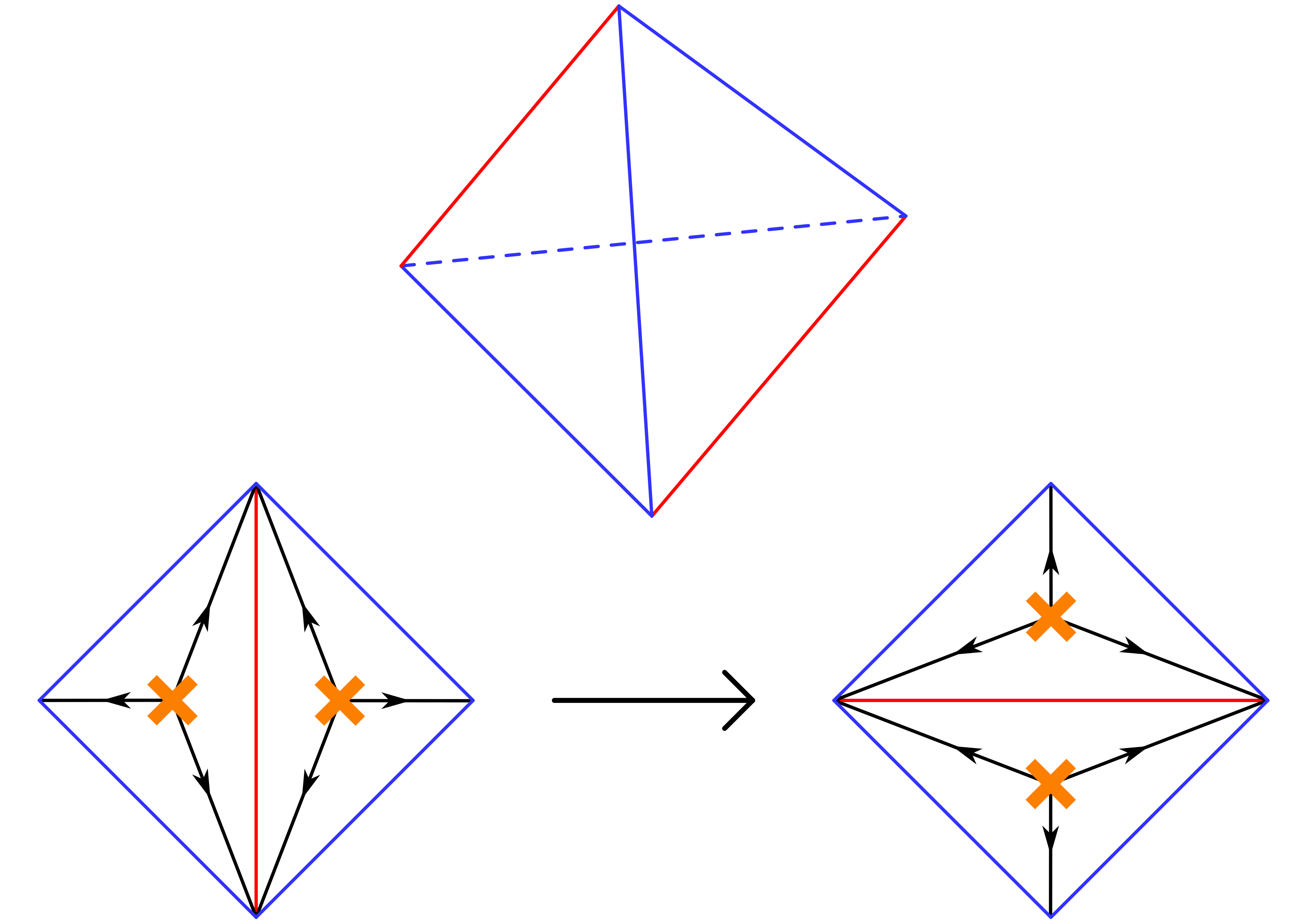}
\put(48,18){$f$}
\put(20,16){\textcolor{Red}{$E$}}
\put(81,18){\textcolor{Red}{$E$}}
\put(33,65){$\triangle$}
\put(33,23){$\subset\scrW(u,\vartheta)$}
\put(95,23){$\subset\scrW(u,\vartheta')$}
\end{overpic}
\caption{\small Each flip $f$ in the transition from $\scrW(u,\vartheta)$ to $\scrW(u,\vartheta')$ corresponds to a tetrahedron $\triangle$ in $M$, where the triangles adjacent to the flipped edge $E$ form the faces of $\triangle$.}
\label{FlipTetrahedron}
\end{figure}

We may also simply change the boundary triangulation $\scrT\to\scrT'$ on the second copy $C'=C$ of the boundary $\partial M=C\sqcup C'$. Let $K=2$. Assume that $\scrT$ and $\scrT'$ are dual to the spectral networks $\scrW=\scrW(u,\vartheta)$ and $\scrW'=\scrW(u,\vartheta')$, respectively, and that $\scrW(u,\vartheta)$ and $\scrW(u,\vartheta')$ are related by a sequence of flips. 
As illustrated in Figure~\ref{FlipTetrahedron}, each flip in the transition from $\scrW(u,\vartheta)$ to $\scrW(u,\vartheta')$ corresponds to a tetrahedron in the decomposition of $M$ into tetrahedra. 
The coordinate change corresponding to such a flip is given by
\begin{align}
\cX_{\gamma_E}^{\scrW'} = \cX_{\gamma_E}^\scrW \qquad \mbox{and} \qquad \cX_\gamma^{\scrW'} = \cX_\gamma^\scrW \, \big(1+\cX_{\gamma_E}^\scrW\big)^{\langle\gamma,\gamma_E\rangle} \ ,
\end{align}
where $\gamma_E$ is the one-cycle on $\Sigma_u$ corresponding to the edge $E$ of $\scrT$ in Figure~\ref{FlipTetrahedron}, and $\gamma$ is any other one-cycle on $\Sigma_u$ (see~\cite[Section~7.6]{gaiotto2013wall}). Let $\gamma$ be any one-cycle with $\langle\gamma,\gamma_E\rangle=1$, and consider the Darboux coordinates \smash{$x=\frac\epsilon{\pi\,\I} \log\big(-\cX_{\gamma_E}^\scrW\big)$}, \smash{$y=\frac1{2\,\epsilon} \log\cX_\gamma^\scrW$} and \smash{$y'=\frac1{2\,\epsilon} \log\cX_\gamma^{\scrW'}$}. The effective twisted superpotential then changes by
\begin{align}\label{eq:twisteddifference}
\widetilde{\cW}{}'^{\,\rm eff}(x;\epsilon) - \Weff(x;\epsilon) 
&= \frac12 \, \int\,\Big(\log\cX_\gamma^{\scrW'}-\log\cX_\gamma^\scrW\Big) \, \de x \nonumber \\[4pt]
&= \frac12 \, \int\,\log\big(1-\E^{\,\pi\,\I\,x/\epsilon}\big) \, \de x \nonumber \\[4pt]
&= -\frac\epsilon{2\pi\,\I} \, {\rm Li}_2(\E^{\,\pi\,\I\,x/\epsilon}) \ .
\end{align}

The difference \eqref{eq:twisteddifference} precisely matches the effective two-dimensional twisted superpotential \smash{$\widetilde{\cW}^{\rm eff}_\triangle(x;\epsilon)$} for the elementary tetrahedron theory ${\sf T}_2[\triangle,\scrT,\varPi_E]$ in the background $\widetilde{S}\,^1\times\R^2$ (the boundary of $D_{R,\epsilon}\times \R^2$). Indeed, the theory ${\sf T}_2[\triangle,\scrT,\varPi_E]$ consists of a single free chiral multiplet $\phi_E$ with charge~$+1$ under the $U(1)$ flavour symmetry associated to the edge $E$, charge~$0$ under the $U(1)_R$ symmetry, and a level $-\frac12$ background Chern-Simons term for the difference of the corresponding gauge fields $A_{\rm f}-A_R$. The result \eqref{eq:twisteddifference} is the one-loop contribution of the chiral field $\phi_E$ with effective twisted mass $\pi\,\I\,x/\epsilon$ to the effective twisted superpotential, which is obtained as a sum over its Kaluza-Klein modes in the compactification to $\R^2$~\cite{Witten:1993yc}. Geometrically it corresponds to the hyperbolic volume of the tetrahedron $\triangle$.

Implementing the modifications to the four-dimensional effective twisted superpotential for a sequence of flips might require us to adapt the polarization on $\Sigma_u$ along the way. This has the effect of an affine symplectic transformation on the three-dimensional $\N=2$ theory ${\sf T}_2[M]$ (see~\cite{Witten:2003ya} and~\cite[Appendix~A]{Dimofte:2013iv}): ``T-type'' transformations add background Chern-Simons couplings for flavour symmetries, while ``S-type'' transformations gauge a flavour symmetry, replacing it with a new topological $U(1)_J$ symmetry. 

This type of generalized boundary condition does not depend on the complex structure of $C$ and hence can only change the one-loop part of the effective twisted superpotential \smash{$\Weff(x;\epsilon)$}. Considering the effective twisted superpotential with respect to the spectral coordinates $\{\cX_A,\cX_B^\pm\}$ corresponding to a resolution of a Fenchel-Nielsen type network, as we did at the end of Section~\ref{ex:recipepure}, is another example of such a generalized boundary condition. Recall that for these choices, we found that the resulting one-loop contribution agrees with the one-loop contribution to \smash{$\Weff(x;\epsilon)$} in the gauge theory or topological string scheme.

So far we have studied two types of more general infrared boundary conditions for the four-dimensional theory ${\sf T}_K[C]$ in the vacuum $u$. By choosing more complicated three-manifolds $M$, we may however also engineer three-dimensional boundary conditions for the theory ${\sf T}_K[C]$ in a different vacuum $u'$ (because of the $Sp(2r,\Z)$ monodromies on the Coulomb branch $\scrB$, the three-manifold $M$ as well as the resulting effective twisted superpotential will depend on a choice of path from $u$ to $u'$ on $\scrB$). In the language of~\cite{Dimofte:2013lba} this means that we should consider three-manifolds with not only non-trivial ``big'' boundary components, but also with non-trivial ``small'' boundary components, corresponding to annuli or tori that stem from truncating the tetrahedra in the decomposition of $M$.\footnote{More accurately, the empty boundary condition $\scrL_{\scrW,\varPi}$ can be constructed, according to~\cite{Dimofte:2013lba}, as a combination of a UV-IR domain wall and an infrared boundary condition. This combination corresponds geometrically to a three-manifold $M$ with two boundary components $C$ and $\overline{C}$. Here $C$ is a ``big'' boundary component together with a triangulation $\scrT$ dual to $\scrW$ and polarization $\varPi$, while $\overline{C}$ is only homeomorphic to $C$ and consists of ``big'' three-punctured spheres glued together with ``small'' annuli. The more general boundary conditions discussed here may then be constructed as a combination of an S-duality wall (which connects $C$ with complex structure $q$ to $C'$ with complex structure $q'$), a UV-IR domain wall (which flows the theory to the vacuum $u'$) as well as an infrared boundary condition (determined by the spectral network $\scrW'$ and polarization $\varPi'$ on $C'$).}

While we do not describe such boundary conditions in detail here, these considerations do motivate the definition of a new object: the {(generalized) effective twisted superpotential \smash{$\Weff_{u,\vartheta,\varPi}(x;\epsilon)$}}. We define this twisted superpotential as the generating function of the brane of $\epsilon$-opers in the spectral coordinates \smash{$\big\{\cX^{\scrW(u,\vartheta)}_{A^i},\cX^{\scrW(u,\vartheta)}_{B_i}\big\}$} obtained from abelianization with respect to \emph{any} spectral network $\scrW(u,\vartheta)$, determined by $u$ and $\vartheta$, as well as the polarization $\varPi$ on the Seiberg-Witten curve $\Sigma_u$. That is, the effective twisted superpotential is extracted from the relations
\begin{align}
\log\Big(-\cX_{A^i}^{\scrW(u,\vartheta)}\big(\nabla_\epsilon^{\rm oper}\big)\Big) &= \frac{\pi\,\I\,x_i}\epsilon \ , \nonumber \\[4pt]
\log\cX_{B_i}^{\scrW(u,\vartheta)}\big(\nabla_\epsilon^{\rm oper}\big) &= 2 \, \frac{\partial\Weff_{u,\vartheta,\varPi}}{\partial x_i} \ .
\end{align}
Physically it corresponds to the effective twisted superpotential of the theory ${\sf T}_K[C]$ in the vacuum $u$ with respect to the boundary condition $\scrL_{\scrW(u,\vartheta),\varPi}$. This superpotential is locally constant on $\scrB\times[0,2\pi)$, and it is analytic in $x$ and $\epsilon$ for fixed $(u,\vartheta)$. It is defined uniquely on $\scrB\times[0,2\pi)$ up to the monodromies of the $Sp(2r,\Z)$ bundle over the Coulomb branch~$\scrB$.

\clearpage
\newpage
\printbibliography


\end{document}